\documentclass[twocolumn]{aastex631}

\usepackage{lipsum}
\usepackage{graphicx}

\received{January 30, 2025}
\revised{September 12, 2025}
\accepted{September 29, 2025}
\submitjournal{ApJS}

\begin{document}

\title{FIELDMAPS Data Release: Far-Infrared Polarization in the ``Bones'' of the Milky Way}
\shorttitle{FIELDMAPS Data Release}
\shortauthors{Coudé et al.}

\author[0000-0002-0859-0805]{Simon Coudé}
\affiliation{Department of Earth, Environment, and Physics, Worcester State University, Worcester, MA 01602, USA}
\affiliation{Center for Astrophysics $\vert$ Harvard \& Smithsonian, 60 Garden Street, Cambridge, MA 02138, USA}

\author[0000-0003-3017-4418]{Ian W. Stephens}
\affiliation{Department of Earth, Environment, and Physics, Worcester State University, Worcester, MA 01602, USA}

\author[0000-0002-2885-1806]{Philip C. Myers}
\affiliation{Center for Astrophysics $\vert$ Harvard \& Smithsonian, 60 Garden Street, Cambridge, MA 02138, USA}

\author[0000-0003-3682-854X]{Nicole Karnath}
\affiliation{Center for Astrophysics $\vert$ Harvard \& Smithsonian, 60 Garden Street, Cambridge, MA 02138, USA}
\affiliation{Space Science Institute, 4765 Walnut St, Suite B Boulder, CO 80301, USA}

\author{Howard A. Smith}
\affiliation{Center for Astrophysics $\vert$ Harvard \& Smithsonian, 60 Garden Street, Cambridge, MA 02138, USA}

\author[0000-0003-0990-8990]{Andrés Guzmán}
\affiliation{Joint Alma Observatory (JAO), Alonso de Córdova 3107, Vitacura, Santiago, Chile}

\author{Jessy Marin}
\affiliation{Center for Astrophysics $\vert$ Harvard \& Smithsonian, 60 Garden Street, Cambridge, MA 02138, USA}
\affiliation{Jodrell Bank Centre for Astrophysics, Department of Physics and Astronomy,\\ The University of Manchester, Manchester, M13 9PL, UK}

\author[0000-0002-2250-730X]{Catherine Zucker}
\affiliation{Center for Astrophysics $\vert$ Harvard \& Smithsonian, 60 Garden Street, Cambridge, MA 02138, USA}

\author[0000-0001-6717-0686]{B-G. Andersson}
\affiliation{McDonald Observatory and Department of Astronomy, University of Texas at Austin, Austin, TX 78712, USA}

\author[0000-0002-7402-6487]{Zhi-Yun Li}
\affiliation{Astronomy Department, University of Virginia, Charlottesville, VA 22904, USA}

\author[0000-0002-4540-6587]{Leslie W. Looney}
\affiliation{Department of Astronomy, University of Illinois, 1002 West Green St, Urbana, IL 61801, USA}

\author[0000-0003-1288-2656]{Giles Novak}
\affiliation{Center for Interdisciplinary Exploration and Research in Astrophysics (CIERA), 1800 Sherman Avenue, Evanston, IL 60201, USA}
\affiliation{Department of Physics \& Astronomy, Northwestern University, 2145 Sheridan Road, Evanston, IL 600208, USA}

\author[0000-0003-2133-4862]{Thushara G. S. Pillai}
\affiliation{Haystack Observatory, Massachusetts Institute of Technology, 99 Millstone Rd., Westford, MA 01886, USA}

\author[0000-0001-7474-6874]{Sarah I. Sadavoy}
\affiliation{Department of Physics, Engineering Physics and Astronomy, Queen's University, Kingston, ON K7L 3N6, Canada}

\author[0000-0002-7125-7685]{Patricio Sanhueza}
\affiliation{Department of Astronomy, School of Science, The University of Tokyo, 7-3-1 Hongo, Bunkyo, Tokyo 113-0033, Japan}

\author[0000-0002-6386-2906]{Archana Soam}
\affiliation{Indian Institute of Astrophysics, II Block, Koramangala, Bengaluru 560034, India}

\correspondingauthor{Simon Coudé}
\email{simoncoude@gmail.com}

\begin{abstract}

Polarization observations of the Milky Way and many other spiral galaxies have found a close correspondence between the orientation of spiral arms and magnetic field lines on scales of hundreds of parsecs. This paper presents polarization measurements at 214~$\mu$m toward ten filamentary candidate ``bones" in the Milky Way using the High-resolution Airborne Wide-band Camera (HAWC+) on the Stratospheric Observatory for Infrared Astronomy (SOFIA).  These data were taken as part of the Filaments Extremely Long and Dark: A Magnetic Polarization Survey (FIELDMAPS) and represent the first study to resolve the magnetic field in spiral arms at parsec scales. We describe the complex yet well-defined polarization structure of all ten candidate bones, and we find a mean difference and standard deviation of $-74^{\circ} \pm 32^{\circ}$ between their filament axis and the plane-of-sky magnetic field, closer to a field perpendicular to their length rather than parallel. By contrast, the 850~$\mu$m polarization data from \textit{Planck} on scales greater than 10~pc show a nearly parallel mean difference of $3^{\circ} \pm 21^{\circ}$. These findings provide further evidence that magnetic fields can change orientation at the scale of dense molecular clouds, even along spiral arms. Finally, we use a power law to fit the dust polarization fraction as a function of total intensity on a cloud-by-cloud basis and find indices between $-0.6$ and $-0.9$, with a mean and standard deviation of $-0.7 \pm 0.1$. The polarization, dust temperature, and column density data presented in this work are publicly available online.
\end{abstract}


\section{Introduction} 
\label{sec:intro}

Far-infrared and radio observations of nearby spiral galaxies have shown a close relationship between the orientation of magnetic field lines and the structure of spiral arms \citep[e.g.,][]{Borlaff2021_SALSA1, Lopez-Rodriguez2021_SALSA2, Borlaff2023_SALSA5, Martin-Alvarez2023_SALSA7}. Generally, spiral arms are hosts to regions of high-mass star formation, which themselves provide feedback (e.g., supernovae) that contributes to maintaining and amplifying the galactic magnetic field through turbulence-driven dynamos \citep[see][and references therein]{Beck2015}{}{}. In turn, the process of star formation, which occurs within filamentary molecular clouds, may itself be regulated by this magnetic field at small scales \citep[e.g.,][]{Andre2014_PPVI, Pattle2023_PPVII}. Characterizing magnetic fields at multiple scales within spiral galaxies can therefore improve our insights into their evolution.  

Within the Milky Way, our own barred spiral galaxy \citep[e.g.,][]{Dame2001,Urquhart2014}, the large-scale magnetic field projected on the plane of the sky is found to be mostly parallel to the Galactic disk as seen by the \textit{Planck} Space Telescope at the scale of the clouds themselves \citep[][]{Planck2015_XIX}. This parallel orientation agrees well with the signatures seen in other spiral galaxies \citep[][]{Jones2020,Borlaff2021_SALSA1}, suggesting that magnetic field lines follow spiral arms. Indeed, such a signature in the plane of the sky can exist at large scales even if, in three dimensions, the field itself has a significant line-of-sight component, such as helical geometry with observable polarity reversals \citep[][]{Tahani2022}. Unfortunately, at the scales probed by \textit{Planck} ($> 5'$ at 850~$\mu$m), it is not possible to resolve the magnetic field within the filaments where star formation occurs, even within the solar neighborhood ($< 500$~pc). Furthermore, our point of view within the Galactic disk complicates the identification of star-forming regions associated with specific spiral arms \citep[e.g.,][]{Zucker2015}.

Infrared dark clouds (IRDCs) in particular are a class of dense and cold molecular clouds that appear to trace spiral arms well. They are opaque in the mid-infrared and often elongated in shape \citep{Perault1996,Egan1998,Carey1998, Carey2000, Hennebelle2001, Price2001, Benjamin2003, Carey2009, Peretto2009}, and they are candidate progenitors of high-mass star formation \citep{Pillai2006b,Jackson2010, Sanhueza2012, Tan2014PPVI, Sanhueza2019, Morii2023}. Based on the observed properties of the best known IRDCs, \citet{Goodman2014} proposed that this category of objects could be the ``bones'' of the Galaxy, which are giant molecular structures closely following the ``spines" of spiral arms in both position and velocity space. Only a few dozen filaments have been determined to be potential bone candidates \citep{Wang2016, Zucker2018}. Bones were further studied using simulations, but it was not possible to recover their observed lengths and widths without including magnetic fields and self-gravity \citep[][]{Zucker2019}.

Since the thermal emission of interstellar dust in IRDCs peaks in the far-infrared \citep{Wang2015}, it is an ideal wavelength regime to characterize their magnetic fields through measurements of dust polarization \citep{Pattle_Fissel2019}. Indeed, in most of the interstellar medium, spinning asymmetric dust grains with paramagnetic properties will have their long axis preferentially aligned to be perpendicular to magnetic field lines, as long as they are within an anisotropic radiation field. This mechanism, known as Radiative Alignment Torques (RATs), is responsible for partially polarizing their thermal emission \citep[][]{Lazarian_Hoang2007, Hoang_Lazarian2008}. Alternative mechanisms, such as radiative alignment without magnetic fields (k-RAT), can also theoretically polarize dust emission in the far-infrared and submillimeter, but evidence for these mechanisms has only been found in a limited number of extreme environments \citep[e.g.,][]{Andersson2024, Pattle2021}. In the typical case of optically thin dust emission in molecular clouds, the plane-of-sky component of the magnetic field can be inferred from observed polarization maps at far-infrared and millimeter wavelengths (approximately 50~$\mu$m to 3~mm) by rotating the polarization vectors by $90^{\circ}$ \citep[e.g.,][]{Andersson2015}.

An important discovery from \textit{Planck} dust polarization maps of nearby star-forming regions is that the Galactic magnetic field structure appears to change direction toward molecular clouds whose hydrogen column density $N_{H_2}$ is above $10^{21}$~cm$^{-2}$ \citep{Planck2016_XXXV}. Indeed, magnetic field lines in these clouds tend to be perpendicular to the length of dense filaments \citep[e.g.,][]{Chapman2011, Doi2020,Kaminsky2023}. This property has been extensively studied using magnetohydrostatic models \citep{Tomisaka2014, Tomisaka2015}. In contrast, lower density striations have been observed to follow magnetic field lines into self-gravitating filaments, potentially funnelling gas toward star-forming cores \citep{Palmeirim2013,Pillai2020}. While fewer observations exist toward IRDCs (and specifically bones), early polarization observations toward these objects indicate that a similar behavior may be occurring in at least some regions of two high-mass filaments \citep[][]{Pillai2015ApJ,Stephens2022}. However, stellar feedback from ongoing massive star formation could also lead to more complex field structures due to expanding HII bubbles \citep{Tahani2023, Beslic2024}.

In this paper, we present the complete set of observations from the Filaments Extremely Long and Dark: A Magnetic Polarization Survey (FIELDMAPS), a legacy program of the Stratospheric Observatory for Infrared Astronomy (SOFIA) to map the polarized dust emission in ten candidate bones of the Milky Way using the High-resolution Airborne Wide-band Camera (HAWC+) at 214~$\mu$m. The primary goal of this paper is to produce a set of systematically reduced and fully comparable polarization maps at a resolution of 18\farcs7 to serve as the foundation for statistically-significant studies of the magnetic field properties in these unique star-forming environments. For this reason, this survey also includes updated data products for observations included in previous studies \citep{Stephens2022,Ngoc2023}. The resulting public archive of high-resolution polarimetric data toward candidate bones of the Milky Way will probe a different regime of high-mass star-forming environments compared to studies of the Central Molecular Zone \citep[see FIREPLACE survey;][]{Butterfield2023_FIRE1,Butterfield2024_FIRE2,Pare2024_FIRE3} and of nearby molecular clouds \citep[see BISTRO survey, e.g.,][]{WardThompson2017}. These data will also increase the number of regions where the relation between filament and magnetic field orientations can be studied \citep[e.g.,][]{Pattle2017, Monsch2018, Doi2020, Arzoumanian2021, Lee2021}.

The contents of this paper are divided as follows: In Section~\ref{sec:observations}, we detail the  acquisition and reduction of the astronomical observations used in this work. In Section~\ref{sec:results}, we present the polarization and magnetic field maps for all ten candidate bones of this survey, we compare the magnetic field structure with \textit{Planck} data, we fit the polarization efficiency for each target, and we discuss the physical significance of our results. Additional technical considerations are also addressed in the Appendices.

\section{Observations} 
\label{sec:observations}

\subsection{SOFIA}
\label{sub:sofia}

SOFIA was an airborne observatory jointly operated by the National Aeronautics and Space Administration (NASA) and the German Aerospace Center (DLR) from 2010 to 2022.\footnote{\href{https://irsa.ipac.caltech.edu/Missions/sofia.html}{https://irsa.ipac.caltech.edu/Missions/sofia.html}} SOFIA was built from a Boeing 747-SP airframe modified to house a Nasmyth–Cassegrain telescope with an effective diameter of 2.5~m and designed to quickly swap instruments between series of flights. The observatory flew at an altitude of 13~km, above 99\% of the water vapor in the atmosphere, which made it capable of observing wavelength regimes between 5~$\mu$m to 600~$\mu$m that are inaccessible from the ground. SOFIA's mobility allowed it to operate from both the Northern and Southern hemisphere, thus making it one of a few observatories capable of accessing the entire night sky.

HAWC+ was a far-infrared polarimetric camera available as a SOFIA facility instrument from 2018 to 2022 \citep{Dowell2010,Harper2018}. Its main components were three cryogenically-cooled 32 by 40 detector arrays composed of Transition-Edge Sensors and separated by a polarizing beam splitter, with two arrays in reflection and one in transmission. In polarization mode, a half-wave plate rotated to four specific angles with a DC~brushless air-coil motor would be introduced into the light path before the beam splitter, and only two arrays were used (one in reflection and one in transmission) to recover the Stokes $Q$ and $U$ parameters. This instrument was capable of observing in five bands: 53~$\mu$m (Band~A), 63~$\mu$m (Band~B), 89~$\mu$m (Band~C), 154~$\mu$m (Band~D), and 214~$\mu$m (Band~E). For each band used in this paper, Table~\ref{tab:bands} provides the Full-Width at Half-Maximum (FWHM) of the telescope's beam, the instrument's instantaneous Field-of-View (FOV) in polarization, the size of the instrument's detectors on the sky, and the final pixel size of the resulting data products.

\begin{table}[] 
    \caption{Characteristics of the HAWC+ observing bands used for this work \citep{Harper2018}.}
    \hspace*{-50pt}
    \begin{tabular}{cccccc}
        \hline
        Band & $\lambda$  & FWHM & Pol. FOV & Detector\tablenotemark{a} & Pixel\tablenotemark{b} \\
         & $\mu$m & \arcsec & \arcmin & \arcsec & \arcsec \\
        \hline
        \hline
        A & 53 & 4.85 & 1.4 x 1.7 & 2.55 & 1.21 \\
        E & 214 & 18.2 & 4.2 x 6.2 & 9.37 & 4.55 \\
        \hline
    \end{tabular}
    \label{tab:bands}
    \tablenotetext{a}{Detector pixel size of the instrument.} 
    \tablenotetext{b}{Map pixel size of the resulting data products.}
\end{table}

FIELDMAPS is a SOFIA Legacy Program (08\_0186, PI: I. Stephens) to map the magnetic field structure in ten candidate bones of the Galaxy using HAWC+ observations of dust polarization at 214~$\mu$m \citep[see also][]{Stephens2022}. The observations for program 08\_0186 were acquired between September~2020 and July~2022 during SOFIA Cycles~8 and 9. This survey also uses archival data at 214~$\mu$m from SOFIA programs 05\_0109 (PI: I. Stephens), 05\_0206 (PI: T. Pillai), and 06\_0027 (PI: I. Stephens). Program 06\_0027 also included 53~$\mu$m data, which are presented in Appendix~\ref{apx:chop_fil5}. The observations used for this paper are all publicly available on the NASA Infrared Science Archive (IRSA). The updated data products created for this data release paper are available online on the FIELDMAPS Dataverse\footnote{\href{https://doi.org/10.7910/DVN/NUXGJE}{https://doi.org/10.7910/DVN/NUXGJE}}. The target selection for the survey is explained in Section~\ref{sub:targets}, the HAWC+\cite{} observing modes used in Appendix~\ref{apx:modes}, and the data acquisition and processing in Section~\ref{sub:acquisition}. A detailed breakdown of observations on a source by source basis is given in Appendix~\ref{apx:obs}.

\subsubsection{Target Selection}
\label{sub:targets}

\begin{table*}[]
\caption{General properties of the 10 Milky Way bones studied for the FIELDMAPS survey.}
\hspace*{-70pt}
\begin{tabular}{lcccccccccccc}
    \hline
    Name & RA & Dec & $l$ & $b$ & Velocity & $D_\odot$ & $D_{Gal}$  & Length  & Width & Mass & $\phi_{F}$ & Arm\\
     & hh:mm:ss & $\pm$dd:mm:ss & $^\circ$ & $^\circ$ & km s$^{-1}$ & kpc & kpc & pc & pc & \(10^3 \, \textup{M}_\odot\) & $^{\circ}$ & \\
    \hline
    \hline
    Filament 1 & 18:41:31 & -05:30:17 & 26.90 & -0.31 & 68 & 4.2 & 4.9 & 20 & 1.5 & 4.0 & 93 & ScN\\
    Filament 2 & 18:38:49 & -07:02:08 & 25.24 & -0.42 & 57 & 3.8 & 5.1 & 38 & 1.1 & 10.5 & 74 & ScN\\
    Filament 4 & 18:30:23 & -10:27:51 & 21.23 & -0.15 & 66 & 4.3 & 4.5 & 19 & 1.8 & 7.1 & 96 & Nor\\
    Filament 5 & 18:25:09 & -12:48:32 & 18.56 & -0.10 & 46 & 4.0 & 4.7 & 61 & 1.3 & 41.5 & 82 & Nor\\
    Filament 6 \tablenotemark{a} & 18:10:32 & -19:21:14 & 11.14 & -0.12 & 31 & 4.1 & 4.3 & 30 & 1.4 & 28.0 & 87 & Nor\\
    Filament 8 & 17:41:05 & -31:13:12 & 357.54 & -0.37 & 4 & 1.3 & 7.0 & 11 & 0.7 & 2.1 & 86 & CrN\\
    Filament 10 & 16:16:27 & -50:52:01 & 332.42 & -0.12 & -49 & 2.9 & 5.9 & 50 & 1.2 & 31.2 & 95 & CtN\\
    G24 & 18:32:57 & -07:48:58 & 23.87 & +0.52 & 96 & 5.3 & 4.0 & 76 & 1.7 & 37.6 & 108 & Nor\\
    G47 & 19:16:40 & +12:45:03 & 47.13 & +0.30 & 58 & 4.2 & 6.2 & 38 & 1.1 & 11.3 & 70 & SgN\\
    G49 & 19:22:37 & +14:10:10 & 49.06 & -0.31 & 68 & 5.2 & 6.3 & 57 & 1.4 & 47.4 & 99 & SgF\\
    \hline
\end{tabular}
\tablenotetext{a}{Also nicknamed the ``Snake'' \citep{Wang2014,Wang2015}, not to be confused with the ``Snake Nebula'' (Barnard~72). \\The coordinates for each cloud were determined using the HAWC+ data in this work, and they are provided in both the ICRS (Right Ascension and Declination) and Galactic (Longitude $l$ and Latitude $b$) reference frames for convenience. $D_\odot$ is the estimated distance of the cloud from the Sun, using the cloud velocities from Table~1 of \citet{Zucker2018} and updated using the model from \citet{Reid2019}, except for Filament~6 and G49 where we respectively use the maser-derived parallax distances from \citet{Li2022} and \citet{Wu2014}. $D_{Gal}$ is the distance of the cloud from the Galactic center, assuming a Sun-Galactic center distance $R_\odot=8.275$~kpc \citep{Gravity2019,Gravity2021} and a solar height $z_\odot=20.8$~pc above the Galactic disk \citep{BennettBovy2019}. The Length, Width, and Mass of each bone are updated from Tables~2 and 3 of \citet{Zucker2018} using the updated distances $D_\odot$ from this table. The widths in particular are the median FWHMs obtained from the \textit{Herschel} dust emission in each bone. The angle $\phi_F$ is the fitted orientation of each filament as measured East of Galactic North (see Section~\ref{sec:results}). The spiral arms associated with each bone are identified as: the Carina near arm (CrN), the Centaurus-Crux near arm (CtN), the Norma arm (Nor), the Sagittarius near (SgN) and far (SgF) arms, and the Scutum near arm (ScN) \citep[][]{Reid2016, Zucker2018, Reid2019}.}
\label{tab:filaments}
\end{table*}

The ten science targets for the FIELDMAPS survey were selected from the list of large-scale filaments collated by \citet{Zucker2018} as candidate bones. From this point, we will refer to these targets as bones instead of candidate bones for simplicity, but we emphasize that the properties of these objects are still actively studied and that their categorization could change in the future. The resulting sample includes bones found near both arm and inter-arm (spur) regions of the Galaxy. Specifically, we observed bones associated with the following six structures defined by \citet{Reid2016}: the Carina near arm (CrN), Centaurus-Crux near arm (CtN), the Norma arm (Nor), the Sagittarius near (SgN) and far (SgF) arms, and the Scutum near arm (ScN). The Scutum near arm in particular is slightly over-represented due to being associated with a majority of the bones identified by \citet{Zucker2015}. The detailed properties of the bones observed for the FIELDMAPS survey are given in Table~\ref{tab:filaments}.

Our target list includes seven bones (Filaments~1, 2, 4, 5, 6, 8, and 10) that each fulfill at least five out of six of the criteria established by \citet{Zucker2015}. The most relevant criteria for this study are that the long axis of these bones are parallel within 30 degrees to the Galactic disk, and they are found to be within 20~pc from the Galactic mid-plane and within 10~km~s$^{-1}$ of the global longitude-latitude-velocity track of their nearest spiral arm. We also include three large-scale \textit{Herschel} filaments (G24, G47, and G49) from \citet{Wang2015} that were part of the subset of contiguous filaments from \citet{Zucker2018}. While these candidates failed one or more of the original bone selection criteria \citep{Zucker2015}, their physical properties otherwise closely resemble those of the seven other bones in this survey. For these reasons, we make no distinction between previously identified bone and large-scale \textit{Herschel} filament, and we simply identify both categories as bones as in \citet{Stephens2022}.

Our target selection was made from the subset compiled by \citet{Zucker2018} of eighteen filaments showing contiguity in \textit{Herschel}-derived column density maps. We used 250~$\mu$m data from the Hi-GAL survey to choose the bones that had the highest predicted brightness contrast between foreground and background. For ten of the bones, most of the filamentary structure was at least 600~MJy~sr$^{-1}$ above the local background emission. At this contrast level, significant polarization could be detected by HAWC+ in the bones with minimal integration time. In comparison, the remaining eight bones would have required much longer exposures, including the prototypical bone candidate Nessie \citep[][]{Jackson2010, Goodman2014}. Nevertheless, the ten selected bones should be a representative sample of the known population based on their physical properties \citep[such as mass and length;][]{Zucker2018} and varying degrees of star formation activity (see Section~\ref{sub:yso}).

Due to their position in the Galaxy, most of the targets in our sample were previously observed in the infrared by surveys of the Galactic plane \citep[e.g.,][]{Perault1996, Price2001, Benjamin2003, Carey2009, Molinari2010}. Follow-up studies using these surveys subsequently identified several of these targets as IRDCs \citep[e.g.,][]{Peretto2009}, Giant Molecular Filaments \citep[GMFs;][]{Ragan2014,Wang2015,AbreuVicente2016,Wang2016}, and bones \citep[][]{Zucker2015,Zucker2018}. Filament~6 in particular is one of the earliest IRDCs found in the Galactic plane \citep[][identified as G11.11-0.12]{Egan1998,Carey1998}, and so it is the most extensively studied object in the FIELDMAPS sample \citep[e.g.,][]{Carey2000,Johnstone2003,Pillai2006a,Pillai2015ApJ,Henning2010,Pillai2015ApJ,Wang2014}.

The distances and associated spiral arms listed in Table~\ref{tab:filaments} have been modified from the original catalog published by \citet{Zucker2018}, which used the Bayesian approach from \citet{Reid2016}. These changes were made to reflect the updated Galactic structure model from the Bar and Spiral Structure Legacy (BeSSeL) Survey \citep{Reid2019},\footnote{\href{http://bessel.vlbi-astrometry.org/node/378}{http://bessel.vlbi-astrometry.org/node/378}} except for Filament~6 and G49 for which we use the most recent maser-derived parallax distances instead (see G011.10-00.11 in Table~1 of \citealt{Li2022} and G049.19-00.34 in Table~3 of \citealt{Wu2014}). The width, length, and mass of each bone were also scaled by the ratio of the new to old distances, and this scaling is squared for the mass. Similarly to \citet{Zucker2018}, we use a far distance probability prior $P_{far}$ of 0.01 for Filaments 1, 2, 4, 5, 6, 8, and 10 since they show unambiguous near-IR extinction features  (see related figures in Section~\ref{sub:yso} and Appendix~\ref{apx:Spitzer}), while the default value of 0.5 is used for G24, G47, and G49. For these last three sources, using $P_{far}$ value of 0.01 or 0.5 does not impact the associated arms. The $P_{far}$ probability is used to disentangle the near/far ambiguity when calculating kinematic distances \citep[][]{Reid2016}. The uncertainties provided by the distance calculator vary between 6~\% and 16~\%, with a median of 9~\% for the sample.

Notable changes from \citet{Zucker2018} include that Filament~5 is slightly more likely to be associated with Norma instead of Scutum near, Filament~8 with Carina instead of Centaurus-Crux near, and G49 with the near arm instead of the far arm of Sagittarius. For Filament~6, the parallax distance of $4.1\pm0.2$~kpc from \citet{Li2022} is larger than the most likely (with a probability~$P=0.61$) kinematic distance of $2.9\pm0.2$~kpc in Scutum near obtained from the BeSSeL calculator. The parallax distance for this filament is instead in better agreement with the second solution from the calculator (with $P=0.39$) at a kinematic distance of $3.8 \pm 0.3$~kpc in Norma. While \citet{Li2022} still refer to Filament~6 as being part of the Scutum arm, we report Norma as its associated arm in Table~\ref{tab:filaments} to remain consistent with the process used for the other targets in the sample. In contrast, for G49, the parallax distance of $5.2 \pm 0.3$~kpc from \citet{Wu2014} is within the uncertainties of the kinematic distance of $5.5 \pm 0.5$~kpc, and so there is no change to the associated nearest arm.

The identification of the closest spiral arm for each bone in Table~\ref{tab:filaments} relies specifically on the model from \citet[][]{Reid2019}, and the results could change when using a different model \citep[e.g.,][]{Vallee2008}. Additionally, new 3D extinction maps of the solar neighborhood with \textit{Gaia} have shown the existence of significant dust structures between spiral arms \citep[][]{Zucker2023}. We reiterate that the targets in in our sample are candidate bones, and that their classification could change in the future. However, even if they existed in inter-arm regions instead, their known properties and relative positions within the disk would still make them uniquely suited to study the role of magnetic fields in dense star-forming filaments of the Galactic plane.

\subsubsection{Data Acquisition}
\label{sub:acquisition}

The characteristics of the two HAWC+ observing modes, Chop-Nod and Scan imaging, are detailed in Appendix~\ref{apx:modes}. In summary, the Chop-Nod observing mode used the secondary mirror of the telescope to alternate between the target and two sky positions to improve the background subtraction, while the Scan mode instead moved the primary mirror in a Lissajous pattern to maximize the integration time on the target \citep{Harper2018}. In each case, when performing polarization observations, the half-wave plate was rotated to four position angles ($5.0^\circ$, $50.0^\circ$, $22.5^\circ$, and $72.5^\circ$) in order to recover the linear Stokes parameters ($I$, $Q$, and $U$; see Section~\ref{sub:equations}).

The observations for the FIELDMAPS legacy program (08\_0186) were originally designed to use the Chop-Nod polarimetry mode of HAWC+ at 214~$\mu$m (Band~E), similarly to those from previous programs also included in this survey (05\_0109, 05\_0206, and 06\_0027), due to the Scan mode not being fully commissioned for the Cycle~8 SOFIA Call for Proposals. The ``Off'' positions required for the initially planned Chop-Nod observing strategy were determined using data from the \textit{Herschel} Hi-GAL survey, but finding symmetrical ``Off'' positions without bright astronomical sources proved challenging due to the proximity of the science targets to the Galactic Disk. In contrast, as discussed in Appendix~\ref{apx:modes}, the Scan mode did not require choosing distinct ``Off'' positions for background subtraction, and it was a more efficient observing mode than Chop-Nod. 

With the advent of shared-risk observing for the Scan polarimetry mode of HAWC+ in Cycle~8, all the proposed observations were redesigned from Chop-Nod to Scan with the help of the HAWC+ Instrument Science team before the program was scheduled to be observed. Additional modifications were made during Cycle~9 (i.e., after 2021/07/01) to add overlapping scans to the remaining sources in order to limit potential edge effects and obtain a more even coverage of the filaments. Two bones in our sample were observed using Chop-Nod, Filament~5 and Filament~6 (the ``Snake''), but only Filament~6 was observed exclusively in this mode as part of SOFIA programs 05\_0109 and 06\_0027. Filament~5 is the sole target in our sample with both Scan and Chop-Nod observations, as a single field was observed in Chop-Nod mode for programs 05\_0206 at 214~$\mu$m and 06\_0027 at 53~$\mu$m (see Appendix~\ref{apx:chop_fil5}).

The detailed descriptions of all observations included in the FIELDMAPS survey are given in Appendix~\ref{apx:obs}. For each bone, we list the observing modes used, the SOFIA flights during which data were obtained, the Level~0 files (raw instrumental data) that had to be discarded before data reduction, if any, as well as the reasons why, and the Astronomical Observations Requests (AORs) used by the HAWC+ instrument software for each planned telescope pointing. These AORs contain most of the information required to write the headers of the  Flexible Image Transport System (FITS) files for the Level~2, 3, and 4 data products created by the reduction pipeline (see Appendix~\ref{apx:obs}).

The typical observing parameters of the Lissajous scans for SOFIA program 08\_0186 were horizontal~$x$ and vertical~$y$ amplitudes of $120\arcsec$, respectively noted as cross-elevation and elevation in the AOR files and the FITS headers. Since these amplitudes already provided a mostly symmetrical coverage, we kept the default range of scan angles from $-30^\circ$ to $+30^\circ$. Although it had no significant impact on our observing strategy, it is worthwhile to note that HAWC+ Lissajous scan angles used an azimuth-elevation reference frame instead of an equatorial one before Cycle~9. The scan duration for each pointing was set to 60~seconds. We aimed for an equal number of repeats for each pointing to provide even coverage of each bone, although in some cases the exact number may have varied slightly depending on the available time during each flight. In practice, each pointing would be repeated at least twice before proceeding to the next, so each filament would be covered sequentially from one end to the other over the course of one or multiple flights. For Filament~6, which was observed exclusively in Chop-Nod mode, AORs for program 05\_0206 and 06\_0027 used nod times of 40 and 50~s, respectively, and both programs used chop amplitudes of $250\arcsec$. Finally, we also include Chop-Nod observations of Filament~5 in both Band~A and E, which we compare to the scan mode observations in Appendix~\ref{apx:chop_fil5}.

\subsubsection{Data Processing}
\label{sub:proc}

We used the SOFIA data reduction pipeline \citep[SOFIA Redux version 1.3.0;][]{clarke_2023_HAWC} to systematically process Level~0 HAWC+ polarization observations downloaded from IRSA into Level~4 data products for the FIELDMAPS survey. The HAWC+ scan mode reduction package used by SOFIA Redux is a Python adaptation of the Java-based \texttt{CRUSH} data reduction software \citep{Kovacs_2008_Crush}, and most of the reduction parameters are shared between the two packages. 

For each bone, all the available Level~0 scan data were loaded into SOFIA Redux using the default reduction parameters provided except for the additional parameters \textit{fixjumps=True} and \textit{rounds=30}. The \textit{fixjumps} parameter identifies and compensates for potentially anomalous flux jumps on individual detectors. The \textit{rounds} parameter controls the number of iterations for the map-making algorithm. A higher number of iterations can improve the recovery of extended emission, but at the cost of increased noise and longer computation times \citep{Kovacs_2008_Crush}. We found that doubling the number of rounds from the default 15 was sufficient to systematically recover the polarized emission across the length of most bones in our sample, but we also note that previous studies of the Central Molecular Zone have reliably recovered extended emission with SOFIA Redux by using a number of rounds as high as~85 \citep{Butterfield2023_FIRE1}. In contrast, we only used the default reduction parameters provided in SOFIA Redux for Chop-Nod observations of Filaments~5 and 6.

When a large set of Level~0 scan files is provided to the HAWC+ reduction pipeline, the software processes up to 12 consecutive files (4 per repeat) sharing the same AOR number and obtained from the same flight into a single Level~2 data product. These Level~2 data products are then calibrated from instrumental counts to Jansky per pixel (Jy/pix) and merged into Level~3 data products for each flight. Appendix~\ref{apx:obs} lists all the files that had to be discarded for each flight, and Appendix~\ref{apx:fix_fil4} describes how the Level~2 files for Filament~4 were corrected for positional errors. SOFIA Redux uses calibration standards based on observations of Uranus, Neptune, and other solar system objects obtained at the start of each HAWC+ flight series during the lifetime of the instrument. The flux calibration accuracy is expected to be within 10\%, which is sufficient for this work as we are mainly interested in the polarization fraction $P$ and angle $\theta$, relative quantities derived from the Stokes~$I$, $Q$, and $U$ parameters (see Section~\ref{sub:equations}).  Finally, the merged Level~3 data products from each flight are combined by the pipeline into a single ``multi-mission'' Level~4 data product. 

The pixel scale of the final data product is $4\farcs55$, and the effective resolution is $18\farcs7$ instead of the diffraction-limited $18\farcs2$ due to the smoothing step during the map-making process. We verified the accuracy of this effective beam by re-reducing calibration observations of Neptune obtained at 214~$\mu$m during Flights~607, 787, and 884 and fitting a two-dimensional Gaussian in the \textsc{CARTA} data analysis tool. We also found that observations of Neptune in Chop-Nod and Scan polarization modes during Flights~607 and 884 show similar beam profiles. However, we could not find Chop-Nod observations of Neptune in Total Intensity mode, and so we cannot directly measure the effect of the smoothing steps between the Chop-Nod and Scan modes reduction pipelines in SOFIA Redux at this time. Nevertheless, for this work, we will refer to $18\farcs7$ as the beam size for HAWC+, although it may be a small overestimation for Chop-Nod observations. 

The resulting Level~4 data products for each bone are FITS files with multiple extensions containing the Stokes~$I$, $Q$, and $U$ parameters and their uncertainties, as well as derived quantities such as the polarized intensity~$I_P$, the polarization fraction~$P$, and the rotation angle $\theta_P$. The calculation of these polarization properties is detailed in Section~\ref{sub:equations}. The complete structure and contents of the Level~4 HAWC+ data products are described by \citet{Gordon2018} and \cite{clarke_2022_Manual}.  

The final step of the data reduction process was to transform the coordinate system of the Level~4 data products from the International Celestial Reference System (ICRS) to the Galactic Coordinate System (GCS). The full description of the transformation from ICRS to GCS is provided in Appendix~\ref{apx:gal_reproj}. The FIELDMAPS data products are provided in both coordinate systems through the survey data repository on the Harvard Dataverse. Since studies of objects within the Galaxy often use the GCS system, this paper will use the GCS data exclusively to present the bones.

\subsection{Herschel}
\label{sub:herschel}

The \textit{Herschel} Space Observatory was a far-infrared mission operated by the European Space Agency (ESA), with support from NASA, from May 2009 to June 2013 \citep{Pilbratt2010}.\footnote{\href{https://irsa.ipac.caltech.edu/Missions/herschel.html}{https://irsa.ipac.caltech.edu/Missions/herschel.html}} The telescope for \textit{Herschel} had an effective diameter of 3.28~m, and the spacecraft carried two imaging cameras: the Photodetecting Array Camera and Spectrometer (PACS) and the Spectral and Photometric Imaging Receiver (SPIRE). For this work, we use observations from the \textit{Herschel} Infrared Galactic Plane Survey \citep[Hi-GAL;][]{Molinari2010} combining five of the continuum bands covered by these two instruments: 70~$\mu$m, 160~$\mu$m, 250~$\mu$m, 350~$\mu$m and 500~$\mu$m. 

The data we used were produced from the raw Hi-GAL observations following the recipe described by \citet{Guzman2015} to obtain continuum maps at a resolution of $36\farcs4$ \citep{Aniano2011}. The resulting data products differ from those of the Hi-GAL survey by using a background subtraction technique that more effectively separates Galactic cirrus or other emission from the molecular clouds themselves; it also corrects for the unknown zero levels of the \textit{Herschel} photometric scales \citep{Guzman2015}. This process emphasizes the accurate measurement of cold, dark filamentary structures like the bones of the FIELDMAPS survey. The continuum data from the Hi-GAL program were obtained in the parallel, fast-scanning mode and reduced with version 9.2.0 of the \textit{Herschel} pipeline. The method convolves the 5-band image set to the lowest available resolution (i.e., the 500~$\mu$m SPIRE beam) using the convolution kernels of \citet{Aniano2011}. The background is subtracted from the images for each field and for each observing band under the assumption that it constitutes a smooth additive component arising from diffuse emission in front and/or behind the filament. 

The background model has two constraints: its value has to be less than the emission at each pixel of the image, within a $2\sigma$ margin, and it cannot vary by more than than $10 \%$ over a scale of $2.5\arcmin$. The iterative algorithm first creates a smoothed image by setting to zero the spatial frequencies corresponding to flux variations on angular scales smaller than $2.5\arcmin$. Whenever a pixel in the smoothed image is larger than the corresponding pixel in the original image by more than $2\sigma$, it is replaced by the value in the original image. The result of the first iteration then replaces the starting image, and the algorithm continues to iterate this process until the difference between two consecutive iterations is below $5\%$ in all pixels. 

An example of the differences between each step of this iterative process across the Galactic disk is illustrated in Figure~1 of \citet{Guzman2015}. We find that the dust emission maps produced in this way are better at removing foreground and background emission than using the \textit{Herschel} pipeline alone, thus enabling a more accurate determination of the dust column densities in each source. Indeed, when fitting the spectral energy distribution (SED) using these new maps, we typically find average dust temperatures~$T_d$ in our filaments that are a few degrees cooler than previous estimates \citep[see Figure~5 in ][]{Guzman2015}. As emphasized by \citet{Zucker2018}, this background subtraction method will also affect the measurement of the filaments' density profiles.

For this work, we use the \textit{Herschel} data described in this section to obtain maps of molecular hydrogen gas column density $N_{H_2}$ and dust temperature $T_d$ for each bone. The parameters used for the SED fit are a gas-to-dust ratio $R_{g/d}$ of 100, a mean molecular weight $\mu$ of 2.8, and a spectral index of emissivity~$\beta$ of 1.75. Additionally, in Appendix~\ref{apx:herschel_comp}, we use these fits to calculate the expected spectral flux densities at 214~$\mu$m  for each bone and compare them to HAWC+ measurements in order to investigate potentially missing large-scale emission in the FIELDMAPS data.

\subsection{Planck}
\label{sub:planck}

The \textit{Planck} Space Observatory was the companion far-infrared mission to \textit{Herschel}, also operated by ESA with support from NASA from May 2009 to October 2013 \citep{Planck2011_I}.\footnote{\href{https://irsa.ipac.caltech.edu/Missions/planck.html}{https://irsa.ipac.caltech.edu/Missions/planck.html}} Equipped with a 1.9~m by 1.6~m elliptical reflector, \textit{Planck's} primary goal was to map the cosmic microwave background (CMB) with continuum and polarization observations using its two onboard instruments: the Low Frequency Instrument (LFI; 30 to 70~GHz) and the High Frequency Instrument (HFI; 100 to 857~GHz). A by-product of the \textit{Planck} mission was an all-sky map of polarized dust emission across the Galaxy. In this paper, we use Public Release~3 polarization maps obtained by the HFI at 353~GHz (850~$\mu$m) with an effective resolution of $5\arcmin$ \citep{Planck2020_I,Planck2020_III}. For each bone, we downloaded $2^\circ$ by $2^\circ$ maps of the Stokes~$I$, $Q_{cmb}$, and $U_{cmb}$ parameters from the IRSA legacy archive. Each map was calibrated from units of cosmic background temperature $K_{cmb}$ to MegaJansky per steradian (MJy~sr$^{-1}$) using the conversion factors from Table~6 of \citet{Planck2014_IX}. We also converted the Stokes~$Q_{cmb}$ and $U_{cmb}$ maps in the COSMO/HEALPix convention \citep[][]{Gorski2005, Planck2015_XIX} to Stokes~$Q_{gcs}$ and $U_{gcs}$ using the International Astronomical Union (IAU) convention \citep[e.g.,][]{HamakerBregman1996} rotated to be in the GCS coordinate system (see Appendix~\ref{apx:gal_reproj}). 

\subsection{Spitzer}

The \textit{Spitzer} space telescope was an infrared mission operated by NASA from August 2003 to January 2020.\footnote{\href{https://irsa.ipac.caltech.edu/Missions/spitzer.html}{https://irsa.ipac.caltech.edu/Missions/spitzer.html}} \textit{Spitzer} was equipped with an 85~cm mirror, and the platform carried two photometric imagers: the Infrared Array Camera \citep[IRAC;][]{Fazio2004} covering four bands at 3.6, 4.5, 5.8 and 8.0~$\mu$m, and the Multiband Imaging Photometer for \textit{Spitzer} \citep[MIPS;][]{Rieke2004} at 24, 70 and 160~$\mu$m. For this work, we use archival data of the Galactic plane from the Galactic Legacy Infrared Midplane Survey Extraordinaire \citep[GLIMPSE;][]{Benjamin2003} and from the MISPGAL survey \citep[][]{Carey2009} survey to create infrared images of each bone.

\section{Results} 
\label{sec:results}

\subsection{Star Formation Activity}
\label{sub:yso}

The list of young stellar objects (YSOs) used for this work was obtained from the catalog produced by \citet{Zhang2019}. In their catalog, the YSOs were identified and classified via multiband photometric catalogs spanning the near- to far-infrared.  The UKIRT Infrared Deep Sky Survey \citep[UKIDSS;][]{Lawrence2007}, the Vía Láctea survey \citep[VVV;][]{Minniti2010}, and the 2MASS point source catalog \citep{Skrutskie200} provided the near-IR identifications. The mid-infrared archival data was obtained from the Spitzer galactic plane surveys, GLIMPSE \citep{Benjamin2003} and MISPGAL \citep{Carey2009}, in addition to ALLWISE \citep{Wright2010} and the Red~MSX Source survey \citep{Lumsden2013}. The far-infrared data used to identify protostellar objects was also obtained from the Hi-GAL survey \citep[][]{Molinari2010}. In this paper, we only include Class~I, Flat Spectrum, Class~II, and Class~III YSOs compiled by \citet{Zhang2019}. We also performed a literature search for Massive Young Stellar Objects (MYSOs), as well as masers\footnote{\href{https://maserdb.net/}{https://maserdb.net/}}, for each bone \citep[e.g.,][]{Svoboda2016,Beuther2019}.

\begin{figure*}[]
\centering
\includegraphics[scale=0.75]{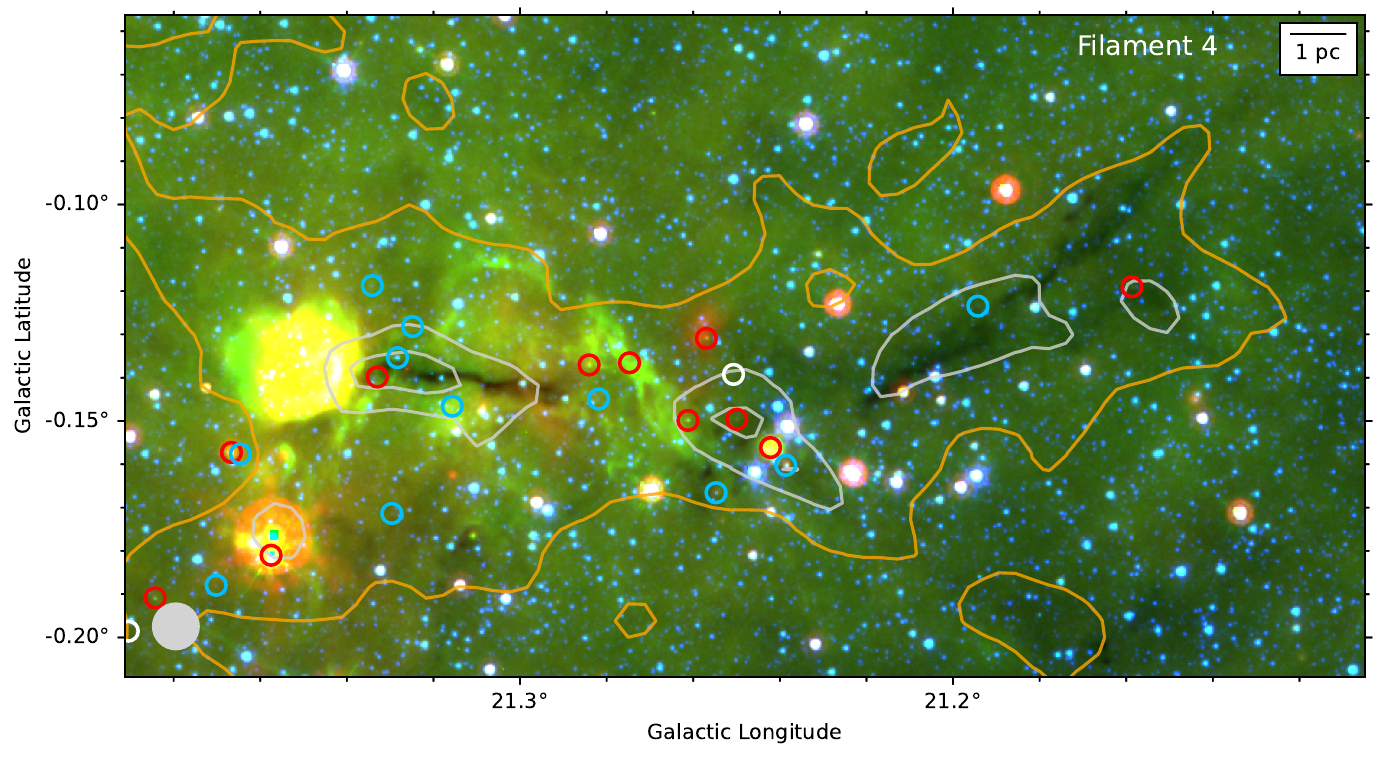}
\caption{\textit{Spitzer} red~(24~$\mu$m), green~(8.0~$\mu$m), and blue~(3.6~$\mu$m) composite image for Filament~4. The color levels were chosen to highlight the infrared-dark features of the bone. Class~I and Flat Spectrum (red), Class~II (blue), and Class~III (white) YSOs from the literature are identified with circles. The three contours trace \textit{Herschel}-derived $N_{H_2}$~column densities of $0.5$, $1.0$, and $1.5 \times 10^{22}$~cm$^{-2}$, with the lowest level identified in orange for clarity. The \textit{Herschel} beam at 500~$\mu$m is given by the gray circle at the bottom left. A reference length of 1~pc is provided at the top right, assuming a distance~$D_\odot$ of 4.5~kpc to the cloud (see Table~\ref{tab:filaments}).
\label{fig:Fil4_RGB}}
\end{figure*}

To illustrate the typical star formation activity in the bones, Figure~\ref{fig:Fil4_RGB} shows the positions of Class~I, II, III, and flat-spectrum YSOs for Filament~4 on a composite red (24~$\mu$m), green (8.0~$\mu$m), and blue~(3.6~$\mu$m) image from the GLIMPSE and MIPSGAL surveys with \textit{Spitzer}. The color levels were chosen to highlight the infrared-dark features of the filament, although the \ion{H}{2} region to the East is saturated as a result. Figure~\ref{fig:Fil4_RGB} also includes contours of \textit{Herschel}-derived $N_{H_2}$~column density from 0.5 to $1.5 \times 10^{22}$~cm$^{-2}$. Filament~4 was chosen as an example because it shows clearly identifiable features associated with IRDCs and signs of ongoing star formation. Appendix~\ref{apx:Spitzer} provides similar images for each bone. While we found no confirmed maser for the region displayed by Figure~\ref{fig:Fil10_G24_RGB}, several have been detected across the FIELDMAPS targets and their positions are plotted in Figures~\ref{fig:Fil1_RGB} to \ref{fig:G47_G49_RGB} in Appendix~\ref{apx:Spitzer}.

\subsection{Derived Polarization Properties}
\label{sub:equations}

Polarization vectors (also called ``pseudo-vectors'' or ``half-vectors'' due to their $180^\circ$ ambiguity) are defined using the Stokes parameters, with Stokes~$I$ being the total intensity of the incoming light, and Stokes~$Q$ and $U$ being the linear components describing the fraction of polarized light $P$ and the rotation angle $\theta$. Following the IAU convention, the polarization angle $\theta$ is measured on the celestial sphere from North to East. The relations between these quantities are given by $Q = P \, I \cos(2\theta)$ and $U = P \, I \sin(2\theta)$.

The full description of how each parameter is calculated during the data reduction process of HAWC+ observations is given by \citet{Gordon2018} and \citet{clarke_2022_Manual}, although we note here that they use a slightly different notation for the polarization fraction. For the FIELDMAPS survey, we provide data products in both the ICRS and GCS reference frames. The process of calculating the updated Stokes parameters~$Q_{gcs}$ and $U_{gcs}$ from Stokes~$Q_{icrs}$ and $U_{icrs}$ is detailed in Appendix~\ref{apx:gal_reproj}. From this point forward, we will refer to Stokes~$Q_{gcs}$ and $U_{gcs}$ as Stokes~$Q$ and $U$ for simplicity, and because the results in this paper were obtained using the data products rotated in the GCS frame. 

We detail below the equations used to derive the polarization properties and their uncertainties from the measured Stokes~$I$, $Q$ and $U$ parameters, assuming the cross-terms of the error covariance matrix are small \citep{Gordon2018}. First, the polarized intensity $I_p$ is simply written as:

\begin{equation}
    I_p = \sqrt{Q^2 + U^2} \, , 
    \label{eq:Ip}
\end{equation}

\noindent with the uncertainty $\delta_{I_p}$:

\begin{equation}
    \delta_{I_p} = \sqrt{ \frac{(Q \, \delta_Q)^2+(U \, \delta_U)^2}{Q^2+U^2}} \, , 
    \label{eq:sigIp}
\end{equation}

\noindent where $\delta_Q$ and $\delta_U$ are the uncertainties for Stokes~$Q$ and $U$, respectively. Although the noise for $Q$ and $U$ can be assumed to be normally distributed, the quadratic form of the polarized intensity $I_p$ means that it instead follows a Rice distribution with a positive bias for low signal-to-noise measurements \citep{Serkowski1958,Wardle1974,Pattle_2019}. This bias needs to be removed to obtain the debiased polarized intensity~$I'_p$:

\begin{equation}
    I'_p = \sqrt{ I_p^2-\delta_{I_p}^2 } .
    \label{eq:debiasedIp}
\end{equation}

\noindent Following this relation, the difference between the biased $I_p$ and debiased $I'_p$ intensities will be less than 6\% for a signal-to-noise ratio of~3 and above. The HAWC+ data reduction pipeline uses a variation of this method to debias~$P$ directly instead of $I_p$ \citep[see Equation~3 from][]{Gordon2018}, but we show in Appendix~\ref{apx:debias} that both methods are equivalent for the signal-to-noise thresholds used in this work. Finally, Equation~\ref{eq:debiasedIp} gives one of the most common approaches to debiasing polarization data, but we note that the accuracy of this process and the use of alternative methods have been studied extensively in the past \citep[e.g.,][]{Naghizadeh1993,Montier2015,Vidal2016,Pattle_2019}. 

From there, we define the debiased polarization fraction~$P$ as:

\begin{equation}
    P = 100 \% \, \frac{I'_p}{I} = 100 \% \, \frac{\sqrt{Q^2 + U^2 - \delta_{I_P}^2}}{I} ,
    \label{eq:pol}
\end{equation}

\noindent which we will express as a percentage here. From this point, we will refer to the debiased polarization fraction $P$ simply as the polarization fraction, as we never use the biased quantity for this work. The uncertainty $\delta_P$ on the polarization fraction is given by:

\begin{equation}
    \delta_P = 100 \% \, \frac{I_p}{I} \, \sqrt{ \left( \frac{\delta_{I_p}}{I_p} \right)^2 + \left( \frac{\delta_I}{I} \right)^2 } ,
    \label{eq:sigpol}
\end{equation}

\noindent where $\delta_I$ is the uncertainty for the Stokes~$I$ total intensity. 

The polarization angle $\theta$ in degrees is given by: 

\begin{equation}
    \theta = \frac{180^\circ}{2 \, \pi} \: \arctan \left( \frac{U}{Q} \right) ,
    \label{eq:theta}
\end{equation}

\noindent with a valid range of -90$^\circ$ to 90$^\circ$, which fully covers the range of possible angles for the polarization vectors described here. Specifically, we use of the two-arguments $\arctan2$ function in Python to solve the quadrant degeneracy of $\arctan$ when dividing the Stokes~$U$ and $Q$ parameters. The uncertainty $\delta_\theta$ is then expressed as:

\begin{equation}
    \delta_\theta = \frac{180^\circ}{2 \, \pi} \: \frac{\sqrt{\left(U \, \delta_Q \right)^2+\left(Q \, \delta_U \right)^2}}{Q^2+U^2} .
    \label{eq:sig_theta}
\end{equation}

Finally, as discussed in Section~\ref{sec:intro}, we obtain the plane-of-sky orientation $\theta_B$ of the magnetic field by rotating the polarization angle $\theta$ by $90^\circ$.

\subsection{Polarization and Magnetic Field Maps}
\label{sub:polmaps}

For the analysis covered in this paper, we used the data products described in Section~\ref{sub:proc} to create catalogs of polarization vectors at 214~$\mu$m that fulfill the following selection criteria: signal-to-noise thresholds of $I/\delta_I > 10$ and $P/\delta_P > 3$, and a maximum polarization fraction $P < 30\%$. The upper limit in polarization fraction and the larger signal-to-noise cutoff for Stokes~$I$ were chosen to remove any vector with an anomalously high polarization fraction, as \textit{Planck} found a maximum polarization fraction $p_{max}$ of $22.5_{-1.5}^{+3.5}$\% in the interstellar medium \citep{Planck2020_XII}. Large values of the polarization fraction~$P$ could be a result of filtering large-scale emission in the Stokes~$I$ map (see Appendix~\ref{apx:herschel_comp}). 

For each bone, Table~\ref{tab:detections} gives the total number vectors $N_{pol}$ in those catalogs, as well as the lower limit $N_{beam}$ of independent beam positions with a detection of polarization (i.e., one measurement per beam area). The number $N_{beam}$ is approximated by multiplying the number of detected polarization vectors $N_{pol}$ by the ratio of the area of a $4\farcs55$ square pixel and the area of a circle with the same diameter as the $18\farcs7$ effective beam FWHM at 214~$\mu$m (resulting in a factor $\sim 0.075$). We also include in Table~\ref{tab:detections} the median uncertainties $M_{\delta_I}$, $M_{\delta_Q}$, and $M_{\delta_U}$ of the Stokes~$I$, $Q$, and $U$ parameters, respectively, for the catalog of polarization vectors for each bone.

\begin{table}[]
\caption{Number of detected polarization vectors and noise characteristics for each bone.}
\hspace*{-45pt}
\begin{tabular}{lccccc}
    \hline
    Name & $N_{pol}$ & $N_{beam}$ & $M_{\delta_I}$ & $M_{\delta_Q}$ & $M_{\delta_U}$ \\
     &  & & \multicolumn{3}{c}{mJy arcsec$^{-2}$} \\
    \hline
    \hline
    Filament 1 & 5,574 & 420 & 0.059 & 0.076 & 0.076 \\
    Filament 2 & 14,459 & 1,089 & 0.077 & 0.105 & 0.104 \\
    Filament 4 & 10,876 & 819 & 0.060 & 0.078 & 0.078 \\
    Filament 5 & 6,266 & 472 & 0.092 & 0.123 & 0.121 \\
    Filament 6 & 6,522 & 491 & 0.081 & 0.118 & 0.119 \\
    Filament 8 & 3,367 & 253 & 0.166 & 0.211 & 0.211 \\
    Filament 10 & 8,881 & 669 & 0.241 & 0.299 & 0.295 \\
    G24 & 26,222 & 1,976 & 0.088 & 0.111 & 0.111 \\
    G47 & 9,553 & 720 & 0.044 & 0.055 & 0.056  \\
    G49 & 12,824 & 966 & 0.215 & 0.243 & 0.240 \\
    \hline
\end{tabular}
\label{tab:detections}
\end{table}

\begin{figure*}[p]
\centering
\includegraphics[scale=0.75]{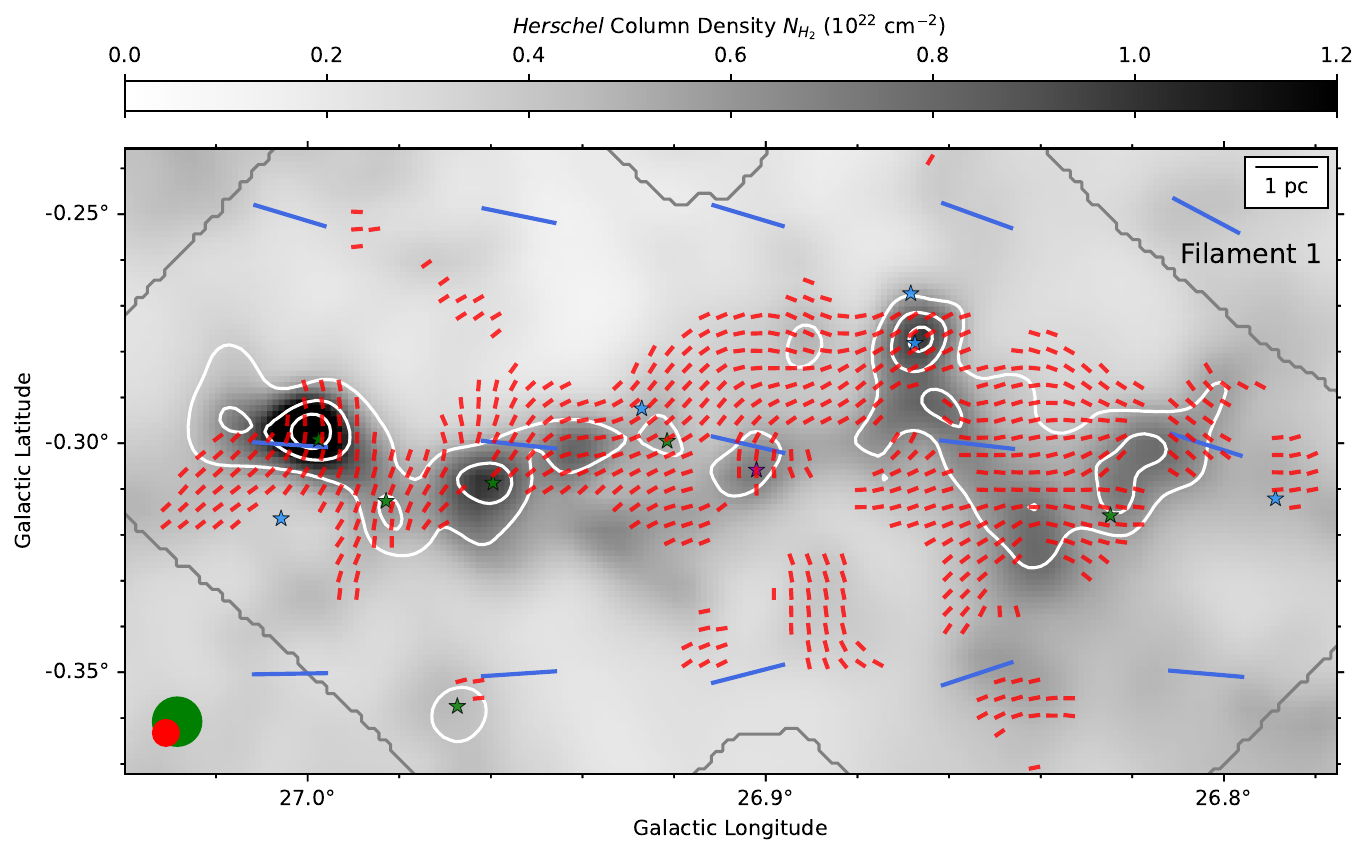}
\includegraphics[scale=0.75]{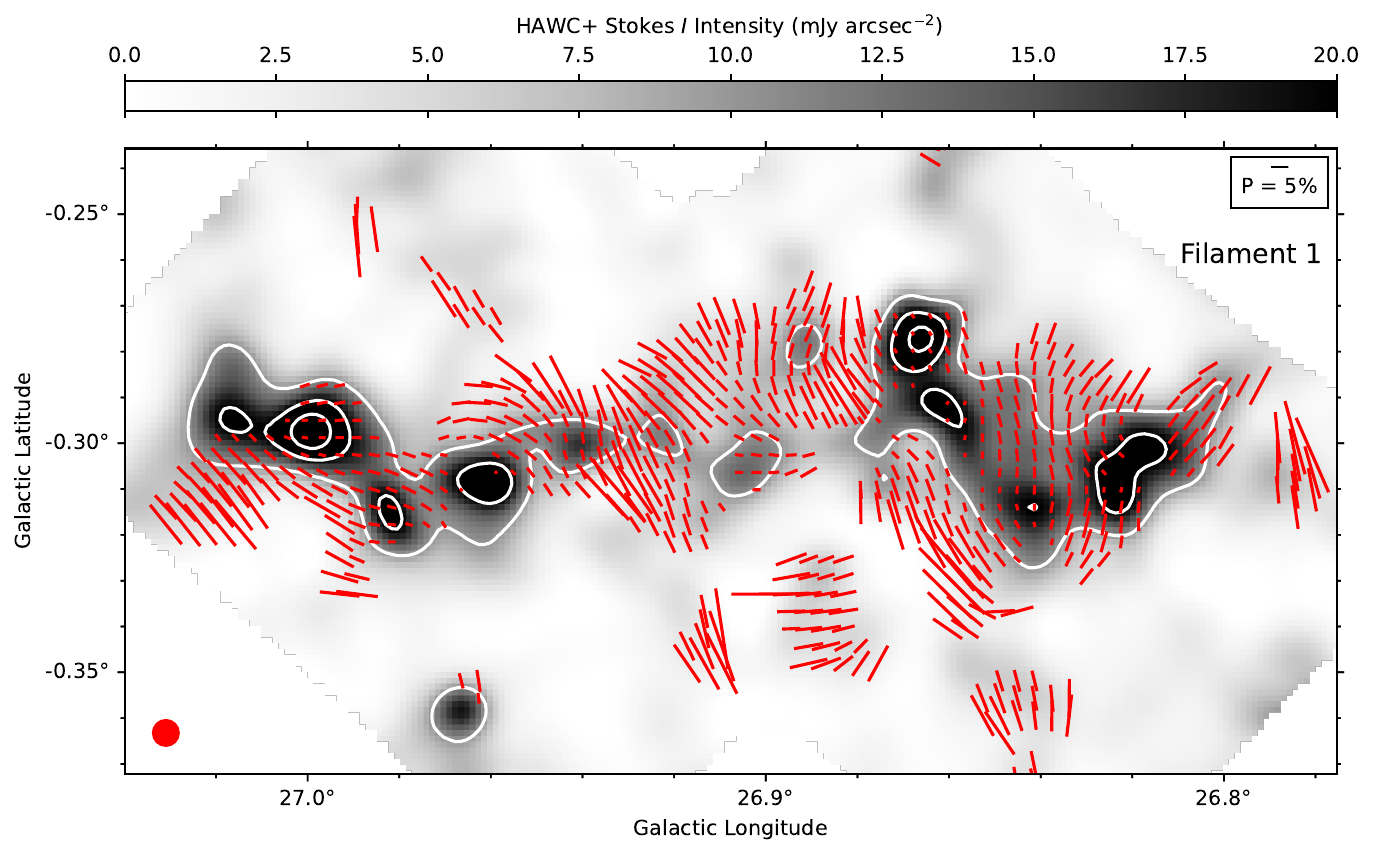}
\caption{\textbf{Filament~1}. \textit{Top}: The inferred magnetic field structure (HAWC+ 214~$\mu$m: red, \textit{Planck} 850~$\mu$m: blue) plotted on the \textit{Herschel}-derived $N_{H_2}$ column density map (gray). The magnetic field angle~$\theta_B$ is obtained from rotating the polarization angle~$\theta$ by $90^\circ$. Class~I (green), Flat Spectrum (violet), and Class~II (blue) YSOs are identified with star symbols. The gray contour denotes the area observed by HAWC+. The circles denote the beam sizes of HAWC+ (red) and \textit{Herschel} (green). \textit{Bottom}: The polarization measurements (red) plotted on the HAWC+ Stokes~$I$ total intensity map (gray) at 214~$\mu$m. The vector length shows the polarization fraction~$P$. For each panel, only every third vector is plotted for both SOFIA and Planck data, and the white contours trace the Stokes~$I$ total intensity from 10 mJy arcsec$^{-2}$ and increasing by steps of 10 mJy arcsec$^{-2}$. See Section~\ref{sub:polmaps} for details.
\label{fig:Fil1_Maps}}
\end{figure*}

\begin{figure*}[p]
\centering
\includegraphics[scale=0.75]{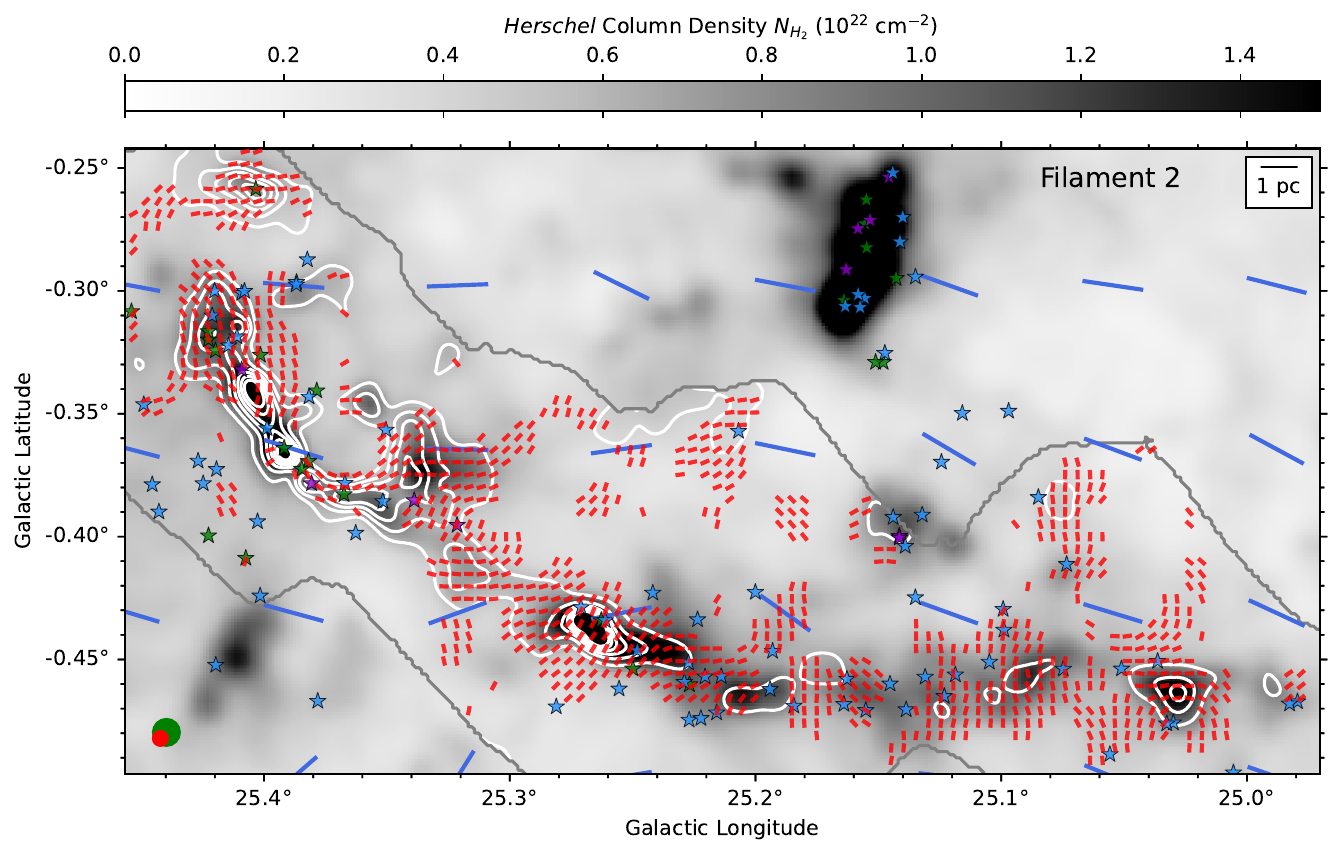}
\includegraphics[scale=0.75]{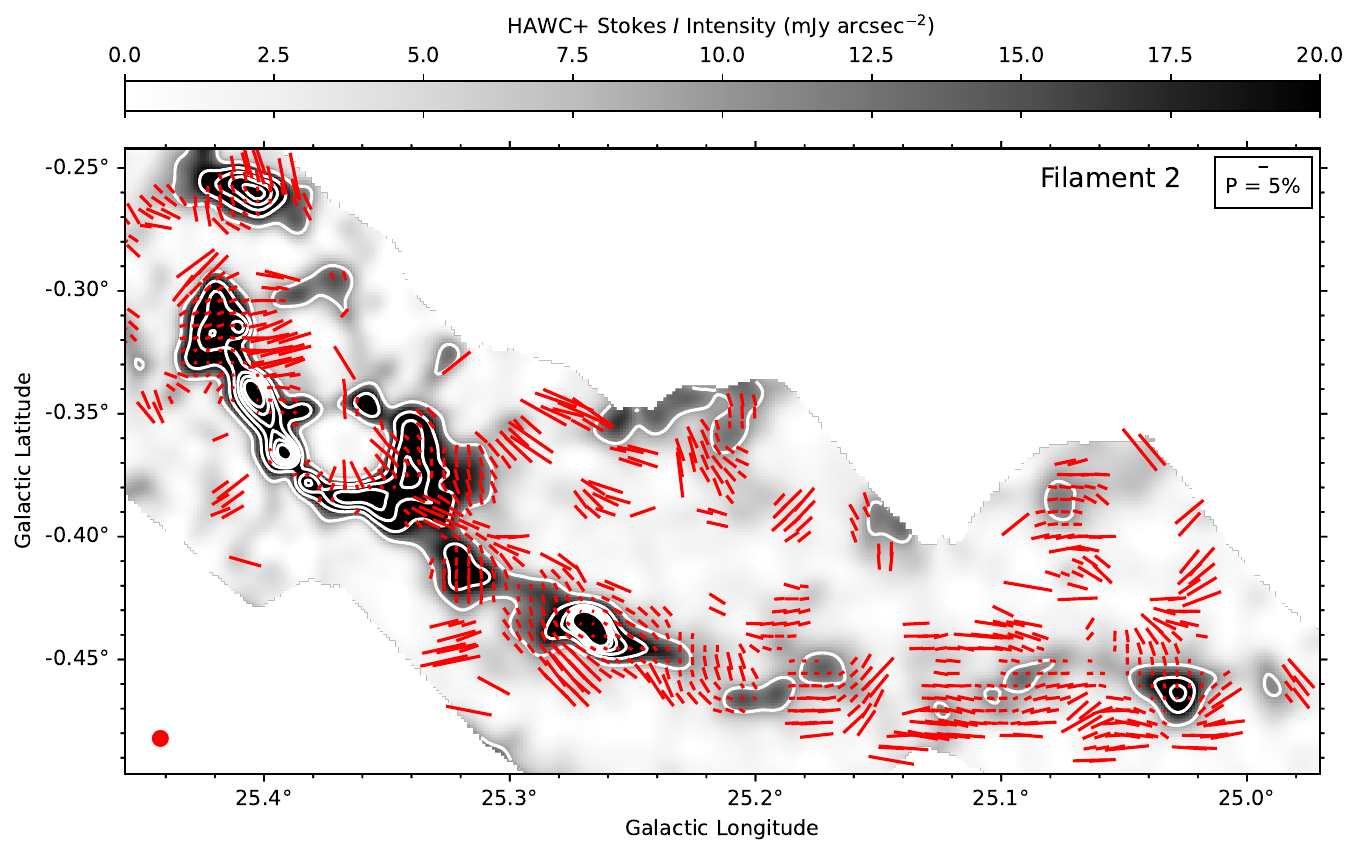}
\caption{\textbf{Filament~2}. Same as Figure~\ref{fig:Fil1_Maps}, with every fourth vector plotted for both SOFIA and \textit{Planck} data. See Section~\ref{sub:polmaps} for details. 
\label{fig:Fil2_Maps}}
\end{figure*}

\begin{figure*}[p]
\centering
\includegraphics[scale=0.75]{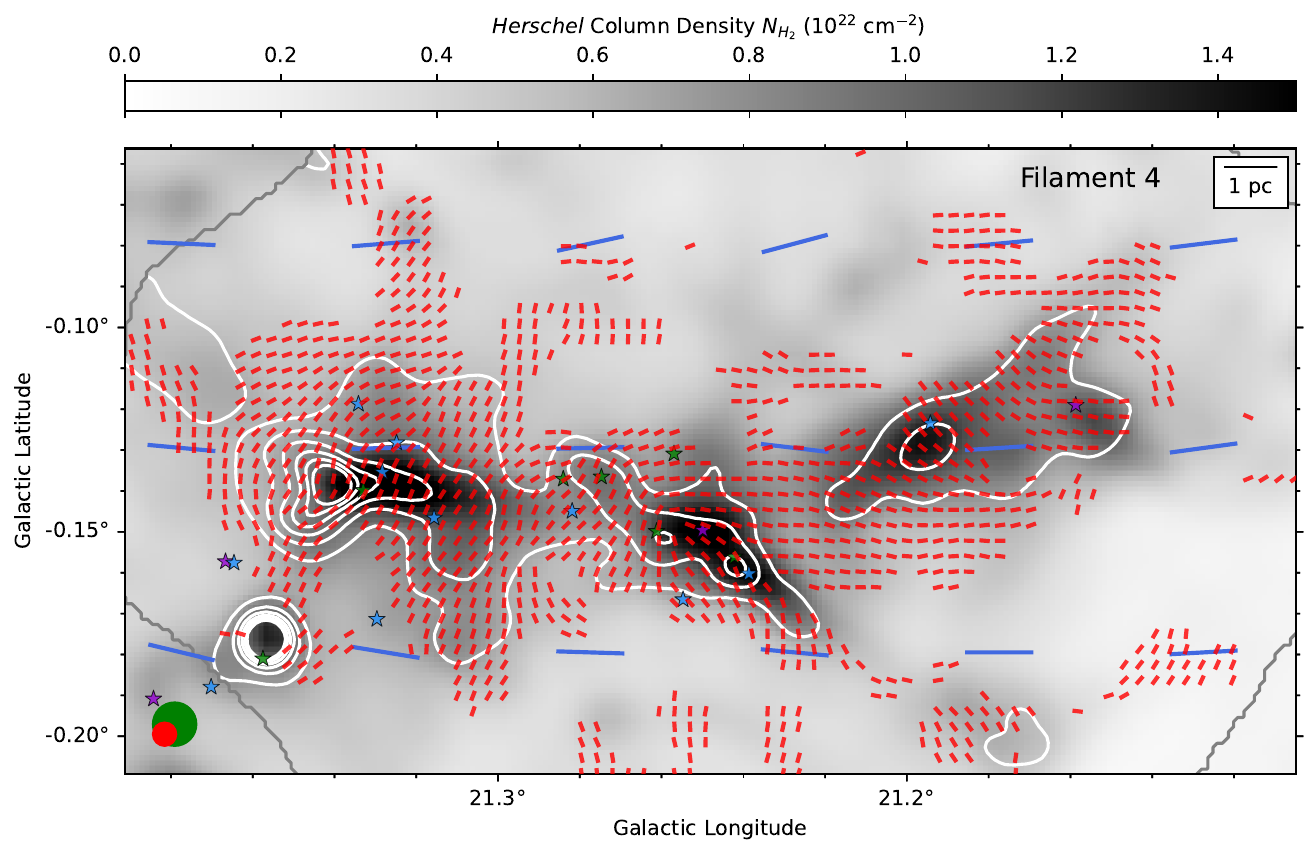}
\includegraphics[scale=0.75]{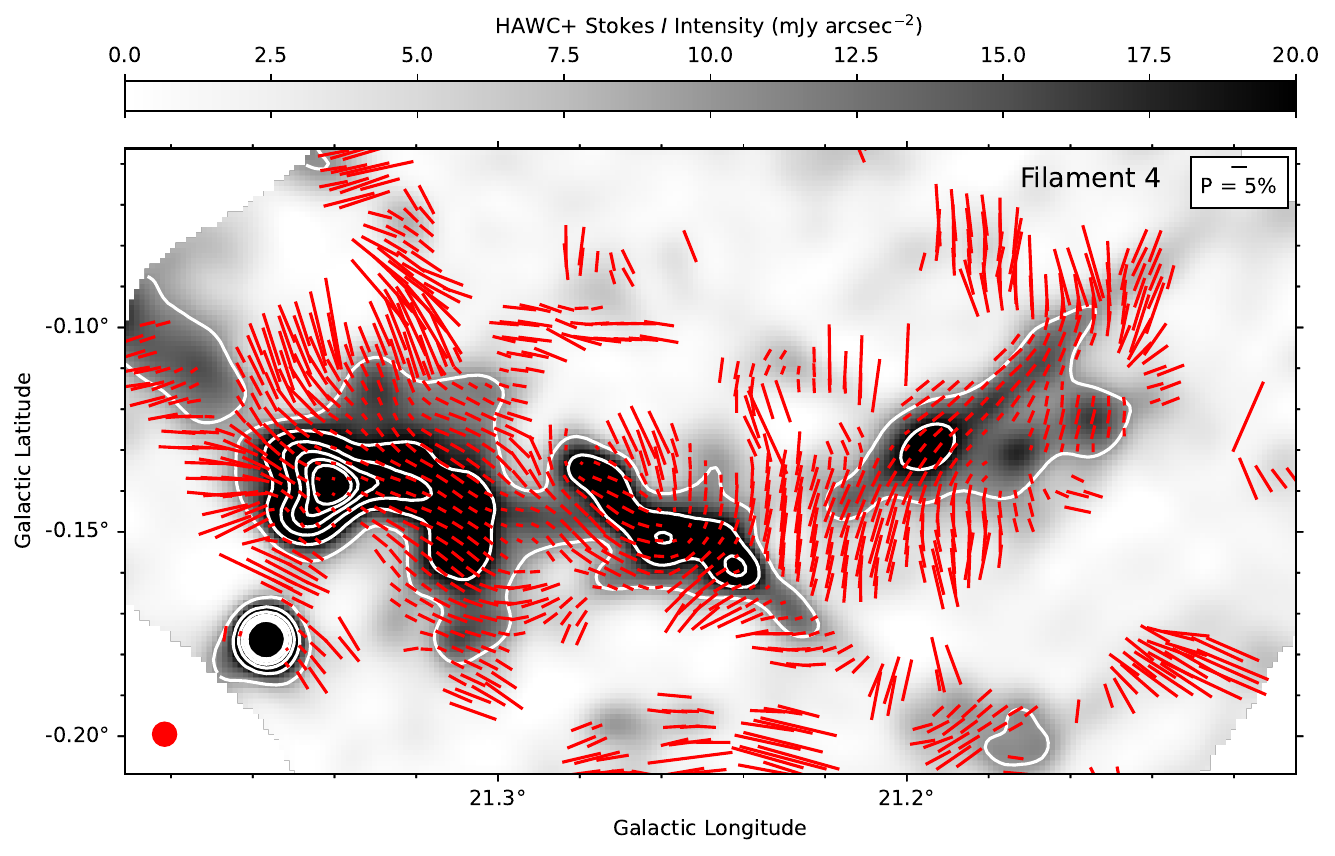}
\caption{\textbf{Filament~4}. Same as Figure~\ref{fig:Fil1_Maps}, with every third vector plotted for both SOFIA and \textit{Planck} data. See Section~\ref{sub:polmaps} for details. This figure displays the same area as Figure~\ref{fig:Fil4_RGB}. 
\label{fig:Fil4_Maps}}
\end{figure*}

\begin{figure*}[p]
\centering
\includegraphics[scale=0.75]{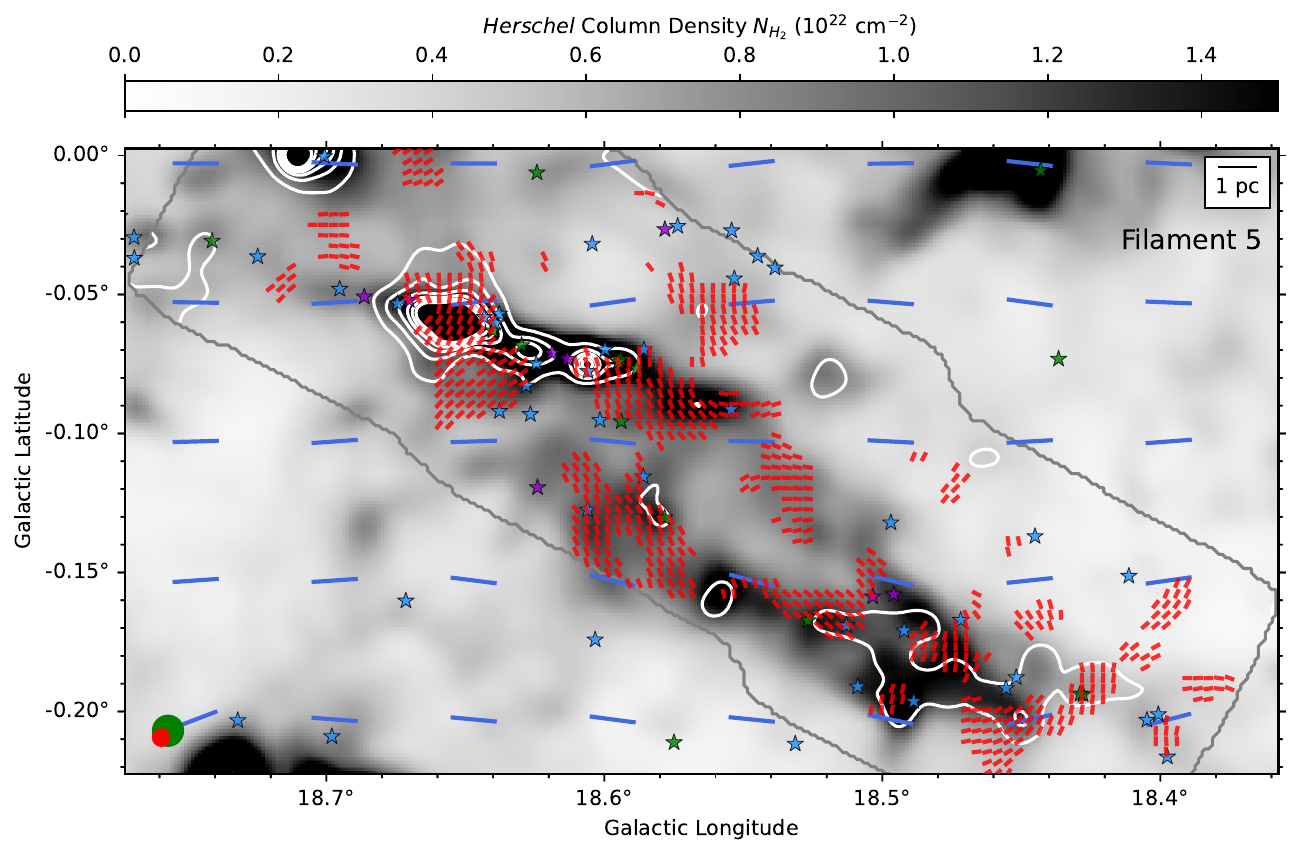}
\includegraphics[scale=0.75]{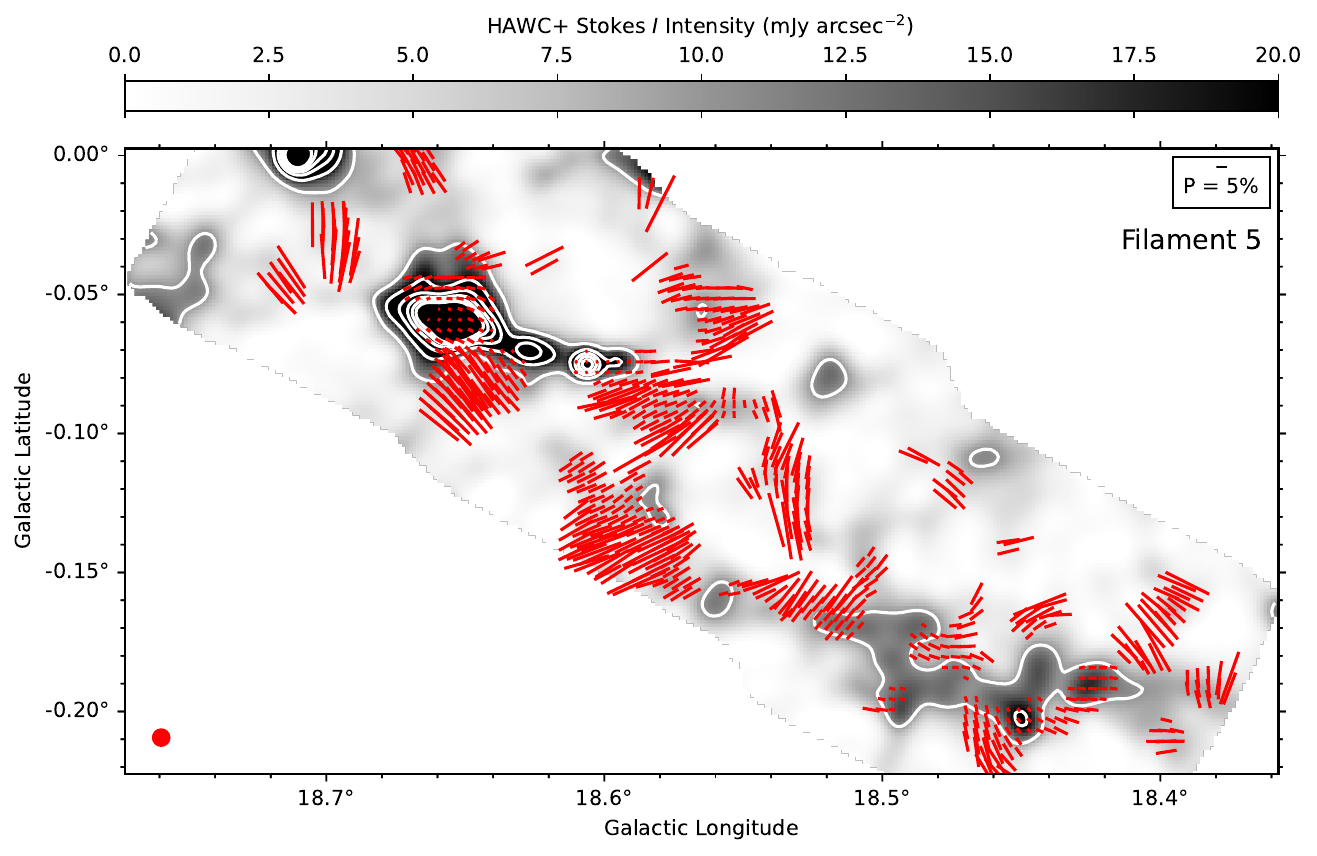}
\caption{\textbf{Filament~5}. Same as Figure~\ref{fig:Fil1_Maps}, with every third vector plotted for both SOFIA and \textit{Planck} data. See Section~\ref{sub:polmaps} for details. 
\label{fig:Fil5_Maps}}
\end{figure*}

\begin{figure*}[p]
\centering
\includegraphics[scale=0.75]{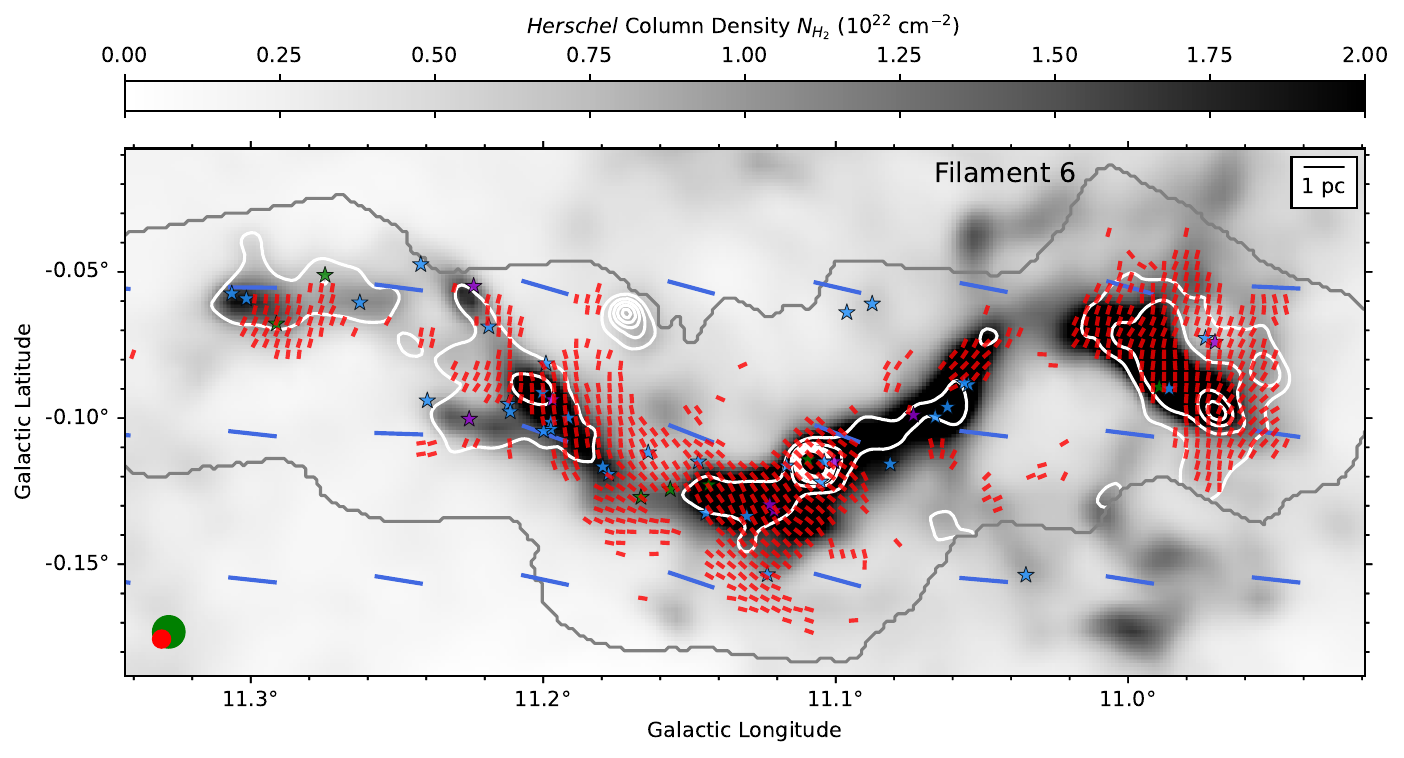}
\includegraphics[scale=0.75]{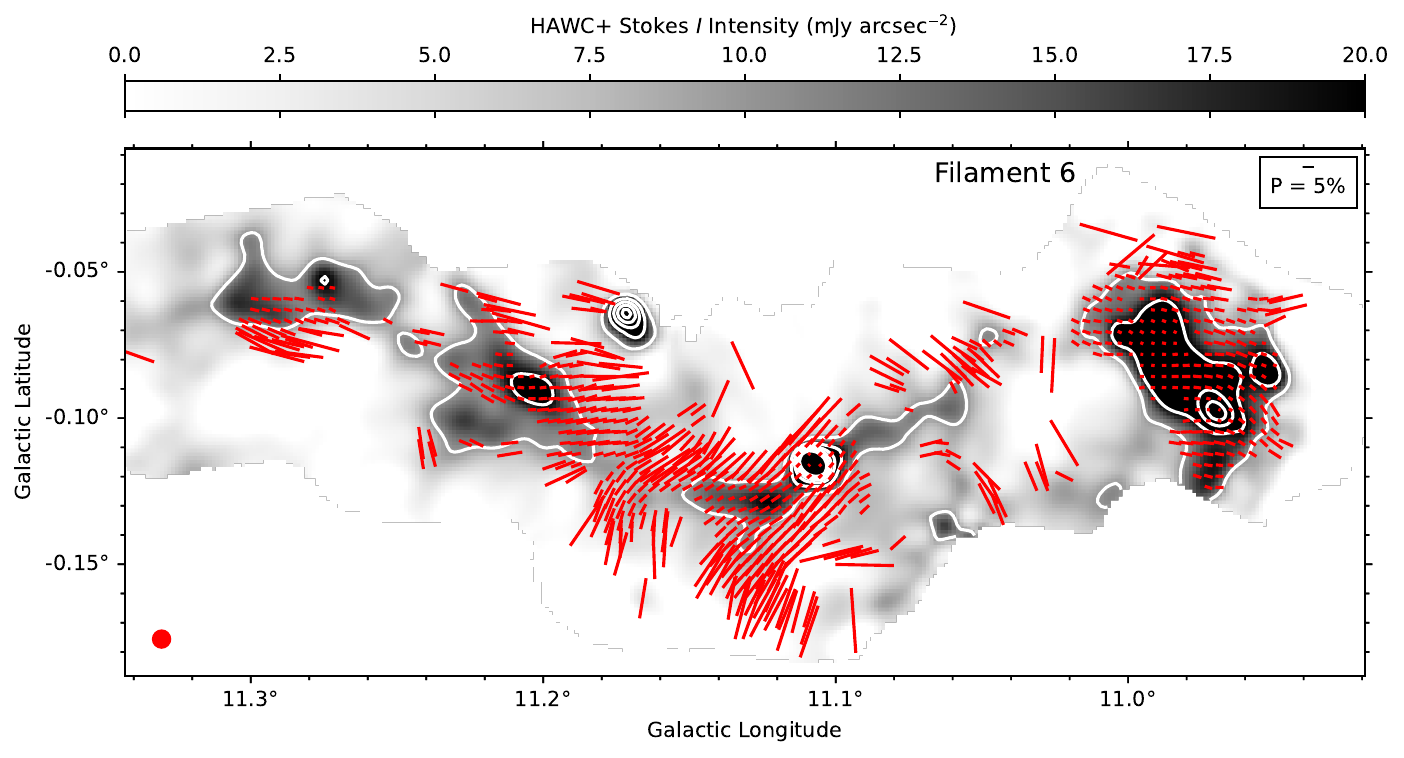}
\caption{\textbf{Filament~6} (i.e., the ``Snake''). Same as Figure~\ref{fig:Fil1_Maps}, with every third vector plotted for both SOFIA and \textit{Planck} data. See Section~\ref{sub:polmaps} for details. 
\label{fig:Snake_Maps}}
\end{figure*}

\begin{figure*}[p]
\centering
\includegraphics[scale=0.75]{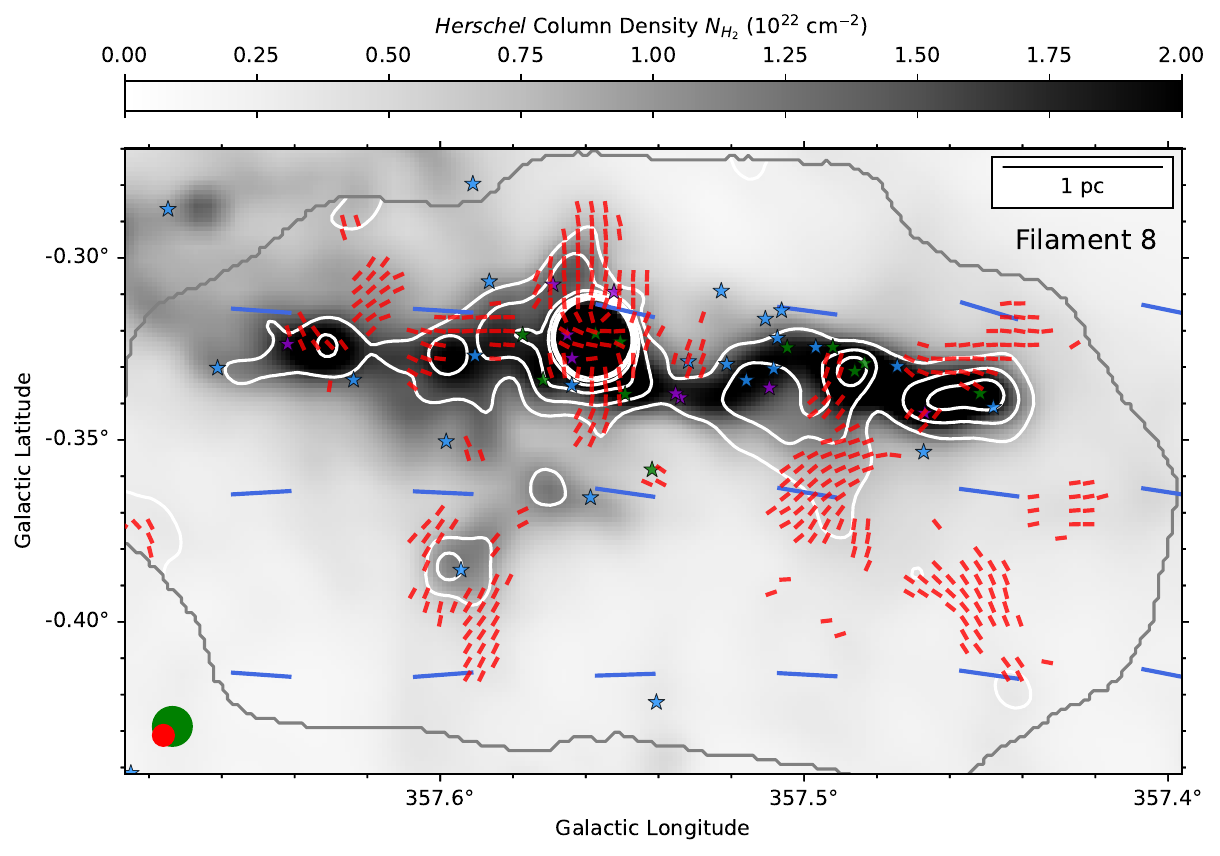}
\includegraphics[scale=0.75]{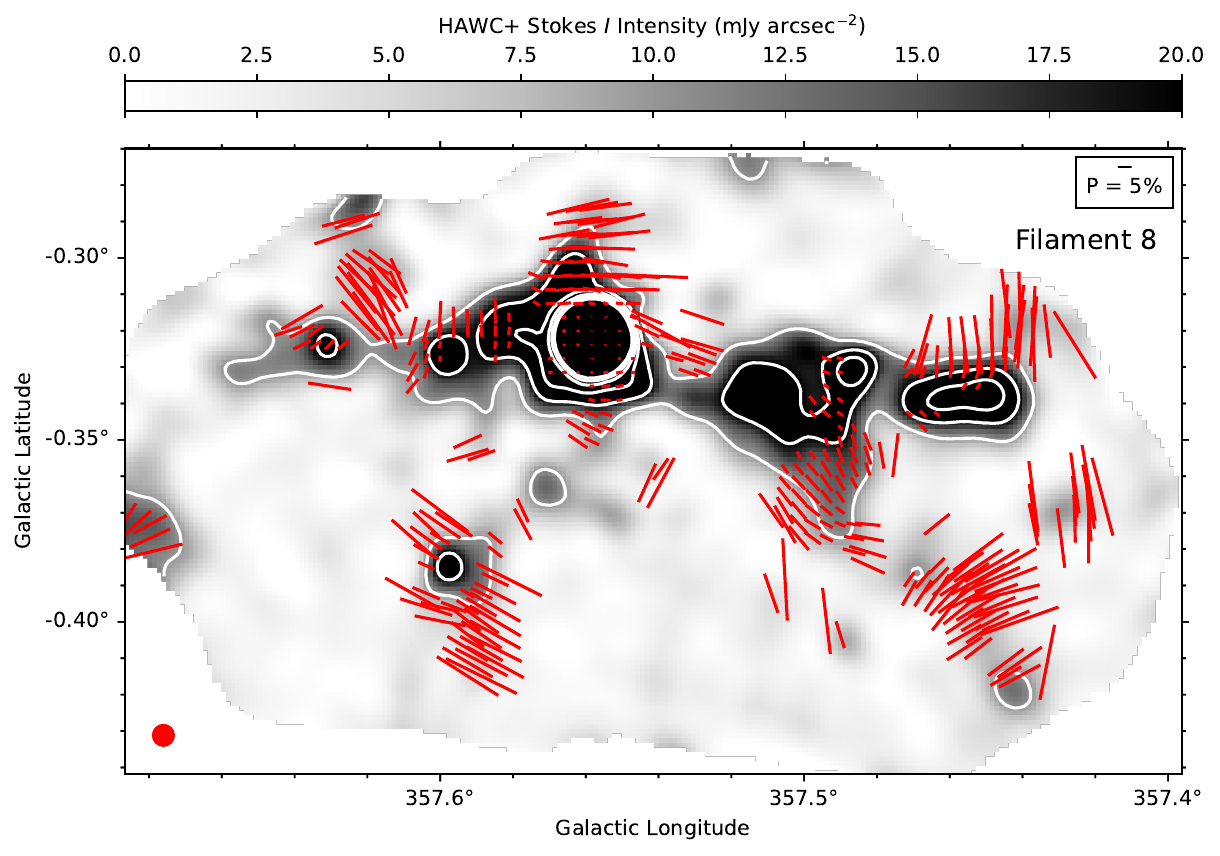}
\caption{\textbf{Filament~8}. Same as Figure~\ref{fig:Fil1_Maps}, with every third vector plotted for both SOFIA and \textit{Planck} data. See Section~\ref{sub:polmaps} for details. 
\label{fig:Fil8_Maps}}
\end{figure*}

\begin{figure*}[p]
\centering
\includegraphics[scale=0.75]{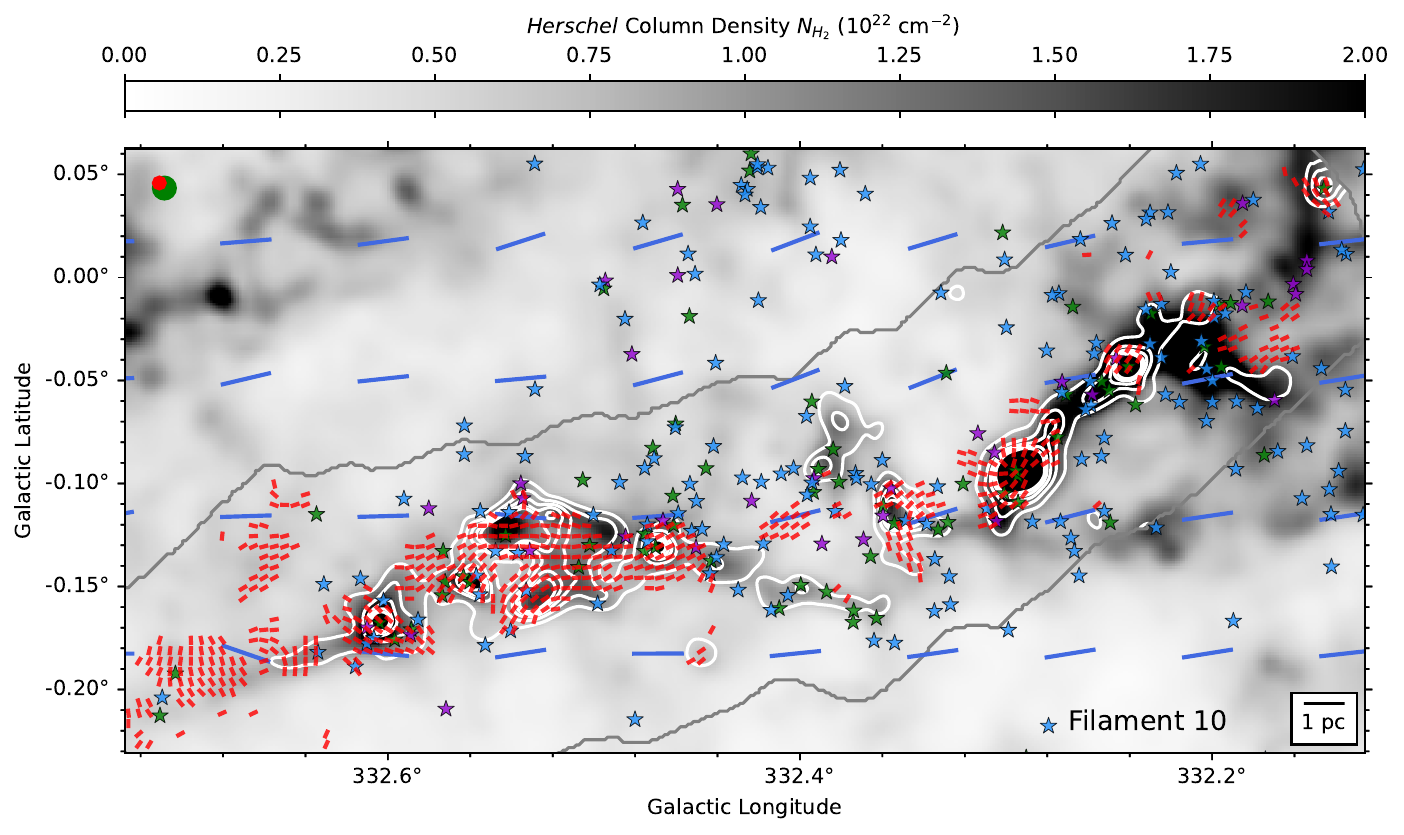}
\includegraphics[scale=0.75]{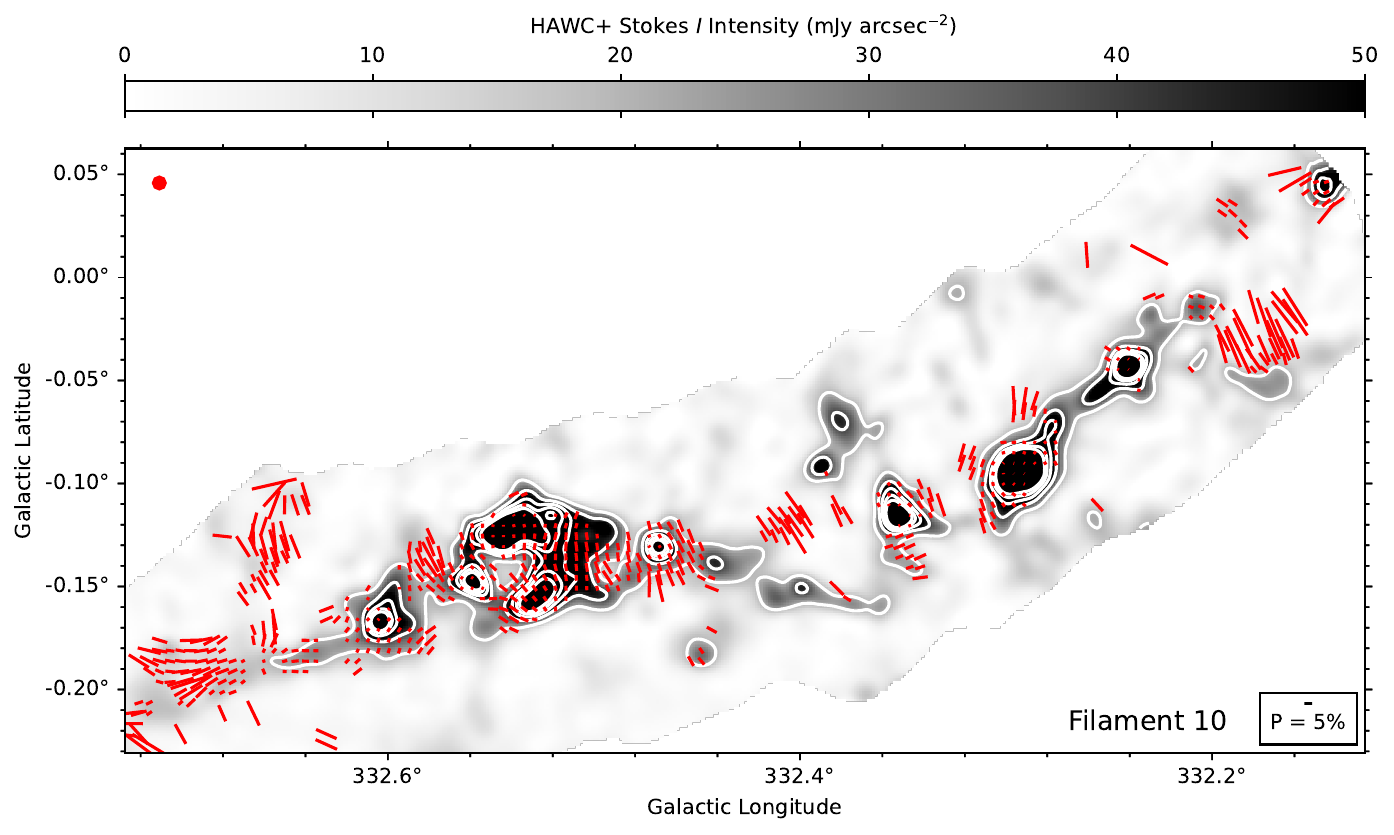}
\caption{\textbf{Filament~10}. Same as Figure~\ref{fig:Fil1_Maps}, with every fourth vector plotted for both SOFIA and \textit{Planck} data. The white contours in the bottom panel also trace the HAWC+ 214~$\mu$m Stokes~$I$ total intensity for the following levels starting at 20 mJy arcsec$^{-2}$ and increasing by steps of 20 mJy arcsec$^{-2}$. See Section~\ref{sub:polmaps} for details. 
\label{fig:Fil10_Maps}}
\end{figure*}

\begin{figure*}[]
\centering
\includegraphics[scale=0.75]{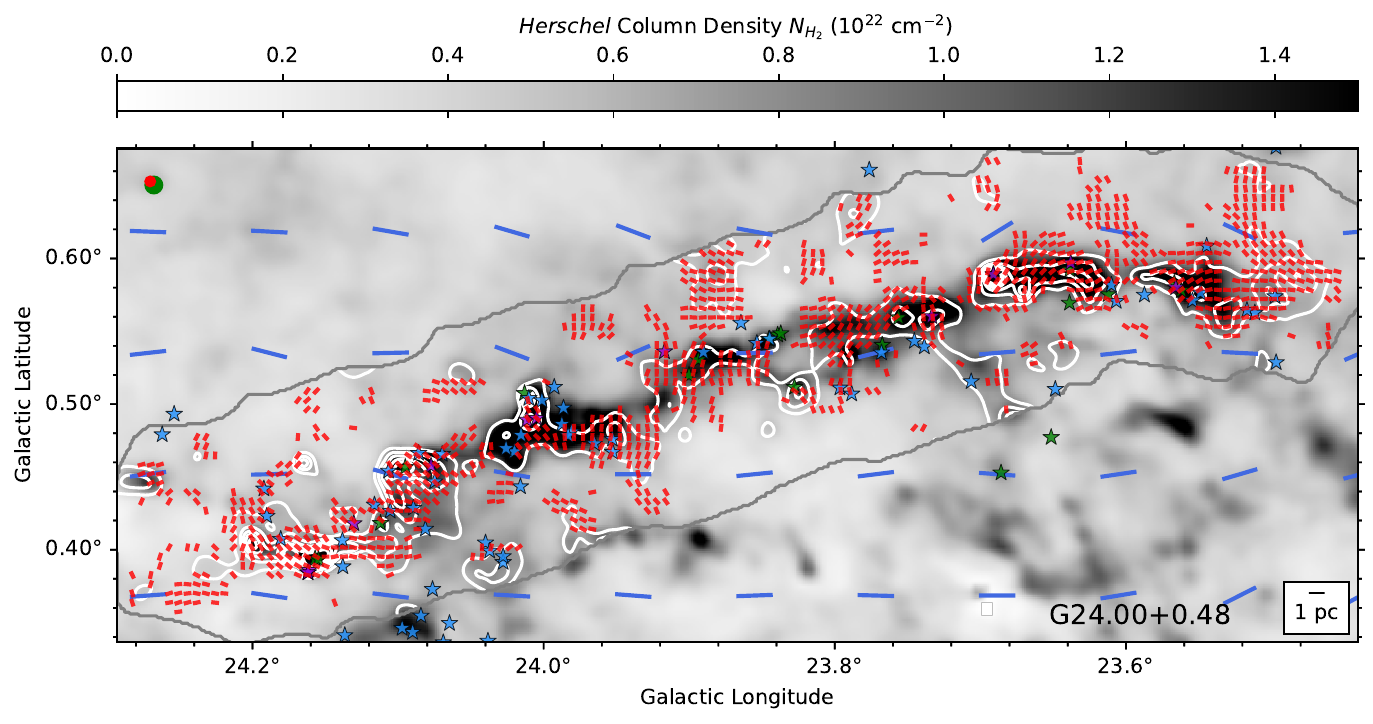}
\includegraphics[scale=0.75]{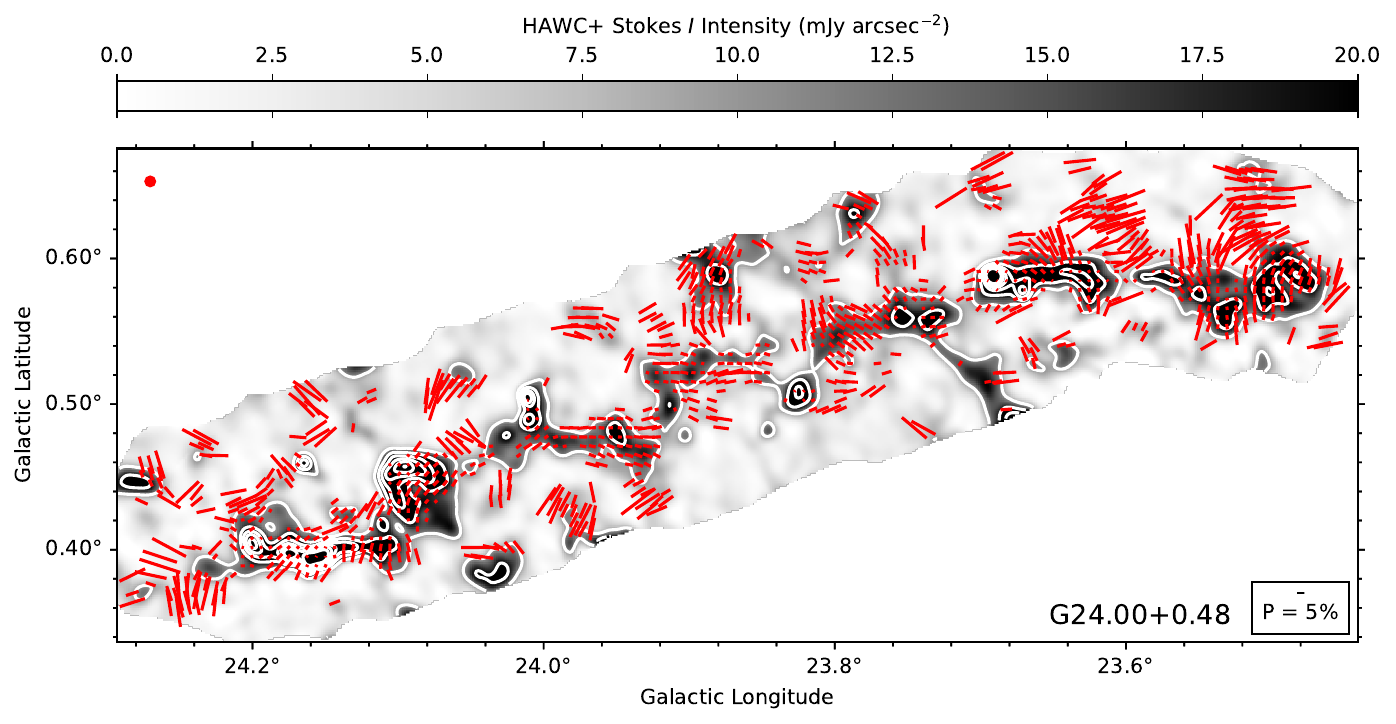}
\caption{\textbf{G24}. Same as Figure~\ref{fig:Fil1_Maps}, with every fifth vector plotted for both SOFIA and \textit{Planck} data. See Section~\ref{sub:polmaps} for details. 
\label{fig:G24_Maps}}
\end{figure*}

\begin{figure*}[p]
\centering
\includegraphics[scale=0.75]{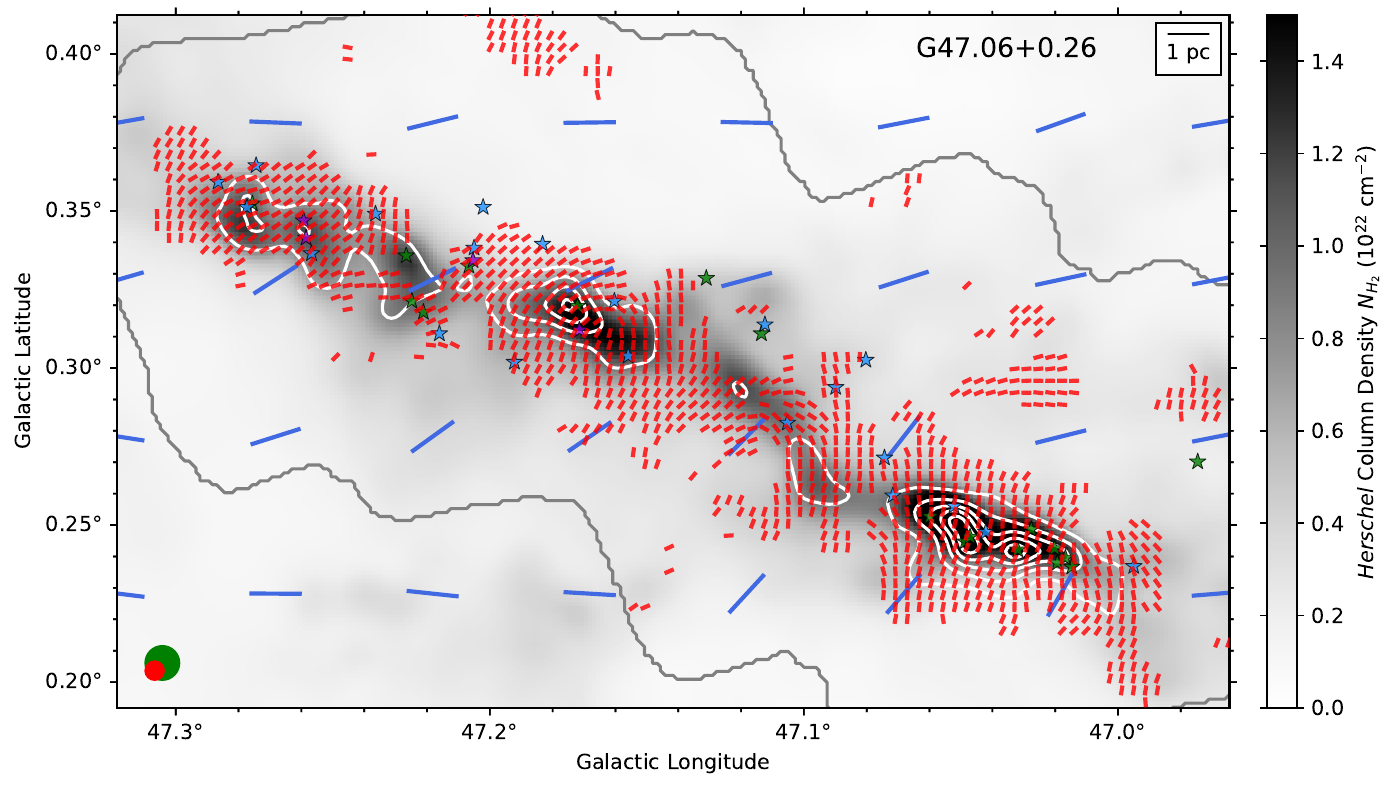}
\includegraphics[scale=0.75]{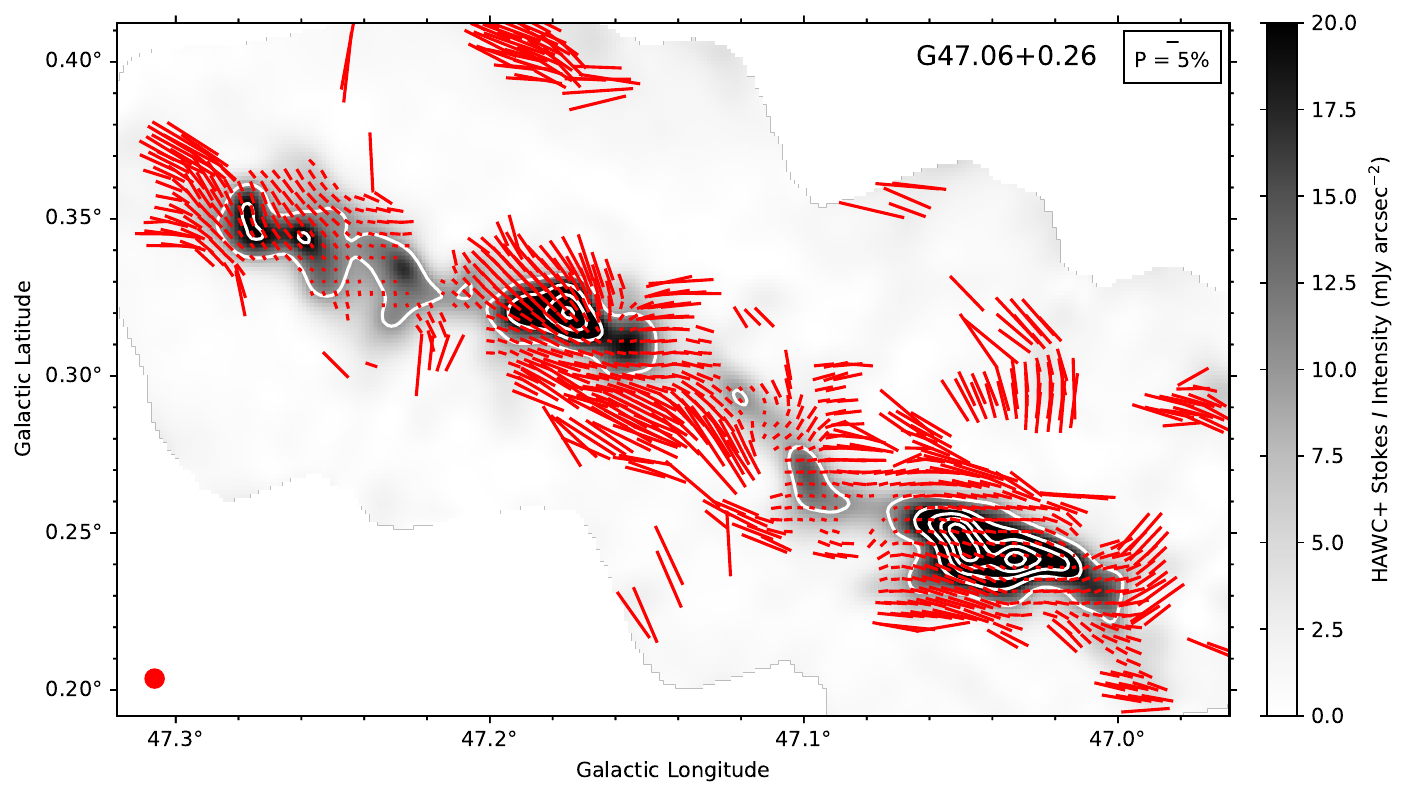}
\caption{\textbf{G47}. Same as Figure~\ref{fig:Fil1_Maps}, with every third vector plotted for both SOFIA and \textit{Planck} data. See Section~\ref{sub:polmaps} for details.
\label{fig:G47_Maps}}
\end{figure*}

\begin{figure*}[p]
\centering
\includegraphics[scale=0.75]{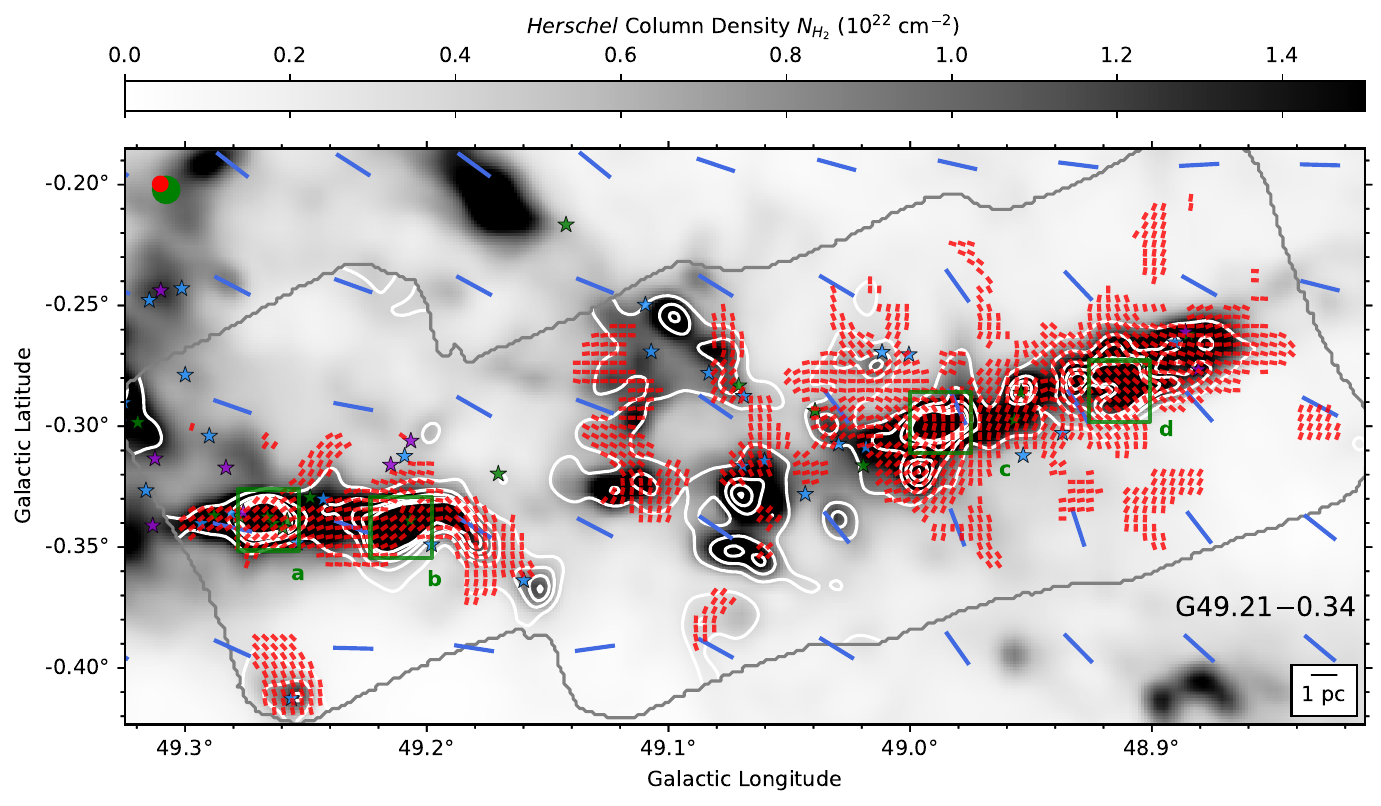}
\includegraphics[scale=0.75]{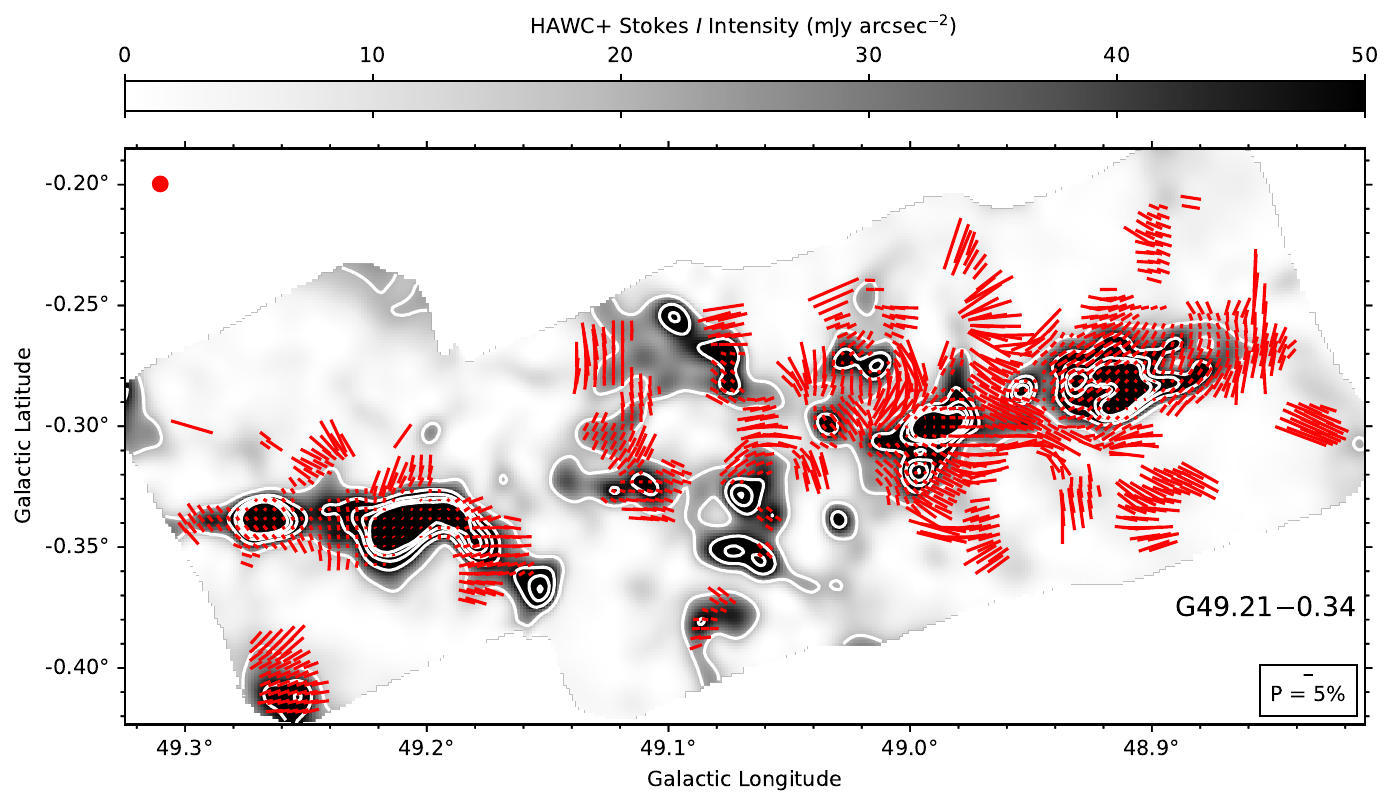}
\caption{\textbf{G49}. Same as Figure~\ref{fig:Fil1_Maps}, with every third vector plotted for both SOFIA and \textit{Planck} data. The white contours trace the HAWC+ 214~$\mu$m Stokes~$I$ total intensity for the following levels starting at 20 mJy arcsec$^{-2}$ and increasing by steps of 30 mJy arcsec$^{-2}$. See Section~\ref{sub:polmaps} for details. Additionally, the green squares in the top panel identify the four regions listed in Table~\ref{tab:G49abcd}.
\label{fig:G49_Maps}}
\end{figure*}

Figures~\ref{fig:Fil1_Maps} through \ref{fig:G49_Maps} present the polarization observations for the FIELDMAPS survey. Each figure is divided into two panels, with the top panel being the map of inferred magnetic field vectors and the bottom panel being the map of measured polarization vectors, as described below: 

\textit{Magnetic field maps:} The top panel of Figures~\ref{fig:Fil1_Maps} through \ref{fig:G49_Maps} shows the plane-of-sky magnetic field structure inferred from both HAWC+ 214~$\mu$m (red vectors) and \textit{Planck} 850~$\mu$m (blue vectors) polarization observations. These magnetic field vectors, with their orientation $\theta_B$ rotated by $90^\circ$ from the derived polarization angle~$\theta$, are plotted over the molecular hydrogen column density $N_{H_2}$ (gray scale) obtained from \textit{Herschel} continuum data (see Section~\ref{sub:herschel}). To facilitate the visualization of the magnetic field structure, the vector lengths are fixed to a single arbitrary length for each map, and we vary the sampling of the plotted HAWC+ and \textit{Planck} vectors depending on the bone (see individual captions). Sampling every fourth polarization vector is equivalent to plotting 1.0 vector per beam element for HAWC+ and 1.25 vector per beam element for \textit{Planck}. The white contours trace the HAWC+ 214~$\mu$m Stokes~$I$ total intensity for the following levels starting at 10 mJy arcsec$^{-2}$ and increasing by steps of 10 mJy arcsec$^{-2}$, unless stated otherwise in the caption. The beam sizes of HAWC+ ($18\farcs7$; red circle) and \textit{Herschel} ($36\farcs4$; green circle) are also given, and, while we do not show it on the plot since it is too large, we note here that the \textit{Planck} beam is ($5\arcmin$). A scale bar is provided to show a reference length of 1~pc for each bone, using the distances listed in Table~\ref{tab:filaments}. Also shown are the Class~I (green stars), Flat Spectrum (violet stars), and Class~II (light blue stars) YSOs identified for this work (see Section~\ref{sub:yso}).

\textit{Polarization maps:} The bottom panel of Figures~\ref{fig:Fil1_Maps} through \ref{fig:G49_Maps} shows the HAWC+ 214~$\mu$m polarization vectors (red) plotted over the Stokes~$I$ map (gray scale). The orientation of each vector is given by the polarization angle~$\theta$ derived from the Stokes~$Q$ and $U$ parameters. The length of each vector is its associated polarization fraction~$P$, and a reference length of $P=5\%$ is provided. To facilitate the comparison between the top and bottom panels of each figure, the white contours in the bottom panel also trace the HAWC+ 214~$\mu$m Stokes~$I$ total intensity for the following levels starting at 10 mJy arcsec$^{-2}$ and increasing by steps of 10 mJy arcsec$^{-2}$, unless stated otherwise in the caption. The HAWC+ beam size ($18\farcs7$) is indicated with a red circle. 

For Filament~5, the results of the Chop-Nod observations are plotted in Figure~\ref{fig:Fil5ChopE} of Appendix~\ref{apx:chop_fil5}.

% Filament 1

% Filament 2

% Filament 4

% Filament 5

% Filament 6 (Snake)

% Filament 8

% Filament 10

%G24

% G47

% G49

\subsection{Distributions of Polarization Angles} 
\label{sub:histograms}

\begin{figure*}
    \centering
    \includegraphics[width=0.495\textwidth]{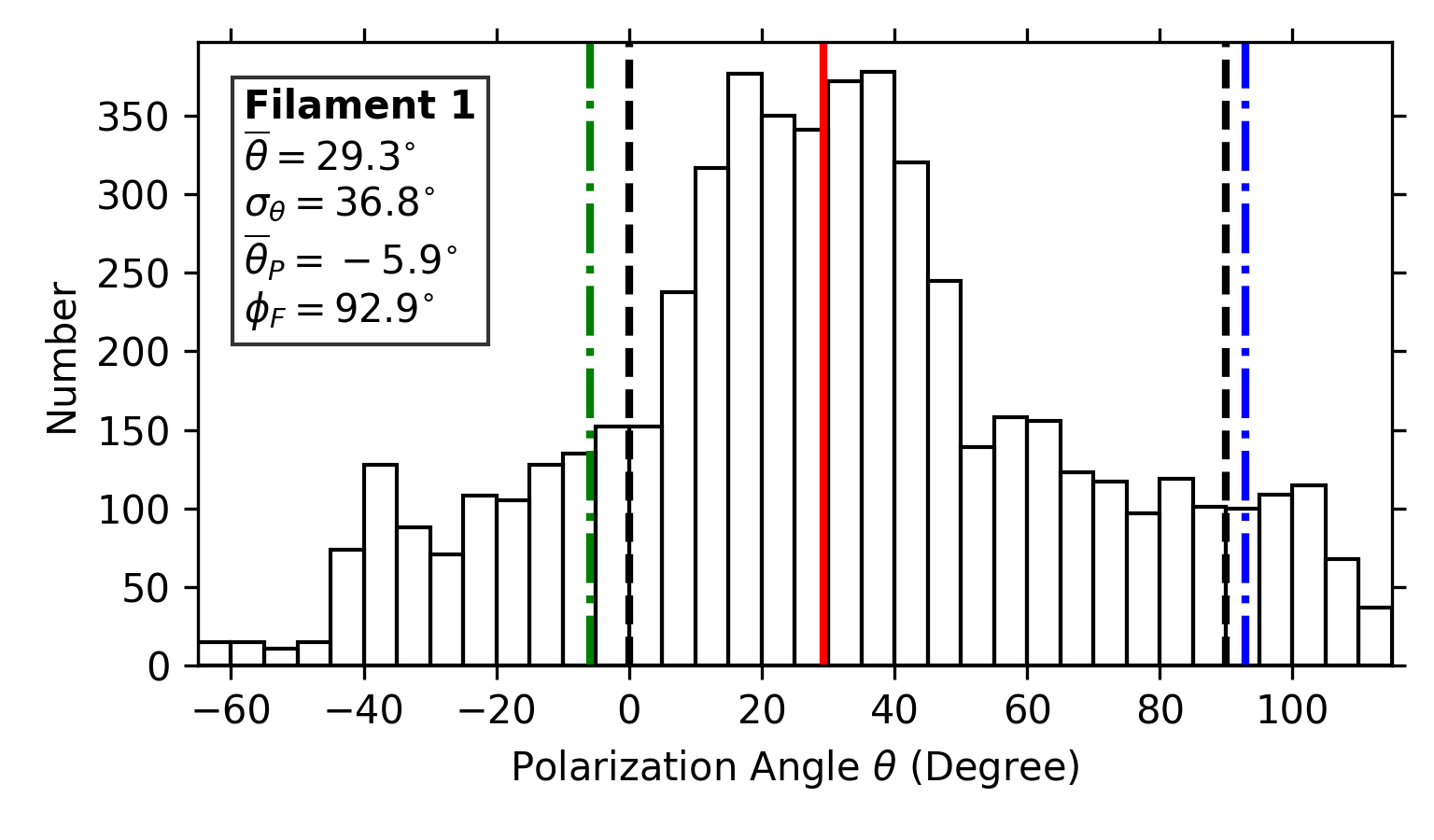}
    \includegraphics[width=0.495\textwidth]{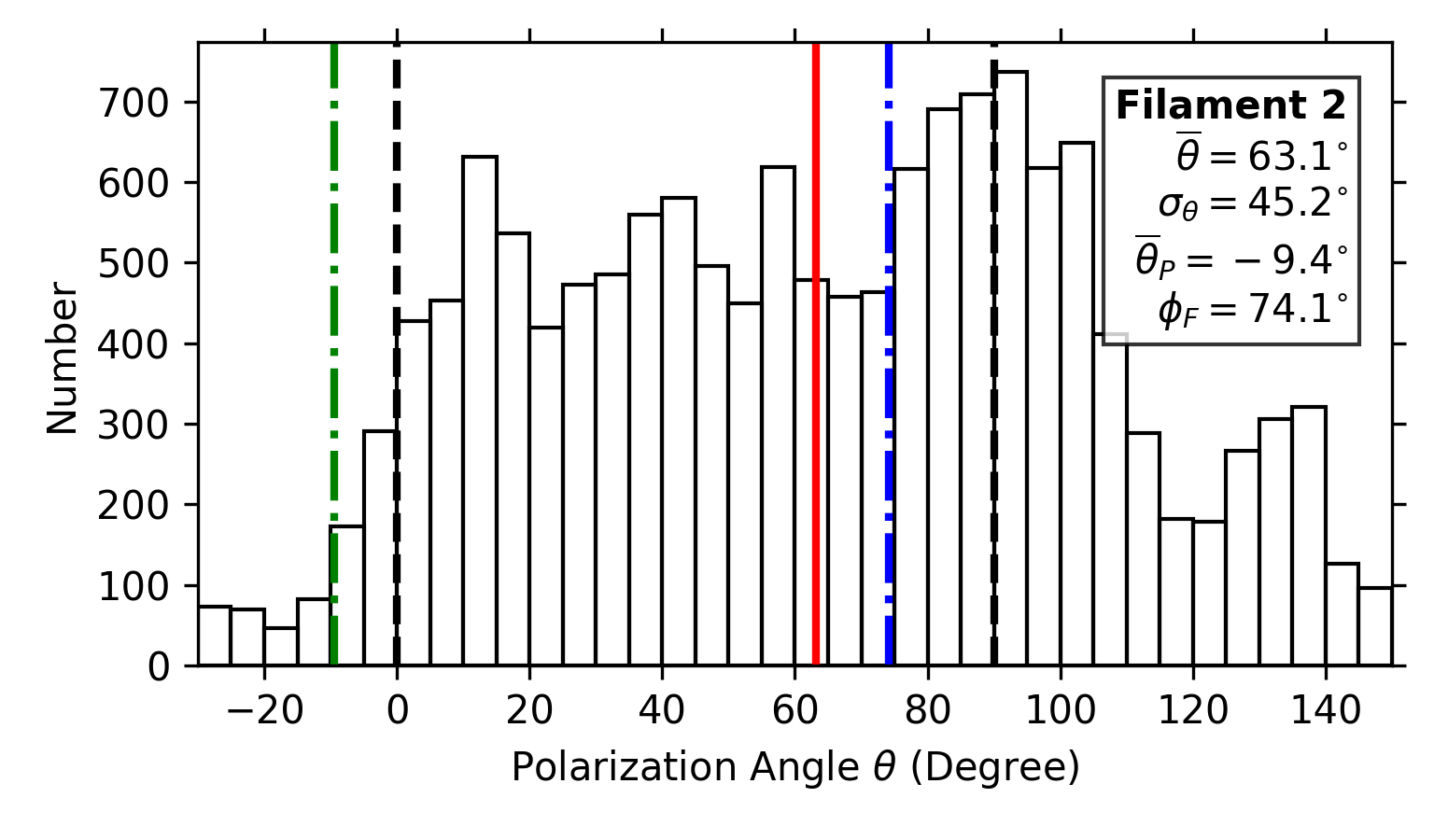}
    \includegraphics[width=0.495\textwidth]{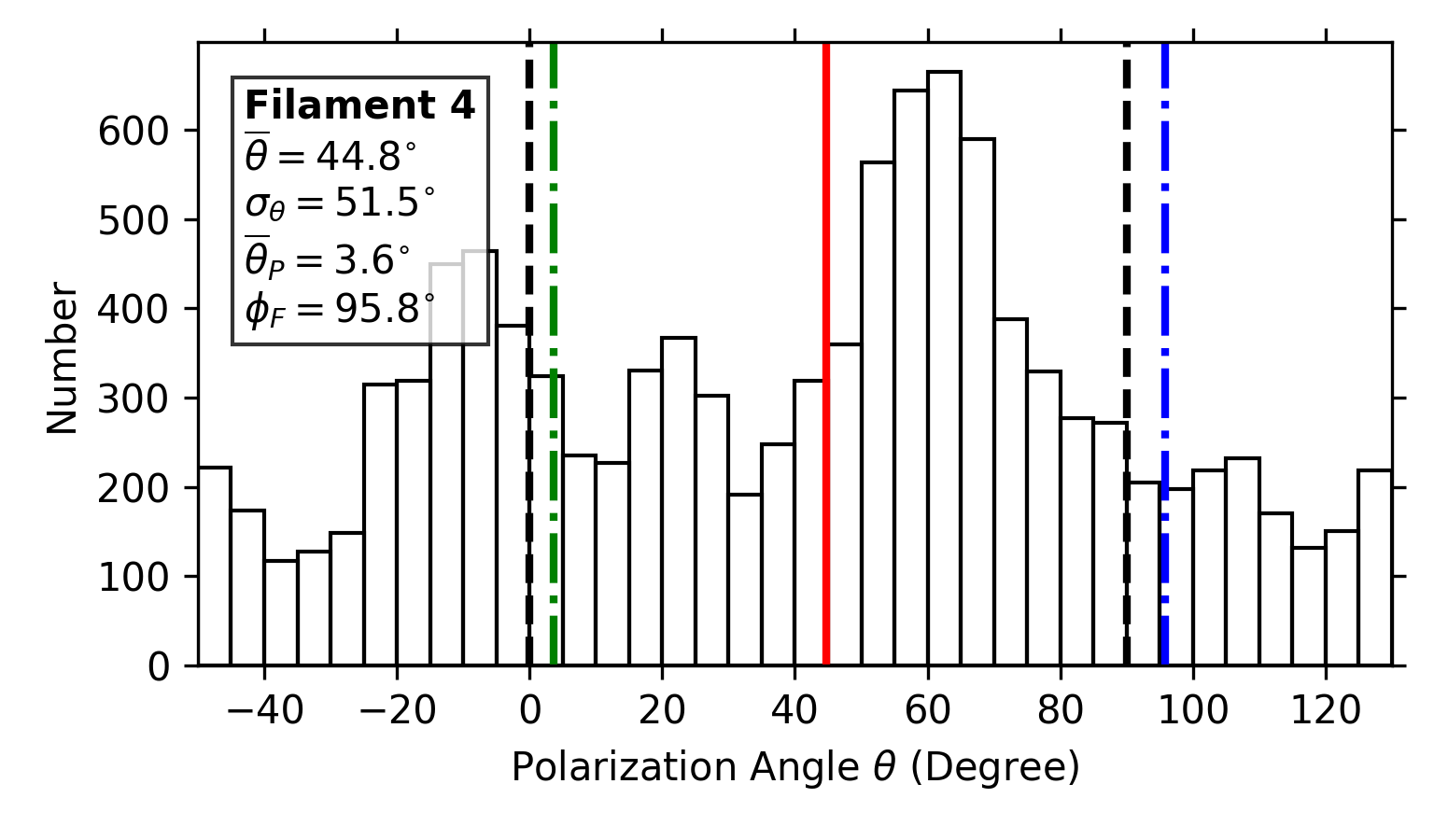}
    \includegraphics[width=0.495\textwidth]{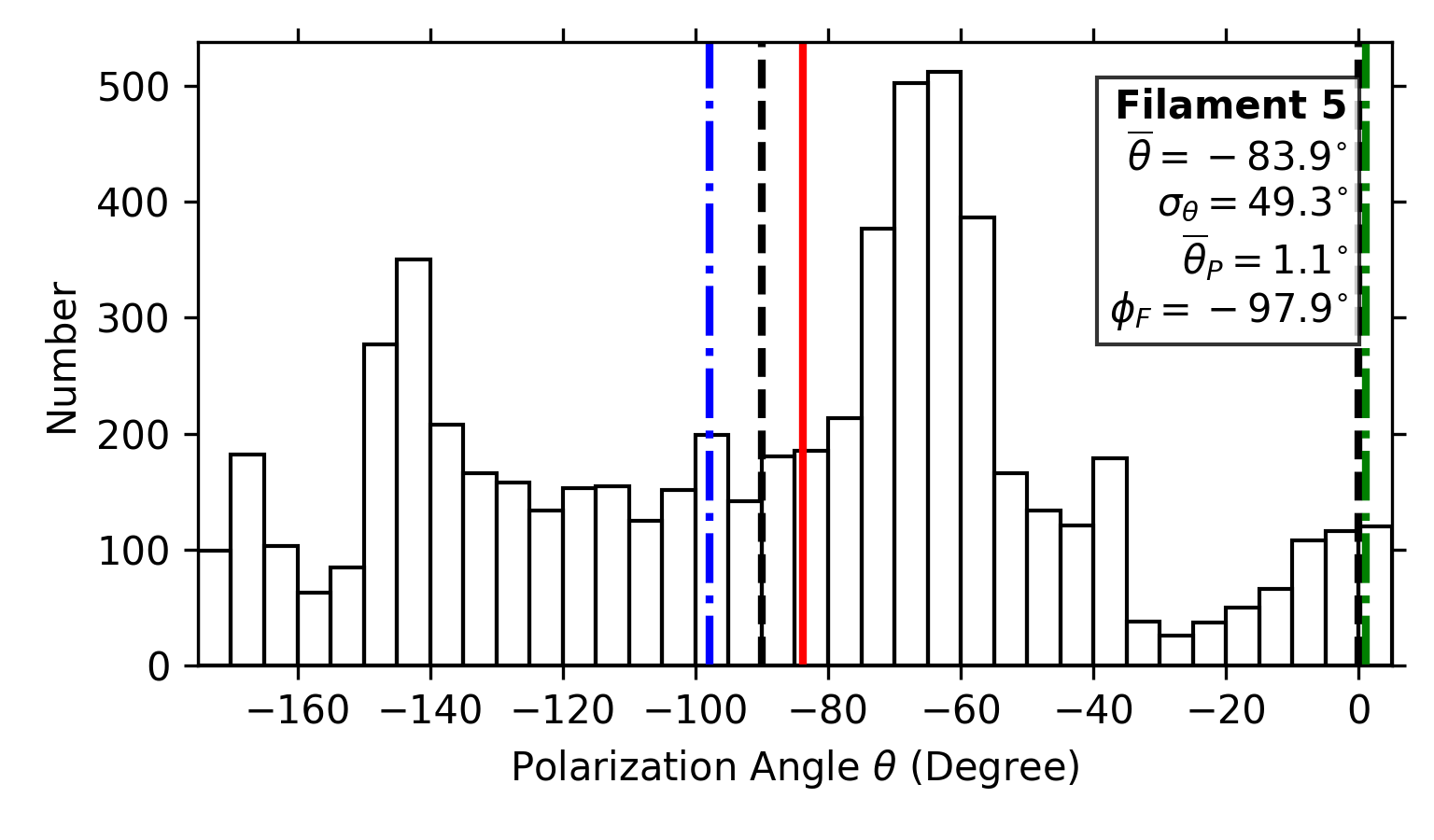}
    \includegraphics[width=0.495\textwidth]{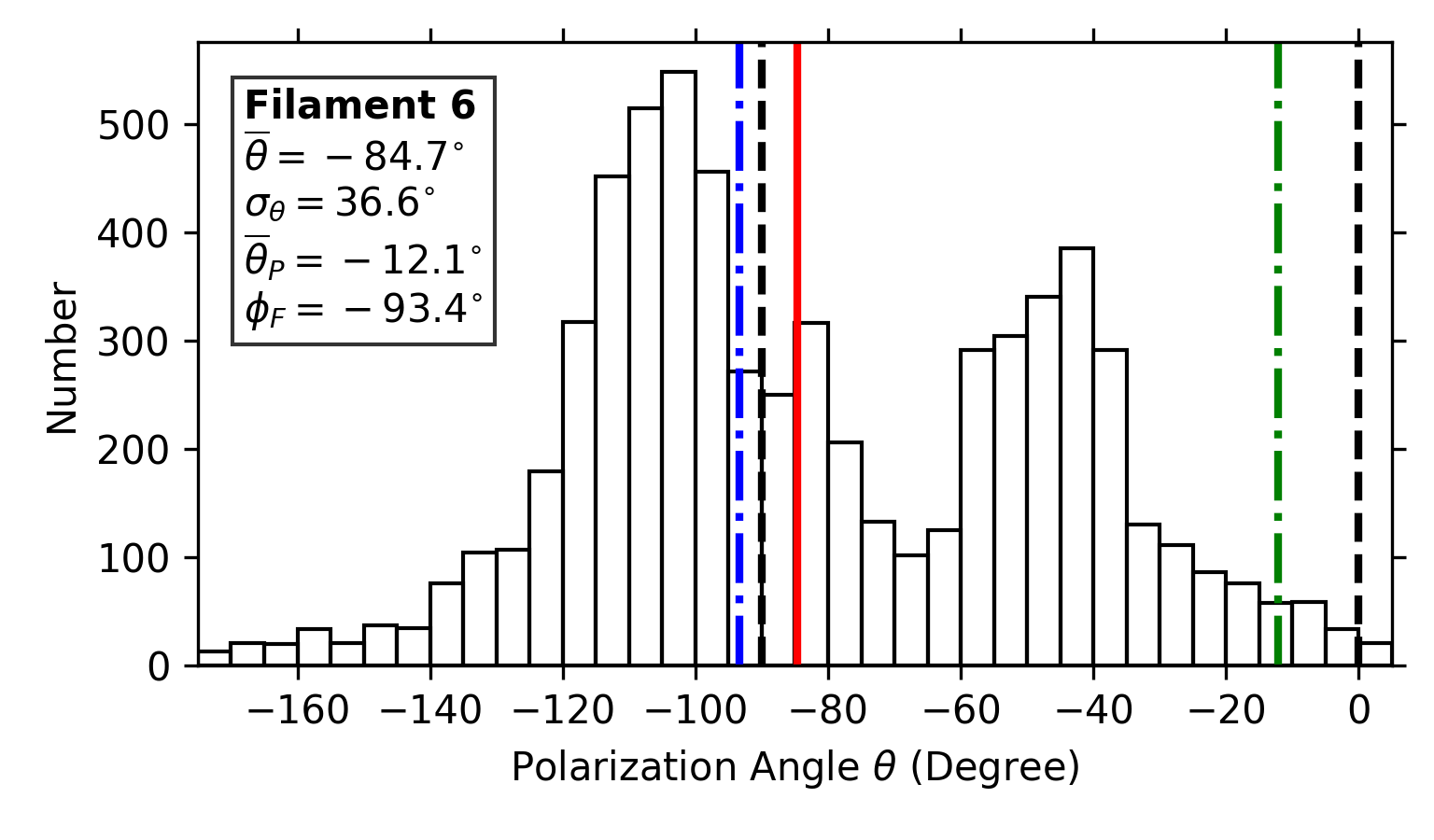}
    \includegraphics[width=0.495\textwidth]{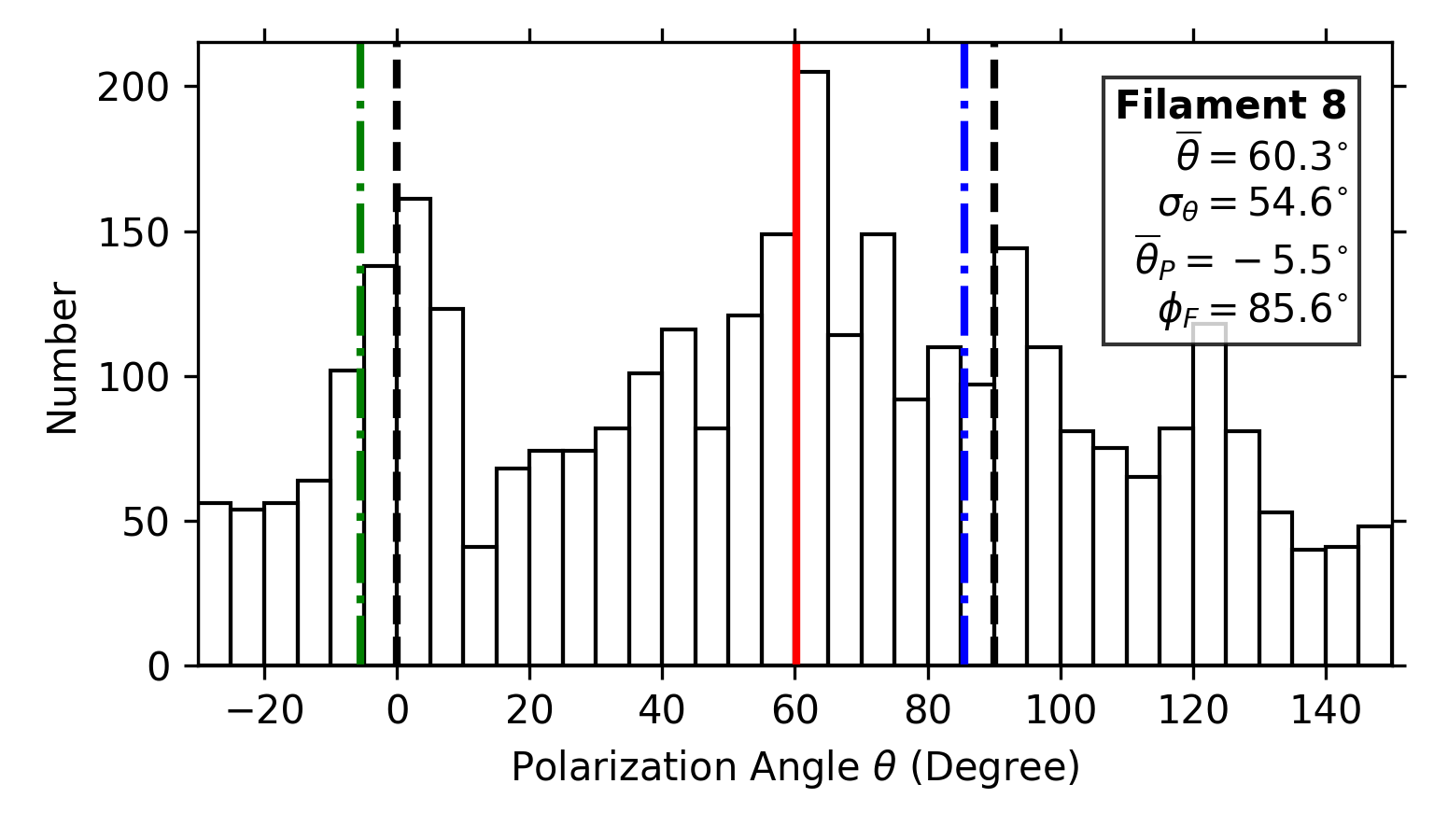}
    \caption{Histograms of polarization angles $\theta$. Continued in Figure~\ref{fig:Fil10-G49_Histo}. All angles are provided relative to Galactic North. The bin size of the histograms is $5^\circ$. For each target in this study, the full red line is the circular mean $\overline{\theta}$ of the distribution, the green dashed-dotted line is the circular mean of the \textit{Planck} polarization angles $\overline{\theta}_P$, and the blue dashed-dotted line is the fitted orientation $\phi_F$ of the filament (see Section~\ref{sub:histograms}). Each histogram is centered on the bin containing the circular mean $\overline{\theta}$. Black dashed lines indicate angles of -90$^\circ$, 0$^\circ$, and 90$^\circ$ for clarity. The measured circular standard deviation $\sigma_\theta$ for each cloud is also given within the insets. The inferred magnetic field orientation $\theta_B$ is obtained from rotating the polarization angle $\theta$ by $90^\circ$.}
    \label{fig:Fil1-5_Histo}
\end{figure*}

\begin{figure*}
    \centering
    \includegraphics[width=0.495\textwidth]{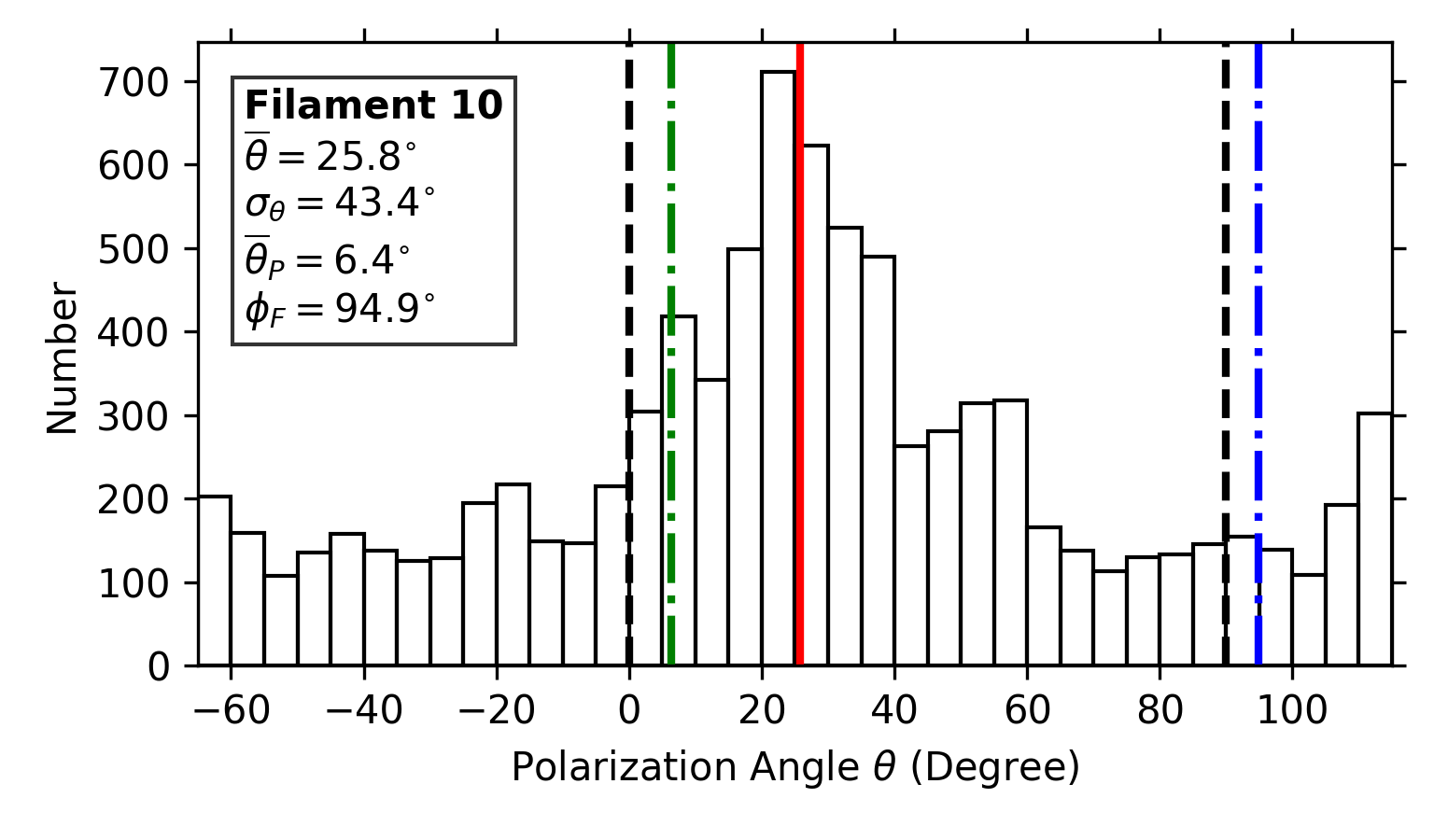}
    \includegraphics[width=0.495\textwidth]{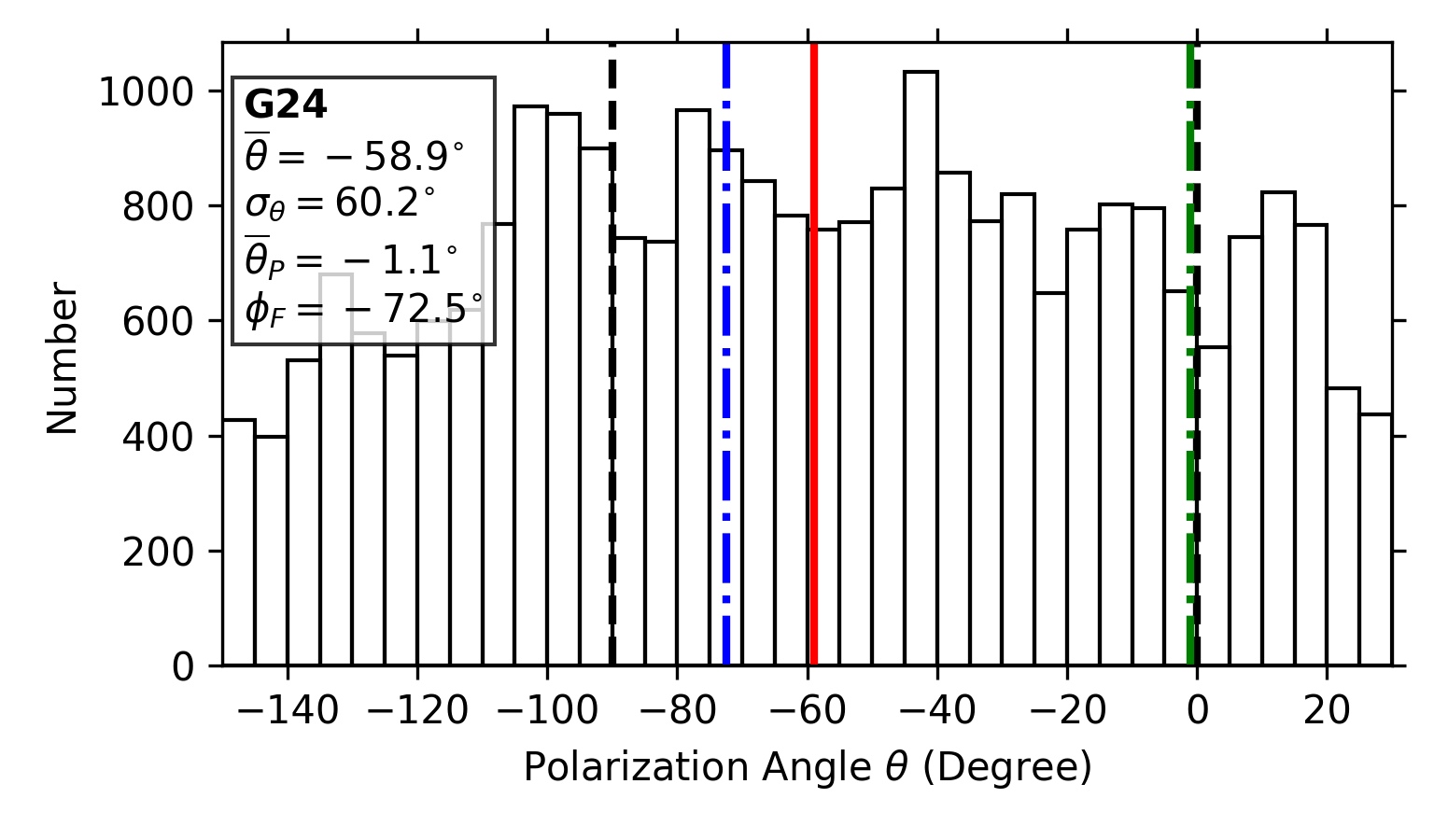}
    \includegraphics[width=0.495\textwidth]{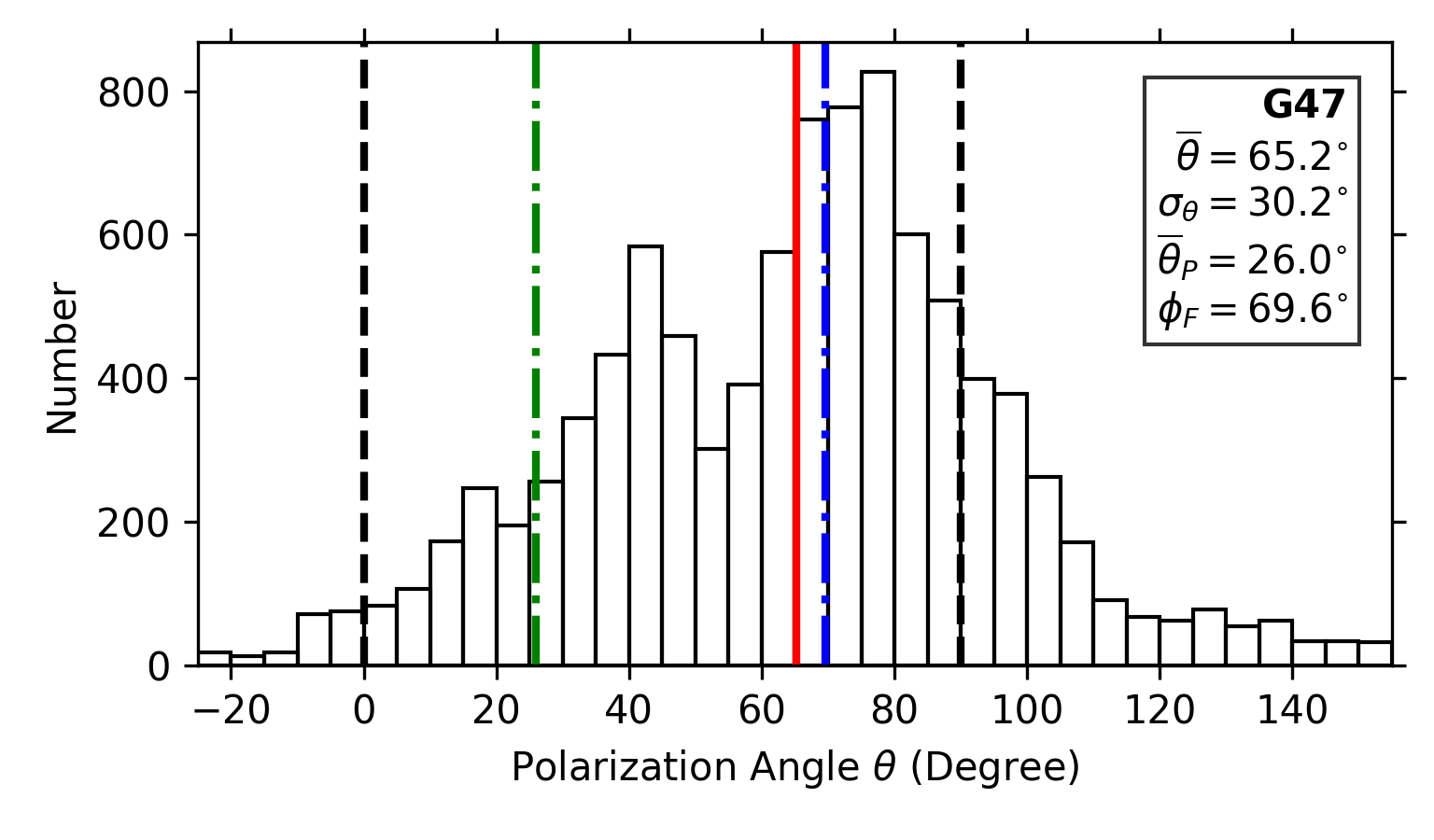}
    \includegraphics[width=0.495\textwidth]{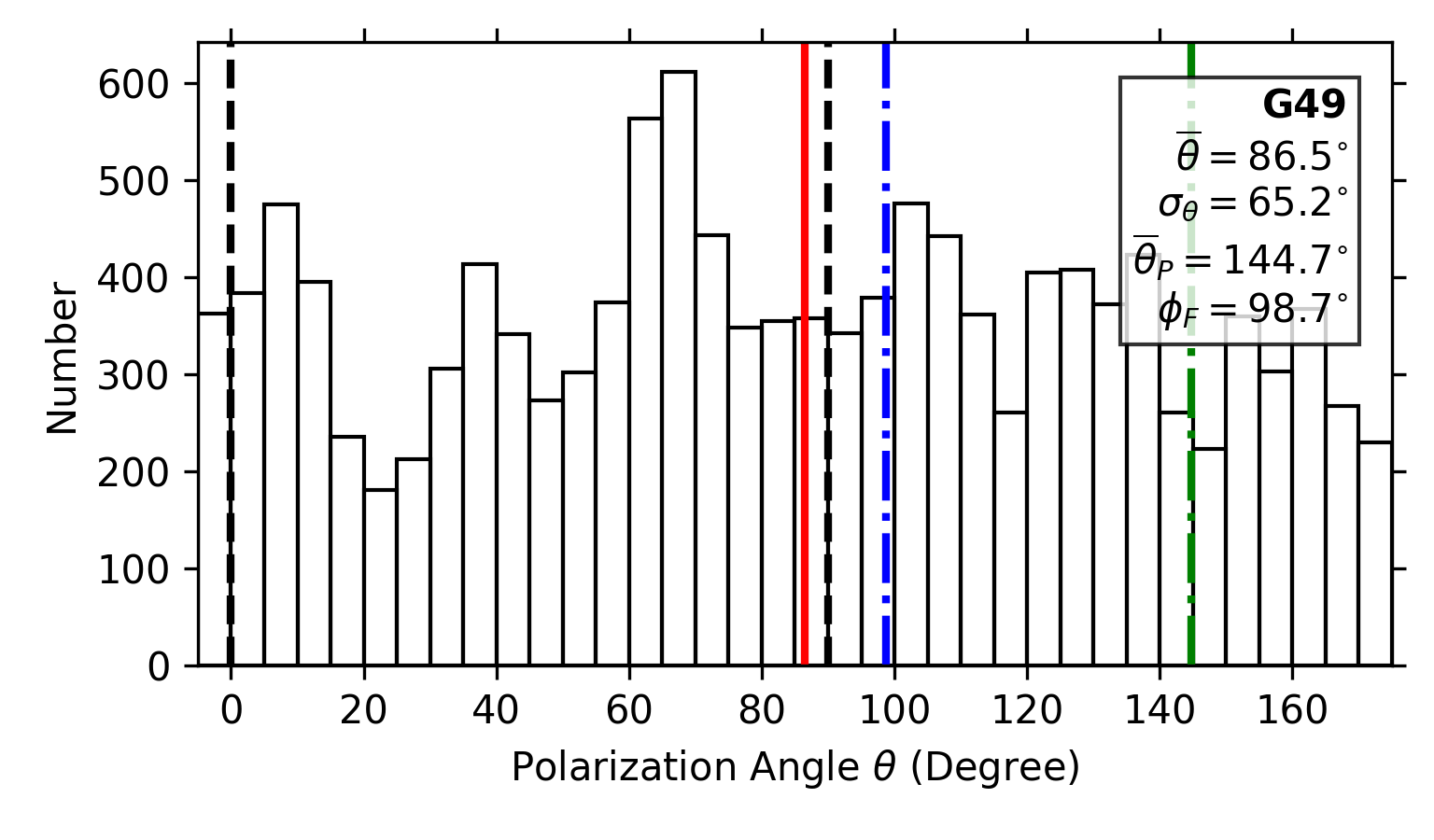}
    \caption{Histograms of polarization angles $\theta$. Continued from Figure~\ref{fig:Fil1-5_Histo}. }
    \label{fig:Fil10-G49_Histo}
\end{figure*}

We compiled the polarization angles $\theta$ from the vector catalogs of each bone as histograms in Figures~\ref{fig:Fil1-5_Histo} and \ref{fig:Fil10-G49_Histo}. The total number of elements in each histogram is given by the $N_{pol}$ parameter in Table~\ref{tab:detections}.Each histogram covers a $180^\circ$ range centered on the circular mean $\overline{\theta}$ of the distribution. This variable range takes into account the $180^\circ$ ambiguity for half-vectors, which means that the lower and upper limits of the x-axis for each histogram are equal. We use directional statistics to find an accurate mean value of the polarization angle distribution \citep[e.g.,][]{Doi2020}. Specifically, the circular mean and standard deviation for each bone were calculated using the library of statistical functions from the SciPy Python package with boundaries of $-90^\circ$ and $90^\circ$. Using the circular mean value calculated for each bone, we find an average polarization angle of $67.9^\circ \pm 30.9^\circ$ for our sample, which is equivalent to $-22.1^\circ \pm 30.9^\circ$ for the magnetic field orientation relative to Galactic North. 

We note that the histogram of a random distribution of angles from $-90^\circ$ to $90^\circ$ will tend toward a continuous uniform distribution, which has a standard deviation of $180^\circ /\sqrt{12} \approx 52.0^\circ$. The circular standard deviation in the limit of a circular uniform distribution is not as well defined, but we nonetheless caution against over-interpreting values~$\sigma_\theta$ greater than $50^\circ$ as a general rule. 

\begin{figure}
    \centering
    \includegraphics[width=0.495\textwidth]{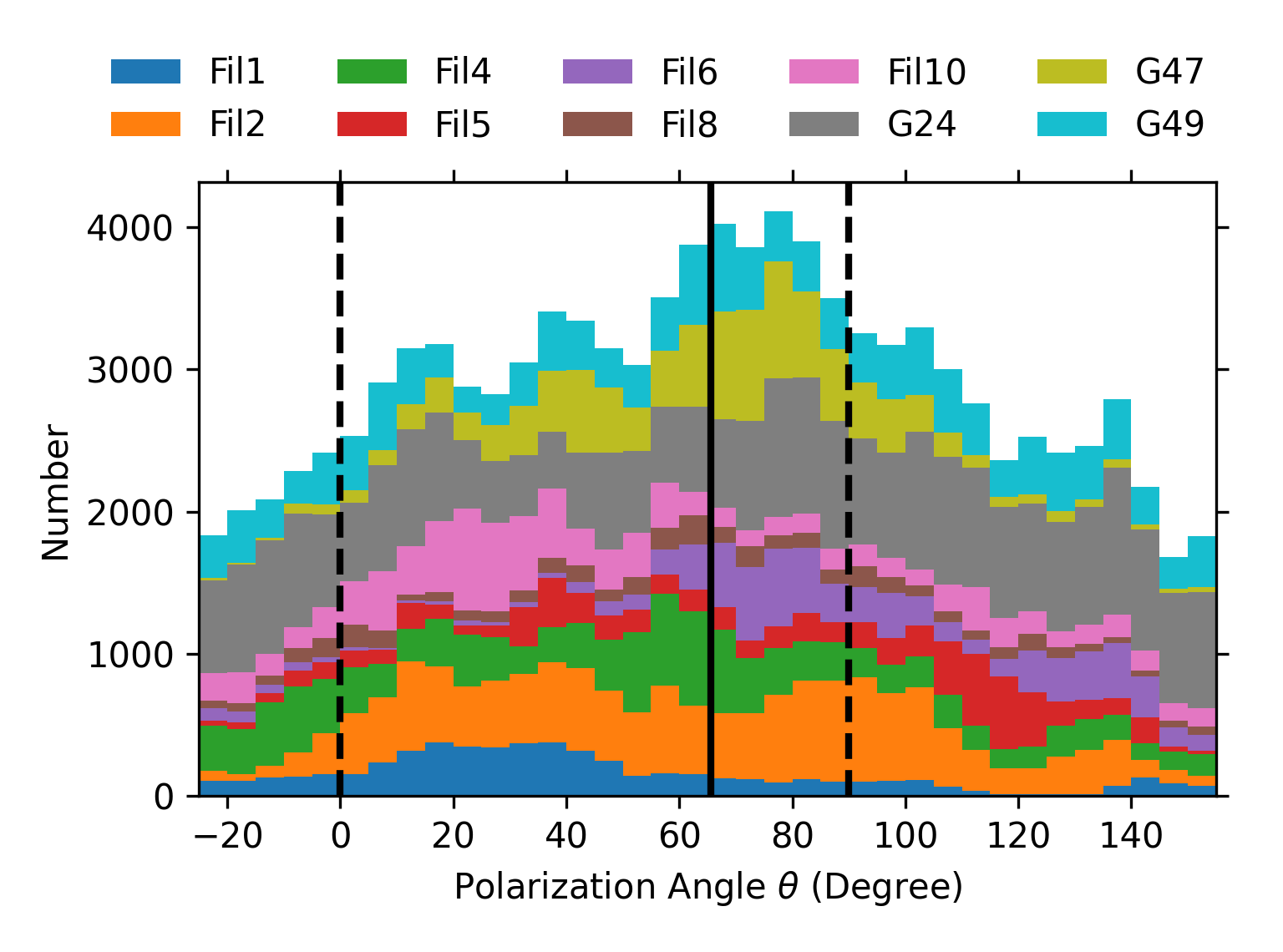}
    \caption{Stacked histogram of polarization angles~$\theta$ for all the bones covered in Figures~\ref{fig:Fil1-5_Histo} and \ref{fig:Fil10-G49_Histo}. The bin size of the histograms is $5^\circ$. Each color represents the contribution from a different bone. The histogram is centered on the bin containing the circular mean $\overline{\theta}$ of the stacked distribution, indicated by the plain black line at $65.6^{\circ}$. Black dashed lines indicate angles of 0$^\circ$ and 90$^\circ$ for clarity.}
    \label{fig:Stacked_Histo}
\end{figure}

Figure~\ref{fig:Stacked_Histo} presents the stacked histogram combining all the distributions shown in Figures~\ref{fig:Fil1-5_Histo} and~\ref{fig:Fil10-G49_Histo}, each identified with a different color. The circular mean and standard deviation of the resulting stacked distribution of polarization angles are $65.6^\circ \pm 56.6^\circ$, which are equivalent to $-24.4^\circ \pm 56.6^\circ$ for the corresponding mean magnetic field angle. Even with a greater weight from the bones with the highest number of detected polarization vectors, the resulting circular mean is within a few degrees from the average of the circular means of each individual bone. Despite the larger circular standard deviation of $56.6^\circ$, the stacked distribution is nevertheless clearly peaked around the mean. As a whole, the mean magnetic field measured with HAWC+ in the bones is therefore closer to perpendicular to the Galactic disk than parallel to it. 

We used \textit{Planck} polarization data to estimate the average Galactic field orientation in the plane of the sky toward the location of each bone. First, we re-gridded their respective HAWC+ Stokes~$I$ maps to share the same footprint and pixel scale as the \textit{Planck} maps described in Section~\ref{sub:planck}. We then used the re-gridded HAWC+ data to mask pixels in the \textit{Planck} maps where the HAWC+ Stokes~$I$ intensity was less than zero or undefined. The polarization angles from the remaining pixels in the \textit{Planck} maps were then used to calculate the circular mean $\overline{\theta}_P$ for each bone, which we identify in Figures~\ref{fig:Fil1-5_Histo} and \ref{fig:Fil10-G49_Histo} as the green dotted lines. Using the circular mean values of each bone as before, we find an average polarization angle of $-3.1^\circ \pm 14.6^\circ$ for our sample, which is equivalent to $86.9^\circ \pm 14.6^\circ$ for the magnetic field orientation relative to Galactic North. These results are in agreement with the expectation of a large-scale magnetic field mostly parallel to the Galactic plane. 

Figures~\ref{fig:Fil1-5_Histo} and \ref{fig:Fil10-G49_Histo} also illustrate the complexity of the plane-of-sky magnetic field structure observed with HAWC+ in each bone. Even Filaments~1, 10, and G47, which are the closest to exhibiting a single-peak distribution, have circular standard deviations of polarization angles greater than $30^\circ$. The large standard deviations are even more evident in the other bones of the sample, and especially for G24 and G49 which each have a circular standard deviation greater than $60^\circ$ and a nearly flat histogram. It would be tempting then to conclude that the magnetic fields in these objects must be weak and disorganized. A closer look at Figures~\ref{fig:Fil1_Maps} through \ref{fig:G49_Maps}, however, shows fields that appear locally well-ordered, but often with smoothly varying orientations between different regions along the length of the bone. Such a behavior has been predicted by Galactic-scale magnetohydrodynamic simulations of giant molecular cloud evolution \citep[][]{Zhao2024}. Additionally, combining multiple components of a magnetic field that is well-ordered locally into a single histogram can lead to distributions with extended wings, which is a well-documented phenomenon in optical polarimetry \citep[e.g.,][]{Goodman1990}.

\begin{table}[]
    \caption{Circular mean and standard deviation of polarization angles in the four brightest regions of G49. }
    \hspace*{-15pt}
    \begin{tabular}{ccccc}
        \hline
        Region & $l$  & $b$ & $\overline{\theta}$ & $\sigma_\theta$ \\
         & $^{\circ}$ & $^{\circ}$ & $^{\circ}$ & $^{\circ}$ \\
        \hline
        \hline
        G49a & 48.913 & -0.285 & $-46.5$ & $7.3$ \\
        G49b & 48.987 & -0.298 & $-89.2$ & $28.1$ \\
        G49c & 49.211 & -0.342 & $28.0$ & $21.9$ \\
        G49d & 49.266 & -0.339 & $40.1$ & $8.9$ \\
        \hline
    \end{tabular}

    \label{tab:G49abcd}
\end{table}

We further investigated the local structure of the magnetic field in G49 by studying the distribution of polarization angles around the four brightest peaks of the bone. We selected G49 because it is the bone with the largest measured circular standard deviation at $65.1^\circ$, as seen in Figure~\ref{fig:Fil10-G49_Histo}. Table~\ref{tab:G49abcd} provides the circular mean~$\overline{\theta}$ and standard deviation~$\sigma_\theta$ for each region, which were calculated from the polarization vectors contained within a $91\arcsec$ by $91\arcsec$ box (about five times the beam size of HAWC+) centered on the listed coordinates. These regions are identified with green squares in the top panel of Figure~\ref{fig:G49_Maps}. The corresponding histograms are shown in Figure~\ref{fig:G49abcd_Histo} of Appendix~\ref{apx:G49_regions}. There are two main takeaways from Table~\ref{tab:G49abcd}: the circular mean~$\overline{\theta}$ differs significantly for each region, and their respective standard deviation~$\sigma_\theta$ is much smaller than the global standard deviation for the bone. The wide distribution of polarization angles for G49 can therefore be explained by the combination of multiple well-ordered magnetic field components across the length of the filament.

\subsubsection{Magnetic Field and Filament Orientations}
\label{sub:fil_orient}

One of the distinctive features of bones is the orientation of their filamentary structure parallel to the Galactic disk \citep[][]{Zucker2015}. As noted in Section~\ref{sec:intro}, the \textit{Planck} team found that the magnetic field in the disk is also oriented mostly parallel to the Galactic plane \citep[][]{Planck2015_XIX}. In contrast, magnetic field lines are generally expected to become orthogonal to the length of dense filaments \citep[e.g.,][]{Tomisaka2014, Planck2016_XXXV, Pattle_Fissel2019}. The FIELDMAPS data provides us with an opportunity to investigate the orientation of magnetic fields in bones relative to their length, and therefore quantify is these fields are closer in average to the Galactic field orientation or to the expected structure in dense filaments.

The spine of the filaments were determined by \citet{Zucker2018} using the \textsc{FilFinder} package \citep{Koch2015}, which produces masks where the ``skeleton'' of a filament is traced by a pixel-wide line. We used a least squares linear regression to find the mean orientations~$\phi_F$ relative to Galactic North (e.g., $\phi_F = 90^\circ$ is parallel to the Galactic Plane). We show the result of these fits for Filament~10 in Figure~\ref{fig:Fil10_Skeleton} of Appendix~\ref{apx:ext_fil10}, and the corresponding figures for each other bone are available on the FIELDMAPS Dataverse. These orientations~$\phi_F$ are identified in Figures~\ref{fig:Fil1-5_Histo} and \ref{fig:Fil10-G49_Histo} as the blue dotted lines, as well as in Table~\ref{tab:BfieldFilOrient} using a range of $0^\circ$ to $180^\circ$ for clarity. We measure the spine position angle of each bone relative to the Galactic plane as $\delta\phi = \phi_F - 90^\circ$, and we find a circular mean and standard deviation of $\langle \delta\phi \rangle= 1.0^{\circ} \pm 10.9^{\circ}$ for the entire sample. While Filament~2, G24, and G47 are slightly more inclined than the rest, all ten bones fulfill the requirement listed in Section~\ref{sub:targets} to be parallel within $30^{\circ}$ from the Galactic plane. 

\begin{table}[]
\caption{Magnetic field and filament orientations for each bone relative to Galactic North.}
\hspace*{-40pt}
\begin{tabular}{lccccc}
    \hline
    Name & $\overline{\theta}_B$ & $\overline{\theta}_{PB}$ & $\phi_F$ & $\delta\overline{\theta}_{B}$ & $\delta\overline{\theta}_{PB}$ \\
     & $^\circ$ & $^\circ$ & $^\circ$  & $^\circ$  & $^\circ$ \\
    \hline
    \hline
    Filament 1 & 119.3 & 84.1 & 93 & -26.3 & 8.9 \\
    Filament 2 & 153.1 & 80.6 & 74 & -79.1 & -6.6 \\
    Filament 4 & 134.8 & 93.6 & 96 & -38.8 & 2.4 \\
    Filament 5 & 6.1 & 91.1 & 82 & 75.9 & -9.1 \\
    Filament 6 & 5.3 & 77.9 & 87 & 81.7 & 9.1 \\
    Filament 8 & 150.3 & 84.5 & 86 & -64.3 & 1.5 \\
    Filament 10 & 115.8 & 96.4 & 95 & -20.8 & -1.4 \\
    G24 & 31.1 & 88.9 & 108 & 75.9 & 18.1 \\
    G47 & 155.2 & 116.0 & 70 & -85.2 & -46.0.  \\
    G49 & 176.5 & 54.7 & 99 & -77.5 & 44.3 \\
    \hline
\end{tabular}
\label{tab:BfieldFilOrient}
\end{table}

\begin{figure}
    \centering
    \includegraphics[width=0.495\textwidth]{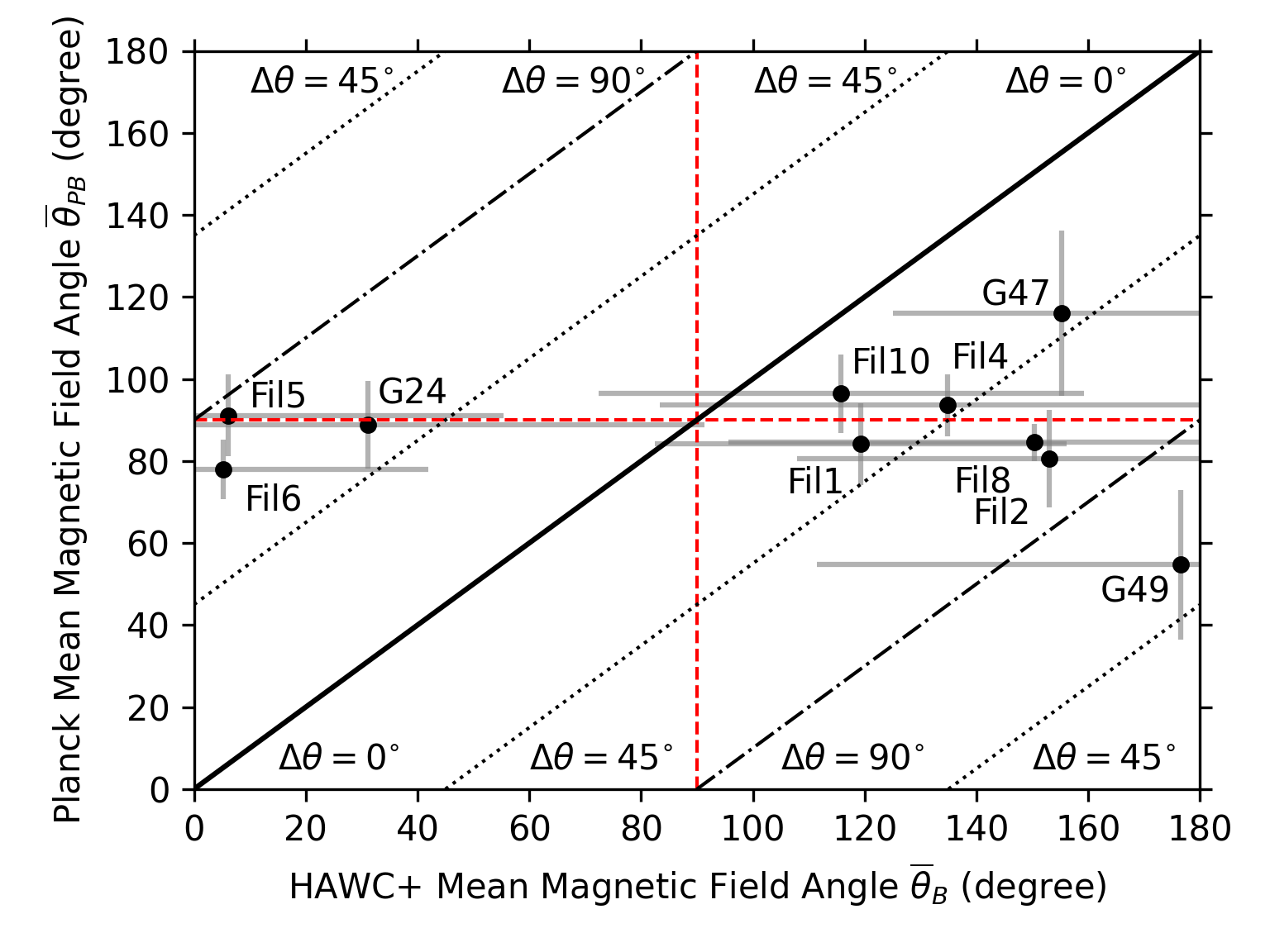}
    \caption{Comparison between the HAWC+ and \textit{Planck} circular means for the inferred magnetic field angle in each filament. The gray error bars trace the circular standard deviation for both HAWC+ and \textit{Planck} data. Each bone is identified with a label. The dashed red lines represent the Galactic disk orientation at $90^{\circ}$. The full, dotted, and dash-dotted lines show the absolute angle differences $\Delta\theta$ of $0^{\circ}$, $45^{\circ}$, and $90^{\circ}$, respectively. The \textit{Planck} mean magnetic field angles for each bone are closer to being parallel to the Galactic disk than those obtained from HAWC+.}
    \label{fig:HP_Means_Comp}
\end{figure}

We then measured the angle differences $\delta\overline{\theta}_B = \overline{\theta}_B - \phi_F$ and $\delta\overline{\theta}_{PB} = \overline{\theta}_{PB} - \phi_F$ of the average magnetic field directions $\overline{\theta}_B$ from HAWC+ and $\overline{\theta}_{PB}$ from \textit{Planck} observations, respectively, relative to the orientation $\phi_F$ of each bone. The values for $\overline{\theta}_B$ and $\overline{\theta}_{PB}$, given in Table~\ref{tab:BfieldFilOrient} using the same $0^\circ$ to $180^\circ$ range as in Figure~\ref{fig:HP_Means_Comp}, were obtained by adding $90^\circ$ to the circular means $\overline{\theta}$ and $\overline{\theta}_P$ calculated previously. The differences $\delta\overline{\theta}_B$ and $\delta\overline{\theta}_{PB}$ are also given in Table~\ref{tab:BfieldFilOrient} using a range of $-90^\circ$ to $90^\circ$. For the HAWC+ data, we find a mean difference $\langle \delta\overline{\theta}_{B} \rangle$ of $-73.7^{\circ} \pm 31.6^{\circ}$ between the filament axis and the plane-of-sky magnetic field. When using the \textit{Planck} data, we instead find a difference $\langle \delta\overline{\theta}_{PB} \rangle$ of $2.8^{\circ} \pm 21.0^{\circ}$. 

On average, the \textit{Planck} data trace a magnetic field parallel to the length of the bones, which themselves are nearly parallel to the Galactic plane. This would be in agreement with a scenario where the magnetic field at large scales closely follows the spiral arms of the Galaxy \citep[e.g.,][]{Borlaff2021_SALSA1}. However, we must note that the magnetic field seen by \textit{Planck} may not be fully representative of the Galactic field at the location of each bone, as the data may be skewed by the integration of multiple components along the line-of-sight \citep[][]{Planck2015_XXI} and by instrumental effects \citep[][]{Planck2015_XIX}. In contrast, the HAWC+ data, which probes smaller scales than \textit{Planck} and are closer, in average, to being perpendicular to the length of bones, supports the expectation that the link between Galactic and local magnetic fields breaks down in dense star-forming regions \citep[e.g.,][]{Stephens2011}. Finally, while these statistics are helpful to glean general trends from the bones as a single population of objects, they may obscure the existence of complex field structures due to feedback from the star formation occurring within them (see Section~\ref{sub:yso}). 

To investigate these potential trends more closely, Figure~\ref{fig:HP_Means_Comp} shows the comparison between the HAWC+ and the \textit{Planck} mean magnetic field directions $\overline{\theta}_{B}$ and $\overline{\theta}_{PB}$ for individual bones. The error bars trace their respective circular standard deviations. The large error bars for the HAWC+ data illustrate the wide spread of polarization angles measured in each filament, as shown in Figures~\ref{fig:Fil1-5_Histo} and \ref{fig:Fil10-G49_Histo}. The two dashed lines indicate the $90^\circ$ orientation of the Milky Way disk relative to Galactic North for both datasets. The diagonals show the difference $\Delta \theta = \vert \overline{\theta}_{B} - \overline{\theta}_{PB} \vert$ for values of $0^\circ$ (full line), $45^\circ$ (dotted line), and $90^\circ$ (dash-dotted line). 

An important trend from Figure~\ref{fig:HP_Means_Comp} is that the \textit{Planck} magnetic field orientations follow the Galactic disk for most bones, while the HAWC+ data covers a wider range of values. Of note, the two bones farthest from the Galactic Center, G47 and G49 (see Table~\ref{tab:filaments}), are also the ones where the \textit{Planck}-derived large-scale magnetic field deviates the most from being parallel to the disk. As shown in Figure~\ref{fig:G47_Maps} and \ref{fig:G49_Maps}, there is a clear shift in the \textit{Planck}-derived field lines toward the spine of these two bones from a field that is mostly parallel to the Galactic plane elsewhere, including for every other bone in the sample. This could be explained by the relative position on the sky of G47 and G49 approximately $50^\circ$ from the Galactic Center, which means that there is likely to be less dust to integrate along the line-of-sight in the \textit{Planck} data compared to the locations of other bones. Indeed, shifts similar to those seen in G47 and G49 in the large-scale field orientation measured by \textit{Planck} have been observed toward dense filaments in less-crowded lines-of-sight \citep[e.g.,][]{Planck2016_XXXV, Doi2020}.

\subsubsection{Relation to Magnetic Field Strengths}
\label{sub:dcf}

The classical approach to measuring the plane-of-sky amplitude $B_{pos}$ of the magnetic field with polarimetry is through the Davis-Chandrasekhar-Fermi (DCF) method \citep[][]{Davis1951,CF1953}. A common form of the DCF equation is given by \citet[][]{Crutcher2004}:

\begin{equation}
    B_{pos} = Q \sqrt{4 \pi \rho} \, \frac{\sigma_V}{\sigma_\theta} \, ,
    \label{eq:DCF}
\end{equation}

\noindent where $\rho$ is the gas density, $\sigma_V$ is the non-thermal velocity dispersion of the gas, $\sigma_\theta$ is the dispersion of polarization angles, and $Q$ is a theoretical correction factor often assumed to be $\sim 0.5$ \citep[][]{Ostriker2001}.

In the context of this paper, the main takeaway from Equation~\ref{eq:DCF} is that the plane-of-sky amplitude $B_{pos}$ is inversely proportional to the angle dispersion $\sigma_\theta$, and so larger angle dispersion will invariably result in weaker field strengths for a given gas velocity dispersion and density. However, the DCF method assumes a single ordered field direction for the entire region considered, which does not appear to be the case across the length of any of the bones considered for this survey. Additionally, simulations have shown that the DCF equation may be unreliable when the dispersion $\sigma_\theta$ is larger than $25^\circ$, as such values would likely be due to super-Alfv\'enic turbulence \citep[][]{Ostriker2001, Stephens2022}. The circular standard deviation values listed in Figures~\ref{fig:Fil1-5_Histo} and \ref{fig:Fil10-G49_Histo} are all above $30^\circ$, which is well beyond that theoretical limit even when accounting for measurement uncertainties. It is therefore necessary to investigate the field strengths in more localized regions of the bones in order to obtain narrower dispersions of angles, as illustrated in Table~\ref{tab:G49abcd}. 

In the specific case of G47, \citet[][]{Stephens2022} measured the deviation $\sigma_\theta$ at different locations along its spine with the use of a rectangular $74\farcs0 \times 55\farcs5$ sliding box. With this approach, they found magnetic field strengths that are generally between 20~$\mu$G to 100~$\mu$G, with a maximum reaching nearly 200~$\mu$G. Most notably, however, is that the resulting mass-to-flux ratio $\lambda$ \citep[see][]{Crutcher2004} for G47 is between 0.2 and 1.7, which means that most regions along the bone are either subcritical (i.e., they are supported against gravitational collapse by the magnetic field when $\lambda < 1$) or critical ($\lambda \approx 1$). The few regions that are supercritical ($\lambda>1$) appear to be sites of active star formation. 

Similarly, \citet[][]{Pillai2015ApJ} used another variation of the DCF technique, based on the structure function (or angular dispersion function) defined by \citet[][]{Houde2009}, to find a magnetic field strength greater than 200~$\mu$G in Filament~6. The resulting mass-to-flux ratio $\lambda$ has an upper limit of 1.1 and so, like G47, the magnetic field in Filament~6 is contributing significantly to supporting the cloud against gravity. 

In each case, far from being weak and disordered, the magnetic field in Filament~6 and G47 is instead dynamically important to support these bones against gravitational collapse. The large angle dispersion values shown in Figures~\ref{fig:Fil1-5_Histo} and \ref{fig:Fil10-G49_Histo} should therefore be seen as an indirect probe of how much the magnetic field's orientation varies along the length of a bone, and not directly as a potential indicator of its strength. With this data release, the magnetic support analysis from \citet[][]{Stephens2022} has been extended to all ten bones observed as part of the FIELDMAPS survey to investigate if the trends seen in Filament~6 and G47 are common for Milky Way bones \citep[][]{Stephens2025}.

\subsection{Polarization Efficiency} 
\label{sub:PvI}

The fractional polarization~$P$ from far-infrared observations can probe, to some extent, the alignment efficiency of a population of interstellar dust grains in a given environment \citep[][]{Andersson2015}. In fact, this alignment efficiency is generally studied by measuring the polarization efficiency $\epsilon_p = P/\tau$, where $\tau$ is the optical depth, and assuming $P \propto \epsilon_p$ for optically thin dust emission in the far-infrared \citep[][]{Jones2015}.

According to the RAT theory of grain alignment in particular, the polarization efficiency~$\epsilon_p$ will depend on the grains' size distribution and the radiation field in their surroundings, among other parameters. In a nutshell, the amount of starlight that reaches the dust grains will directly impact how well they will align with magnetic field lines, and larger dust grains are also predicted to have a better alignment. For this reason, it may be more helpful to plot the polarization fraction~$P$ against the visual extinction $A_V$, as was done historically with optical and near-infrared observations. We note that there is a theoretical upper limit for large grains where the internal alignment through the Barnett effect becomes inefficient, but we do not expect it to be significant in the case of the typical Mathis-Rumpl-Nordsieck (MRN) distributions of sizes in molecular clouds \citep[][]{MRN1977,Hoang_Lazarian2008}. 

However, measuring $A_V$ from far-infrared and submillimeter observations of molecular clouds requires an accurate determination of the emissivity spectral index $\beta$ and of the reddening factor $R_V$ \citep[e.g.,][]{Jones2015,Coude2019}. Since we assumed a fixed~$\beta$ of $1.75$ (see Section~\ref{sub:herschel}), new SEDs with~$\beta$ as a free parameter will need to be fitted for each bone in order to obtain far-infrared opacity maps that could be converted into extinction maps. Furthermore, this future work would benefit from the addition of longer-wavelength data, such as observations at 850~$\mu$m from the JCMT, to improve the determination of the spectral index~$\beta$ \citep[e.g.,][]{Sadavoy2013}. 

\begin{figure*}[ht]
    \centering
    \includegraphics[width=0.495\textwidth]{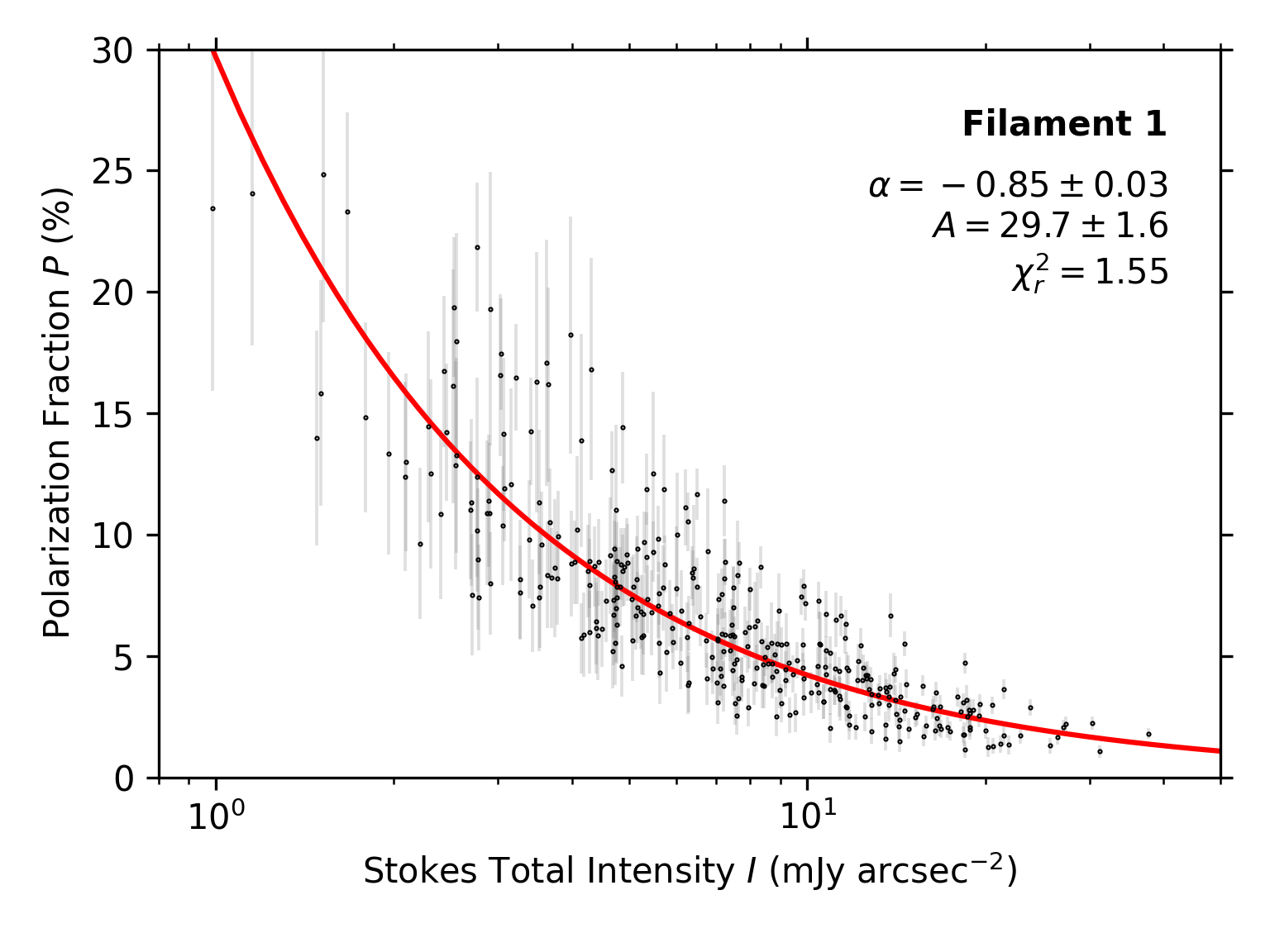}
    \includegraphics[width=0.495\textwidth]{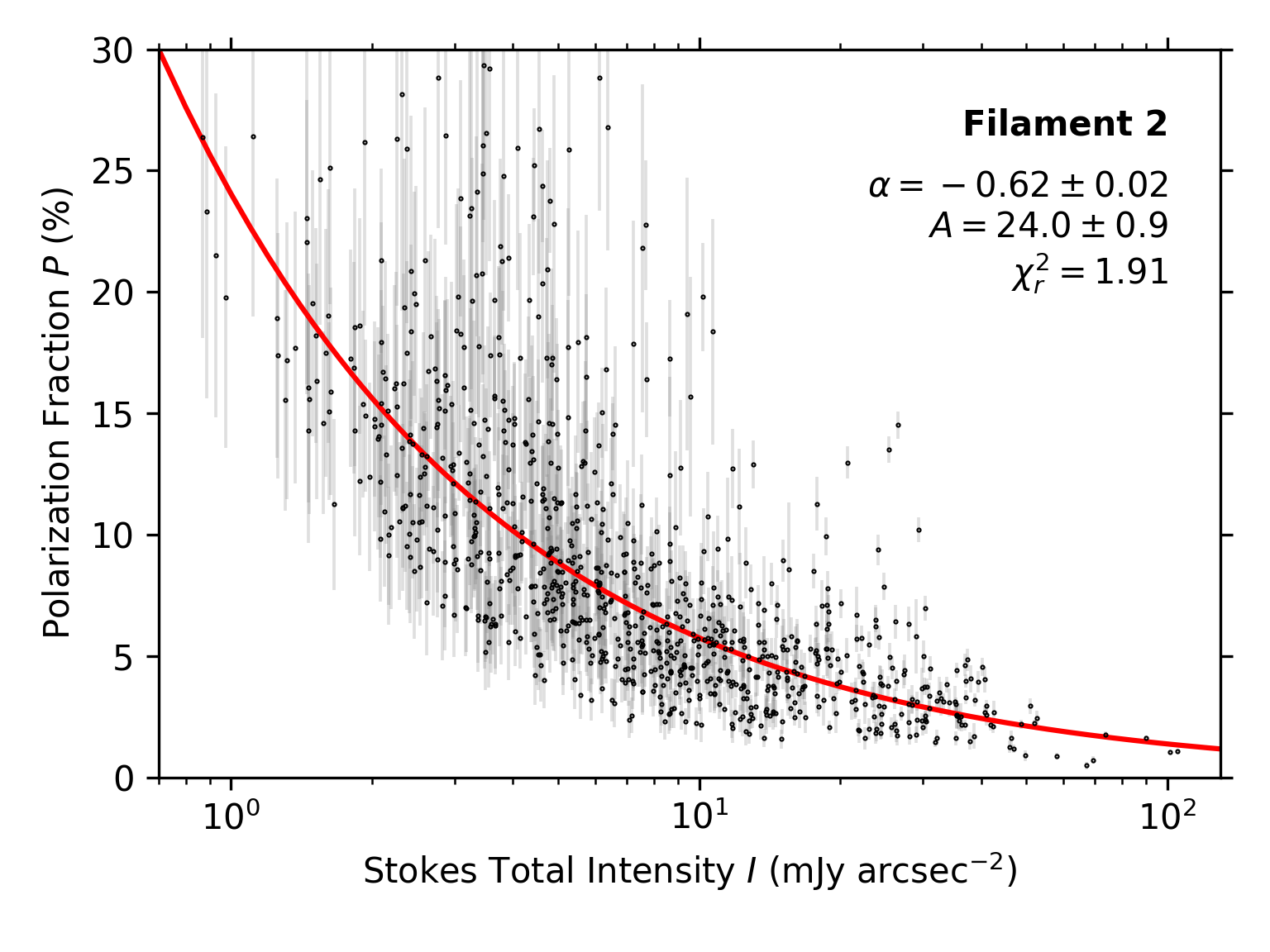}
    \includegraphics[width=0.495\textwidth]{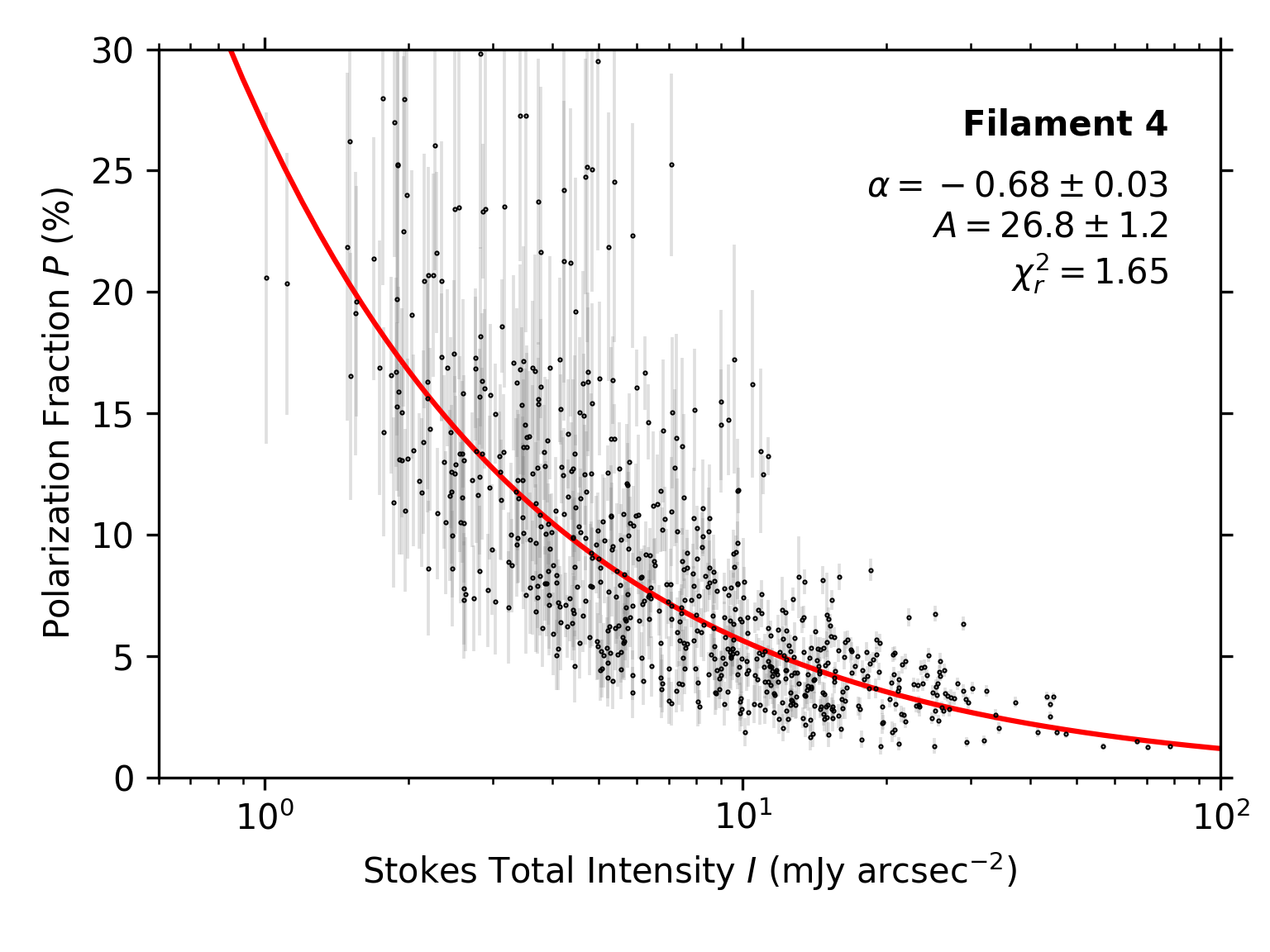}
    \includegraphics[width=0.495\textwidth]{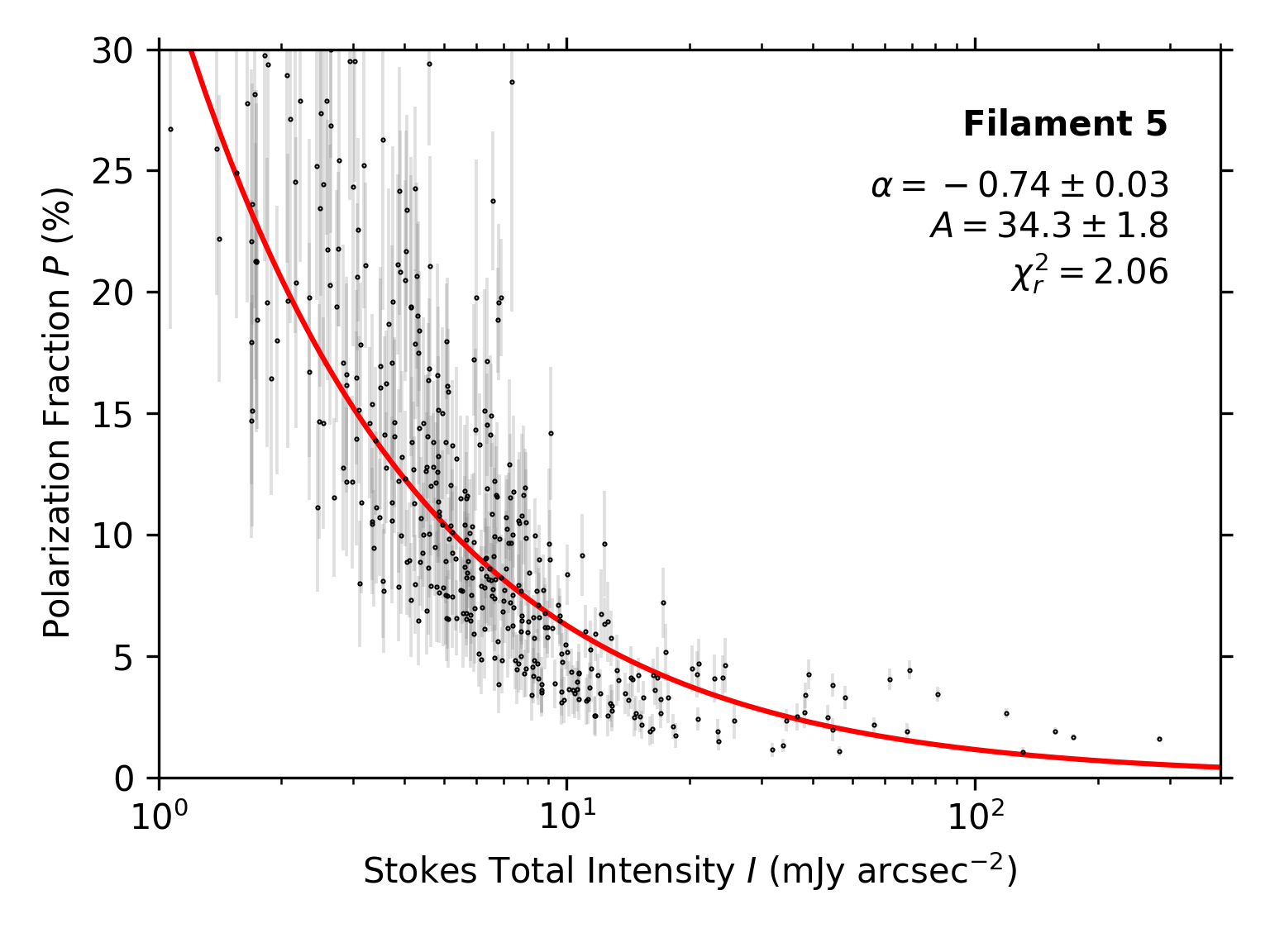}
    \includegraphics[width=0.495\textwidth]{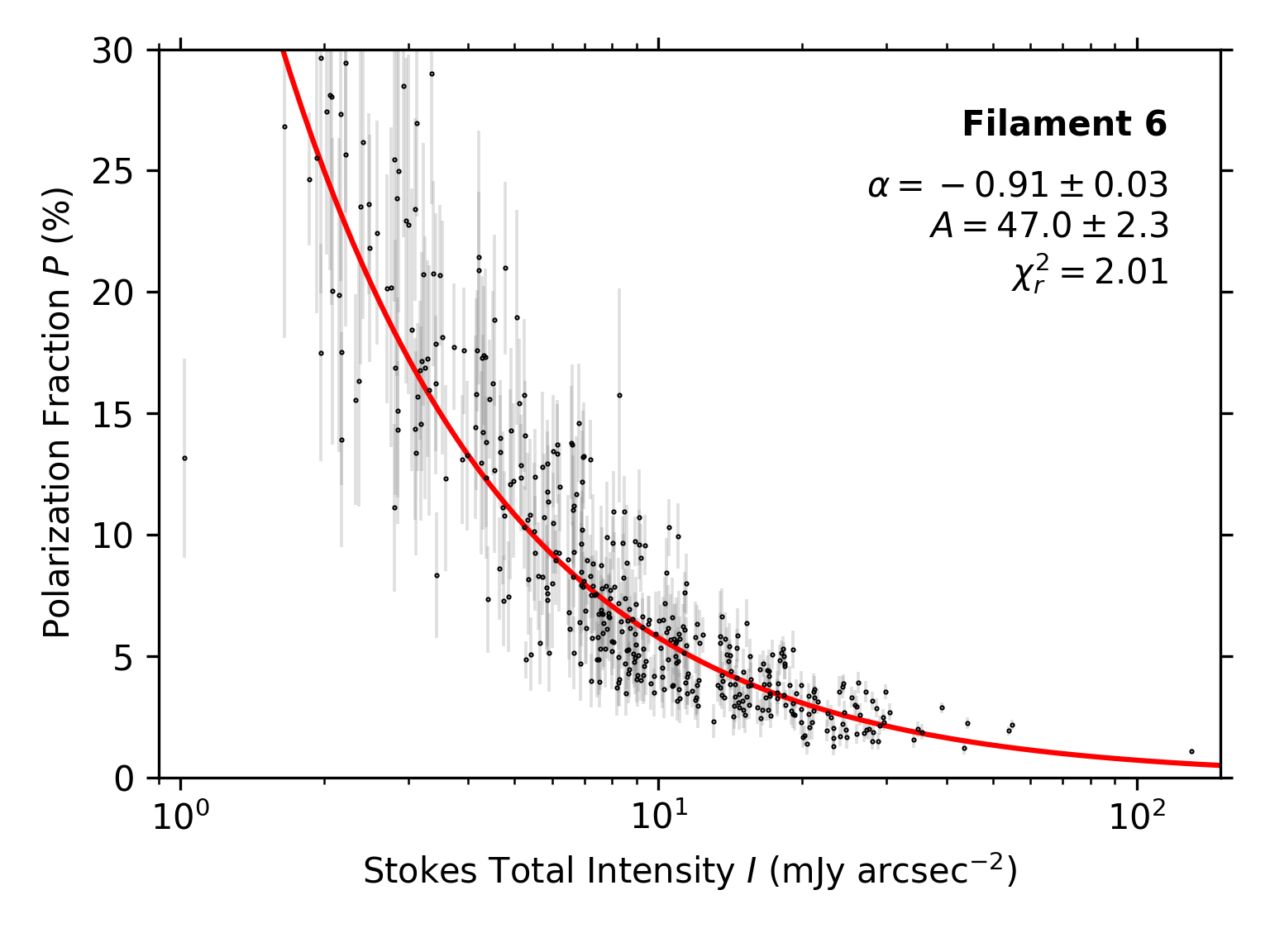}
    \includegraphics[width=0.495\textwidth]{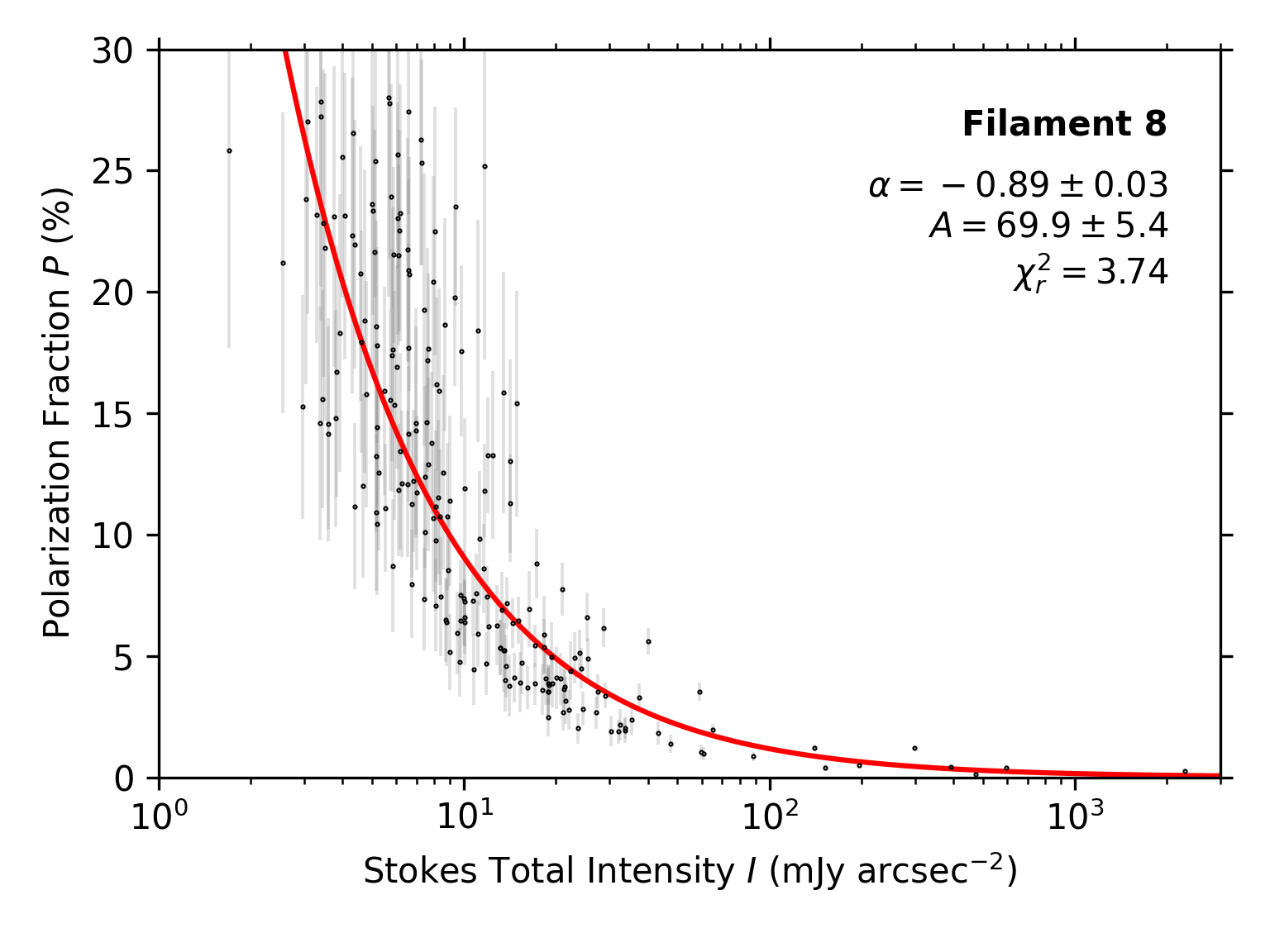}
    \caption{Debiased polarization fraction $P$ as a function of the Stokes~$I$ total intensity for the vectors plotted in Figures~\ref{fig:Fil1_Maps} through \ref{fig:Fil8_Maps}. The intensity $I$ is shown on a logarithmic scale to better visualize the spread of $P$ for lower values of $I$. The best fit to the power law given in Equation~\ref{eq:PvI} is shown as a red solid line for each bone, and the associated parameters $A$ and $\alpha$ are provided directly on the plot. Only every fourth vector is plotted and used for the fit, and the uncertainty in $P$ is shown as gray lines. Continued in Figure~\ref{fig:Fil10-G49_PvI}.}
    \label{fig:Fil1-5_PvI}
\end{figure*}

\begin{figure*}[ht]
    \centering
    \includegraphics[width=0.495\textwidth]{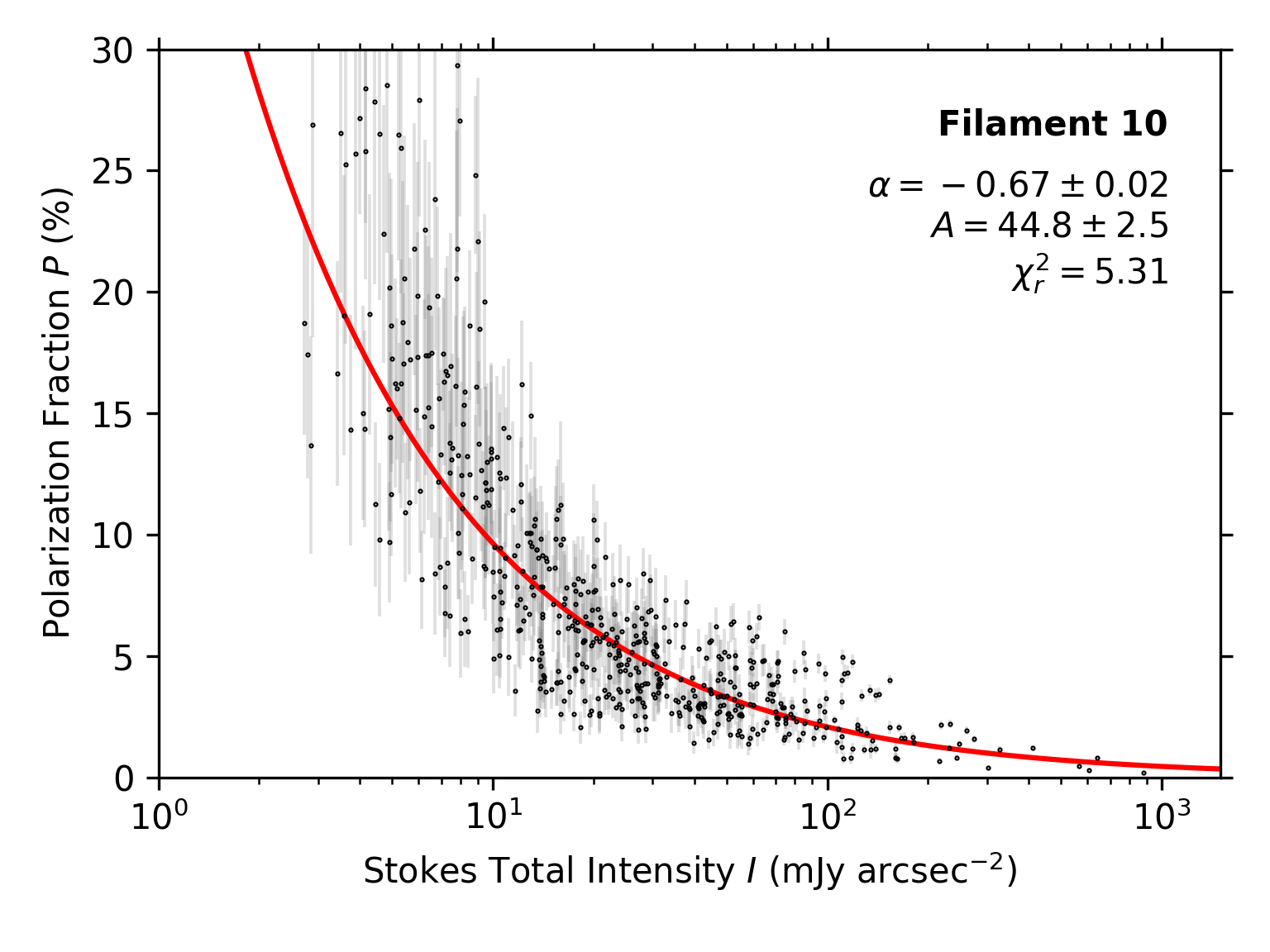}
    \includegraphics[width=0.495\textwidth]{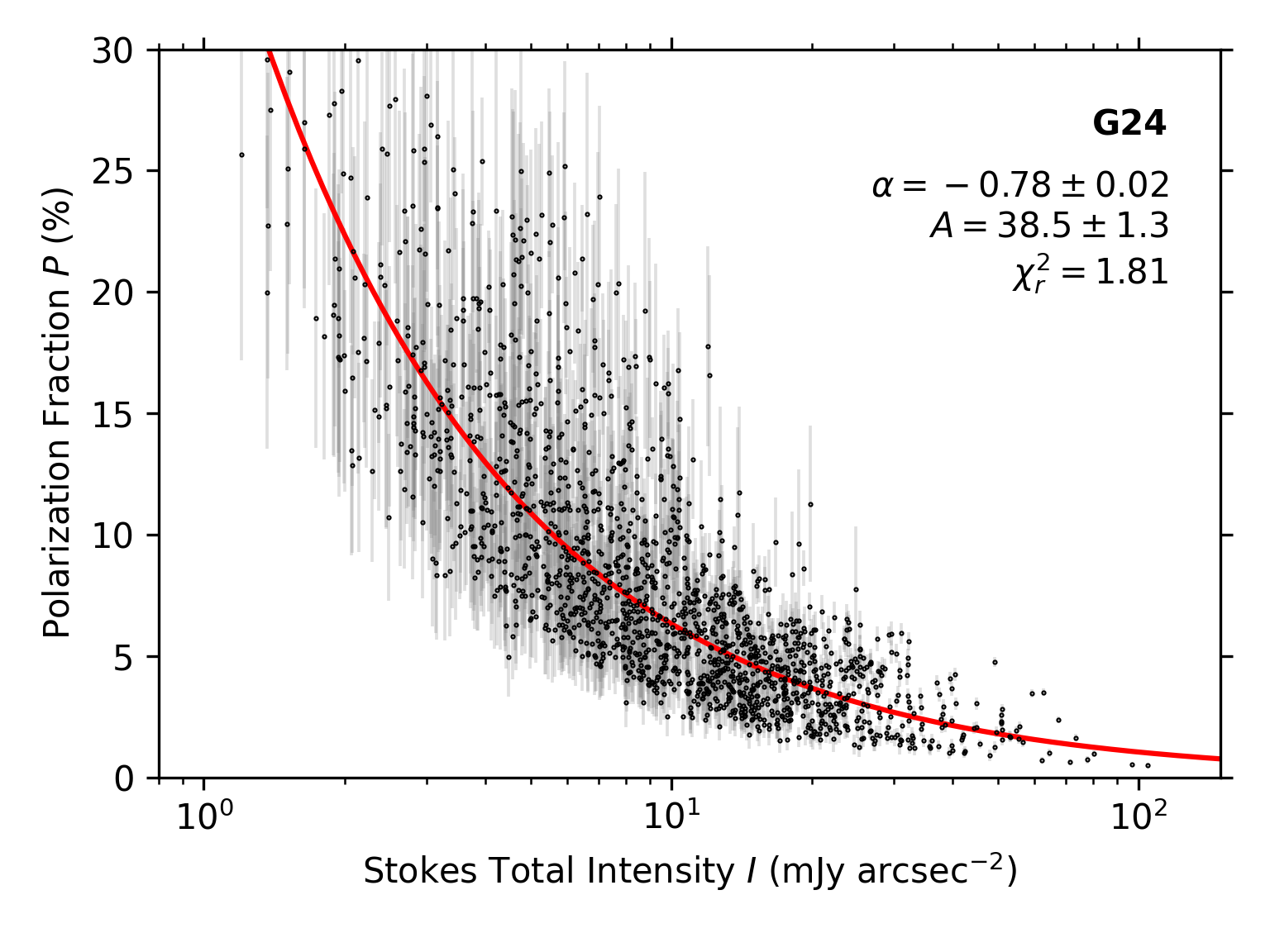}
    \includegraphics[width=0.495\textwidth]{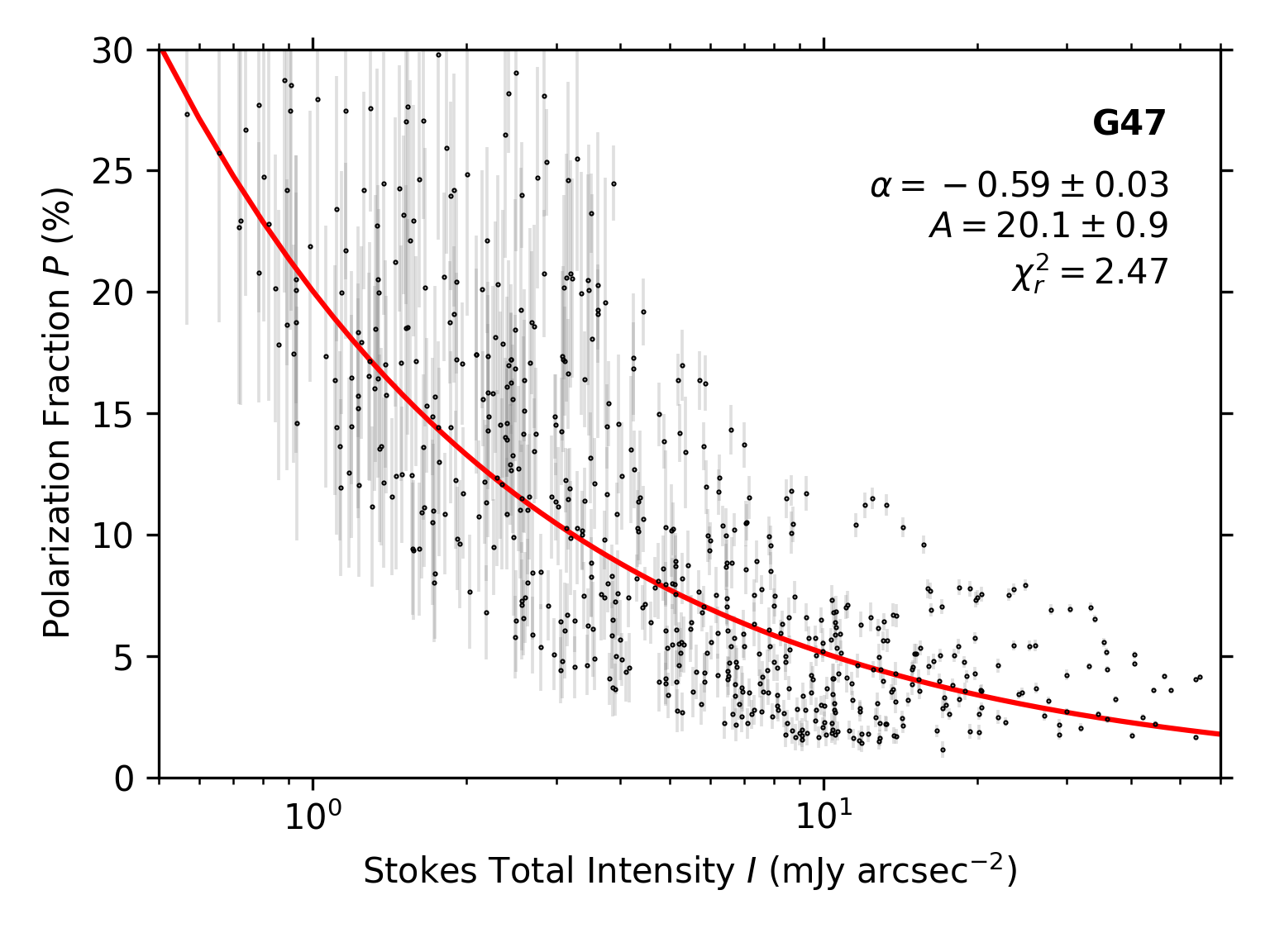}
    \includegraphics[width=0.495\textwidth]{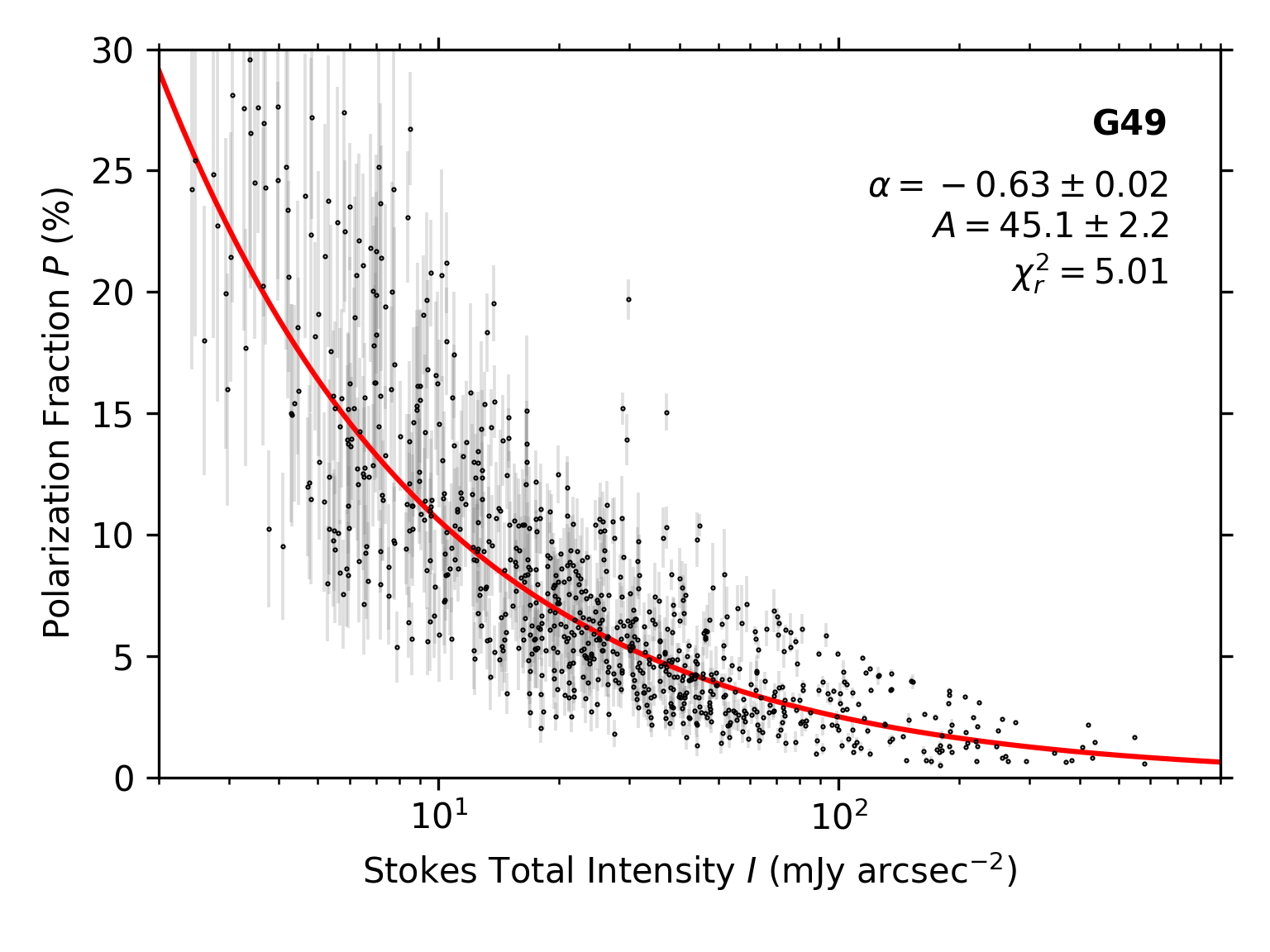}
    \caption{Debiased polarization fraction $P$ as a function of the Stokes~$I$ total intensity for the vectors plotted in Figure~\ref{fig:Fil10_Maps} through \ref{fig:G49_Maps}.The intensity $I$ is shown on a logarithmic scale to better visualize the spread of $P$ for lower values of $I$. The best fit to the power law given in Equation~\ref{eq:PvI} is shown as a red plain line for each bone, and the associated parameters $A$ and $\alpha$ are provided directly on the plot. Only every fourth vector is plotted and used for the fit, and the uncertainty in $P$ is shown as gray lines.  Continued from Figure~\ref{fig:Fil1-5_PvI}.}
    \label{fig:Fil10-G49_PvI}
\end{figure*}

Due to the challenges of calculating the visual extinction~$A_V$ from data at longer wavelengths, the Stokes~$I$ total intensity is sometimes used as an imperfect proxy for the optical depth. The relation between the debiased polarization fraction $P$ defined in Equation~\ref{eq:pol} and the Stokes~$I$ total intensity is plotted in Figures~\ref{fig:Fil1-5_PvI} and \ref{fig:Fil10-G49_PvI} for the vector catalogs plotted in Figures~\ref{fig:Fil1_Maps} through \ref{fig:G49_Maps}. For each bone, we used least squares to fit the logarithmic form of the following power law:
\begin{equation}
    P = A \, I^{\alpha} \, ,
    \label{eq:PvI}
\end{equation}
\noindent where $A$ is a proportionality factor and $\alpha$ is the power-law index. We find an average power-law index $\overline{\alpha}$ of $-0.7 \pm 0.1$ for the entire sample when sampling every fourth vector (i.e., the width of the beam's FWHM), with the full range spanning from $-0.59$ to $-0.91$. The results are mostly unchanged when sampling every vector in each data set instead, and the related figures are also available on Dataverse. These indices are comparable to those found in nearby star-forming regions (approximately $-0.8$) using 850~$\mu$m polarization data from the James Clerk Maxwell Telescope (JCMT) \citep[e.g., from the BISTRO survey;][]{Soam2018, Kwon2018, Coude2019}. 

The relation between the polarization fraction~$P$ and Stokes~$I$ in Figures~\ref{fig:Fil1-5_PvI} and \ref{fig:Fil10-G49_PvI} reproduces the well-known phenomenon of depolarization, or polarization ``hole'', seen in high density environments both in observations \citep[e.g.,][]{Dotson1996,Matthews2002,Girart2006,Liu2013,Alves2014,Hull2014} and theory \citep[e.g.,][]{Padoan2001,Bethell2007,Pelkonen2009}. A steeper power-law index $\alpha$ is expected for inefficient grain alignment in a cloud. With values below $-0.5$ for $\alpha$, it is possible that dust grains are poorly aligned within the densest regions of bones. However, \citet[][]{Pattle_2019} found that the Ricean behavior of the noise $\delta_{I_p}$ for the polarization intensity $I_p$ could artificially lead to a steeper power-law index $\alpha$. 

We calculated the reduced chi-square $\chi^2_r$ for each fit, which yielded values between $1.5$ and $5.3$. A value $\chi^2_r > 1$ could be understood as an imperfect fit to the data, but the values we obtain are not surprising considering the large spread of data points in Figures~\ref{fig:Fil1-5_PvI} and \ref{fig:Fil10-G49_PvI}. We also provide the fitting uncertainties rounded up to the second decimal for the power-law index in each figure, but we caution against over-interpreting them due to the data spread. As a comparison with the Barnard~1 clump, a dark cloud of the Perseus molecular cloud complex, \citet[][]{Coude2019} found a power index $\alpha = -0.85$ and $\chi^2_r = 3.4$, which they argued indicates that the data cannot be modeled accurately by a single power law. In this situation, a region-by-region analysis may be preferable in order to offset the changing physical conditions across the cloud. 

% Removed the following sentence due to changing how the data is sampled:
%Additionally, some measurements within a beam size from each other are likely to be correlated, which may lead to an overestimation of the degrees of freedom, and thus an underestimation of $\chi^2_r$. We also provide the fitting uncertainties rounded up to the second decimal for the power-law index in each figure, but we caution against over-interpreting them due to the data spread.

Indeed, one of the reasons for the spread of data points in the relation between the polarization fraction~$P$ and the Stokes~$I$ total intensity, and the difficulty to fit them accurately with a single power law, is that we are combining regions with different physical conditions within each bone.  Since all the bones in our sample show signs of past or current star formation, it is likely that stellar feedback has affected dust properties across different regions (e.g., see Figure~\ref{fig:Fil10_PvTmap} in Appendix~\ref{apx:ext_fil10} for variations of dust temperatures~$T_d$ in Filament~10), which in turn would have an impact on the measured polarization~$P$. Additionally, the dust emission in some regions could be optically thick at 214~$\mu$m, but not at 850~$\mu$m, which would complicate the comparison of the polarization fraction~$P$ between the two wavelengths. 

In the far-infrared, the optical depth~$\tau_\nu$ at a frequency~$\nu$ should be proportional to the hydrogen column density~$N_{H_2}$ when $\beta$ is fixed \citep[see Equation~A2 from][]{Jones2015}. We therefore investigated the relation between the polarization fraction~$P$ and the column density~$N_{H_2}$ for each bone, using the column density maps described in Section~\ref{sub:herschel}. We provide an example of this relation for Filament~10 in the left panel of Figure~\ref{fig:Fil10_PvN_PvT}, using HAWC+ Stokes~$I$, $Q$, and $U$ data smoothed and re-projected to the same resolution and pixel scale as the \textit{Herschel}-derived maps. When fitting a power-law of the form $P \propto N_{H_2}^\gamma$, we find indices $\gamma$ that are generally steeper than the indices $\alpha$ from Figures~\ref{fig:Fil1-5_PvI} and \ref{fig:Fil10-G49_PvI}, even going as low as $\gamma = -1.40$. This result does not change if we instead use the same non-smoothed vector catalogs as Figures~\ref{fig:Fil1-5_PvI} and \ref{fig:Fil10-G49_PvI}. However, as exemplified in the left panel of Figure~\ref{fig:Fil10_PvN_PvT} in Appendix~\ref{apx:ext_fil10}, there is typically a larger spread of data points in each case compared to the relation between the polarization fraction~$P$ and Stokes~$I$. These results suggest that the column density~$N_{H_2}$ measurements used in this work are not a better proxy for the visual extinction~$A_V$ than Stokes~$I$.

Additionally, the polarization fraction~$P$ measured with HAWC+ could be overestimated in lower density regions due to the filtering of large-scale dust emission in the data reduction pipeline. Figures~\ref{fig:Fil1-5_IvIest} and \ref{fig:Fil10-G49_IvIest} in Appendix~\ref{apx:herschel_comp} show that the measured HAWC+ spectral flux densities systematically underestimate those predicted with \textit{Herschel} data at low intensities. Since the atmosphere is unpolarized, the effect of the background subtraction on the recovered low-level flux could be more pronounced for Stokes~$I$ than for $Q$ and $U$ due to how these parameters are individually treated by the HAWC+ pipeline. Following Equation~\ref{eq:pol}, the resulting polarization fraction~$P$ would then be overestimated.

To explore this hypothetical effect on the power law given in Equation~\ref{eq:PvI}, we considered the unrealistic scenario where Stokes~$Q$ and $U$ are fully recovered for large spatial scales by the HAWC+ pipeline, but not Stokes~$I$. We correct for such a discrepancy by replacing the HAWC+ Stokes~$I$ measurements with the \textit{Herschel}-predicted 214~$\mu$m flux densities from Appendix~\ref{apx:herschel_comp}. While the resulting polarization fractions are systematically lower than those reported in Figures~\ref{fig:Fil1-5_PvI} and \ref{fig:Fil10-G49_PvI}, the effect on the spectral index~$\alpha$ is not as conclusive. Only four bones, Filaments~2, 5, 10, and G49, show new indices below $-0.5$, as illustrated in the right panel of Figure~\ref{fig:Fil10_PvI_PvIest} for Filament~10. If differences exist between the large-scale filtering of Stokes $I$, $Q$, and $U$ in the HAWC+ pipeline, they may be insufficient to account for the trends shown in Figures~\ref{fig:Fil1-5_PvI} and \ref{fig:Fil10-G49_PvI}, at least not for all the bones in our sample. In any case, these results highlight the importance of developing instruments and observing strategies capable of reliably recovering fainter, more extended polarized dust emission in order to accurately measure the polarization fraction over a larger range of densities within molecular clouds. Space missions in particular, such as the proposed ``PRobe far-Infrared Mission for Astrophysics'' \citep[PRIMA; e.g.,][]{Moullet2023,Dowell2024}, do not need to remove the contribution of Earth's atmosphere to the data and therefore can more reliably recover large-scale polarized flux.

An additional mechanism to decrease the polarization fraction~$P$ is the ``disruption'' (or destruction) of dust grains by radiative torques \citep[RATD;][]{Hoang2019}. Indeed, in the RAT paradigm of grain alignment, large dust grains could theoretically be torn apart by centrifugal forces if the ambient radiation field is strong enough. Assuming that dust grains are heated by the incoming starlight, the relation between the polarization fraction~$P$ and the dust temperature~$T_d$ can therefore be used to probe this effect. The alignment efficiency of the grains should increase with increasing dust temperature~$T_d$ until it reaches a threshold temperature, at which point the polarization fraction~$P$ will drop noticeably. Using this relation, \citet[][]{Tram2021a,Tram2021} found evidence for grain disruption near the high-mass star Oph~S1 of the star-forming region $\rho$~Ophiuchi~A and in the starburst region of the 30~Doradus complex in the Large Magellanic Cloud. We also investigated the relation between the polarization fraction~$P$ and the dust temperature~$T_d$ for the FIELDMAPS bones, and we provide an example for Filament~10 in the right panel of Figure~\ref{fig:Fil10_PvN_PvT} using the same smoothed vector catalog as described previously for the gas column density~$N_{H_2}$. While there are signs of decreasing polarization fraction~$P$ above $26$~K for some bones, conclusive results would require detailed modeling that is beyond the scope of this paper. 

Finally, it is important to state that the polarization fraction~$P$ will also be affected by potentially unresolved magnetic field structures within the beam of the telescope \citep[e.g.,][and references therein]{Andersson2015,Pattle_Fissel2019}. For example, this could be the result of tangled field lines due to turbulent motions in the gas, which would lead to a polarization signal that partially cancels itself when integrated along the line-of-sight, thus reducing of the polarization fraction. The inclination of field lines relative to the line-of-sight can also influence the observed fraction of polarization in filaments \citep[][]{Doi2021b}. 

Multi-wavelengths observations in the far-infrared and millimeter regimes are required to lift the degeneracy of these competing effects on the polarization fraction~$P$, as the observed grain alignment efficiency is expected to have a wavelength dependence while the integrated field directions should not \citep[e.g.,][]{Michail2021,Fanciullo2022,Tram2024}. With radio observations of Faraday rotation, it also becomes possible to measure the line-of-sight component of magnetic fields near giant molecular filaments, and thus have a measure of their 3D structure \citep[e.g.,][]{Tahani2022}. In addition, combining optical polarimetry with distance measurements from the \textit{Gaia} space telescope can identify the foreground components of dust polarization along the line-of-sight toward molecular clouds \citep[][]{Doi2021a,Doi2024}. 

\section{Discussion} 
\label{sec:discussion}

In Section~\ref{sub:fil_orient}, we compared the magnetic field structure observed in the FIELDMAPS bones to the orientation of each filament and to the large-scale field probed with the \textit{Planck} telescope. We found a mean difference and standard deviation of $-74^{\circ} \pm 32^{\circ}$ between the plane-of-sky magnetic field directions measured with HAWC+ at 214~$\mu$m and the filament orientations. The difference measured with \textit{Planck} at 850~$\mu$m is $3^{\circ} \pm 21^{\circ}$ instead. The magnetic field orientations measured with HAWC+ are closer to perpendicular to the length of the filaments, which are themselves parallel to the Galactic plane. In comparison, the \textit{Planck} data traces a field that is nearly parallel on average to the bones. As a reminder, the effective resolutions of HAWC+ at 214~$\mu$m and \textit{Planck} at 850~$\mu$m are respectively 18\farcs7 and 5\arcmin.

When looking at the Galaxy as a whole with both dust polarization and synchrotron data, the \textit{Planck} team determined that the Galaxy's magnetic field was homogeneously parallel to the disk up to Galactic latitudes of $\pm 10^\circ$ in most regions \citep[][]{Planck2015_XIX}. However, this picture is complicated by the relatively low values of dust polarized intensity measured across the plane, which at the scales probed by \textit{Planck} are likely due to unresolved field structures and integrated signal along the line-of-sight. The FIELDMAPS data confirms, at least in bones, that there can be significant changes in the field structure at higher resolutions, as illustrated by the histograms in Figures~\ref{fig:Fil1-5_Histo} and~\ref{fig:Fil10-G49_Histo}. Similarly, in the more turbulent Central Molecular Zone of the Galaxy, the FIREPLACE survey observed a complex magnetic field structure with a bi-modal distribution split between parallel and perpendicular to the Galactic plane \citep[][Par\'e et al., in prep.]{Pare2024_FIRE3}.

In addition, \citet{Planck2016_XXXV} found that magnetic field orientations tend to become more perpendicular to structures with hydrogen column densities~$N_H$ above $10^{21}$~cm$^{-2}$ in star-forming regions of the Gould belt. The mean field orientations measured in FIELDMAPS data indicate that such a change in orientation must be occurring in bones, although with significant variations along the length of the filaments. In some bones, there is even evidence for a density threshold at which a perpendicular alignment is unambiguously preferred \citep[][]{Stephens2025}. However, HAWC+ observations are generally only sensitive to higher column densities than those probed by \textit{Planck}. If the transition in field orientation happens at similar densities in bones as in nearby star-forming regions, then the current \textit{Planck} data may not be capable of identifying it due to the telescope's resolution and the position of these filaments within the Galactic plane. Higher resolution observations of dust polarization probing the same densities as \textit{Planck} are needed, such as those that would be achieved with PRIMA \citep[e.g.,][]{Moullet2023,Dowell2024}.

Nevertheless, we can already make useful comparisons with observations of magnetic fields in other spiral galaxies. The ``Survey on extragALactic magnetiSm with SOFIA'' (SALSA) in particular found that magnetic fields are closely aligned to spiral arms in both far-infrared and radio data, although with some differences in pitch angle and polarization levels \citep[][]{Borlaff2021_SALSA1,Lopez-Rodriguez2022_SALSA4,Lopez-Rodriguez2023_SALSA6}. When viewed from within, the resulting plane-of-sky magnetic field would likely exhibit a strong component parallel to the disk, consistent with patterns inferred from the Planck polarization measurements of our own Galaxy \citep[e.g.,][]{Beck2015,Planck2015_XIX,Borlaff2023_SALSA5}.

The SALSA survey also found lower polarization levels in spiral arms than in inter-arm regions, with the lowest values associated with the locations of active star formation in the arms \citep[][]{Lopez-Rodriguez2022_SALSA4}. Similarly to the \textit{Planck} polarization measurements discussed previously, this discrepancy can be explained by unresolved field structures, such as a turbulent field in the denser star-forming material probed in the arms \citep[][]{Borlaff2021_SALSA1, Borlaff2023_SALSA5}. Stellar feedback and turbulence from high-mass star formation activity is believed to contribute to the dynamo maintaining the magnetic field of spiral galaxies at large-scales \citep[][]{Beck2015}. If the FIELDMAPS bones are co-located with the spiral arms of our Galaxy (see Table~\ref{tab:filaments}), then they may be representative of comparable structures within the arms of other galaxies. In that case, the complex magnetic field morphologies observed in the FIELDMAPS polarization data at 214~$\mu$m may provide indirect evidence for unresolved structures in far-infrared observations of extragalactic sources.

In Galactic-scale simulations, the magnetic field associated with kiloparsec-long filaments in the cold neutral medium of spiral arms is generally parallel to their axis, in agreement with both \textit{Planck} and extra-galactic results for comparable column densities \citep[][]{Pillsworth2025}. Furthermore, models with magnetic fields have been found to significantly increase the dense star-forming gas in filaments when compared to purely hydrodynamic models, with perpendicular fields increasing the accretion rate and line masses of the filaments \citep[][]{Pillsworth2024}. In some cases, magnetic fields can also extend the lifetime of star-forming filaments by dampening the effect of stellar feedback at small scales, allowing more than one episode of star formation to occur \citep[][]{Hix2023}. The star formation tracers seen in the \textit{Spitzer} data from Figure~\ref{fig:Fil4_RGB} and Appendix~\ref{apx:Spitzer} may hint at such a scenario happening in bones. Finally, complex field structures akin to those observed along the length of the FIELDMAPS filaments may also be occurring in multi-scales simulations as a result of a number of factors, such as turbulence from supernovae feedback and compression by the expansion of super-bubbles \citep[][]{Zhao2024}.

\section{Summary} 
\label{sec:conclusion}

In this paper, we provided the full data release of the FIELDMAPS legacy survey, which mapped the polarized dust thermal emission in ten candidate bones of the Milky Way, referred here as bones for simplicity, at 214~$\mu$m using the HAWC+ camera aboard SOFIA. The primary goal of this survey is to characterize the magnetic field structure within these bones, some of the densest filamentary structures associated with the spiral arms of the Galaxy. From these HAWC+ polarization observations, we created maps of the plane-of-sky magnetic field~$B_{pos}$ structure across each bone, compared to the gas column density~$N_{H_2}$ derived from \textit{Herschel} observations and the large-scale field structured inferred from 850~$\mu$m measurements of polarization with \textit{Planck}.

We quantified the distribution of polarization angles in all the FIELDMAPS targets. In each case, we identified the mean polarization directions from the 214~$\mu$m HAWC+ data and the 850~$\mu$m \textit{Planck data}, as well as the orientation of the bone's spine. The average inclination of the bones is nearly parallel to the galactic disk at $1^{\circ} \pm 11^{\circ}$. We find a mean difference of $-74^{\circ} \pm 32^{\circ}$ between their filament axis and the plane-of-sky component of the magnetic field seen by HAWC+. In comparison, the difference between the filament orientation and the large-scale field traced by \textit{Planck} is $3^{\circ} \pm 21^{\circ}$. From these polarization data, we find that these large-scale fields change from being parallel to the Galactic disk to being closer to perpendicular, in average, to the long axis of each bone. At smaller scales, the magnetic field directions within bones are locally well-defined, but can change significantly across their length. 

We also produced maps of the polarization vectors for all the bones of the FIELDMAPS survey, as well as plots of the relation between the polarization fraction~$P$ and the Stokes~$I$ total intensity. We fitted a power-law in each plot to investigate the polarization efficiency in the bones, and we found an average power-law index and standard deviation of $\alpha = -0.7 \pm 0.1$ across the entire sample. We find similarly steep indices when plotting the polarization fraction~$P$ against the gas column density~$N_{H_2}$ instead. These fitted indices are comparable to those measured in nearby molecular clouds, and they may point to poor grain alignment in the densest regions of bones. However, several additional factors could contribute to artificially lowering the indices measured from our HAWC+ data, and follow-up studies using polarization observations at multiple wavelengths are needed to fully explore the alignment efficiency of dust grains in bones.

Finally, all the data products, analysis codes, and figures produced for this data release paper are made publicly available online\footnote{\href{https://doi.org/10.7910/DVN/NUXGJE}{https://doi.org/10.7910/DVN/NUXGJE}}. With its uniform sample of ten bones, the FIELDMAPS data release is an excellent starting point for follow-up studies using both current and future polarimeters.

\vspace{12pt}

%\begin{acknowledgments} % commented to allow the acknowledgements to be split over multiple pages
The NASA/DLR Stratospheric Observatory for Infrared Astronomy (SOFIA) was jointly operated by the Universities Space Research Association (USRA), under NASA contract NNA17BF53C, and the Deutsches SOFIA Institut (DSI) under DLR contract 50 OK 0901 to the University of Stuttgart. Financial support for S.C. and I.W.S. was provided by NASA through award 08\_0186 issued by USRA.
L.W.L. acknowledges support from NSF AST-1910364 and NSF AST-2307844. P.S. was partially supported by a Grant-in-Aid for Scientific Research (KAKENHI Number JP22H01271 and JP23H01221) of JSPS.

We thank Atilla Kov\'acs for insightful discussions on the data reduction processes for polarization measurements and the intricacies of the CRUSH software (https://www.sigmyne.com/crush/). We also thank Jens Kauffmann for discussions on the HAWC+ data reduction processes and Kate Pattle for discussions on depolarization.

We thank the HAWC+ instrument science team for the planning and acquisition of the data used in this project: Ryan Arneson, Peter Ashton, Melanie Clarke, Simon Coud\'e, Curtis DeWitt, Sareh Eftekharzadeh, Michael Gordon, Kyle Kaplan, Nicole Karnath, Enrique Lopez-Rodriguez, Steve Mairs, Ed Montiel, Dennis O'Flaherty, Sachin Shenoy, and Bill Vacca. Finally, we also thank everyone who contributed to SOFIA during its lifetime for their dedication to the success of this unique mission.

We also thank the anonymous referee for their insightful comments and suggestions.

\textit{Herschel} was an ESA space observatory with science instruments provided by European-led Principal Investigator consortia and with important participation from NASA. This work was also based on observations obtained with Planck (http://www.esa.int/Planck), an ESA science mission with instruments and contributions directly funded by ESA Member States, NASA, and Canada. 

This research has made use of the NASA/IPAC Infrared Science Archive, which is funded by NASA and operated by the California Institute of Technology, of NASA’s Astrophysics Data System, and of APLpy, an open-source plotting package for Python (https://aplpy.github.io/).
%\end{acknowledgments}

\facility{SOFIA, \textit{Herschel}, \textit{Planck}, IRSA}
\software{APLpy \citep{RobitailleBressert2012}, Astropy \citep{Astropy2013,Astropy2018,Astropy2022}, Photutils \citep{photutils_2024}, reproject \citep{Robitaille2020}, CARTA \citep{CARTA_2021}}

\appendix

\section{Observing Modes}
\label{apx:modes}

\paragraph{Nod-Match-Chop}

The initial observing mode for HAWC+ was Nod-Match-Chop, or Chop-Nod for short, a technique described in detail by \citet{Hildebrand2000}. In this mode, the telescope array is moved, or ``nodded'', between two ``Nod'' positions each found symmetrically on a line centered on the science target. At each ``Nod'' position, the secondary mirror alternates, or ``chops'', at a frequency of $10.2$~Hz between an ``On'' (target) and an ``Off'' (sky) position. This process of chopping and nodding effectively creates two ``Off'' positions that are are used as time-dependent references to remove the sky emission during data reduction \citep{Harper2018, clarke_2022_Manual}. The Chop amplitude and angle are chosen so that the ``Off'' positions avoid astronomical sources of emission, although it is not always possible in crowded regions such as the Galactic Disk.  A typical Chop amplitude for SOFIA was between 3.5' and 4.2'. Additionally, the telescope array is dithered to four positions during a Chop-Nod observing sequence to limit the effect of bad pixels on the reduced observations. To be complete, a set of Chop-Nod observations also requires a pair of internal calibration files (IntCal) at the start and at the end in order to measure the phase delay needed for demodulating the chopping time stream, and to apply a flat-field correction accounting for detector gains and bad pixels. Overall, Chop-Nod observing was less efficient than On-The-Fly mapping due to the time spent off-source and the large overheads, but the sky removal for this technique is generally considered to be more reliable if no astronomical sources are found in the ``Off'' positions. 

\paragraph{Scan Imaging} 

The second observing mode for HAWC+ was Scan imaging, also often called On-The-Fly mapping. Scan imaging was the primary mode for total intensity observations, and it was made available to the community for polarization observations as shared-risk starting in Cycle~8. The main scanning strategy used for HAWC+ was to perform a Lissajous pattern with the telescope array while keeping the secondary mirror fixed \citep{Harper2018}. A Lissajous pattern is a combination of two sinusoidal functions, one horizontal $x$ and one vertical $y$, that is non-repeating when the ratio of frequencies is irrational (in the case of HAWC+, $\omega_y/\omega_x \simeq \sqrt{2}$). Since such a Lissajous scan never returns to the exact same position, it minimizes the effect of various sources of noise in the data, such as bad pixels and the $1/f$ noise \citep{Kovacs_2008_Scans}. For a given reference frame, the parameters that could be tweaked for HAWC+ scans are the horizontal and vertical amplitudes, the range of initial scan angles, the scan rate (typically 100''/s or 200''/s), and the duration. A caveat of using a Lissajous scan is that amplitudes that are significantly larger than the detector array's field-of-view can lead to lower sensitivities in the center of a map due to the larger amount of time spent at the edges.

\paragraph{Polarization} 

When performing polarization observations with any of the previously described observing modes, the half-wave plate for the selected filter was first inserted into the optical path of the instrument, and then it was rotated to four distinct angles during observing: 5.0$^{\circ}$, 50.0$^{\circ}$, 27.5$^{\circ}$, and 72.5$^{\circ}$ \citep[see][]{Harper2018}. These four angles are sufficient to recover the linear Stokes parameters ($I$, $Q$, and $U$; see Section~\ref{sub:equations}) during data reduction \citep{clarke_2022_Manual}. Each HAWC+ band had an associated half-wave plate, except for Bands~A and B which shared the same. 

In Chop-Nod mode, the half-wave plate was rotated after each Nod pair (ABBA; see \citealt{Harper2018}) for a given dither position. While this approach increased overheads, Chop-Nod observing with HAWC+ generated a single file for each dither position that contains data for all of the four required half-wave plate angles. A typical set of Chop-Nod polarization observations therefore consists of four files, one for each dither position, sandwiched between two internal calibration files at the start and at the end. Since each file has all the required angle information, the data reduction pipeline can proceed even if some dither positions are missing (for example, if a set had to be interrupted for any reason).

In Scan mode, the half-wave plate was rotated after individual scans. Each scan generated a single file, and so each file contains information for a single half-wave plate position angle. A full polarization observation in Scan mode therefore requires sets of four files, one for each position angle, in order to be processed by the data reduction pipeline. This method led to lower overheads since the half-wave plate was not rotated as often as when observing in Chop-Nod mode, but at the risk of having to discard an entire set of scans if even one was missing as all four scans had to be completed between line-of-sight rewinds of the telescope array. These rewinds were a necessary component of the stabilization mechanism used to negate the rotation of the sky on the detector array during observing, and this concept is explained in Section~1.2.4 in the SOFIA Observer's Handbook. Overall, polarization observations in Scan mode were shown to be an improvement of a factor~$\sim 1.8$ in sensitivity relative to Chop-Nod for the same on-source time \citep{Lopez-Rodriguez2022_SALSA3}. 

\section{Details of data acquisition}
\label{apx:obs}

Astronomical Observations Requests (AORs) were Extensible Markup Language (XML) files created and modified using the Unified SOFIA Proposal and Observation Tool (USPOT). They contained all the configuration information required by the instrument and the telescope array to perform the requested observations, and they were used as the reference to write the headers of the resulting Flexible Image Transport System (FITS) files. If needed, accurate information for individual AORs can be accessed through the headers of Level~2 (corrected data) and Level~3 (calibrated data) files, while the headers in Level~3 (merged data) and Level~4 (merged and multi-mission data) files contain information from only one of the AORs used for the mosaicked data. The ambiguous nomenclature for Level~3 HAWC+ data is an historical quirk of the data reduction pipeline, but the two types can be differentiated on IRSA by the Product Type parameters \textit{calibrate} and \textit{merge}, respectively. We only provide Level~4 data products as part of the FIELDMAPS survey data repository, and so the FITS headers of corrected Level~2 and calibrated Level~3 data products on IRSA remain the best reference for exact information on the configuration of every observation performed for this survey. 

In cases where the pipeline would crash during data reduction, we proceeded to inspect each individual polarization sets that were obtained across all flights for the affected bones. We identified several anomalous Level~0 files as the source of these crashes, and thus discarded them before re-running the reduction pipeline described previously. As mentioned in Section~\ref{sub:acquisition}, we list all the problematic Level~0 files for each bone in this appendix. Otherwise, the most significant correction to the data was fixing the observations of Filament~4 taken during Flight~684. Due to a communication error between the instrument and the telescope array, there is a shift in the World Coordinate System (WCS) information recorded in the data for Flight~684. Appendix~\ref{apx:fix_fil4} explains how the Level~2 data products for this flight were corrected to sub-pixel accuracy before being processed into Level~3 and 4 data products.

\paragraph{Filament 1} was observed with HAWC+ in scan mode on Flights~726 (2021/05/05) and 729 (2021/05/11). Files 90 and 111 from Flights~726 were discarded as they are calibration files erroneously associated with this program. Furthermore, two sets of four files from Flight~726 were discarded due to generating errors during data reduction: files 162 to 165 and 186 to 189, inclusively. The final Level~4 data product was obtained by mosaicking two pointings with overlapping fields of view (AORs~\#60 and 61). 

\paragraph{Filament 2} was observed with HAWC+ in scan mode on Flights~727 (2021/05/06) and 879 (2022/06/01). The final Level~4 data product was obtained by mosaicking four pointings with overlapping fields of view (AORs~\#62, 63, 64, 65, 104, and 105). 

\paragraph{Filament 4} was observed with HAWC+ in scan mode on Flights~684 (2020/09/10) and 685 (2020/09/11). Due to a communication error between the instrument and the telescope array, there is a shift in the WCS information for observations taken on Flight~684. Appendix~\ref{apx:fix_fil4} describes how this shift was corrected manually during the data reduction process. Two sets of four files from Flight~684 were discarded: files 25 to 28 due to exhibiting a larger WCS shift than other observations during that night, and 74 to 77 for exhibiting an additional drift in WCS coordinates within a single set of halfwave plate positions. The final Level~4 data product was obtained by mosaicking two pointings with overlapping fields of view (AORs~\#66, 67, 92, 93, and 94). 

\paragraph{Filament 5} was observed with HAWC+ in scan mode on Flights~728 (2021/05/07), 732 (2021/05/14), and 734 (2021/05/19). Files 97 and 138 on Flight~732, and 100, 172, and 174 on Flight~734, were discarded due to generating errors during data reduction. All files from Flight~728 were also removed because they showed clear discrepancies with the data obtained during Flights~732 and 734. Based on the flight summary for Flight~728, the calibration of the data obtained for Filament~5 may have been affected by attempts to extend the duration of HAWC+'s cryogenic system during the last leg of the flight.\footnote{See \href{https://irsa.ipac.caltech.edu/data/SOFIA/docs/proposing-observing/flight-plans/index.html}{https://irsa.ipac.caltech.edu/data/SOFIA/docs/proposing-observing/flight-plans/index.html}} The large instrumental phase offsets observed when performing internal calibration tests on that night did not affect the quality of the observations for G49 compared to Flights~726 and 733, and so these offsets are unlikely to have independently impacted the data for Filament~5. The final Level~4 data product was obtained by mosaicking three pointings with overlapping fields of view (AORs~\#68, 69, and 70).

Additionally, Filament~5 was also observed at 214~$\mu$m with HAWC+ in Chop-Nod mode on Flight~441 (2017/10/18) for program 05\_0109. Unfortunately, even when using a chop distance of $7.5'$, one of the two ``Off'' positions used for background subtraction contained a far-infrared source, and so a fake negative source is found at $l=$+18:35:54.456 and $b=$-0:04:52.005. Only one pointing was observed for this program (AOR~\#4). These observations were followed up at 53~$\mu$m on Flight~485 (2018/07/12) for program 06\_0206 with a single pointing (AOR~\#8).

\paragraph{Filament 6 (The Snake)} was observed with HAWC+ in Chop-Node mode for program 05\_0206 on Flight~394 (2017/05/12), and for program 06\_0027 on Flight~486 (2018/07/13). The final Level~4 data product was obtained by mosaicking seven pointings with overlapping fields of view, two from program 05\_0206 (AORs~\#1, 2) and five from program 06\_0027 (AORs~\#1, 2, 3 , 4, and 5).

\paragraph{Filament 8} was observed with HAWC+ in scan mode on Flight~892 (2022/06/30). The final Level~4 data product was obtained by mosaicking three pointings with overlapping fields of view (AORs~\#73, 74, and 118).

\paragraph{Filament 10} was observed with HAWC+ in scan mode on Flight~894 (2022/07/05). The final Level~4 data product was obtained by mosaicking eight pointings with overlapping fields of view (AORs~\#75, 76, 77, 78, 112, 113, 114, and 115).

\paragraph{G24} was observed with HAWC+ in scan mode on Flights~880 (2022/06/02), 881 (2022/06/03), 882 (2022/06/04), and 886 (2022/06/11). The final Level~4 data product was obtained by mosaicking twelve pointings with overlapping fields of view (AORs~\#79, 80, 81, 82, 83, 84, 106, 107, 108, 109, 110, and 111).

\paragraph{G47} was observed with HAWC+ in scan mode on Flights~687 (2020/09/15), 689 (2020/09/23), and 690 (2020/09/24). The final Level~4 data product was obtained by mosaicking four pointings with overlapping fields of view (AORs~\#85, 86, 87, and 100). These observations were previously published by \citet{Stephens2022}.

\paragraph{G49} was observed with HAWC+ in scan mode on Flights~726 (2021/05/05), 728 (2021/05/07), and 733 (2021/05/18). File~160 from Flight~733 was discarded as it is a calibration file erroneously associated with this program. Contrary to Filament~5, observations of G49 during Flight~728 do not show any obvious discrepancy with those obtained during Flights~726 and 733. The final Level~4 data product was obtained by mosaicking four pointings with overlapping fields of view (AORs~\#88, 89, 90, and 91).

\section{Extended analysis of the polarization data in Filament 5}
\label{apx:chop_fil5}

\begin{figure*}
    \centering
    \includegraphics[width=0.495\textwidth]{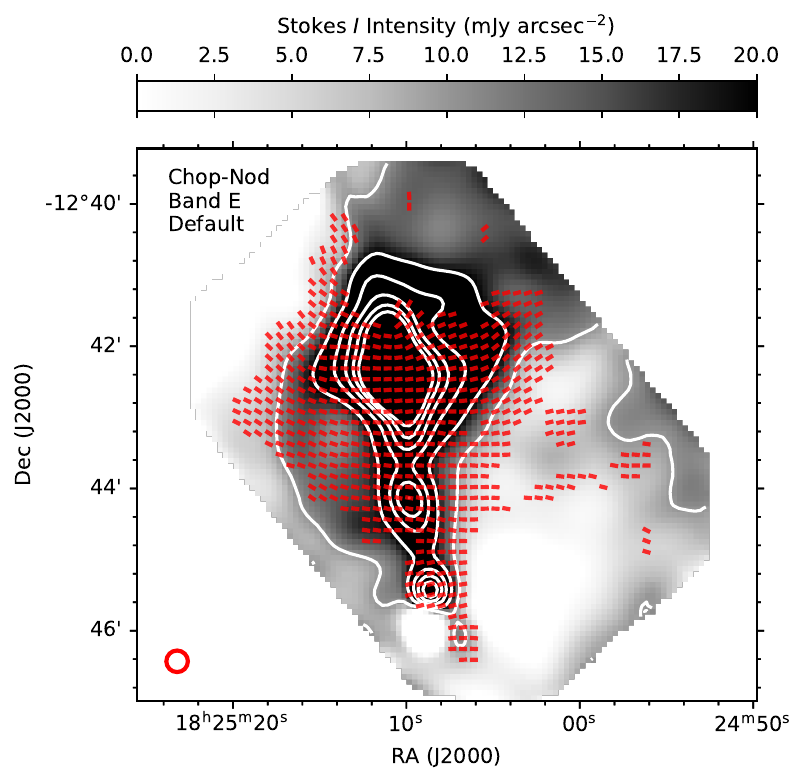}
    \includegraphics[width=0.495\textwidth]{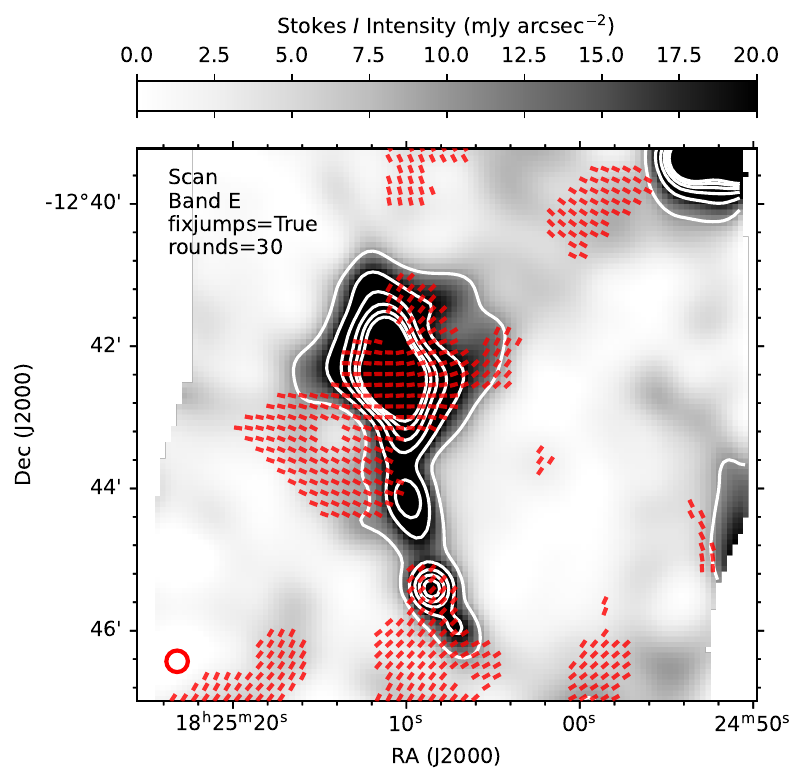}
    \includegraphics[width=0.495\textwidth]{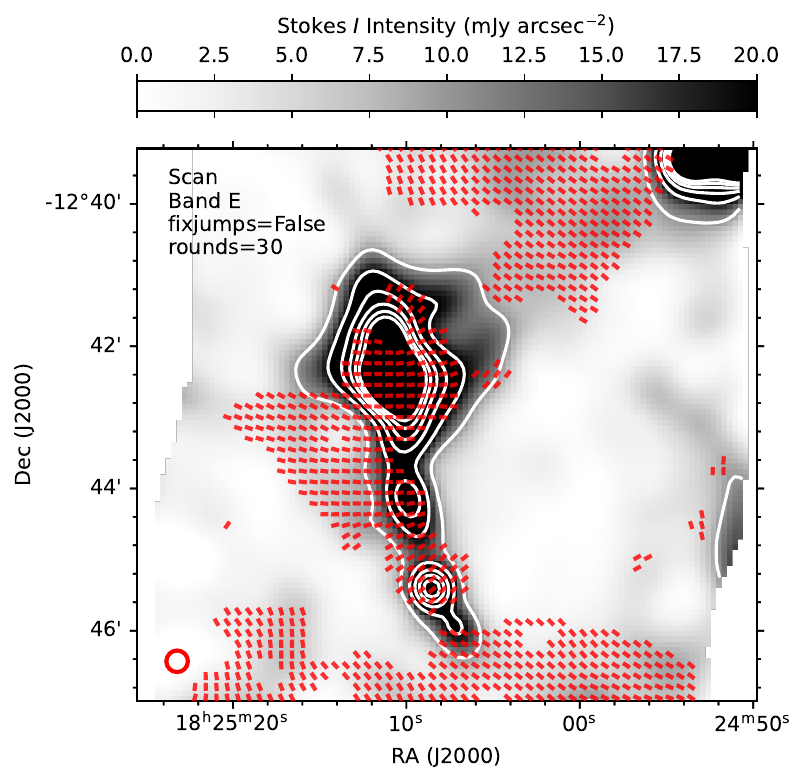}
    \includegraphics[width=0.495\textwidth]{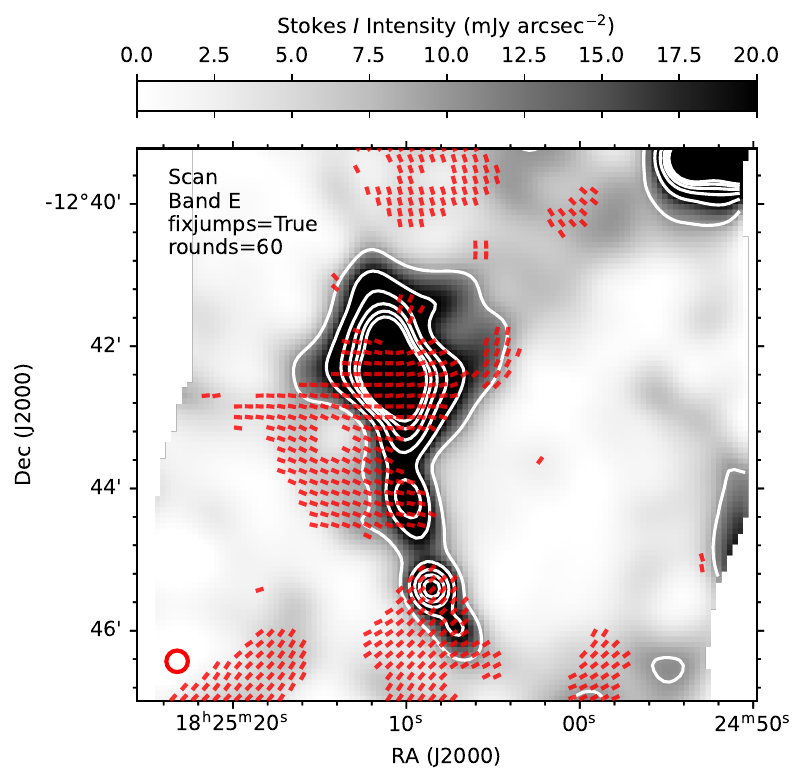}
    \caption{The magnetic field structure (red single-length vectors) inferred from HAWC+ 214~$\mu$m (Band~E) observations for four different data reduction procedures for Filament~5. Every other vector is plotted. The gray scale is the measured 214~$\mu$m Stokes~$I$ total intensity, and the white contours trace levels of 10 to 60~mJy~arcsec$^{-2}$ by steps of 10~mJy~arcsec$^{-2}$. The red circle shows the beam of HAWC+ at 214~$\mu$m. \textit{Top left}: Chop-Nod data reduced using the default pipeline parameters. \textit{Top right}: Scan mode data reduced using \textit{rounds=30} and \textit{fixjumps=True} (see also Figure~\ref{fig:Fil5_Maps} for full map). \textit{Bottom left}: Scan mode data reduced using \textit{rounds=30} and \textit{fixjumps=False}. \textit{Bottom right}: Scan mode data reduced using \textit{rounds=60} and \textit{fixjumps=True}.}
    \label{fig:Fil5ChopE}
\end{figure*}

As noted in Section~\ref{sub:targets} and Appendix~\ref{apx:obs}, Filament~5 was the only bone in our sample observed with HAWC+ in polarization using both Chop-Nod and Scan modes. It is therefore the only target where we can effectively quantify the differences between these two modes. We also used the data for Filament~5 to investigate the effect of the SOFIA Redux Scan mode reduction parameters \textit{fixjumps} and \textit{rounds}. Finally, Filament~5 is also the only bone with observations at 53~$\mu$m (Band~A), which we also present in this appendix. 

We focus first on the differences in the Stokes~$I$ total intensity and the polarization angle~$\theta$ between the two observing modes.  As a reminder, the polarization angle~$\theta$ is a function of both Stokes~$Q$ and $U$ (see Equation~\ref{eq:theta}), and it is the crucial quantity to characterize the plane-of-sky magnetic field structure in the bones. For this analysis, we use the data in the ICRS reference frame as it is the default format used by the reduction pipeline. We therefore use Stokes~$Q_{icrs}$ and $U_{icrs}$ instead of Stokes~$Q_{gal}$ and $U_{gal}$, and the resulting polarization angle~$\theta$ is defined following the IAU convention from Celestial North to East. In Figure~\ref{fig:Fil5ChopE}, we use the magnetic field angle $\theta_B$, which is simply $\theta_B=\theta + 90^\circ$, to better relate the different data reduction results with those from Figures~\ref{fig:Fil1_Maps} to \ref{fig:G49_Maps}. 

\begin{figure*}
    \centering
    \includegraphics[width=0.495\textwidth]{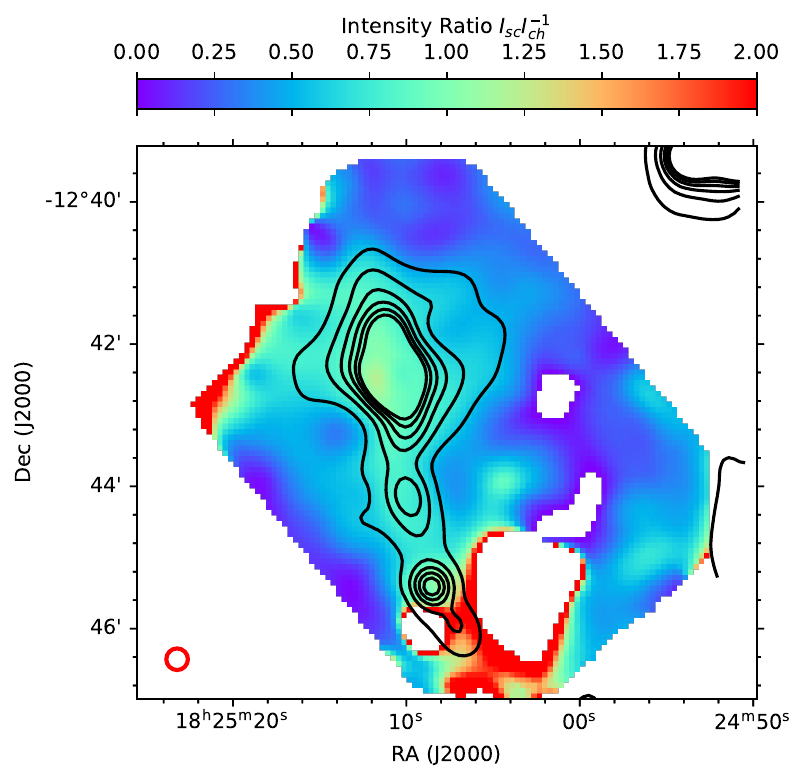}
    \includegraphics[width=0.495\textwidth]{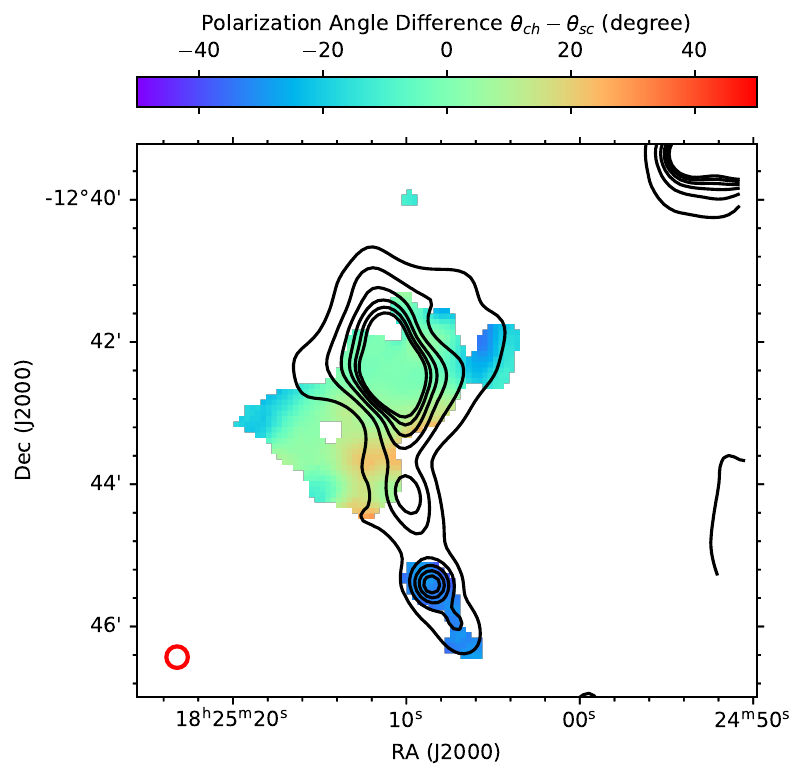}
    \caption{Comparison between Chop-Nod and Scan mode data for Filament~5. The Stokes~$I$, $Q$, and $U$ data from the Scan mode observations were re-gridded to match the Chop-Nod data. The black contours trace the Stokes~$I$ total intensity from the Scan mode map from 10 to 60 mJy arcsec$^{-1}$ by steps of 10 mJy arcsec$^{-1}$. The HAWC+ beam is given as a red circle in each panel. \textit{Left:} The intensity ratio $I_{sc} I^{-1}_{ch}$ is shown on a rainbow scale for the pixels where $I>0$ in both maps. \textit{Right:} The difference of the polarization angles $\theta_{ch}-\theta_{sc}$ is shown on a rainbow scale for the pixels where $P/\delta_P > 3$, $P < 30 \%$, and $I/\delta_I > 10$ in both maps.}
    \label{fig:Fil5IvIOvO}
\end{figure*}

Figure~\ref{fig:Fil5ChopE} presents the inferred magnetic field maps of Filament~5 for different data reduction products in Chop-Nod and Scan modes. As illustrated by the intensity contours in each panel, the Chop-Nod data recovers more low-level extended emission than the Scan mode data. However, the Chop-Nod data also displays significant negative flux that does not exist in the Scan mode data (see also the left panel of Figure~\ref{fig:Fil5IvIOvO}), which is indicative of bright sources in one or both of the ``Off'' positions for the Chop-Nod observations. These characteristics fit the expected limitations of the Chop-Nod and Scan observing modes, as discussed in Appendix~\ref{apx:modes}. Considering that all the bones observed for FIELDMAPS are in crowded regions of the sky in the far-infrared, the existence of significant negative sources in the Chop-Nod data for Filament~5 is further evidence supporting the choice of Scan mode observing for this survey.

Figure~\ref{fig:Fil5IvIOvO} shows the comparison in Stokes~$I$ total intensity and in polarization angle~$\theta$ between Chop-Nod and Scan mode data for Filament~5. The Stokes~$I$ intensity and the polarization angle~$\theta$ are respectively identified as $I_{ch}$ and $\theta_{ch}$ for the Chop-Nod data, and $I_{sc}$ and $\theta_{sc}$ for the Scan data. These maps were obtained by re-projecting the Scan data to the same grid as the Chop-Nod data using the \textsc{reproject} package. As noted previously, the intensity ratio map in the left panel of Figure~\ref{fig:Fil5IvIOvO} exhibits regions with negative flux close to the source, which also impacts the surrounding areas. The region with the largest angle difference in the right panel is also the one closest to this negative flux in the South of the map.

We fitted a linear relation $I_{sc} = a \, I_{ch} +b$ to the intensities measured in the Chop-Nod and Scan mode data for the pixels plotted in the left panel of Figure~\ref{fig:Fil5IvIOvO}. We found a coefficient $a = 0.95$ and a constant $b = -4.5$~mJy arcsec$^{-1}$. While some additional low-level flux is recovered in the Chop-Nod data, the linear trend between the data products is only 5~\% away from a perfect 1:1 correlation, which is comparable to the calibration uncertainties for HAWC+ \citep{Harper2018}. For the difference of polarization angles between the two modes for the pixels shown in the right panel of Figure~\ref{fig:Fil5IvIOvO}, we find an average value $\langle \theta_{ch} - \theta_{sc} \rangle$ of $2.8^\circ$ and a standard deviation of $15.0^\circ$. As a reference, the median angle uncertainty $\delta_{\theta}$ for each mode in these pixels is approximately $5^\circ$, with a minimum of $1^\circ$ and a maximum of $9^\circ$. Overall, we find a reasonably good agreement between the Chop-Nod and Scan data for the pixels that can be directly compared in Filament~5.

To investigate the effect of the two observing modes on the quantities derived in Section~\ref{sub:PvI} for Filament~5, we fitted a power-law following Equation~\ref{eq:PvI} to the relation between the polarization fraction~$P$ and the Stokes~$I$ intensity. We specifically limited our analysis to the region displayed in Figure~\ref{fig:Fil5ChopE}. For the Chop-Nod data, we find a power-law index $\alpha = -0.78 \pm 0.01$, while the Scan data gives an index $\alpha = -0.69 \pm 0.01$. Of note, using a subset of the Scan data for Filament~5 gives a shallower result for the index $\alpha$ than reported in Figure~\ref{fig:Fil1-5_PvI}. However, the difference $\Delta \alpha = 0.09$ between the Chop-Nod and Scan data is larger than the fit uncertainties of $0.01$. This could be an indication that the statistical uncertainties reported in Section~\ref{sub:PvI} for the power-law fits may be underestimating the real measurement uncertainties when including the entire bone. Additionally, an underestimation of Stokes~$I$ from extended emission in the Scan data could lead to a significantly different index $\alpha$, as seen in the right panel of Figure~\ref{fig:Fil10_PvI_PvIest} in Appendix~\ref{apx:ext_fil10}.

We also tested Scan mode data reduction products using differing parameters when running SOFIA Redux. Figure~\ref{fig:Fil5ChopE} shows the results for the data product used in this paper (top right panel), the data product obtained when using $fixjumps=False$ (bottom left panel), and the data product obtained when using $rounds=60$ (bottom right panel). While not pictured, we also ran a reduction with $fixjumps=True$ and $rounds=15$. Increasing the number of iterations in the pipeline from 30 to 60 appears to help recover additional vectors in the middle clump of Filament~5, which agree with those from the Chop-Nod data, but otherwise improvements elsewhere are minor. The largest increase in the number of recovered vectors is when we use $fixjumps=False$. However, there is a nearly $90^\circ$ difference in the polarization angles measured in the southern part of the plots in Figure~\ref{fig:Fil5ChopE} between  the $True$ and $False$ products. Critically, the Chop-Nod map shows a closer agreement with the Scan mode data using $fixjumps=True$. Increasing the number of detections is meaningless if the recovered angles cannot be trusted in parts of the map, and so a conservative approach using the $fixjumps=True$ parameter is warranted in our systematic reduction of the bones for FIELDMAPS.

In general, the $fixjumps$ parameter is useful when a HAWC+ detector displays an anomalously high flux relative to its neighbors, which usually appear as a ``worm'' on the Stokes~$I$ map of the final data reduction product \citep{clarke_2023_HAWC}. Otherwise, the resulting polarization maps should be mostly comparable between $fixjumps=False$ and $True$. Several factors may have contributed to seeing such a strong difference between these two values in our data for Filament~5. First, the region shown in Figure~\ref{fig:Fil5ChopE} lies at the interface between two distinct pointings, which could make it more susceptible to potential edge effects, such as higher noise. Additionally, the coverage between the two fields is uneven due to losing the data from Flight~728 (see Appendix~\ref{apx:obs}), with the northernmost field showing a higher noise level by approximately a factor 2. A greater emphasis was put on acquiring more even coverage of the bones in subsequent observations for the survey, but unfortunately the SOFIA mission concluded before Filament~5 could be revisited. 

Interestingly, the polarization vectors seen in the brightest region of Figure~\ref{fig:Fil5ChopE} agree somewhat between the $fixjumps=False$ and $True$ data products. Since the dust emission is much weaker South of this region (see the bottom panel of Figure~\ref{fig:Fil5_Maps}), a plausible explanation would be that a detector flux jump while this pointing was observed may have had a greater effect on the background subtraction in the Stokes parameters due to the relatively low signal-to-noise emission. An inspection of each data product, all available in the online repository, reveals that this is indeed the location where the Stokes~$U$ maps differ the most, which would explain the observed changes in polarization angle. 

\begin{figure*}
    \centering
    \includegraphics[width=0.495\textwidth]{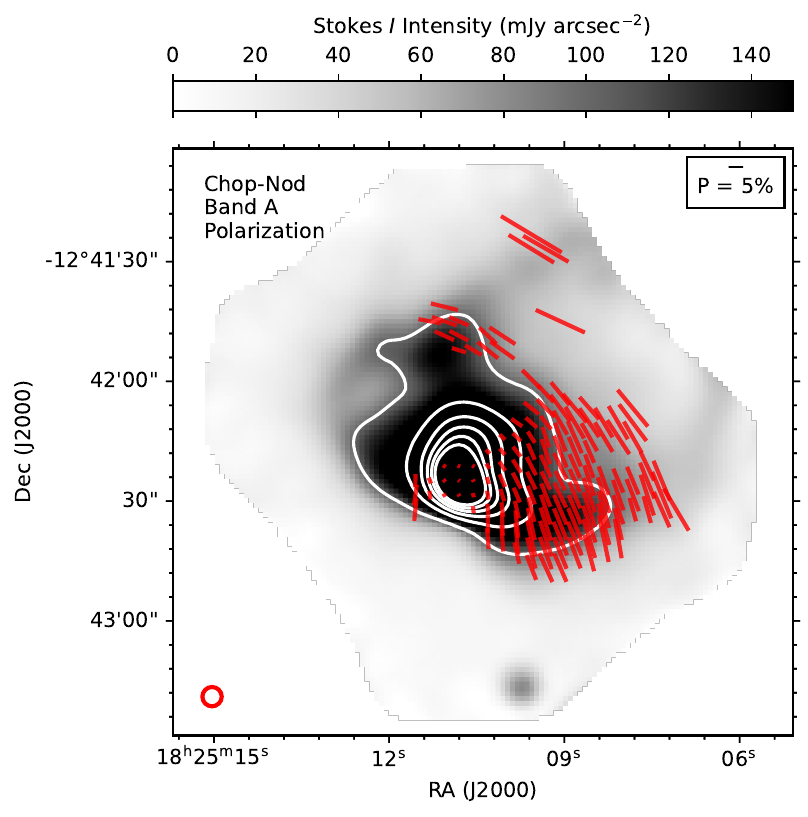}
    \includegraphics[width=0.495\textwidth]{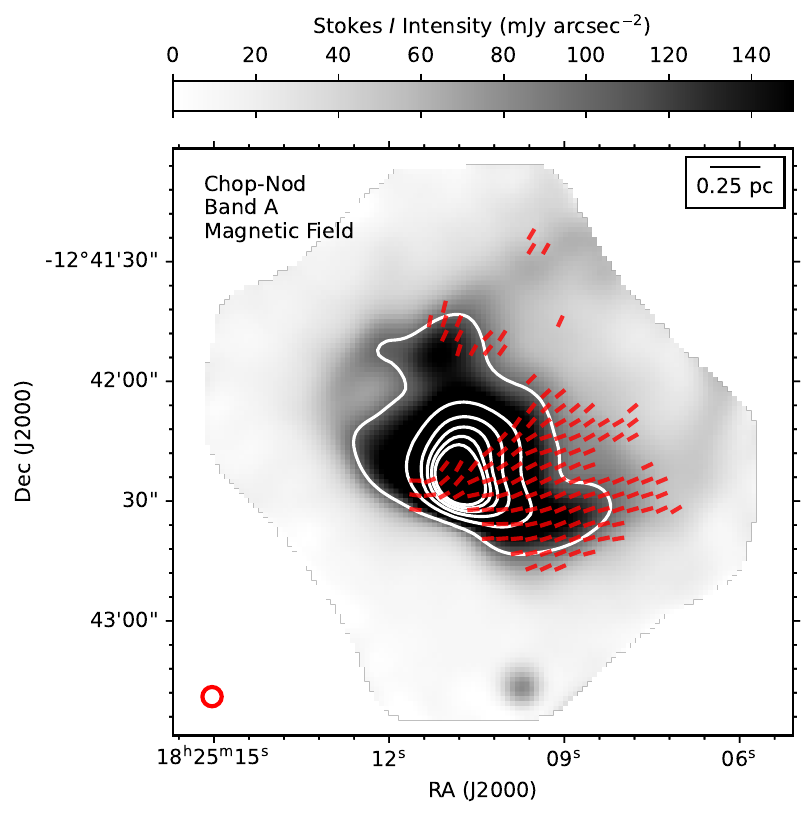}
    \caption{Chop-Nod observations of Filament~5 with HAWC+ at 53~$\mu$m (Band~A). The gray scale is the measured 53~$\mu$m Stokes~$I$ total intensity, and the white contours trace levels of 100 to 600~mJy~arcsec$^{-2}$ by steps of 100~mJy~arcsec$^{-2}$. The red circle shows the beam of HAWC+ at 53~$\mu$m. \textit{Left:} Polarization map shown as red vectors with the length representing the polarization fraction~$P$. \textit{Right:} Inferred magnetic field map shown as red single-length vectors. A scale length of 0.25~pc is given for reference.}
    \label{fig:Fil5ChopA}
\end{figure*}

Finally, we present in Figure~\ref{fig:Fil5ChopA} the polarization and plane-of-sky magnetic field maps of Filament~5 derived from 53~$\mu$m (Band~A) Chop-Nod observations with HAWC+. The smaller field-of-view in Band~A covers only the brightest, northernmost clump shown in Figure~\ref{fig:Fil5ChopE}. This region is associated with the massive IRAS 18223-1243 young stellar object \citep{Chan1996} and its compact HII region \citep{Sewilo2004}, making it a clear site of high-mass star formation. In order to compare the polarization angles between the two wavelengths, we re-projected the Scan data at 214~$\mu$m to the same grid as the Chop-Nod data at 53~$\mu$m. We find a mean angle difference $\langle \theta_{53} - \theta_{214} \rangle$ of $-13.3^\circ$ and a standard deviation of $12.0^\circ$. While the 214~$\mu$m observations is significantly over-sampled, this comparison shows that the 53~$\mu$m polarization data exhibits more structure at smaller scales while still being mostly aligned in the same direction overall. Following Equation~\ref{eq:PvI}, we also fitted a power-law function to the relation between the polarization fraction~$P$ and the Stokes~$I$ total intensity at 53~$\mu$m, and we find a power-law index $\alpha$ of $-0.88 \pm 0.01$.

\section{Reference frame correction in Filament 4}
\label{apx:fix_fil4}

Due to a communication error between the instrument and the telescope array during SOFIA Flight 684 (2020/09/10), all the observations taken for Filament~4 on that night had incorrect WCS information in their headers. Instead of discarding the data from this flight, which represented nearly half of the total observations for Filament~4, we attempted to measure this coordinate shift and apply a correction directly to the headers of the Level~2 data before the calibration and merging steps described in Section~\ref{sub:proc}. 

The HAWC+ coverage for Filament~4 contains a candidate massive YSO associated with an HII region that appears as a near-perfect circle for the SOFIA beam at 214~$\mu$m \citep[G021.3570-00.1795 from][also IRAS 18279-1024]{Urquhart2009}. This fortunate occurrence provided us with a reference source that could be easily fitted with a 2D Gaussian, thus helping to accurately determine its centroid position in data from both Flights~684 and 685. 

Before we proceeded to fix the WCS information, however, we first completed a visual inspection of each HAWC+ polarization data set obtained during Flight~684, and compared them with those from Flight~685. As mentioned in Appendix~\ref{apx:obs}, we removed files 25 to 28 due to displaying a seemingly larger coordinate shift than the rest of the data. We could not measure this shift accurately since the field-of-view for these files did not include G021.3570-00.1795, and so removing them served as a precaution to avoid including a potential pointing error in the final data product. We also removed files 74 to 77, as they showed an obvious pointing instability between individual half-wave plate positions, leading to duplicated sources in the data. 

Reducing the remaining files for Flight~684 created six Level~2 data products, three of which contained G021.3570-00.1795 in their field-of-view. We used the \textsc{centroid\_2dg} function from the \textsc{Photutils} package to fit a 2D Gaussian to this source and determine its centroid position for each Level~2 file. The pointing error was measured with the standard deviation of the fitted centroid positions in each file. For Flight~684, we found a pointing error of 0.42~pixel in Right Ascension and 0.16~pixel in Declination. For Flight~685, the pointing error of the four Level~2 data products containing G021.3570-00.1795 is 0.05~pixel in Right Ascension and 0.04~pixel in Declination. While the pointing accuracy for Flight~685 is superior, finding a sub-pixel pointing accuracy (or less than a quarter of the beam size) for Flight~684 nevertheless reassured us that a single positional shift could be implemented to fix all six Level~2 products from this flight.

We calculated the required positional correction by taking the difference of the average centroid position of G021.3570-00.1795 between Flights~684 and 685, which is $-0.049397^\circ$ in Right Ascension and $0.130828^\circ$ in Declination. The six corrected Level~2 data products for Flight~684 were then generated with new headers where these values were added to the original CRVAL1 and CRVAL2 keywords. Finally, all the corrected Level~2 data products from Flight~684 were combined with those from Flight~685 in the SOFIA Redux pipeline to obtain updated Level~3 and Level~4 products for Filament~4. For future reference, the Level~2 files and the Python codes used for this reference frame correction in Filament~4 are included in the online data repository for this paper. 

\section{Re-projection to Galactic Coordinates}
\label{apx:gal_reproj}

\subsection{SOFIA}
\label{apx:gal_reproj_SOFIA}

After the FIELDMAPS data has been reduced following the procedure described in Section~\ref{sub:proc}, each Level~4 data product was transformed from the International Celestial Reference System (ICRS) to the Galactic Coordinate System (GCS). In this Appendix, we describe in detail the steps of this conversion process. The codes used for the re-projection to Galactic coordinates for each bone are included as part of the FIELDMAPS data release. 

First, for each extension of the original data products, a copy of the FITS header is used as a template where only specific keywords are modified. The CTYPE1, CTYPE2, and WCSNAME keywords are updated to reflect the new GCS reference frame. The CRVAL1 and CRVAL2 values are converted from ICRS to GCS using Astropy's SkyCoord class. The NAXIS1 and NAXIS2 values are increased to allow for the rotation of the field-of-view without cropping any astronomical data. 

These modified FITS headers are then used as the reference when re-projecting each extension of the Level~4 data products from ICRS to GCS. Specifically, we use the reproject\_exact function from the \textsc{reproject} Python package. These re-projected extensions are then recombined into a new FITS container, including a copy of the three tabular extensions described in Table~2 of \citet{Gordon2018}. 

In order to complete the conversion to the GCS reference frame, the Stokes~$Q_{icrs}$ and $U_{icrs}$ parameters have to be recalculated to obtain Stokes~$Q_{gcs}$ and $U_{gcs}$. To achieve this, we must first determine the rotation of the polarization angle from $\theta_{icrs}$ to $\theta_{gcs}$ on the plane on the sky for each pixel position. This rotation is described by the following equation \citep{Appenzeller1968,Panopoulou2016b,Panopoulou2016a}: 

\begin{equation}
    \theta_{gcs} - \theta_{icrs} = \frac{180^\circ}{\pi} \: \arctan \left( \frac{\sin(l_{cnp}-l)}{\tan b_{cnp}  \, \cos b -\sin b \, \cos(l_{cnp}-l)} \right) ,
    \label{eq:galproj}
\end{equation}

\noindent where $l$ and $b$ are the Galactic longitude and latitude in radians of a given pixel. We use the two-arguments $\arctan2$ function in Python to lift the quadrant degeneracy of $\arctan$ when dividing the Stokes~$U$ and $Q$ parameters. The coordinates $l_{cnp}$ and $b_{cnp}$ of the Celestial North Pole are 2.145~radians ($122.9^\circ$) and 0.473~radians ($27.1^\circ$), respectively. 

The polarization intensity $I_p$ should remain the same in both reference frames, so we have:

\begin{equation}
    I_p = \sqrt{Q_{icrs}^2 + U_{icrs}^2} = \sqrt{Q_{gcs}^2 + U_{gcs}^2} \, .
\end{equation}

\noindent Knowing the rotated polarization angle $\theta_{gcs}$, we can then use the definition of the Stokes parameters from Section~\ref{sub:equations} to find $Q_{gcs} = I_p \cos \left( 2 \, \theta_{gcs}\right)$ and $U_{gcs} = I_p \sin \left( 2 \, \theta_{gcs} \right)$. The Stokes~$I$ total intensity also does not depend on the reference frame ($I_{gcs} = I_{icrs}$), and the same can be assumed for the error maps of each Stokes parameter ($\delta I$, $\delta Q$, $\delta U$). 

With the Stokes~$I_{gcs}$, $Q_{gcs}$, and $U_{gcs}$ parameters, we can use Equations~\ref{eq:pol} through \ref{eq:sig_theta} to recalculate the polarization properties of the bones, such as the de-biased polarization fraction~$P$, the polarization angle~$\theta$, the magnetic field angle~$\theta_B$, and their related uncertainties. While they are not directly used in this paper, we nonetheless run the original data products in the ICRS reference frame through the same equations to provide data products with polarization properties calculated following an identical procedure in each coordinate system. These files are all available through the survey's data repository. 

\subsection{Planck}
\label{apx:gal_reproj_Planck}

Polarization angles from \textit{Planck} observations are measured clockwise from Galactic North to West \citep[][]{Planck2015_XIX} in a similar manner to the COSMO/HEALPix convention \citep{Gorski2005}, which is opposite to the International Astronomical Union's (IAU) convention of measuring polarization counter-clockwise from North to East \citep[e.g.,][]{HamakerBregman1996}. While this work uses the same GCS frame as the \textit{Planck} data, we follow the IAU convention for measuring polarization angles from Galactic North to East. This results in a sign difference between the Stokes~$U_{cmb}$ parameters from the \textit{Planck} archival data and the Stokes~$U_P$ parameters from this work, assuming the units for the \textit{Planck} data have already been converted using the values from Table~6 of \citet{Planck2014_IX}. 

As before, the polarized intensity~$I_p$ is unaffected by the choice of reference frame: 

\begin{equation}
    I_p = \sqrt{Q_{cmb}^2 + U_{cmb}^2} = \sqrt{Q_P^2 + U_P^2} ,
\end{equation}

\noindent and the polarization angle~$\psi$ for \textit{Planck} in the COSMO convention is defined using the Stokes~$Q_{cmb}$ and $U_{cmb}$ parameters as follows: 

\begin{equation}
    \psi = \frac{180^\circ}{2\pi} \arctan \left( \frac{Q_{cmb}}{U_{cmb}} \right) .
\end{equation}

\noindent Since the only difference between the two reference frames is the direction from which the angle is measured relative to Galactic North, the relation between the polarization angle $\psi$ in the COSMO convention and the polarization angle $\theta_P$ in the GCS reference frame for \textit{Planck} is simply given by: 

\begin{equation}
    \theta_P = \frac{180^\circ}{\pi}  \arctan \left( \frac{-\sin \psi}{\cos \psi} \right) ,
\end{equation}

\noindent with the Stokes parameters $Q_P = I_p \cos 2 \, \theta_P$ and $U_P = I_p \sin 2 \, \theta_P$. Using these equations, we easily verify that only the sign of Stokes~$U$ changes between the two measurement conventions. In each case, as with Equation~\ref{eq:galproj}, we use the $\arctan2$ function in Python to avoid the inherent quadrant degeneracy of $\arctan$. Finally, with Stokes~$Q_P$ and $U_P$ in the GCS frame, the polarization properties for the \textit{Planck} data are recalculated using Equations~\ref{eq:pol} through \ref{eq:sig_theta}.

\section{Comparison with \textit{Herschel} data}
\label{apx:herschel_comp}

As noted in Section~\ref{sub:herschel} and in Appendix~\ref{apx:modes}, the removal of the atmospheric background from HAWC+ observations creates a discrepancy between the background fluxes measured by SOFIA and \textit{Herschel}. In this appendix, we use the maps of hydrogen column density~$N_{H_2}$ and dust temperature~$T_d$ derived from \textit{Herschel} observations for the FIELDMAPS bones to quantify this difference in the measured flux densities at 214~$\mu$m between the two observatories.  

The flux density~$I_\nu$ at a frequency~$\nu$ for dust thermal emission is obtained from the following modified blackbody equation:

\begin{equation}
    I_\nu = N_d \, \kappa_\nu  \, B_\nu (T_d) \, ,
    \label{eq:Ipredicted} 
\end{equation}

\noindent where $N_d$ is the dust column density, $\kappa_\nu$ is the dust opacity at frequency~$\nu$ \citep[e.g.,][]{Ossenkopf1994}, and $B_\nu (T_d)$ is Planck's law for a blackbody for a dust temperature~$T_d$. In the far-infrared, the dust opacity~$\kappa_\nu$ can be expressed as a power-law \citep[][]{Hildebrand1983}:

\begin{equation}
    \kappa_\nu = \kappa_0 \left( \frac{\nu}{\nu_0} \right)^\beta \, ,
    \label{eq:opacity} 
\end{equation}

\noindent where $\kappa_0$ is the opacity at the reference frequency~$\nu_0$, and $\beta$ is the spectral index of emissivity. The relation between the column densities of the dust~$N_d$ (in g cm$^{-2}$) and the molecular hydrogen~$N_{H_2}$ (in cm$^{-2}$) is given by:

\begin{equation}
    N_d = \frac{\mu \, m_H \, N_{H_2}}{R_{g/d}} \, ,
    \label{eq:dustcolumn}
\end{equation}

\noindent where $\mu$ is the mean molecular weight of the gas, $m_H$ is the mass of the hydrogen atom, and $R_{g/d}$ is the gas-to-dust ratio. For this work, we use the same parameters as Appendix~B of \citet{Wang2015}, which are a mean molecular weight~$\mu$ of~2.8 \citep[][]{Kauffmann2008}, a gas-to-dust ratio~$R_{g/d}$ of~100, a reference dust opacity~$\kappa_0$ of~4.0~cm$^{2}$~g$^{-1}$ at a frequency~$\nu_0$ of 505~GHz, and a spectral index of emissivity~$\beta$ of~1.75.

\begin{figure*}[ht]
    \centering
    \includegraphics[width=0.495\textwidth]{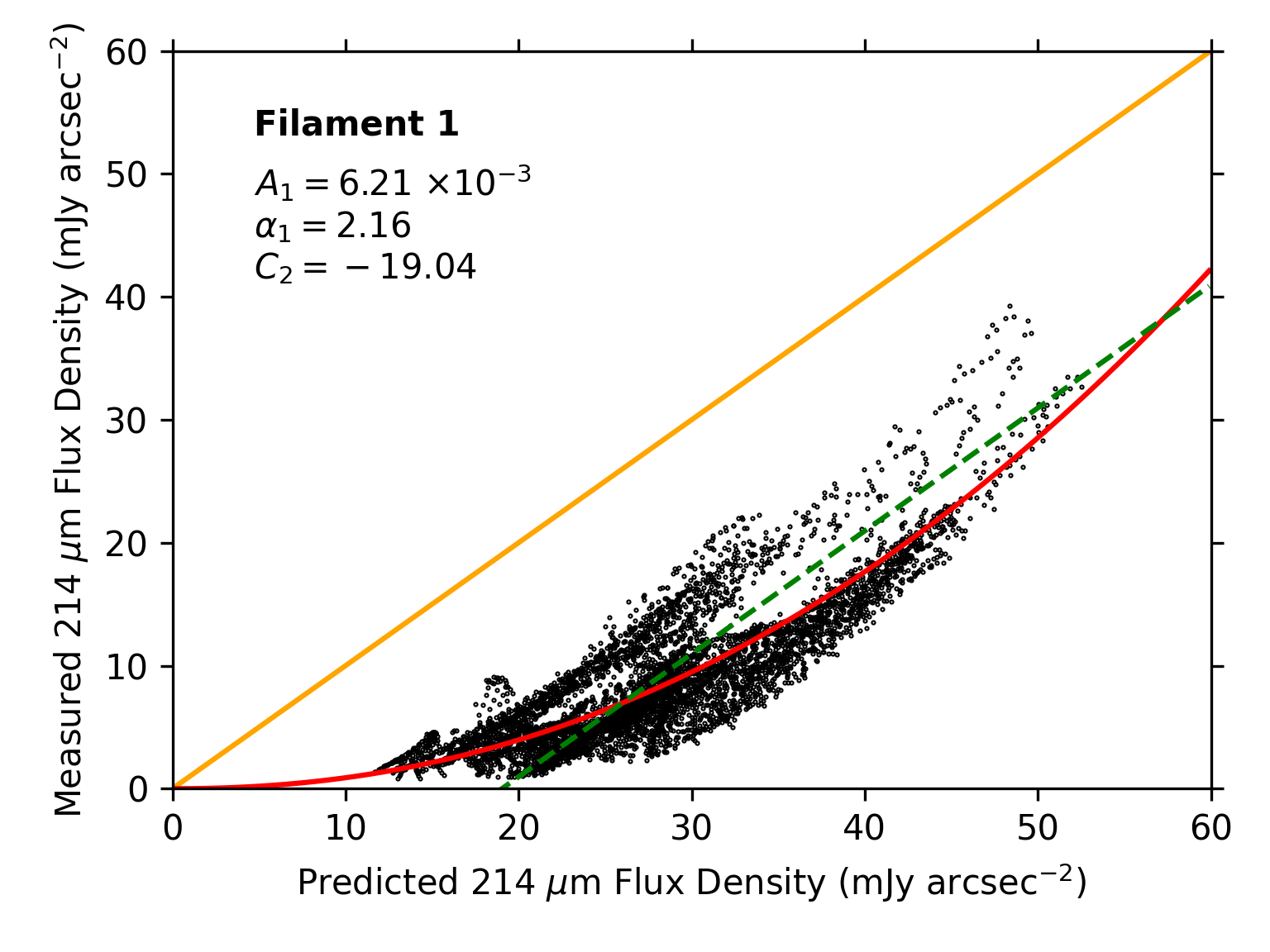}
    \includegraphics[width=0.495\textwidth]{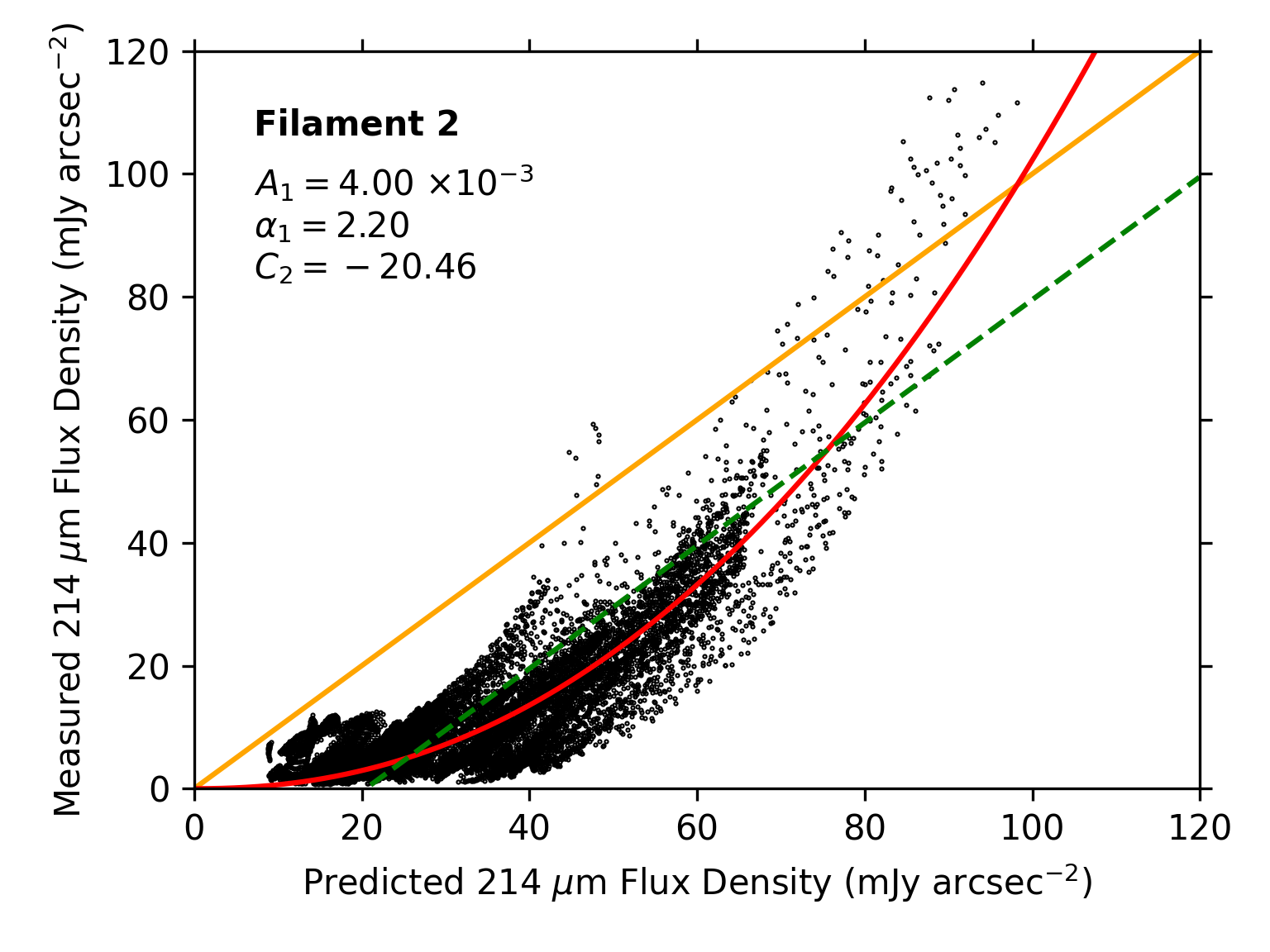}
    \includegraphics[width=0.495\textwidth]{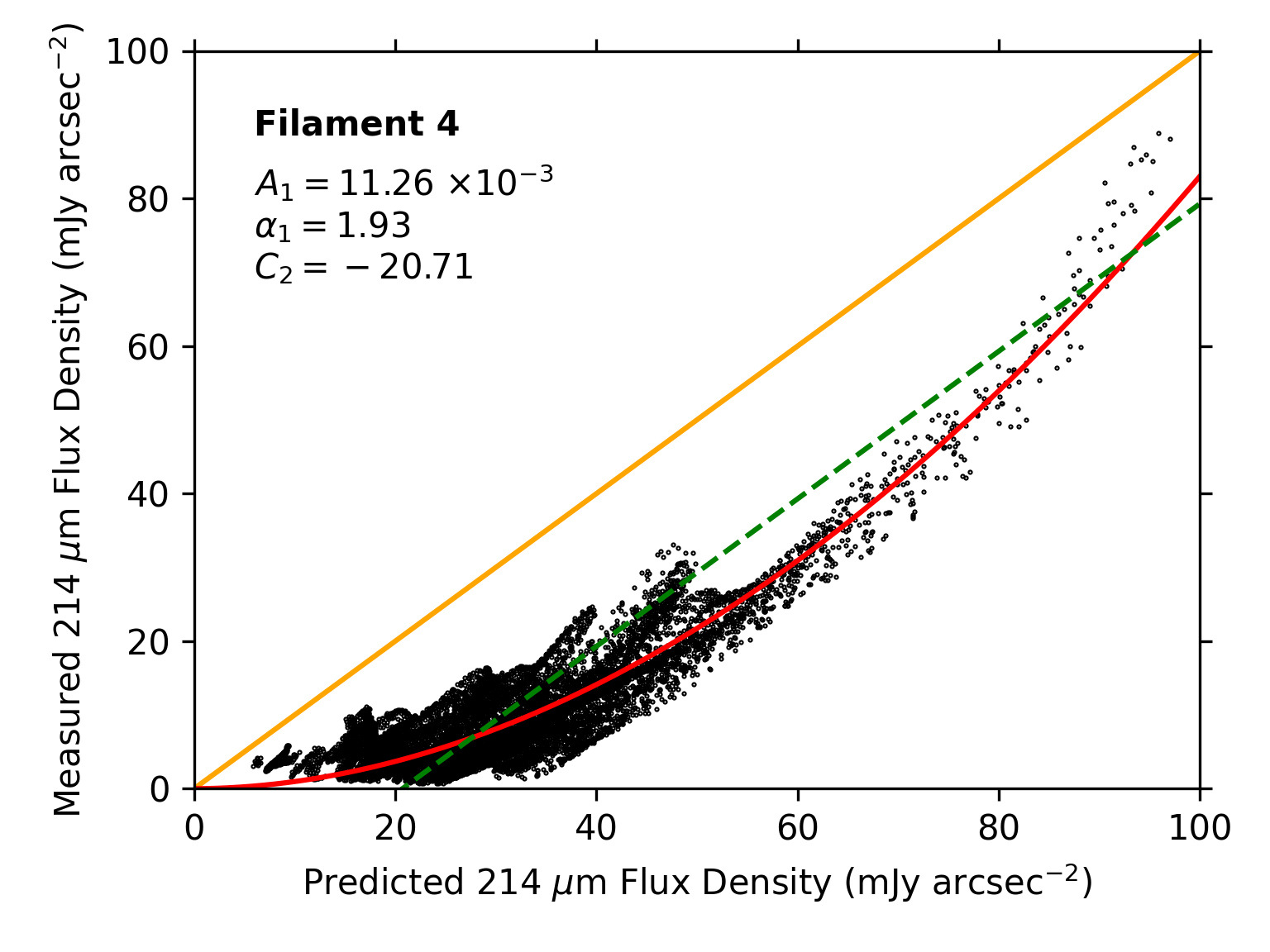}
    \includegraphics[width=0.495\textwidth]{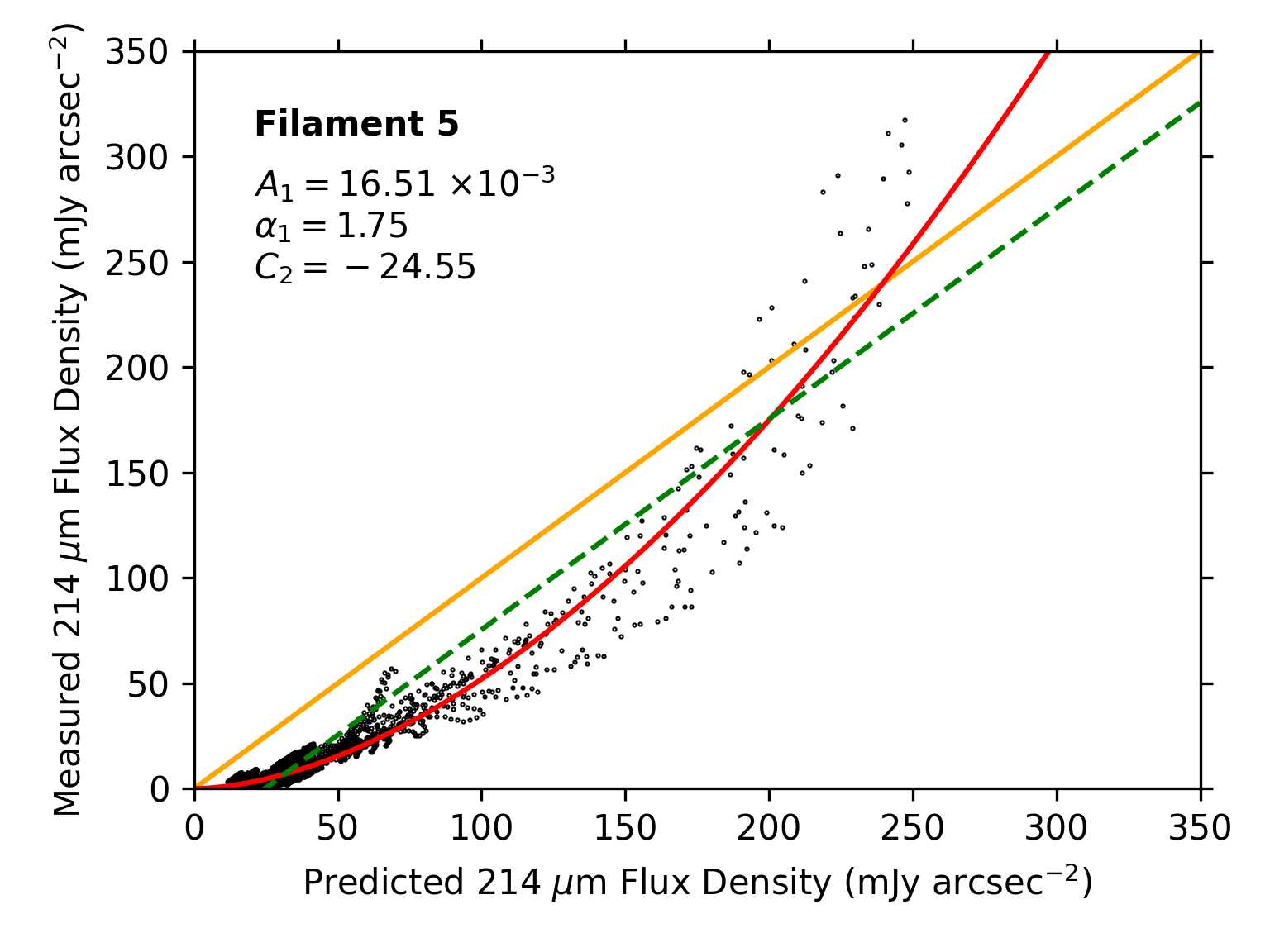}
    \includegraphics[width=0.495\textwidth]{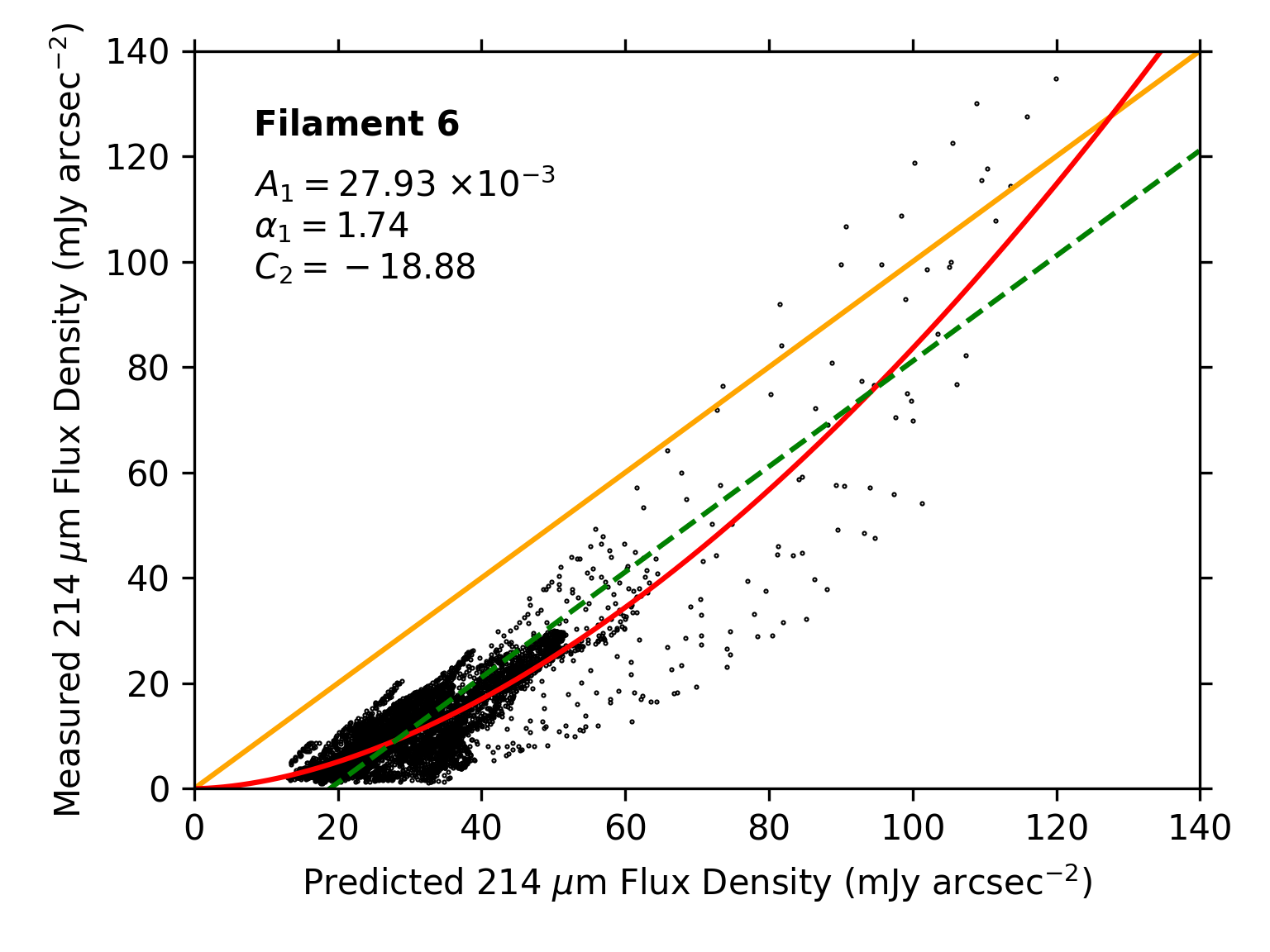}
    \includegraphics[width=0.495\textwidth]{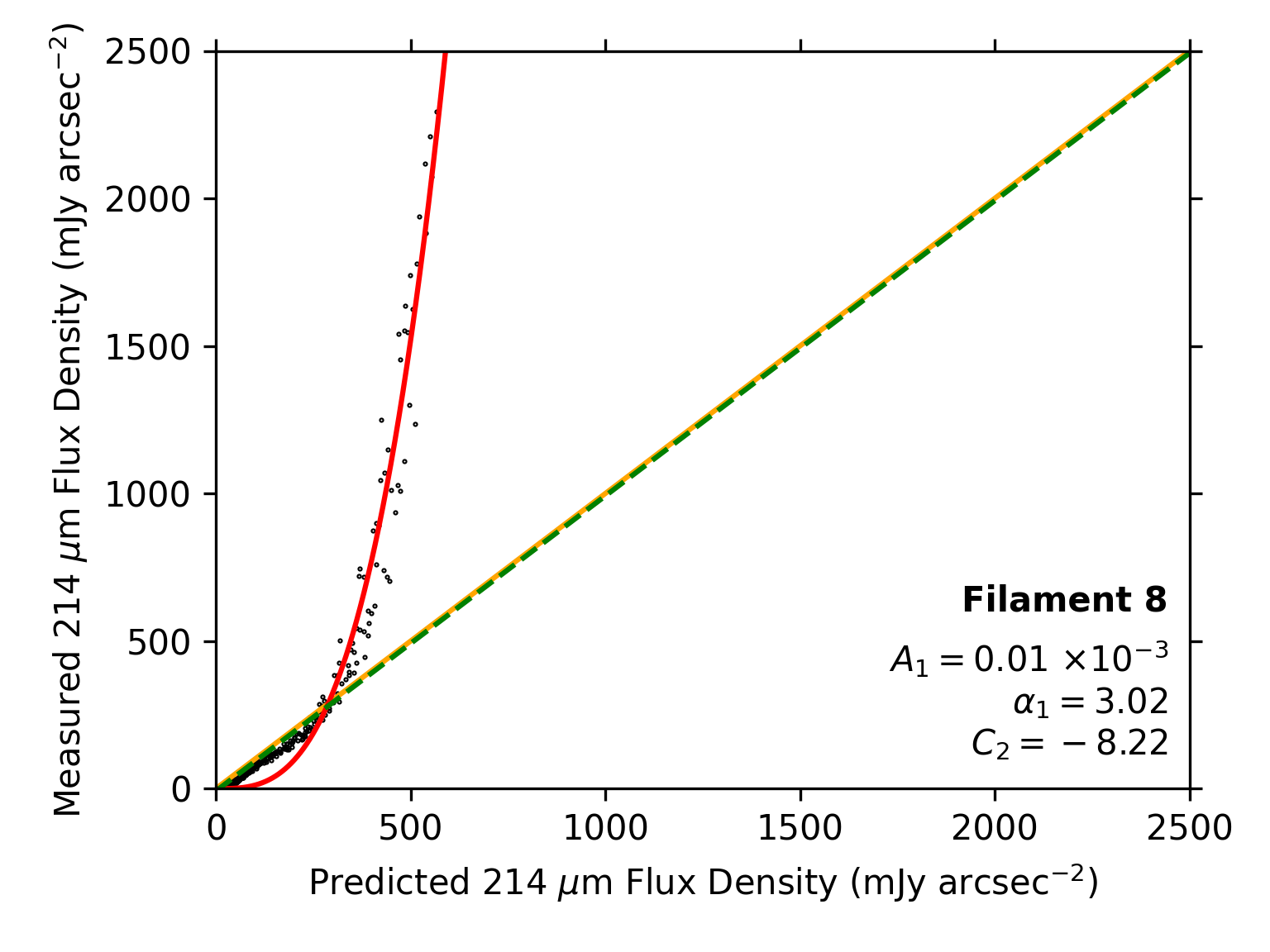}
    \caption{Comparison between the measured and predicted flux densities at 214~$\mu$m for each FIELDMAPS bone. Continued on Figure~\ref{fig:Fil10-G49_IvIest}. The measured flux densities are taken from the HAWC+ Stokes~$I$ total intensity maps shown in the bottom panels of Figures~\ref{fig:Fil1_Maps} to \ref{fig:G49_Maps}. The predicted flux densities are calculated from Equation~\ref{eq:Ipredicted} using the \textit{Herschel}-derived maps of hydrogen column density~$N_{H_2}$ and dust temperature $T_d$. The full orange line shows a 1:1 relation between the predicted~$I_{est}$ and measured~$I_m$ flux densities, the full red line shows a power-law fit of the form $I_{m} = A_1 \, I_{est}^{\alpha_1}$, and the dashed green line shows a constant linear regression $I_{m} = I_{est} + C_2$.}
    \label{fig:Fil1-5_IvIest}
\end{figure*}

\begin{figure*}[ht]
    \centering
    \includegraphics[width=0.495\textwidth]{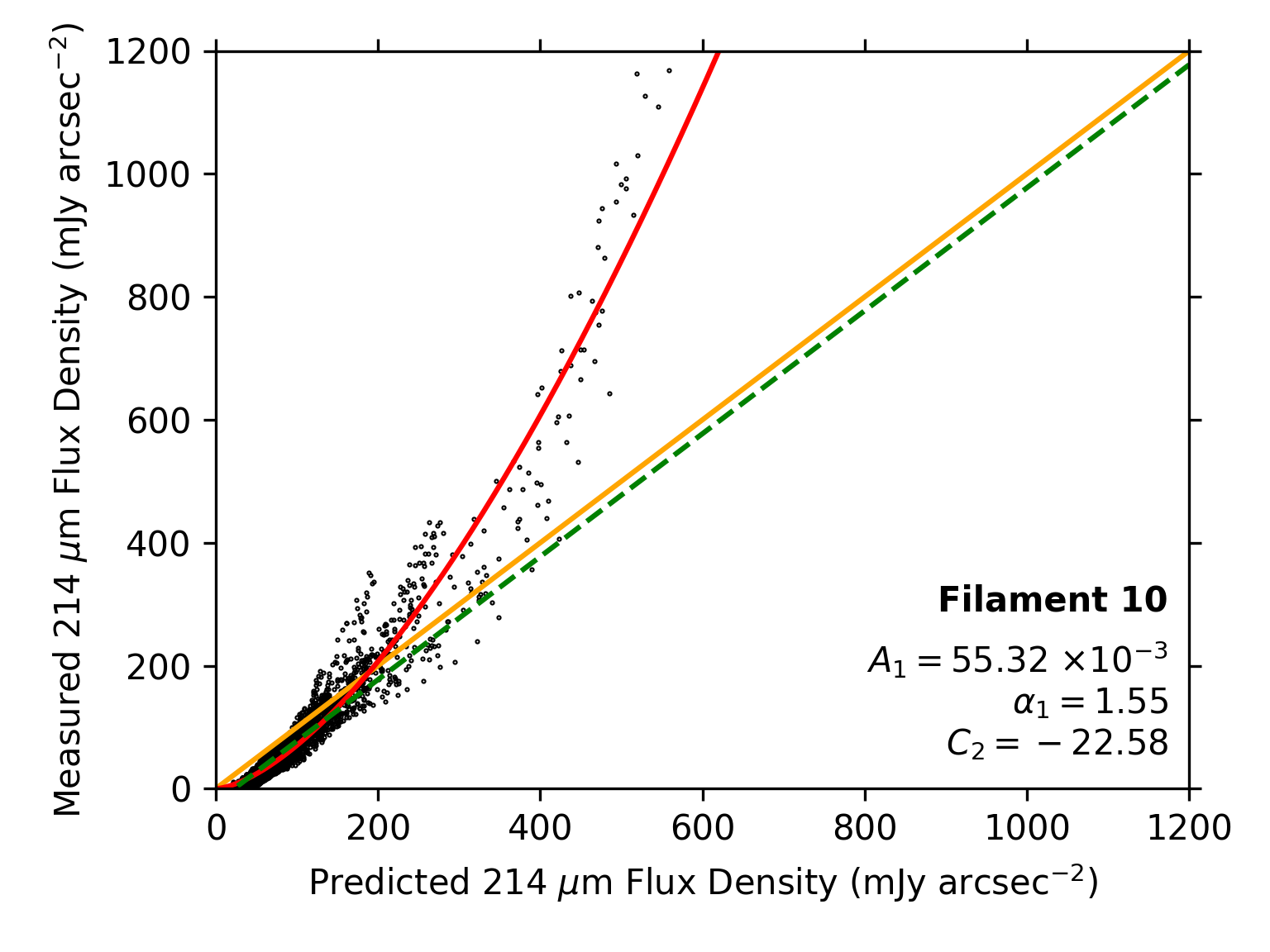}
    \includegraphics[width=0.495\textwidth]{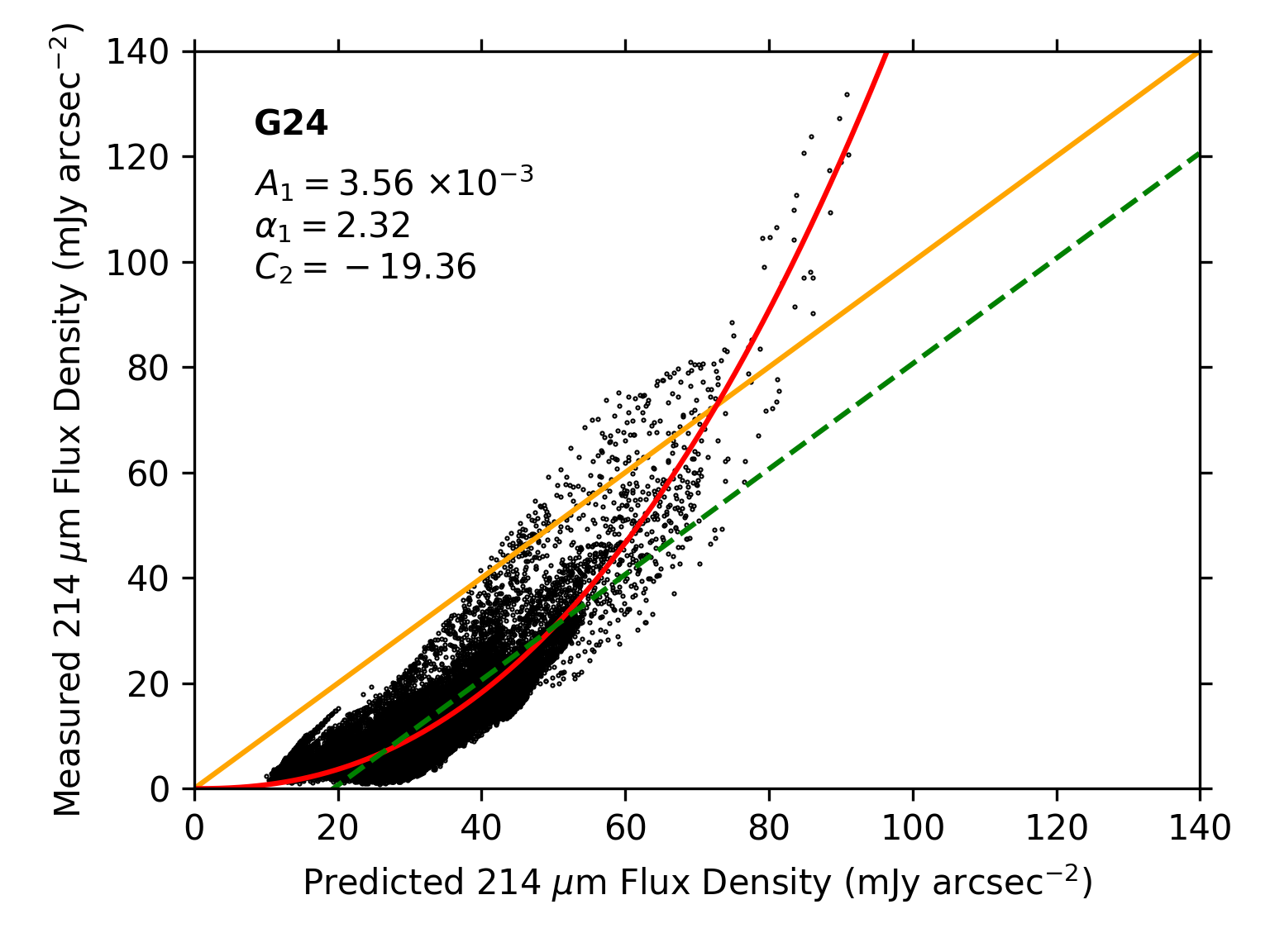}
    \includegraphics[width=0.495\textwidth]{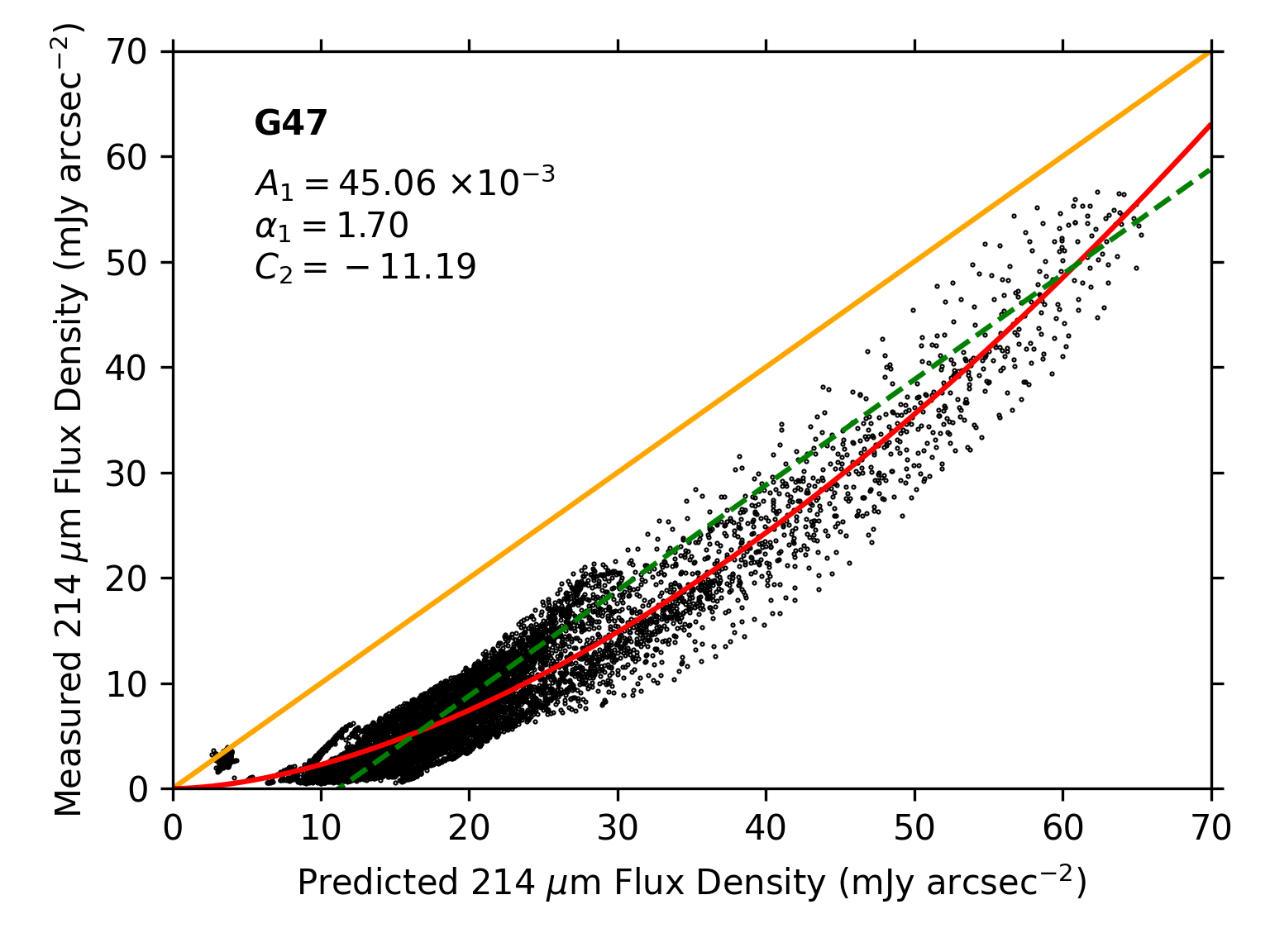}
    \includegraphics[width=0.495\textwidth]{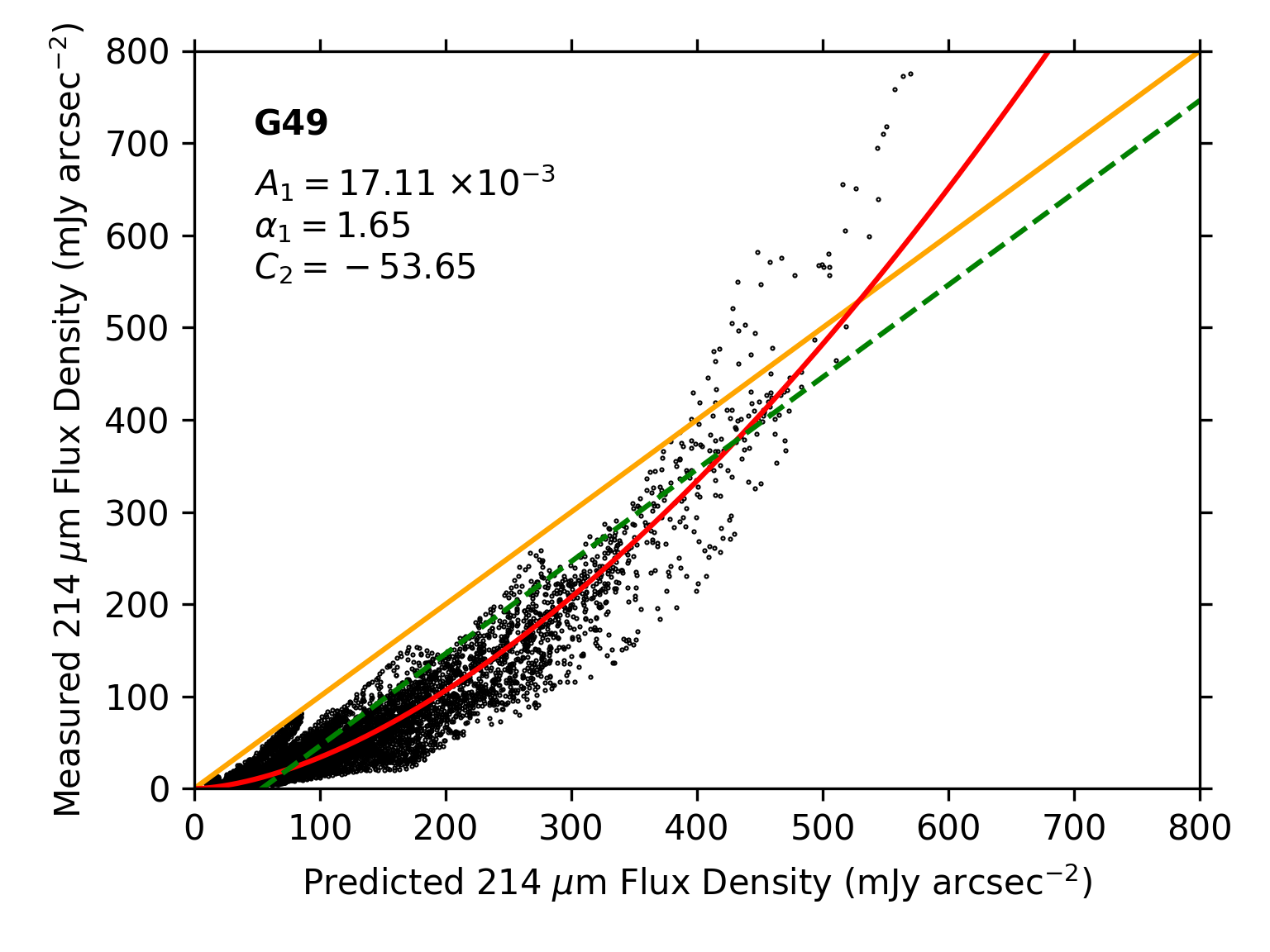}
    \caption{Comparison between the measured and predicted flux densities at 214~$\mu$m for each FIELDMAPS bone. Continued from Figure~\ref{fig:Fil1-5_IvIest}. The measured flux densities are taken from the HAWC+ Stokes~$I$ total intensity maps shown in the bottom panels of Figures~\ref{fig:Fil1_Maps} to \ref{fig:G49_Maps}. The predicted flux densities are calculated from Equation~\ref{eq:Ipredicted} using the \textit{Herschel}-derived maps of hydrogen column density~$N_{H_2}$ and dust temperature $T_d$. The full orange line shows a 1:1 relation between the predicted~$I_{est}$ and measured~$I_m$ flux densities, the full red line shows a power-law fit of the form $I_{m} = A_1 \, I_{est}^{\alpha_1}$, and the dashed green line shows a constant linear regression $I_{m} = I_{est} + C_2$. }
    \label{fig:Fil10-G49_IvIest}
\end{figure*}

For each bone, using Equations~\ref{eq:Ipredicted} to \ref{eq:dustcolumn}, we create a map of predicted flux densities at 214~$\mu$m from the \textit{Herschel}-derived maps of hydrogen column density~$N_{H_2}$ and dust temperature~$T_d$. These maps are re-gridded using the \textsc{reproject} Python package to match the Stokes~$I$ maps of the HAWC+ observations. Figures~\ref{fig:Fil1-5_IvIest} and \ref{fig:Fil10-G49_IvIest} show the comparison between the HAWC+ measured and \textit{Herschel} predicted flux densities at 214~$\mu$m for each pixel with a detection of polarization (i.e., $I/\delta_I > 10$, $P/\delta_P > 3$, and $P<30\%$). The pixels that pass the selection criteria for polarization are the most relevant in the context of this work, but they may not fully cover the range of differences between SOFIA and \textit{Herschel} flux measurements.

In each panel of Figures~\ref{fig:Fil1-5_IvIest} and \ref{fig:Fil10-G49_IvIest}, we plot three relations between the measured and predicted flux densities. First, we show the 1:1 relation for reference, a power-law fit, and a linear relation assuming a constant offset between the two sets of data. Overall, we find that the predicted \textit{Herschel} flux densities tend to be larger than those measured with HAWC+ at lower values, as expected for a difference in the background levels. However, a constant flux offset is insufficient to fully describe the differences between the data, as the flux densities are closer together at higher levels. In some cases, the flux densities measured with HAWC+ are even significantly larger than those predicted from the \textit{Herschel} data. This behavior could be explained by the dilution of sources with high fluxes by the larger \textit{Herschel} beam ($36\farcs4$, see Section~\ref{sub:herschel}), or by the inaccuracy of fitting the dust thermal emission using the properties of a single dust population and a fixed spectral index~$\beta$. Additionally, while a power law can reproduce the relation between SOFIA and \textit{Herschel} at higher flux densities, there is significant variation in the measured spectral indices~$\alpha_1$, which means that such a simple relation is also insufficient to systematically describe the discrepancies between the two observatories. 

\section{Debiasing HAWC+ Data}
\label{apx:debias}

In Section~\ref{sub:equations}, we note the need to debias the polarized intensity $I_p$. The polarized intensity $I_p$ follows a Rice distribution that has a positive bias, which is more significant when the signal-to-noise ratio is low \citep{Serkowski1958,Wardle1974,Pattle_2019}. From Equation~\ref{eq:debiasedIp}, we therefore expressed the debiased polarized intensity $I'_P$ as:

\begin{equation}
    I'_p = \sqrt{ I_p^2-\delta_{I_p}^2 } = \sqrt{Q^2+U^2-\delta_{I_p}^2 } , 
    \label{Apxeq:debiasIp1}
\end{equation}

\noindent with the uncertainty $\delta_{I_p}$:

\begin{equation}
    \delta_{I_p} = \sqrt{ \frac{(Q \, \delta_Q)^2+(U \, \delta_U)^2}{Q^2+U^2}} \, , 
    \label{Apxeq:sigIp}
\end{equation}

\noindent where $\delta_Q$ and $\delta_U$ are the uncertainties for Stokes~$Q$ and $U$, respectively, and assuming a propagation of non-correlated errors. 

However, an alternative approach is given by \citet{Gordon2018} and \citet{clarke_2022_Manual} for debiasing HAWC+ data. Specifically, using the notation from \citet{Gordon2018}, the biased percent polarization~$p$ is given by:

\begin{equation}
    p = 100 \% \, \sqrt{\left(\frac{Q}{I} \right)^2 + \left( \frac{U}{I} \right)^2} \, ,
    \label{Apxeq:p}
\end{equation}

\noindent with uncertainty $\delta_p$:

\begin{equation}
    \delta_p = \frac{100 \%}{I} \, \sqrt{\left[ \frac{(Q \, \delta_Q)^2+(U \, \delta_U)^2}{Q^2+U^2} \right]  + \left[\left(\frac{Q}{I} \right)^2 + \left( \frac{U}{I} \right)^2 \right] \delta_I^2} = 100 \% \: \frac{I_p}{I} \, \sqrt{ \left(\frac{\delta_{I_p}}{I_p}\right)^2  +  \left(\frac{\delta_I}{I} \right)^2} \, ,
    \label{Apxeq:sigp}
\end{equation}

\noindent with $\delta_I$ being the uncertainty on Stokes~$I$, and assuming the cross-terms $\delta_{QU}$, $\delta_{QI}$, and $\delta_{UI}$ are negligible. The debiased percent polarization $p'$ is then:

\begin{equation}
    p' = \sqrt{p^2-\delta_p^2} \, ,
    \label{Apxeq:debiasp1}
\end{equation}

\noindent and, following the notation from \citet{Gordon2018}, the resulting debiased polarized intensity $I_{p'}$ is: 

\begin{equation}
    I_{p'} = \frac{p' \, I}{100 \%} \, ,
    \label{Apxeq:debiasIp2}
\end{equation}

\noindent which is not identical to Equation~\ref{Apxeq:debiasIp1}, due in part to the presence of the signal-to-noise ratio $\delta_I/I$ for the total intensity in Equations~\ref{Apxeq:sigp} and \ref{Apxeq:debiasp1}. 

The extensions of the FITS files produced directly by the HAWC+ pipeline follow the debiasing from Equation~\ref{Apxeq:debiasp1} and \ref{Apxeq:debiasIp2}, while the data products for FIELDMAPS follow the debiasing from Equation~\ref{Apxeq:debiasIp1}. We argue here that both methods are essentially equivalent for the polarization detection thresholds used for this work: $p/\delta_p > 3$, $I/\delta_I > 10$, and $p < 30\%$.

First, the polarization fraction $p$ for dust thermal emission in the Galaxy is generally expected to be below 30~\%, with \textit{Planck} finding a maximum $p_{max}$ of $22.5_{-1.5}^{+3.5}$~\% \citep{Planck2020_XII}, and thus $I_p \leq 0.3 \, I$. Furthermore, due to how Stokes~$I$, $Q$, and $U$ are measured \citep{clarke_2022_Manual}, the uncertainties $\delta_Q$ and $\delta_U$ tend to be slightly larger than $\delta_I$ (see Table~\ref{tab:detections}), which following Equation~\ref{Apxeq:sigIp} leads to $\delta_{I_p} \ge \delta_I$. The $\delta_{I_p}/I_p$ relation can then be approximated by:

\begin{equation}
    \frac{\delta_{I_p}}{I_p} \geq \frac{\delta_I}{0.3 \, I} \, .
\end{equation}

\noindent In this example, if we use lower the limit of 10 for $I/\delta_I$, then we find $I_p/\delta_{I_p} \leq 3$. When squared, the inverse ratio of these quantities gives: 

\begin{equation}
   \left( \frac{\delta_I}{I} \right)^{2} \left( \frac{\delta_{I_p}}{I_p} \right)^{-2}  \leq 0.09 \, .
\end{equation}

\noindent In this extreme example, we find that $(\delta_{I_p}/I_p)^2$ is at least ten times larger than $(\delta_I/I)^2$. However, more generally, lower values of $p$ and a higher signal-to-noise $I/\delta_I$ will exacerbate this difference, and so we can confidently write: 

\begin{equation}
   \left( \frac{\delta_{I_p}}{I_p} \right)^2 \gg \left( \frac{\delta_I}{I} \right)^2 \, .
   \label{Apxeq:gg}
\end{equation}

\noindent Using Equation~\ref{Apxeq:gg}, we can now simplify Equation~\ref{Apxeq:sigp} to become:

\begin{equation}
    \delta_p \approx 100 \% \: \frac{I_p}{I} \, \sqrt{ \left(\frac{\delta_{I_p}}{I_p}\right)^2} = 100 \% \: \frac{\delta_{I_p}}{I} \, ,
\end{equation}

\noindent which when substituted in Equation~\ref{Apxeq:debiasp1} leads to: 

\begin{equation}
    p' = \frac{100 \%}{I} \sqrt{I_p^2-\delta_{I_p}^2} = 100 \% \: \frac{I_p'}{I} = P ,
    \label{Apxeq:debiasp3}
\end{equation}

\noindent where $P$ is the debiased polarization fraction defined in Equation~\ref{eq:pol}. Through the approximation given in Equation~\ref{Apxeq:gg}, we have recovered both the debiased polarized intensity from Equations~\ref{eq:debiasedIp} and \ref{Apxeq:debiasIp1}, but also an equation of the same form as the relation between $p'$ and $I_{p'}$ from Equation~\ref{Apxeq:debiasIp2}. We therefore conclude that, at least for the signal-to-noise thresholds used in this work, both debiasing methods are nearly indistinguishable. 

\section{Angle Dispersion in G49}
\label{apx:G49_regions}

\begin{figure*}
    \centering
    \includegraphics[width=0.495\textwidth]{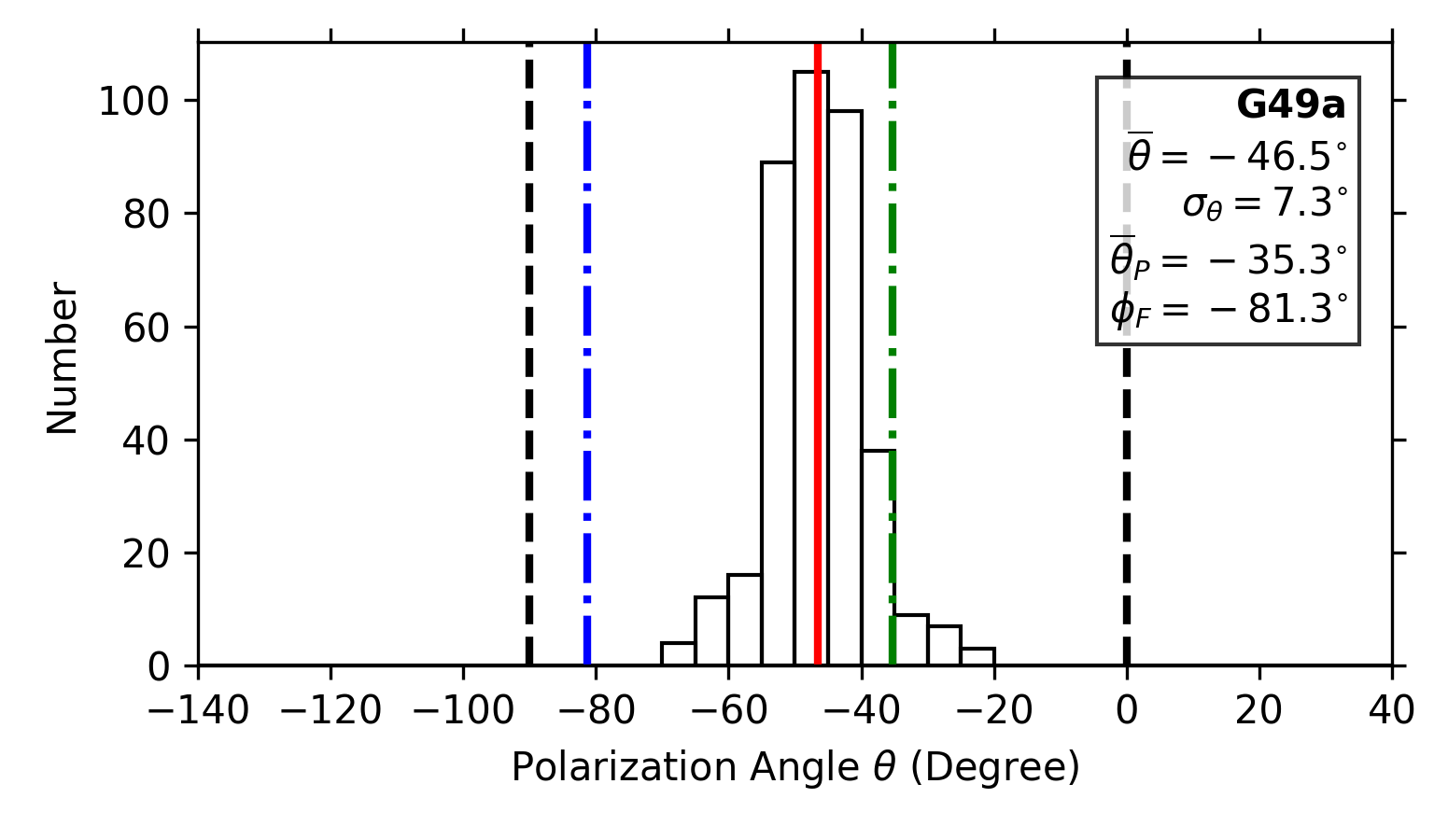}
    \includegraphics[width=0.495\textwidth]{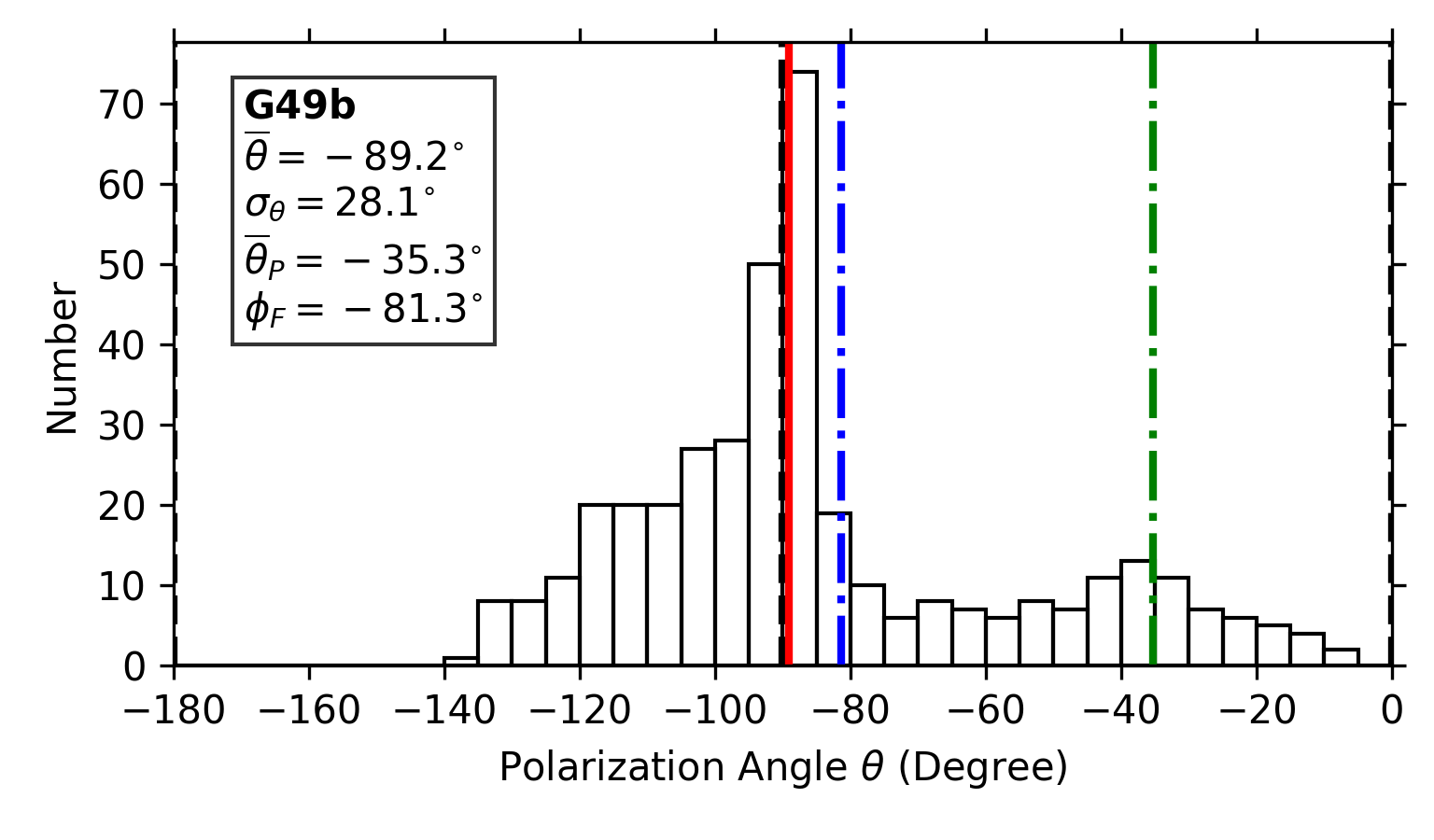}
    \includegraphics[width=0.495\textwidth]{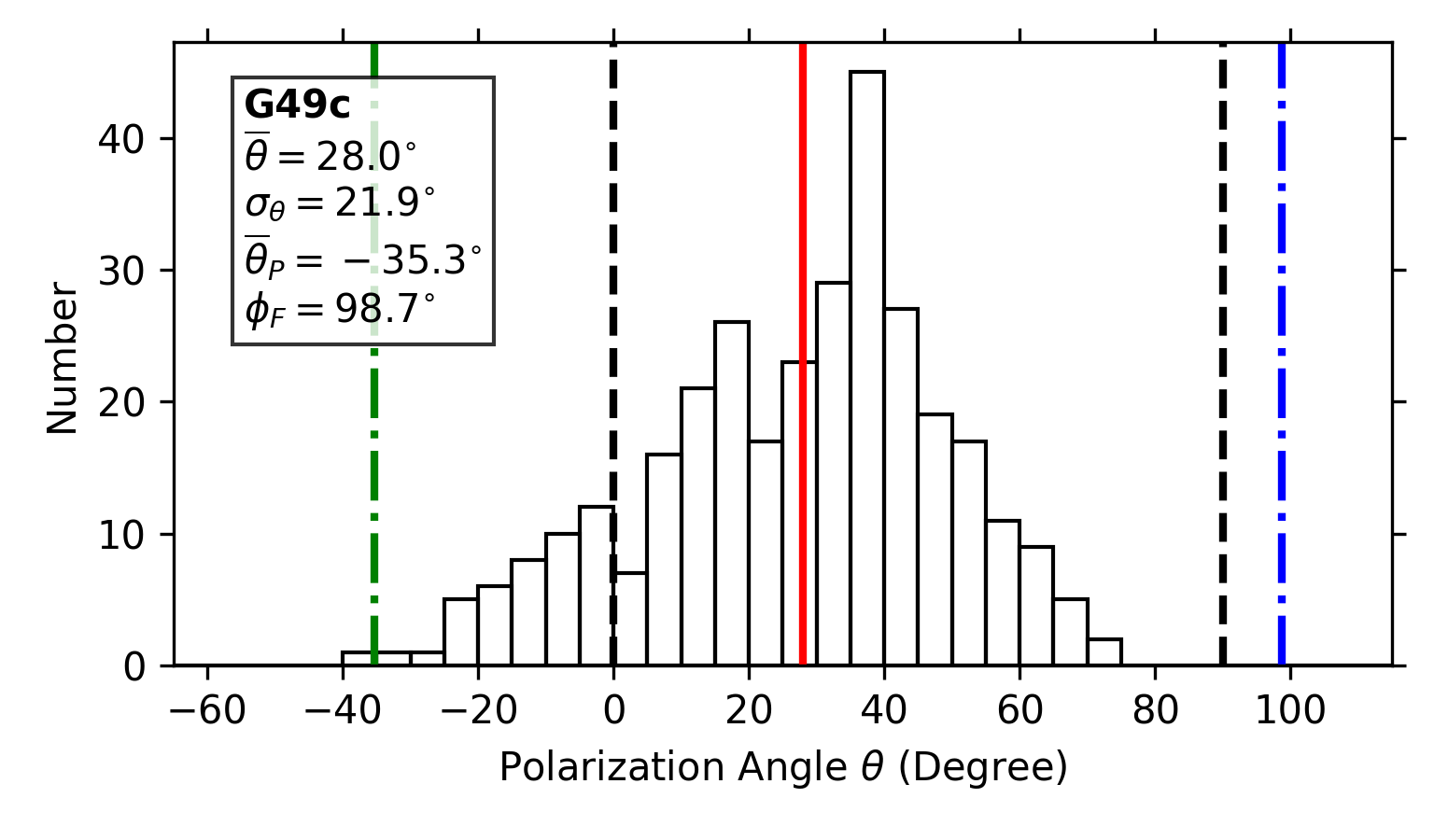}
    \includegraphics[width=0.495\textwidth]{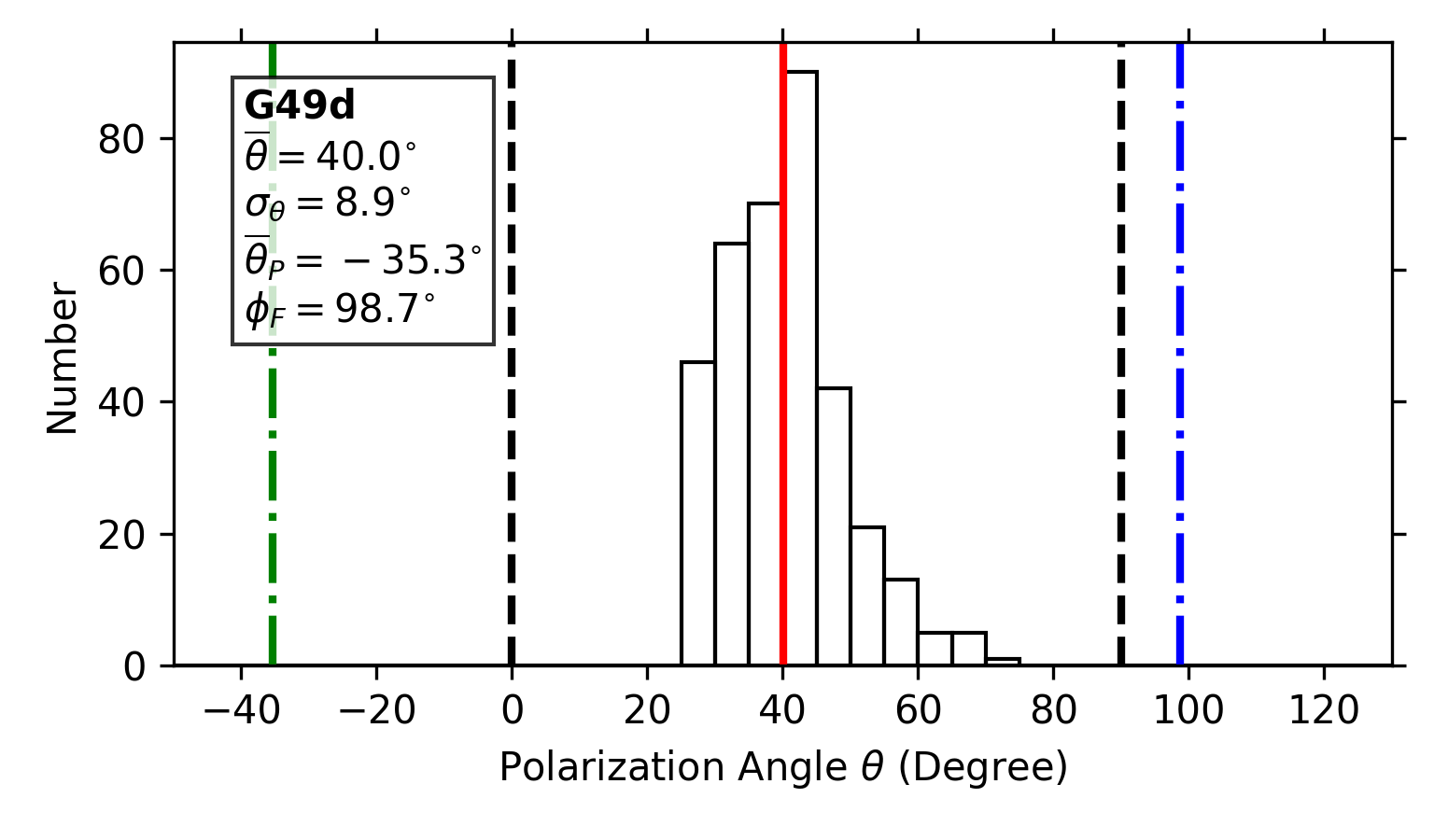}
    \caption{Histograms of polarization angles $\theta$ for the four regions of G49 listed in Table~\ref{tab:G49abcd}. All angles are provided relative to Galactic North. The bin size of the histograms is $5^\circ$. The full red line is the circular mean $\overline{\theta}$ of the distribution, the green dashed-dotted line is the circular mean of the \textit{Planck} polarization angles $\overline{\theta}_P$, and the blue dashed-dotted line is the fitted orientation $\phi_F$ of the filament (see Section~\ref{sub:histograms}). Each histogram is centered on the bin containing the circular mean $\overline{\theta}$. Black dashed lines indicate angles of -90$^\circ$, 0$^\circ$, and 90$^\circ$ for clarity. The measured circular standard deviation $\sigma_\theta$ for each cloud is also given within the insets. The inferred magnetic field orientation $\theta_B$ is obtained from rotating the polarization angle $\theta$ by $90^\circ$.}
    \label{fig:G49abcd_Histo}
\end{figure*}

Figure~\ref{fig:G49abcd_Histo} provides the four histograms used to calculate the circular means and standard deviations given in Table~\ref{tab:G49abcd} of Section~\ref{sub:histograms} for four regions in G49. These regions are $91\arcsec$ by $91\arcsec$ boxes positioned on the four brightest peaks in G49 at 214~$\mu$m, as displayed on the top panel of Figure~\ref{fig:G49_Maps}. The histograms follow the same format as those shown in Figures~\ref{fig:Fil1-5_Histo} and \ref{fig:Fil10-G49_Histo}.

\section{Additional figures}
\label{apx:ext_fil10}

\begin{figure*}
\centering
\includegraphics[scale=0.75]{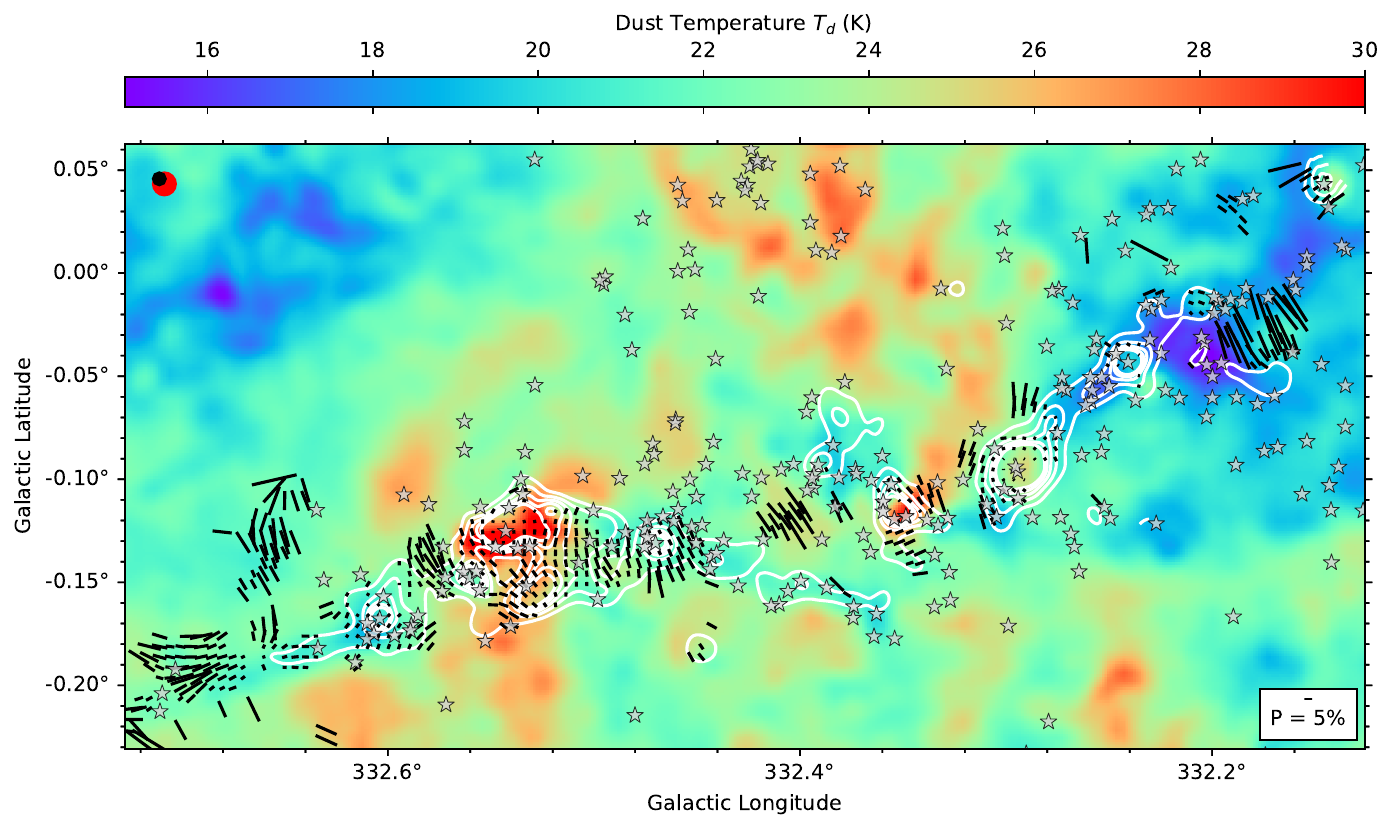}
\caption{The polarization vectors (black) plotted on a color map of dust temperatures~$T_d$ for Filament~10. The vector length shows the polarization fraction~$P$, and only every fourth vector is plotted. The white contours trace reference levels of the Stokes~$I$ total intensity. Known YSOs are identified with star symbols. The circles denote the beam sizes of HAWC+ (black) and \textit{Herschel} (red).}
\label{fig:Fil10_PvTmap}
\end{figure*}

\begin{figure*}
    \centering
    \includegraphics[width=0.495\textwidth]{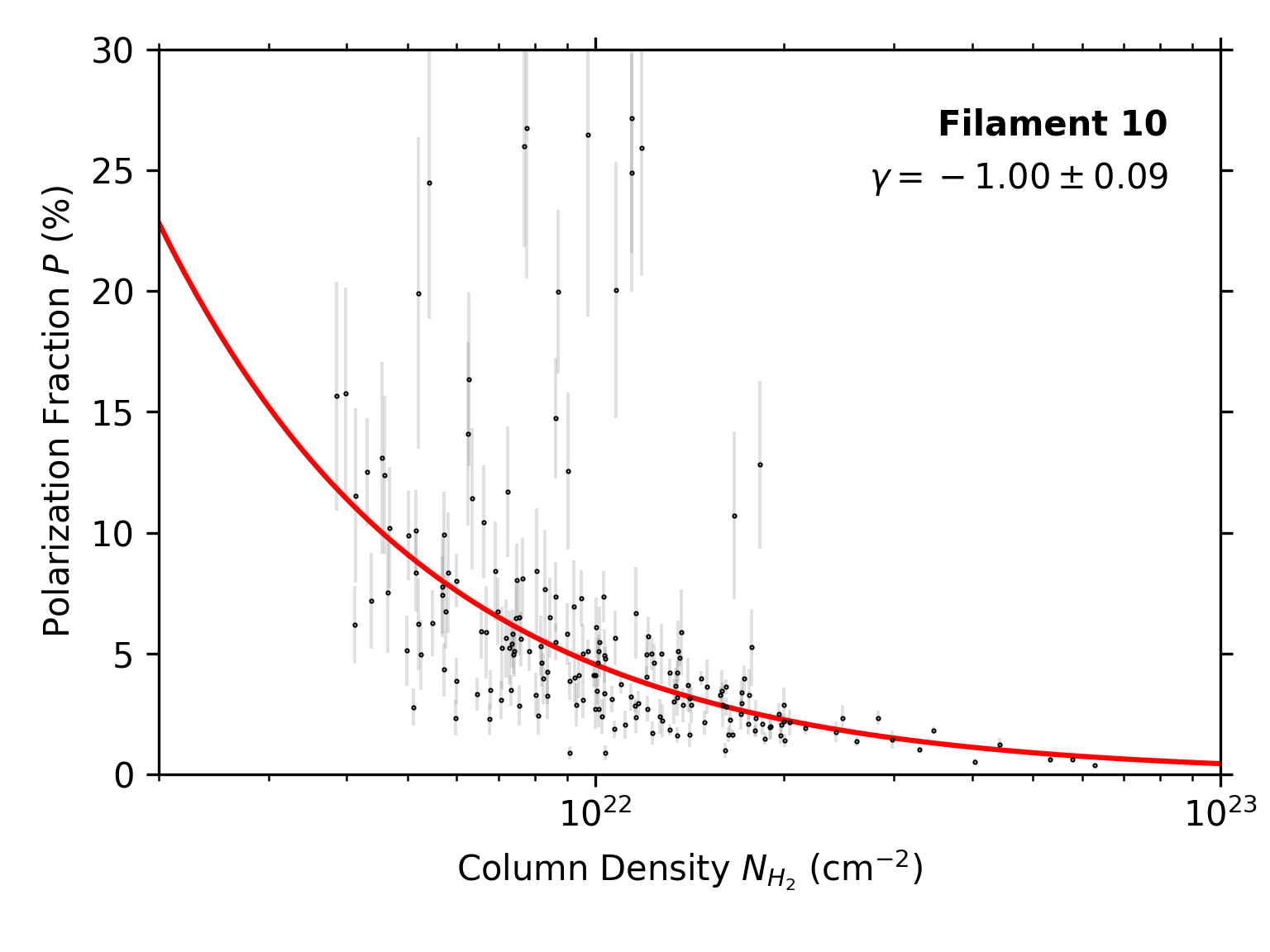}
    \includegraphics[width=0.495\textwidth]{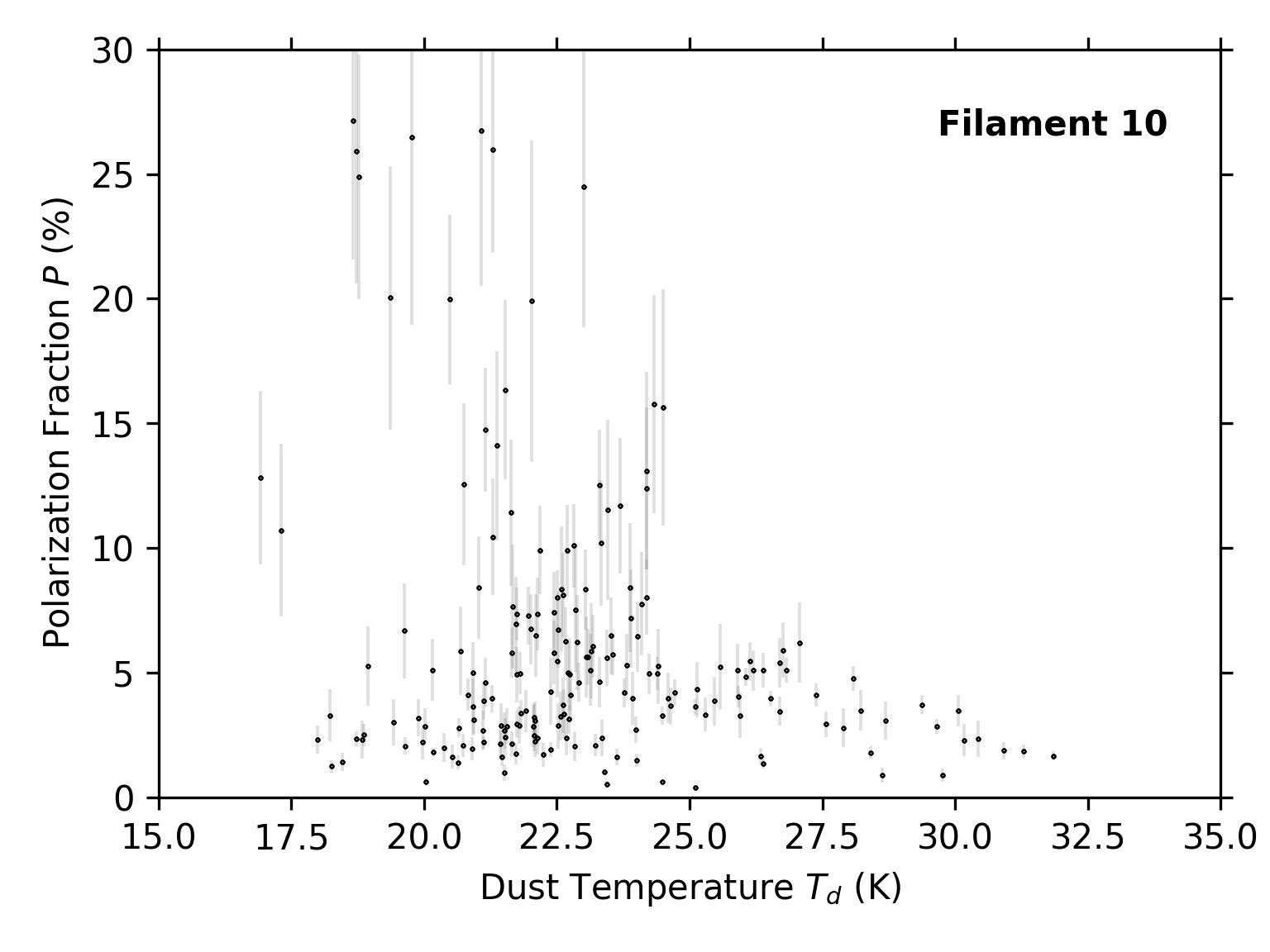}
    \caption{Polarization fraction~$P$ as a function of the gas column density~$N_{H_2}$ (left panel) and the dust temperature~$T_d$ (right panel) for Filament~10. The full red line traces a power-law fit of the form $P \propto N_{H_2} \! ^\gamma$. The HAWC+ data was smoothed and re-projected to share the same resolution and pixel scale as the \textit{Herschel} column density and temperature maps described in Section~\ref{sub:herschel}. Only every other vector from the resulting smoothed catalog is plotted and used for the fit, and the uncertainty in $P$ is shown as gray lines.}
    \label{fig:Fil10_PvN_PvT}
\end{figure*}

\begin{figure*}
    \centering
    \includegraphics[width=0.495\textwidth]{Fil10_PvI_plot_Step4.png}
    \includegraphics[width=0.495\textwidth]{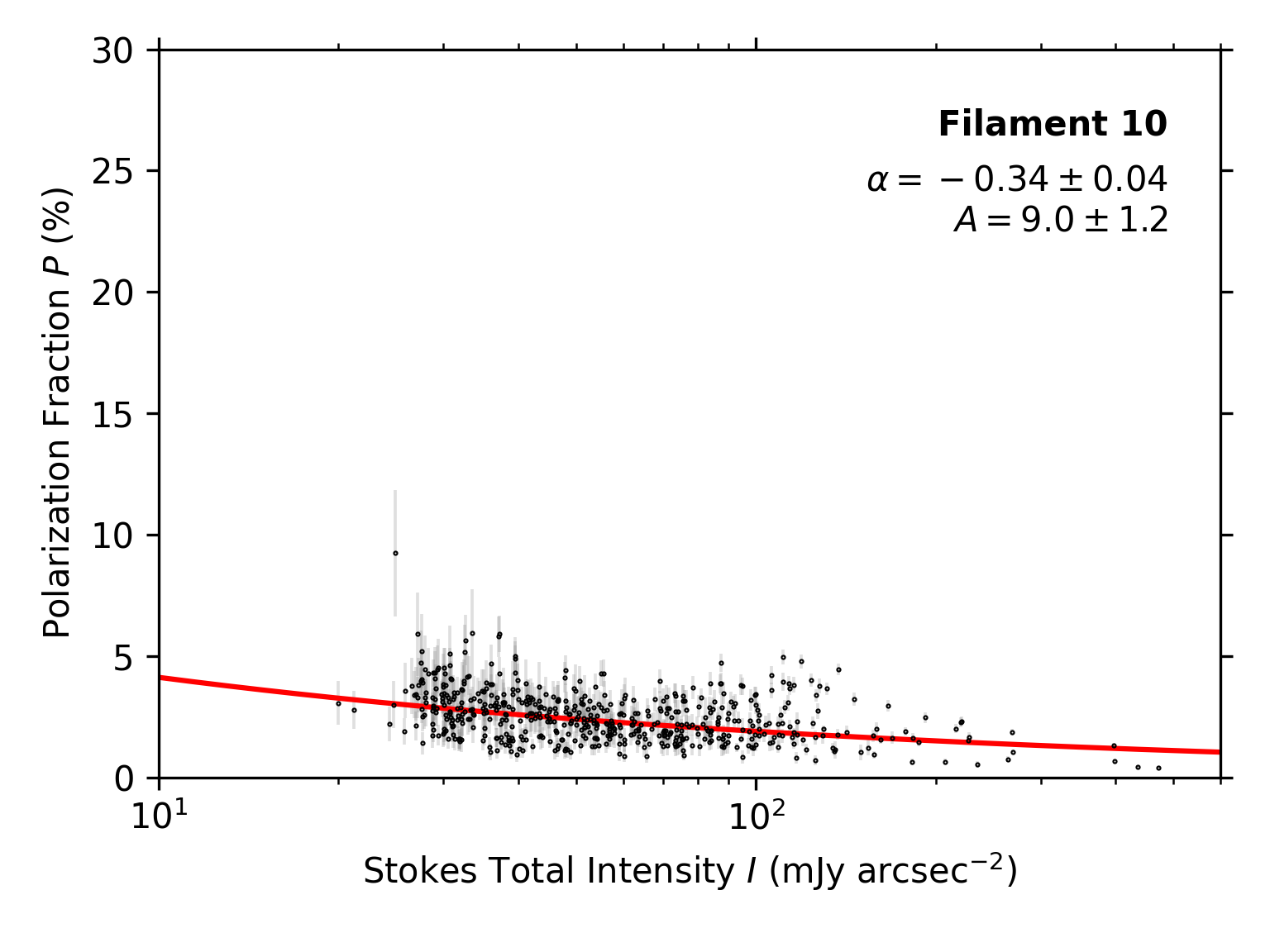}
    \caption{Polarization fraction~$P$ as a function of the Stokes~$I$ total intensity for Filament~10. The left panel shows the same relation as in Figure~\ref{fig:Fil10-G49_PvI} as reference, while the right panel substitutes the Stokes~$I$ total intensity by the \textit{Herschel}-derived flux densities at 214~$\mu$m from Appendix~\ref{apx:herschel_comp}. In each panel, the data is fitted with a power law of the form $P = A \, I^\alpha$. Only every fourth vector is plotted and used for the fit, and the uncertainty in $P$ is shown as gray lines.}
    \label{fig:Fil10_PvI_PvIest}
\end{figure*}

\begin{figure*}
\centering
\includegraphics[scale=0.75]{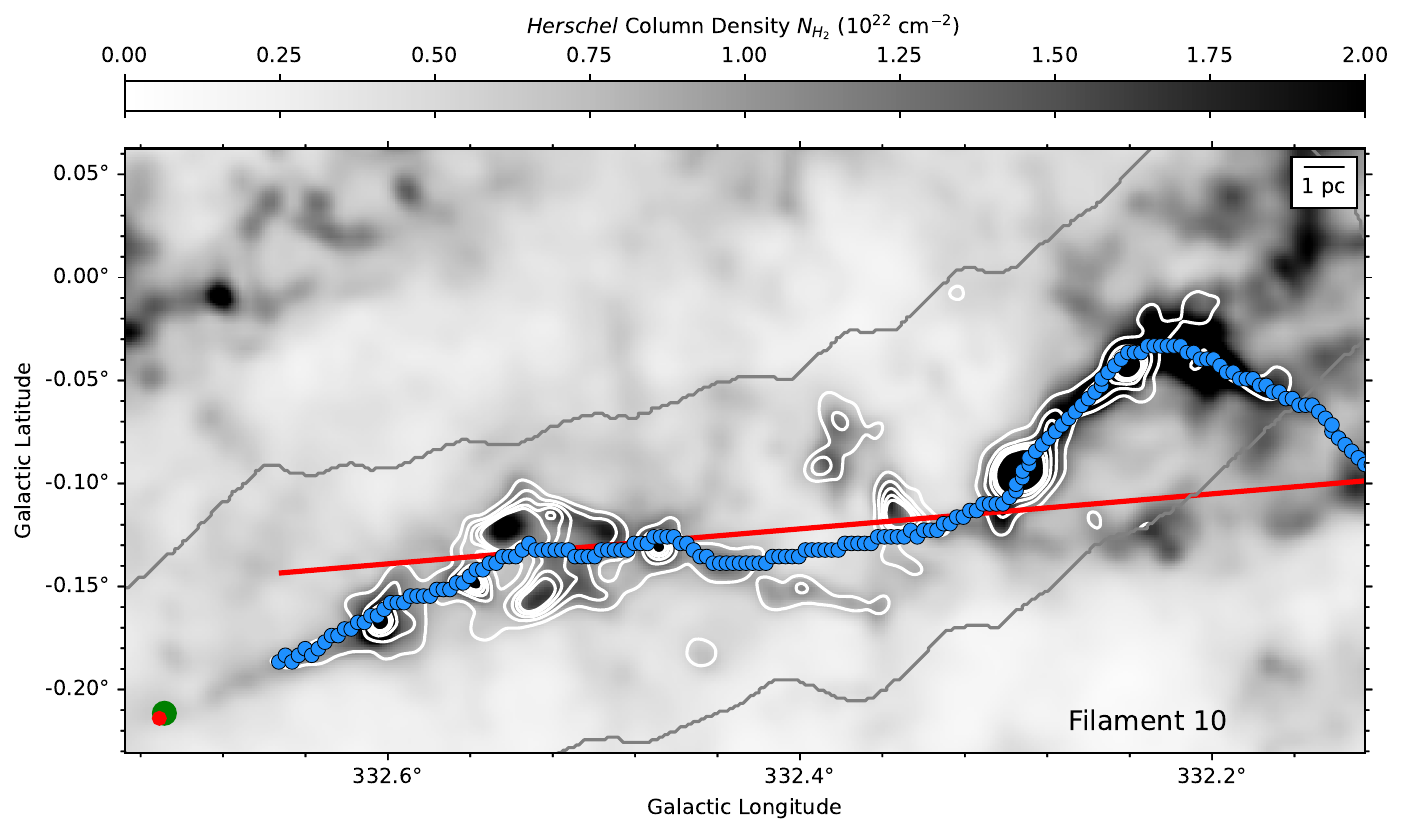}
\caption{Filament orientation for Filament~10. The blue circles show the result the \textsc{FilFinder} algorithm for the spine of the filament. The red line is the linear regression used to calculate the filament orientation~$\phi_F$.}
\label{fig:Fil10_Skeleton}
\end{figure*}

Additional figures for each bone can be found in the online repository for the FIELDMAPS survey. While we could not include all these figures in the paper, we present in this Appendix those created for Filament~10 as an example of their contents. Figure~\ref{fig:Fil10_PvTmap} shows the polarization vectors from the bottom panel of Figure~\ref{fig:Fil10_Maps}, but plotted on the \textit{Herschel}-derived map of dust temperatures~$T_d$ instead. In a similar manner to Figure~\ref{fig:Fil10-G49_PvI}, Figure~\ref{fig:Fil10_PvN_PvT} shows the relations between the polarization fraction~$P$ and the gas column density~$N_{H_2}$, and between the polarization fraction~$P$ and the dust temperature~$T_d$, for Filament 10. In each case, the HAWC+ data was smoothed and re-projected to the resolution and pixel scale of the column density and dust temperature maps. Alternative figures using the same non-smoothed vector catalogs as Figures~\ref{fig:Fil1-5_PvI} and \ref{fig:Fil10-G49_PvI} are also available on Dataverse. A power law of the form $P \propto N_{H_2} \! ^\gamma$ if fitted for the column density. Figure~\ref{fig:Fil10_Skeleton} displays the result of the \textsc{FilFinder} algorithm for Filament~10, as well as the linear fit described in Section~\ref{sub:fil_orient} for the filament orientation~$\phi_F$. In Figure~\ref{fig:Fil10_PvI_PvIest}, we compare the effect of using HAWC+ and \textit{Herschel} flux densities (see Appendix~\ref{apx:herschel_comp}) for the relation between the polarization fraction~$P$ and the Stokes~$I$ total intensity. The left panel of Figure~\ref{fig:Fil10_PvI_PvIest} is a reproduction of the panel for Filament~10 in Figure~\ref{fig:Fil10-G49_PvI}. In each case, we fit a power law of the form $P = A \, I^\alpha$, as described in Section~\ref{sub:PvI}. 

\section{\textit{Spitzer} Images}
\label{apx:Spitzer}

\begin{figure*}
\centering
\includegraphics[scale=0.75]{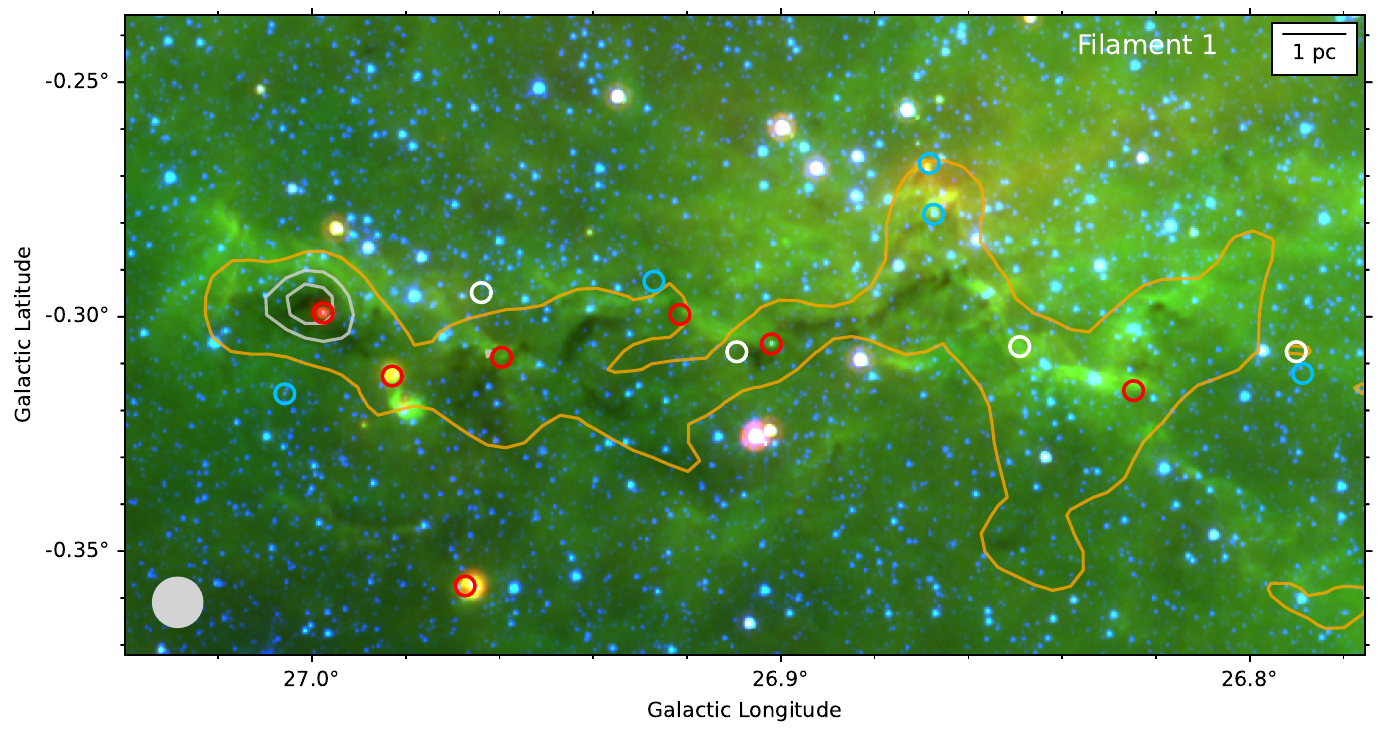}
\caption{\textit{Spitzer} red~(24~$\mu$m), green~(8.0~$\mu$m), and blue~(3.6~$\mu$m) composite image for Filament~1. The color levels were chosen to highlight infrared-dark features. Class~I and Flat Spectrum (red), Class~II (blue), and Class~III (white) YSOs from the literature are identified with circles. The three contours trace \textit{Herschel}-derived $N_{H_2}$~column densities of $0.5$, $1.0$, and $1.5 \times 10^{22}$~cm$^{-2}$, with the lowest level identified in orange for clarity. The \textit{Herschel} beam is given by the gray circle at the bottom left.
\label{fig:Fil1_RGB}}
\end{figure*}

\begin{figure*}
\centering
\includegraphics[scale=0.75]{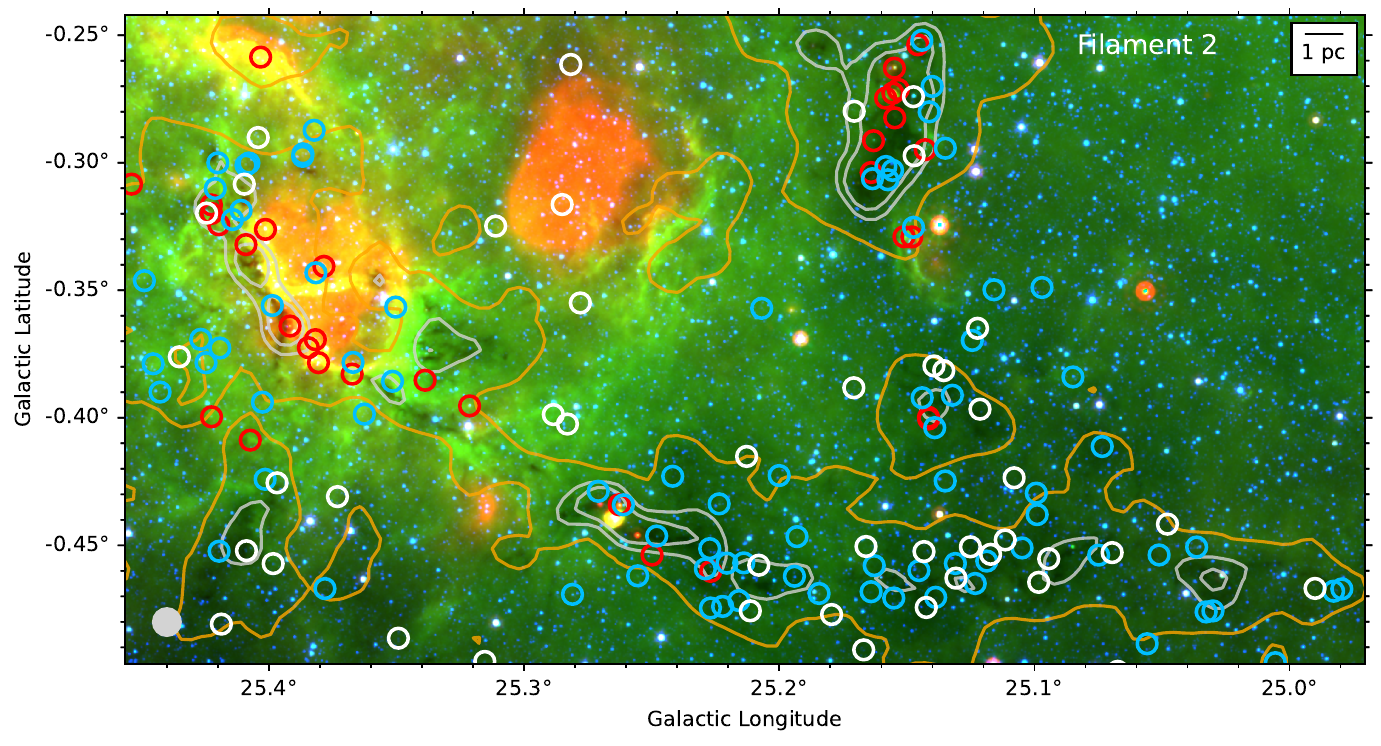}
\includegraphics[scale=0.75]{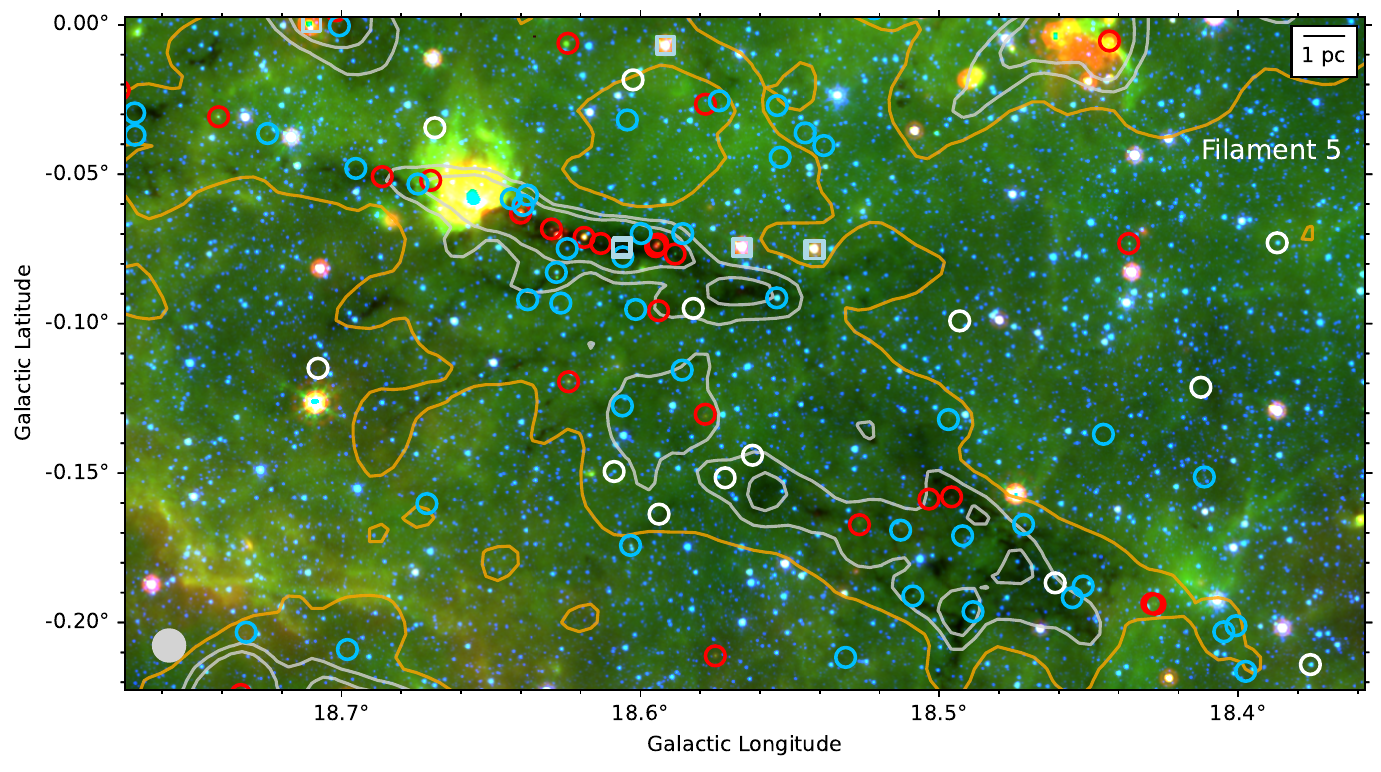}
\caption{\textit{Spitzer} red~(24~$\mu$m), green~(8.0~$\mu$m), and blue~(3.6~$\mu$m) composite image for Filament~2 (top) and Filament~5 (bottom). The color levels were chosen to highlight infrared-dark features. Class~I (green), Flat Spectrum (violet), Class~II (blue), and Class~III (red) YSOs from the literature are identified with star symbols. Masers are identified with squares when available.  The three contours trace \textit{Herschel}-derived $N_{H_2}$~column densities of $0.5$, $1.0$, and $1.5 \times 10^{22}$~cm$^{-2}$, with the lowest level identified in orange for clarity. The \textit{Herschel} beam is given by the gray circle at the bottom left.
\label{fig:Fil2_5_RGB}}
\end{figure*}

\begin{figure*}
\centering
\includegraphics[scale=0.75]{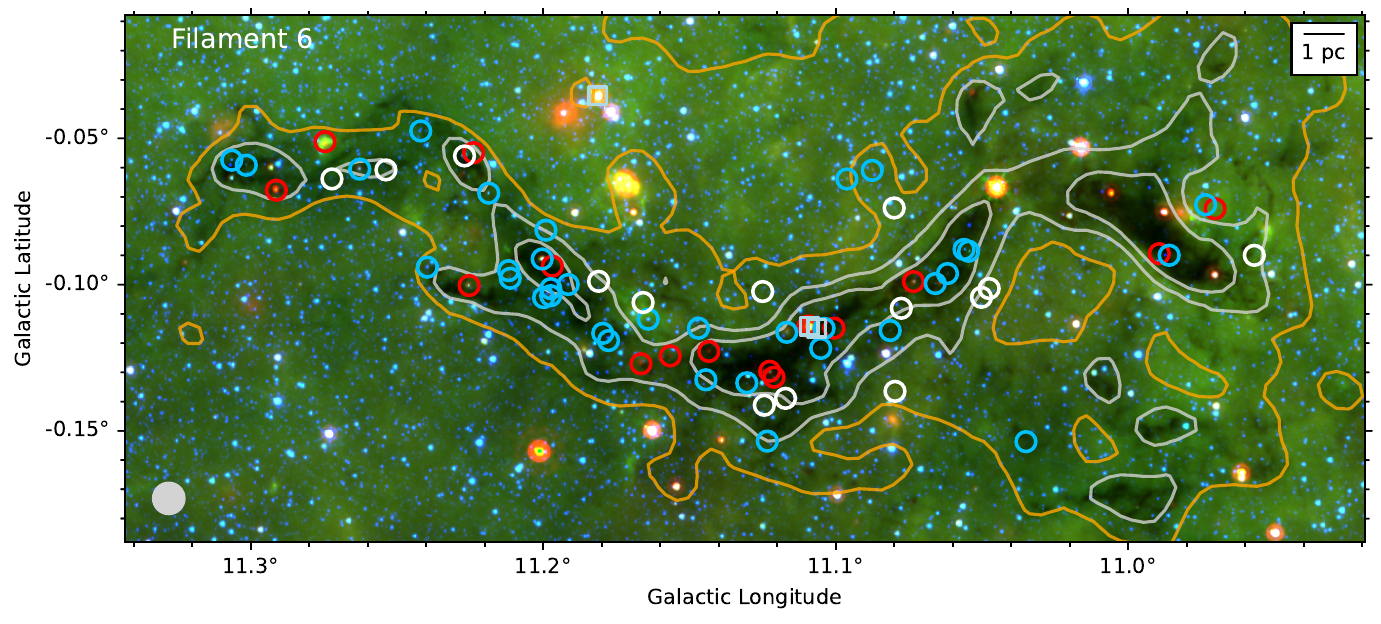}
\includegraphics[scale=0.75]{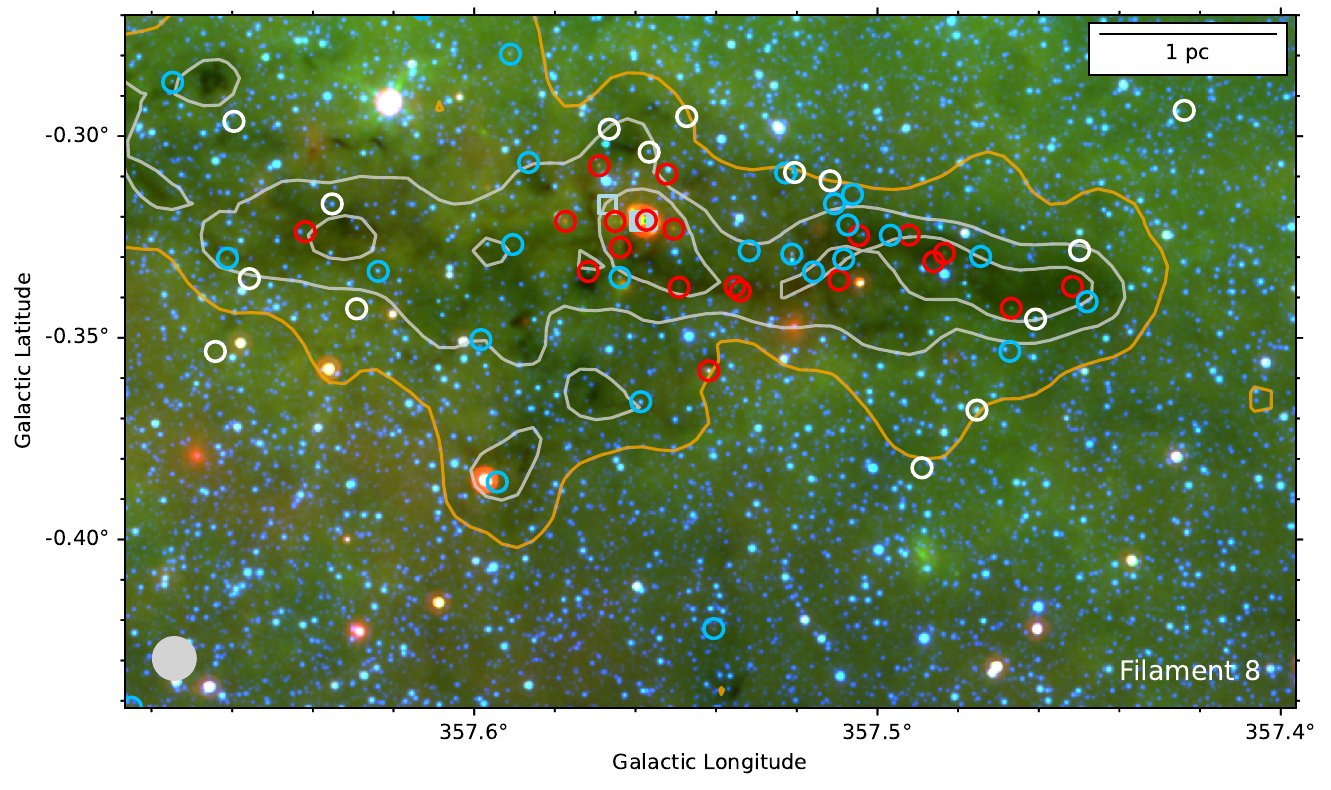}
\caption{\textit{Spitzer} red~(24~$\mu$m), green~(8.0~$\mu$m), and blue~(3.6~$\mu$m) composite image for Filament~6 (top) and Filament~8 (bottom). The color levels were chosen to highlight infrared-dark features. Class~I and Flat Spectrum (red), Class~II (blue), and Class~III (white) YSOs from the literature are identified with circles. Masers are identified with squares when available.  The three contours trace \textit{Herschel}-derived $N_{H_2}$~column densities of $0.5$, $1.0$, and $2.0 \times 10^{22}$~cm$^{-2}$, with the lowest level identified in orange for clarity. The \textit{Herschel} beam is given by the gray circle at the bottom left.
\label{fig:Fil6_8_RGB}}
\end{figure*}

\begin{figure*}
\centering
\includegraphics[scale=0.75]{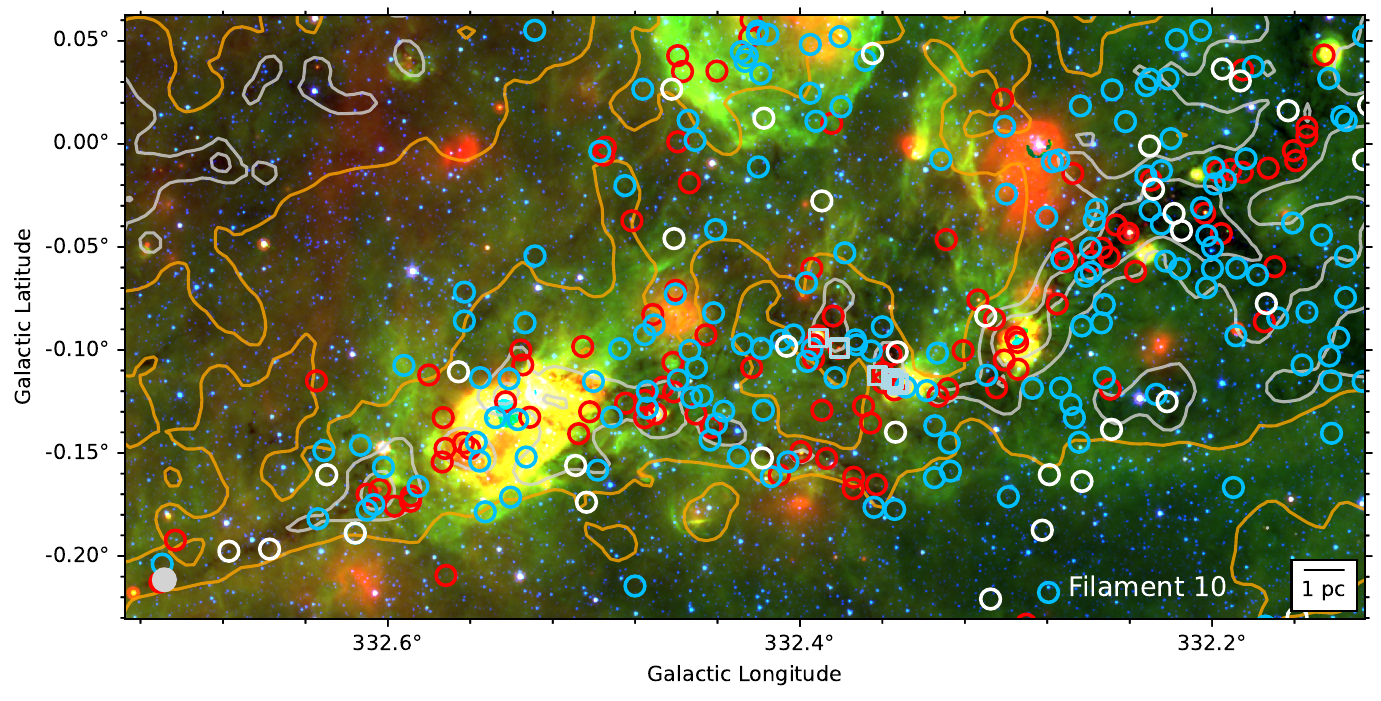}
\includegraphics[scale=0.75]{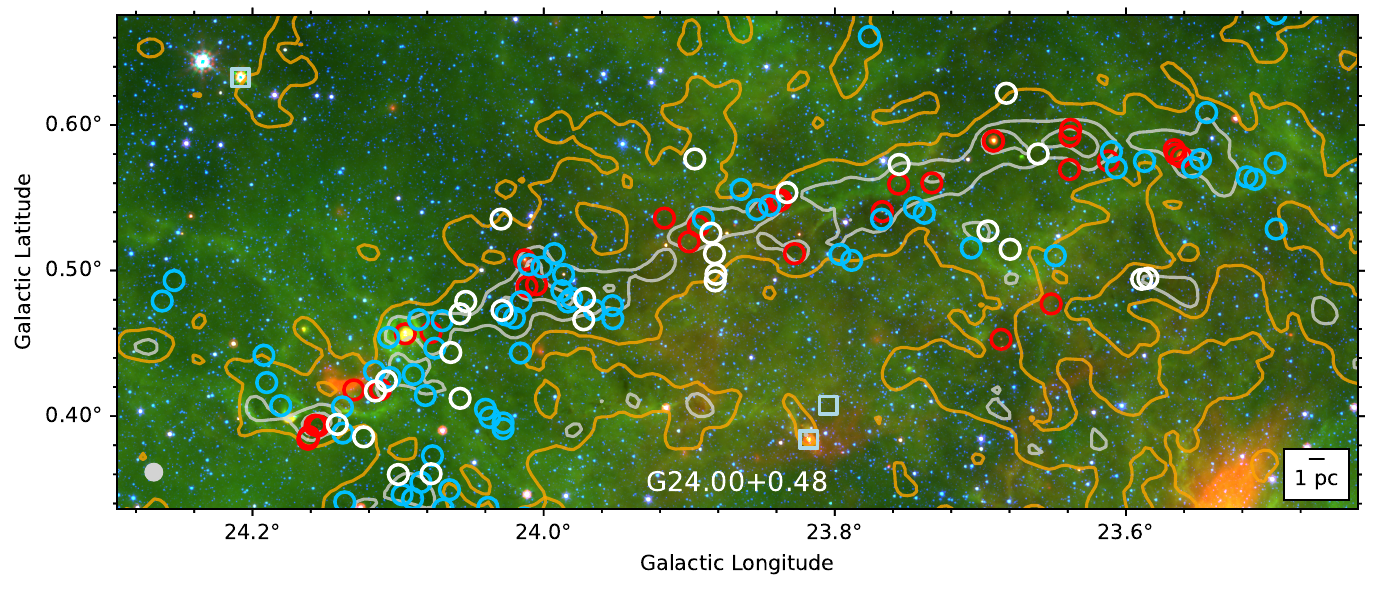}
\caption{\textit{Spitzer} red~(24~$\mu$m), green~(8.0~$\mu$m), and blue~(3.6~$\mu$m) composite image for Filament~10 (top) and G24 (bottom). The color levels were chosen to highlight infrared-dark features. Class~I (green), Flat Spectrum (violet), Class~II (blue), and Class~III (red) YSOs from the literature are identified with star symbols. Masers are identified with squares when available.  The three contours trace \textit{Herschel}-derived $N_{H_2}$~column densities of $0.5$, $1.0$, and $2.0 \times 10^{22}$~cm$^{-2}$, with the lowest level identified in orange for clarity. The \textit{Herschel} beam is given by the gray circle at the bottom left.
\label{fig:Fil10_G24_RGB}}
\end{figure*}

\begin{figure*}
\centering
\includegraphics[scale=0.75]{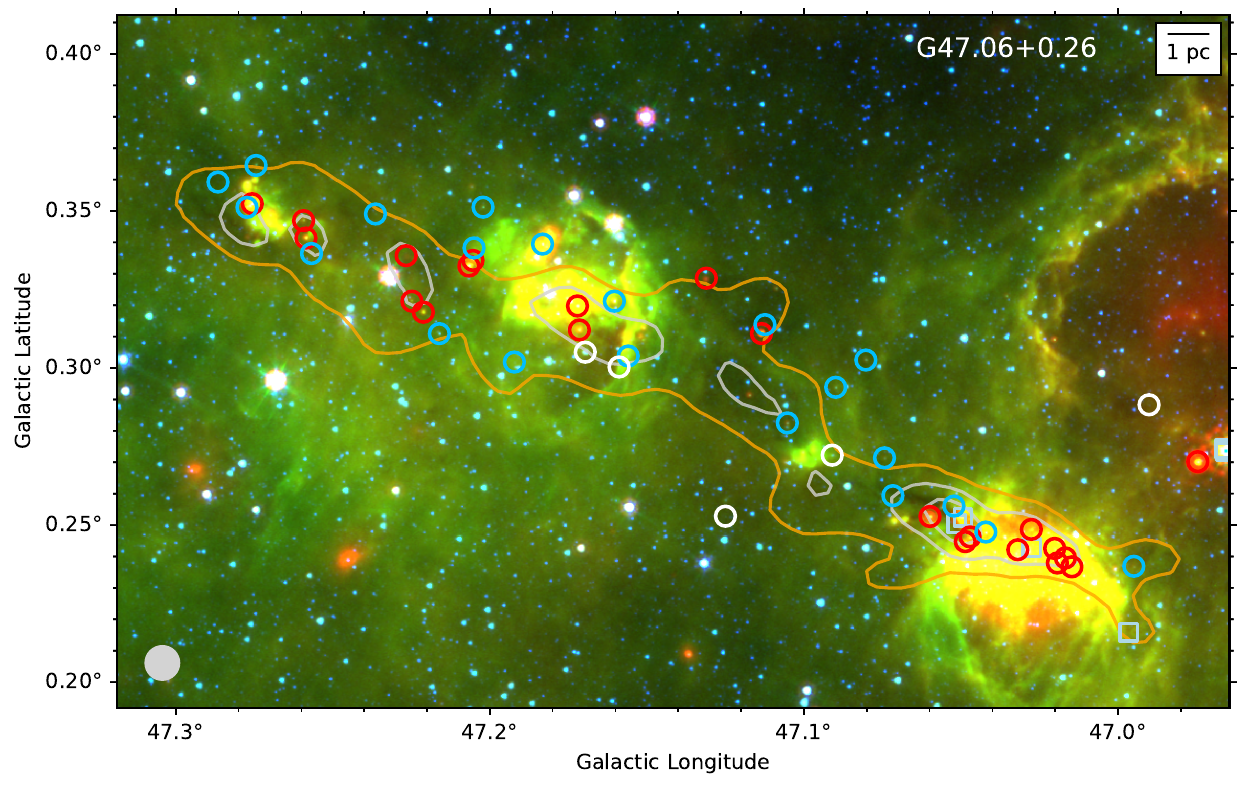}
\includegraphics[scale=0.75]{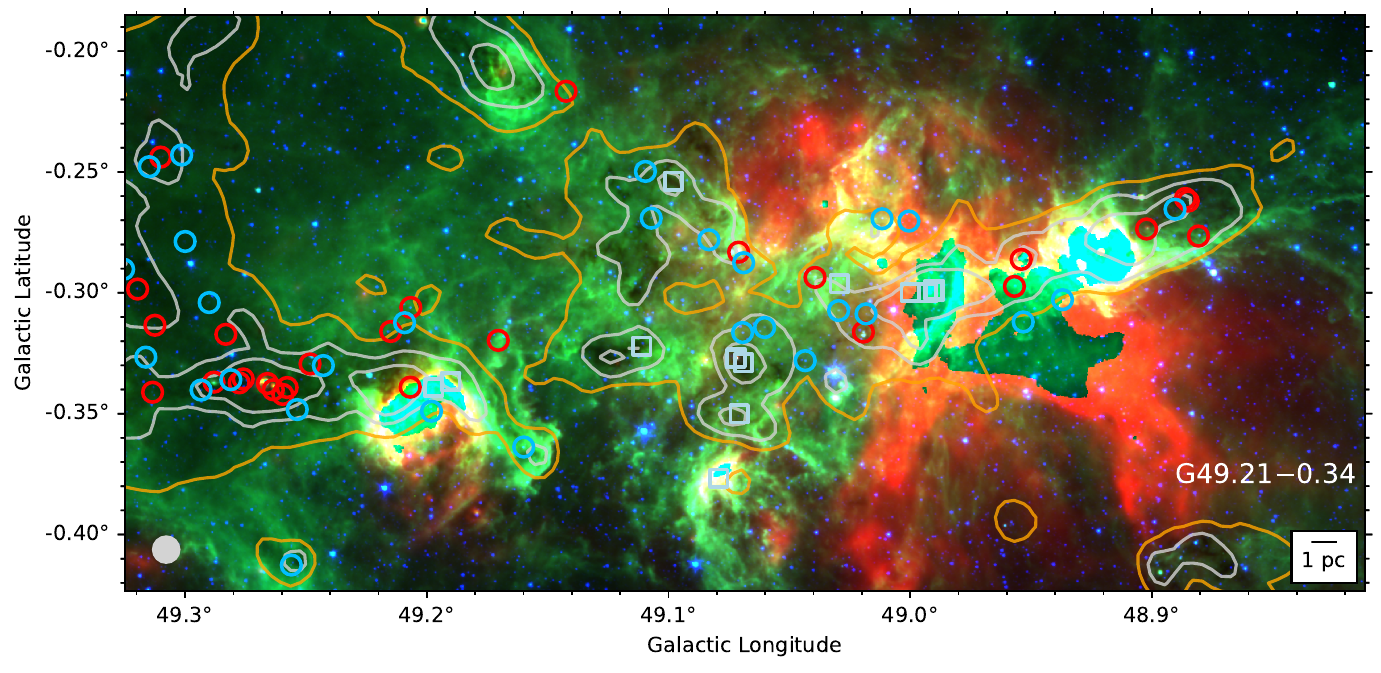}
\caption{\textit{Spitzer} red~(24~$\mu$m), green~(8.0~$\mu$m), and blue~(3.6~$\mu$m) composite image for G47 (top) and G49 (bottom). The color levels were chosen to highlight infrared-dark features, but we note that G49 is heavily saturated in these \textit{Spitzer} bands. Class~I and Flat Spectrum (red), Class~II (blue), and Class~III (white) YSOs from the literature are identified with circles. Masers are identified with squares when available.  The three contours trace \textit{Herschel}-derived $N_{H_2}$~column densities of $0.5$, $1.0$, and $2.0 \times 10^{22}$~cm$^{-2}$, with the lowest level identified in orange for clarity. The \textit{Herschel} beam is given by the gray circle at the bottom left.
\label{fig:G47_G49_RGB}}
\end{figure*}

As noted in Section~\ref{sub:yso}, we have created composite red (24~$\mu$m), green (8.0~$\mu$m), and blue (3.6~$\mu$m) images of the FIELDMAPS targets from data obtained by the GLIMPSE \citep{Benjamin2003} and MIPSGAL \citep{Carey2009} surveys with \textit{Spitzer}. Figures~\ref{fig:Fil1_RGB} through \ref{fig:G47_G49_RGB} present the resulting images for all the FIELDMAPS bones for the regions displayed in Figures~\ref{fig:Fil1_Maps} through Figure~\ref{fig:G49_Maps}, except for Filament~4 which is already shown in Figure~\ref{fig:Fil4_RGB}. Each figure identifies the positions of Class~I, II, III, and flat-spectrum YSOs, as well as masers when available. Contours show \textit{Herschel}-derived $N_{H_2}$~column densities to more easily identify the denser regions of the filaments. As in Figure~\ref{fig:Fil4_RGB}, the color levels were chosen to highlight the infrared-dark features of the bones. Howdver, this process was complicated by the fact that some bones are saturated in the \textit{Spitzer} bands, with G24 being the most extreme example. Reference lengths of 1~pc are also provided for each bone using the distances listed in Table~\ref{tab:filaments}. These composite images illustrate the wide range of star-forming conditions covered by the FIELDMAPS bones.

% Fil 1

% Fil 2 and 5

% Fil 6 and 8

% Fil 10 and G24

% G47 and G49

\bibliographystyle{aasjournal}
\bibliography{references.bib}

\end{document}